\documentclass{cernrep}
\pdfoutput=1
\usepackage{color, siunitx, float}
\usepackage[dvipsnames]{xcolor}
\usepackage{tabularx}
\usepackage{subfigure}
\usepackage[flushleft]{threeparttable}
\usepackage{nicefrac}
\usepackage[unicode=true,pdfusetitle,
 bookmarks=true,bookmarksnumbered=false,bookmarksopen=false, breaklinks=false, pdfborder={0 0 1}, citecolor=Maroon, backref=false, colorlinks=true, linkcolor=blue, linktoc=page]{hyperref}
\makeatletter
\usepackage{hyperref}
\newcommand\footnoteref[1]{\protected@xdef\@thefnmark{\ref{#1}}\@footnotemark}
\makeatother
\usepackage{ifthen} 

\newboolean{articletitles}
\setboolean{articletitles}{true} 

\DeclareSIUnit\barn{b}
 
\newcommand{\eqn}{equation}
\newcommand{\lam}{\lambda}
\newcommand{\GeV}{{\ensuremath\rm GeV}}
\newcommand{\fb}{{\ensuremath\rm fb}}
\newcommand{\TeV}{{\ensuremath\rm TeV}}

\newcommand{\pb}{{\ensuremath\rm pb}}

\newcommand{\pt}{\ensuremath{p_{\rm{T}}}\xspace}
\renewcommand{\Pg}{\ensuremath{\mathrm{g}}\xspace}
\renewcommand{\PH}{\ensuremath{\mathrm{H}}\xspace}
\newcommand{\PV}{\ensuremath{\mathrm{V}}\xspace}
\renewcommand{\PW}{\ensuremath{\mathrm{W}}\xspace}
\renewcommand{\PZ}{\ensuremath{\mathrm{Z}}\xspace}
\newcommand{\PQ}{\ensuremath{\mathrm{q}}\xspace} 
\newcommand{\PAQ}{\ensuremath{\overline{\mathrm{q}}}\xspace} 
\renewcommand{\PQb}{\ensuremath{\mathrm{b}}\xspace} 
\renewcommand{\PAQb}{\ensuremath{\overline{\mathrm{b}}}\xspace}
\renewcommand{\PQc}{\ensuremath{\mathrm{c}}\xspace} 
\renewcommand{\PAQc}{\ensuremath{\overline{\mathrm{c}}}\xspace}
\newcommand{\bbbar}{\PQb{}\PAQb\xspace}
\newcommand{\bbbb}{\bbbar{}\bbbar\xspace}
\newcommand{\ccbar}{\PQc{}\PAQc\xspace}
\newcommand{\qqbar}{\PQ{}\PAQ\xspace}
\newcommand{\mSD}{\ensuremath{m_\mathrm{SD}}\xspace}

\newcommand{\kt}{\ensuremath{k_{\mathrm{T}}}\xspace}
\renewcommand{\GeV}{\ensuremath{\,\text{Ge\hspace{-.08em}V}}\xspace}
\renewcommand{\TeV}{\ensuremath{\,\text{Te\hspace{-.08em}V}}\xspace}

\newcommand{\lb}{\left(}
\newcommand{\rb}{\right)}
\newcommand{\newc}{\newcommand}
\newcommand{\be}{\begin{equation}}
\newcommand{\ee}{\end{equation}}
\newcommand{\bea}{\begin{eqnarray}}
\newcommand{\eea}{\end{eqnarray}}
\newc{\ol}{\overline}
\newc{\wt}{\widetilde}
\renewcommand{\bs}{\boldsymbol}
\newc{\m}{\mathcal}
\newc{\la}{\langle}
\newc{\ra}{\rangle}

\newcommand{\beq}{\begin{eqnarray}}
\newcommand{\eeq}{\end{eqnarray}}
\newcommand{\bpmatrix}{\begin{pmatrix}}
\newcommand{\epmatrix}{\end{pmatrix}}
\newcommand{\ba}{\begin{array}}
\newcommand{\ea}{\end{array}}

\usepackage{titlepage}
\addauthor{HHH Whitepaper}{}

\title{HHH Whitepaper}
\abstract{
We here report on the progress of the HHH Workshop, that took place in Dubrovnik in July 2023. After the discovery of a particle that complies with the properties of the Higgs boson of the Standard Model, all Standard Model (SM) parameters are in principle determined. However, in order to verify or falsify the model, the full form of the potential has to be determined. This includes the measurement of the triple and quartic scalar couplings. 

We here report on ongoing progress of measurements for multi-scalar final states, with an emphasis on three SM-like scalar bosons at 125 \GeV, but also mentioning other options. We discuss both experimental progress and challenges as well as theoretical studies and models that can enhance such rates with respect to the SM predictions.

}

\begin{document}
\titlepage

\tableofcontents
\clearpage
\section*{Introduction }~\label{sec:introduction}
{\sl V. Brigljevic, D. Ferencek, G. Landsberg, T. Robens, M. Stamenkovic, T. Susa }

In October 2022 we were contacted and asked about the possibility
to hold the first triple Higgs workshop in Dubrovnik in the
Summer of 2023. A few months earlier, in June of 2022, Dubrovnik
hosted the 2022 Higgs Pairs Workshop (\href{https://indico.cern.ch/e/HH2022}{https://indico.cern.ch/e/HH2022}).
While adding only one letter (H) to the workshop topic,
this seemed a daring step forward in many respects. The HHH process
itself seemed quite beyond the LHC reach. While the expected Standard Model (SM) cross section for HH production ($\sim 30$ fb) makes it
barely observable with the full expected luminosity
at the LHC, the expected SM cross section for HHH production ($\sim 0.1$ fb)
is lower by more than two orders of magnitude, making its
observation completely beyond reach at the LHC. The study of
Double Higgs production at the LHC is also a very well
established research topic, with many theoretical and experimental
results available, which was manifested in the rich
program of the 2022 workshop filling almost five full days
of interesting talks and discussions. On the other side,
the study of HHH production is still a largely uncharted territory.
While several theoretical studies and calculations already exist,
there is no experimental result on searches for such processes yet.

There are, however, several common aspects closely connecting HH
and HHH production and their studies: both processes provide unique
handles to explore the Higgs potential and in particular
the Higgs self-couplings. Also, many of the analysis tools and techniques
developed for non-resonant or resonant HH analyses are expected to be of
great importance for tackling HHH final states.
Both are also sensitive to similar Beyond the Standard Model (BSM) models and in particular to extensions
of the SM scalar sector, which could in some cases largely enhance their
production cross sections and make even HHH production experimentally
reachable at the LHC.

Bringing together interested theorists and experimentalists to discuss
this  very new topic represented an exciting challenge. As the
local HEP community in Croatia was itself directly involved, both
on the experimental and theoretical side, in some of the first
HHH studies,  accepting to host it became an easy decision, resulting
in the organization of the first HHH workshop in Dubrovnik from
July 14 to 16 2023 at the Inter-University Centre Dubrovnik.

We certainly did not regret the decision as the workshop really
provided a very stimulating atmosphere and exchange of ideas.
We would like to thank all participants for contributing
with excellent talks and lively and very open discussions.
We have good hope that it will serve as a catalyst for
future work on HHH production, hopefully soon leading
to the first experimental search results on HHH production.

The Inter-University Centre Dubrovnik provided an excellent environment
and infrastructure and we would like to acknowledge the friendly and
very efficient support of their staff, notably Nada, Nikolina and Tomi.
They greatly contributed to making the workshop a success and a very pleasant
experience for all participants. They have enthusiastically welcomed all
CERN-related academic events and let us feel very welcome, leading us to
come back again and again to IUC as our preferred venue for the organization
of scientific meetings in Croatia. Of course, the city of Dubrovnik
with its rich history and unique old town and natural surroundings
did its part too.

At the end of the workshop there was a clear consensus among workshop
participants that the discussions started in Dubrovnik should continue
and that this should only be the first HHH workshop. Consensus
also emerged to reconvene in Dubrovnik with the
hope to see the first experimental results from HHH searches
at the LHC. This second workshop is likely to take place in 
the fall of 2025. Stay tuned!

To facilitate and trigger further work, the participants agreed
that a written track of the presented results and ideas should be kept,
resulting in the decision to write this White Paper. We hope it will
serve as a useful overview of current results and a catalyst for
both theoretical and experimental future work on HHH production.

{

This manuscript is structured loosely around the contributions to the HHH workshop, where we tried to group similar topics into common sections. For some topics, we also enhanced the content presented in talks in order to render a more complete overview on the current state of the art as well as future prospects and challenges.\\

This manuscript is organized as follows. In section \ref{sec:thover}, we give a short overview on the current status of theory predictions in both the Standard Model and New Physics scenarios, with relevant references to subsequent sections if feasible. Section \ref{sec:qcd} addresses the important topic of predictions within the SM, with a focus on topics arizing from QCD corrections, jet definitions, and scale uncertainties. In section \ref{sec:exp-hh}, we discuss possible lessons that can be learned from the investigation of DiHiggs final states for the study of multi scalar final states. This is further elaborated on in section \ref{sec:chall}, where we discuss current experimental prospects and challenges for such searches. In section \ref{sec:hh_ml}, we present an example of a study that addresses diHiggs final states using machine learning. Section \ref{sec:flavotag} we give an overview on the current state of the art for flavour tagging, focussing on ATLAS and CMS. In section \ref{sec:thst_models}, we touch upon the broad range of new physics models that can render enhanced triple scalar final states. We address the simulation of such new physics scenarios in a possible simplified approach in section \ref{sec:simp}. Conclusions and outlook are presented in section \ref{sec:summ}.\\

}

\noindent
Vuko Brigljevic, RBI (Zagreb) \\
on behalf of the local organizers: \\
V.B., Bhakti Chitroda, Dinko Ferenček, Tania Robens and Tatjana Šuša

\clearpage{}
\section{{\bfseries A window on Standard Model physics and beyond with triple-Higgs production}}
\label{sec:thover}
{\sl B. Fuks}

The discovery of a Higgs boson with a mass of about 125GeV at the LHC~\cite{ATLAS:2012yve, CMS:2012qbp} has been one of the most important developments in high-energy physics over the last decade. It provided the first crucial insights into the nature of the electroweak symmetry breaking mechanism, the generation of fermion masses, as well as into establishing the Standard-Model nature of the observed new state. Since then, extensive efforts have been undertaken to unravel its properties. In particular, both the ATLAS~\cite{ATLAS:2022vkf} and CMS~\cite{CMS:2022dwd} collaborations have meticulously investigated its tree-level Yukawa couplings with third-generation fermions and weak gauge bosons, as well as its loop-induced couplings with gluons and photons. Measurements have consistently shown excellent agreement with the predictions of the Standard Model, albeit within the present experimental and theoretical uncertainties.

However, to definitely ascertain whether the observed Higgs state aligns with the predictions of the Standard Model, it is imperative to gather information on the shape of the Higgs potential. This necessitates independent measurements of the Higgs cubic, quartic and even higher-order self-couplings. Presently, available data only loosely constrains some of these parameters, allowing for the possibility of significant deviations from the Standard Model~\cite{ATLAS:2022jtk, CMS:2022dwd}. This is especially motivating for new physics scenarios incorporating an extended scalar sector with additional scalar fields. Moreover, understanding the intricacies of the Higgs potential is crucial for the exploration of the mechanisms underlying the electroweak phase transition and the matter-antimatter asymmetry in the universe. Therefore, regardless of a potential discovery of physics beyond the Standard Model in the future, measuring the Higgs cubic and quartic self-couplings stands out as one of the primary objectives of the physics programme at current and future high-energy colliders~\cite{Contino:2016spe, FCC:2018byv, Cepeda:2019klc, Azzi:2019yne}.

In the Standard Model, the Higgs potential reads
\be
  V(\Phi) = -\mu^2 \Phi^\dag \Phi + \lambda \big(\Phi^\dag \Phi\big)^2\,,
\ee
where $\Phi$ represents the weak Higgs doublet, and $\mu$ and $\lambda$ denote the typical Higgs quadratic and quartic interaction terms, respectively. After electroweak symmetry breaking, the neutral component of the Higgs doublet acquires a vacuum expectation value $v$. Consequently, the potential can be reformulated in terms of the physical Higgs field, $h$, as
\be\label{eq:Vh}
  V(h) = \frac12 m_h^2 h^2 + \lambda_{hhh} v h^3 + \frac14 \lambda_{hhhh} h^4
  \qquad\text{with}\qquad
 \lambda_{hhh} =  \lambda_{hhhh} = \frac{m_h^2}{2 v}\,.
\ee
The Higgs self-couplings $\lambda_{hhh}$ and $\lambda_{hhhh}$ are thus inherently linked to both the Higgs mass $m_h$ and the vacuum expectation value $v$. While predictions for these couplings can be derived from existing experimental knowledge, ($v\simeq \SI{246}{\giga\electronvolt}$ and $m_h\simeq \SI{125}{\giga\electronvolt}$), direct measurements are crucial for independent confirmation. Legacy LHC measurements are anticipated to provide an $\mathcal{O}(1)$ estimate of the triple-Higgs coupling $\lambda_{hhh}$ relative to its Standard Model value~\cite{Cepeda:2019klc, Azzi:2019yne}. However, significant direct information on $\lambda_{hhhh}$ is not expected~\cite{Plehn:2005nk, Binoth:2006ym}. Therefore, substantial deviations from these values may persist for the foreseeable future, particularly in scenarios where all other Higgs properties align with Standard Model predictions. 

Accordingly, various studies have explored the potential of both existing and proposed proton-proton colliders to constrain the two Higgs self-couplings through potentially innovative strategies. These investigations typically interpret their findings following one of two approaches, and utilise either the so-called `$\kappa$-framework'~\cite{LHCHiggsCrossSectionWorkingGroup:2012nn, LHCHiggsCrossSectionWorkingGroup:2013rie} or well-defined models of physics beyond the Standard Model. The latter usually incorporate an extended scalar sector with additional weak Higgs singlets and doublets~\cite{Profumo:2007wc, Branco:2011iw, Espinosa:2011ax, Profumo:2014opa, Robens:2015gla, Akeroyd:2016ymd, Robens:2019kga}, and hence rely on a possibly complex parameter space and a very different scalar potential embedding a Standard-Model-like component. In contrast, the kappa framework represents the simplest and most effective method to include new physics effects into the Higgs potential, and it relies on the introduction of two new physics parameters, $\kappa_3$ and $\kappa_4$. These quantities act as modifiers of the cubic and quartic Higgs couplings from their Standard model values. Consequently, the potential~\eqref{eq:Vh} can be expressed as
\be
  V(h) = \frac12 m_h^2 h^2 + \lambda_{hhh} (1+\kappa_3) v h^3 + \frac14 \lambda_{hhhh} (1+\kappa_4) h^4\,,
\ee
with the Standard Model configuration defined by $\kappa_3=\kappa_4=0$.

The first step in the exploration of the Higgs potential involves the study of the trilinear Higgs self-coupling. A primary avenue for investigating this coupling is through the production of Higgs-boson pairs at hadron colliders~\cite{DiMicco:2019ngk}. In the Standard Model, this process is associated with substantial cross section reaching approximately $\sigma_{hh} \simeq \SI{31}{\femto\barn}$ and \SI{38}{\femto\barn} for LHC centre-of-mass energies of $\sqrt{s}=\SI{13}{\tera\electronvolt}$ and $\SI{14}{\tera\electronvolt}$, respectively, and increasing to $\SI{4.4}{\pico\barn}$ at $\sqrt{s}=\SI{100}{\tera\electronvolt}$. These cross sections, that reach a percent-level precision, correspond to state-of-the-art predictions that incorporate next-to-next-to-next-to-leading-order corrections in QCD and soft-gluon resummation at the next-to-next-to-next-to-leading-logarithmic accuracy~\cite{AH:2022elh}. Such a high production rate, that could even be higher in new physics scenarios less sensitive to destructive interference between diagrams, allows for the investigation of various final states to probe the Higgs cubic coupling, with the most promising signatures arising from final state systems composed of four $b$-jets, or a pair of photons combined with either a pair of $b$-jets or a pair of tau leptons~\cite{ATLAS:2021tyg, ATLAS:2022hsp, CMS:2022cju}. {Modern machine-learning techniques have been proven to be highly efficient to extract the signal from the overwhelming background, and their prospects for the future is quite encouraging, as quantitatively assessed in section~\ref{sec:hh_ml}.} Additionally, the triple Higgs coupling indirectly influences single Higgs production, where it arises through self-energy and vertex higher-order loop-corrections~\cite{Degrassi:2016wml, Gorbahn:2016uoy, Bizon:2016wgr, Maltoni:2017ims}. Recently, the ATLAS collaboration exploited this and jointly utilised measurements originating from both di-Higgs and single-Higgs studies to impose the most stringent constraints to date on $\kappa_3$~\cite{ATLAS:2022jtk}, which must satisfy:
\be
   \kappa_3 \in  [-0.4, 6.3]\,.
\ee
{The impact of the different final states relevant to di-Higgs production at the LHC Run~2 is discussed in further detail in section~\ref{sec:exp-hh}.}

\begin{figure}
  \centering
  \includegraphics[width=.23\textwidth]{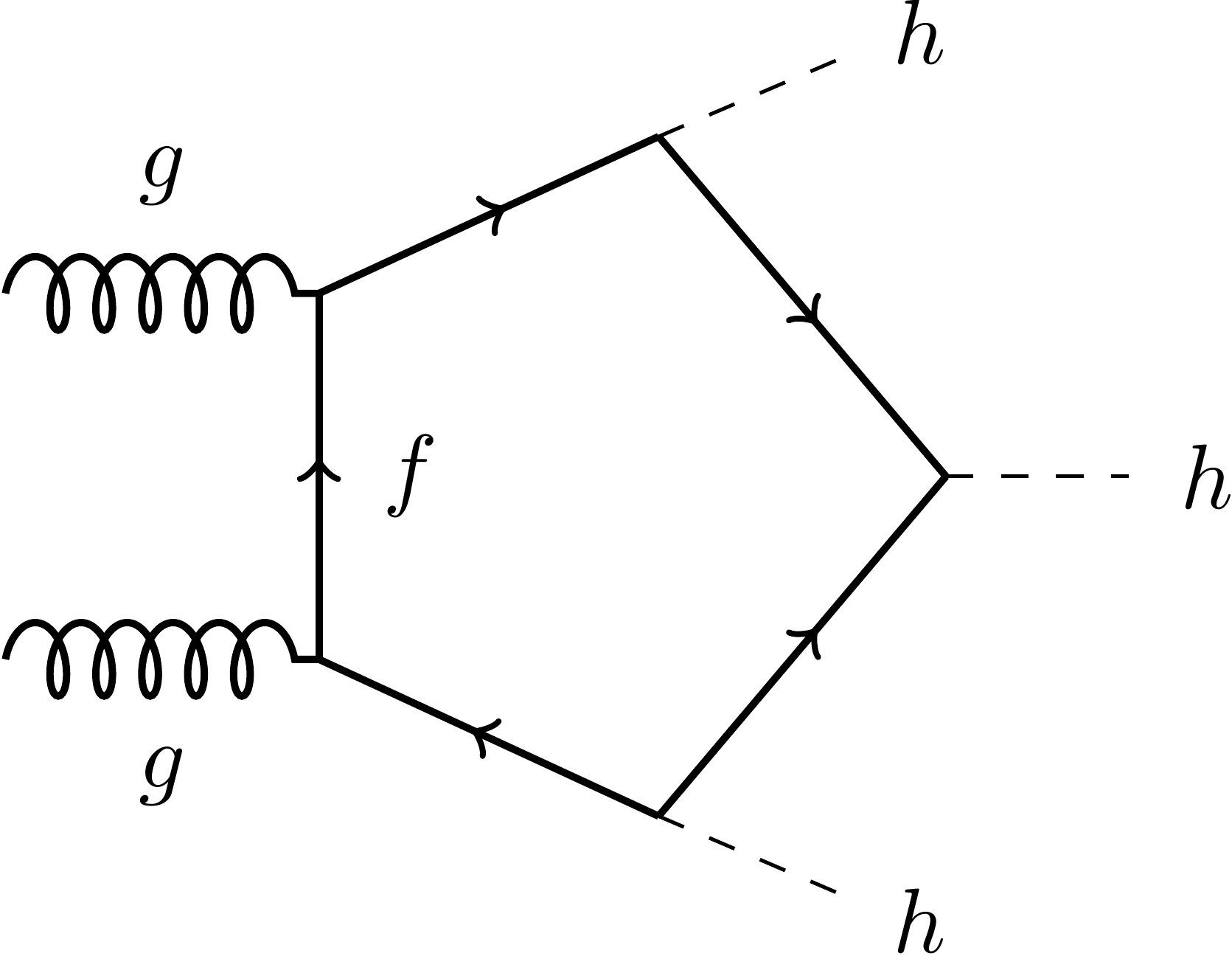}
  \includegraphics[width=.23\textwidth]{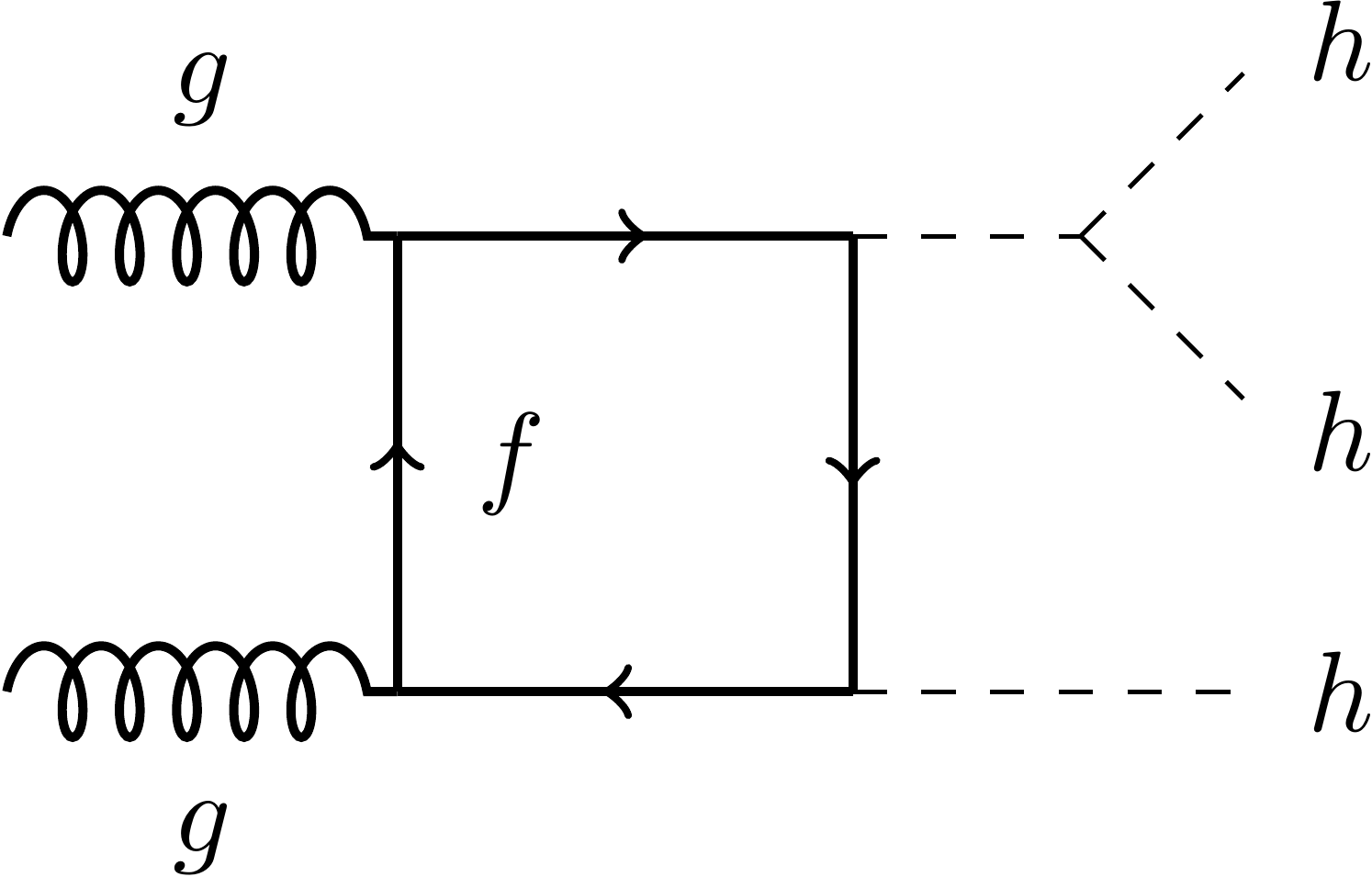}
  \includegraphics[width=.23\textwidth]{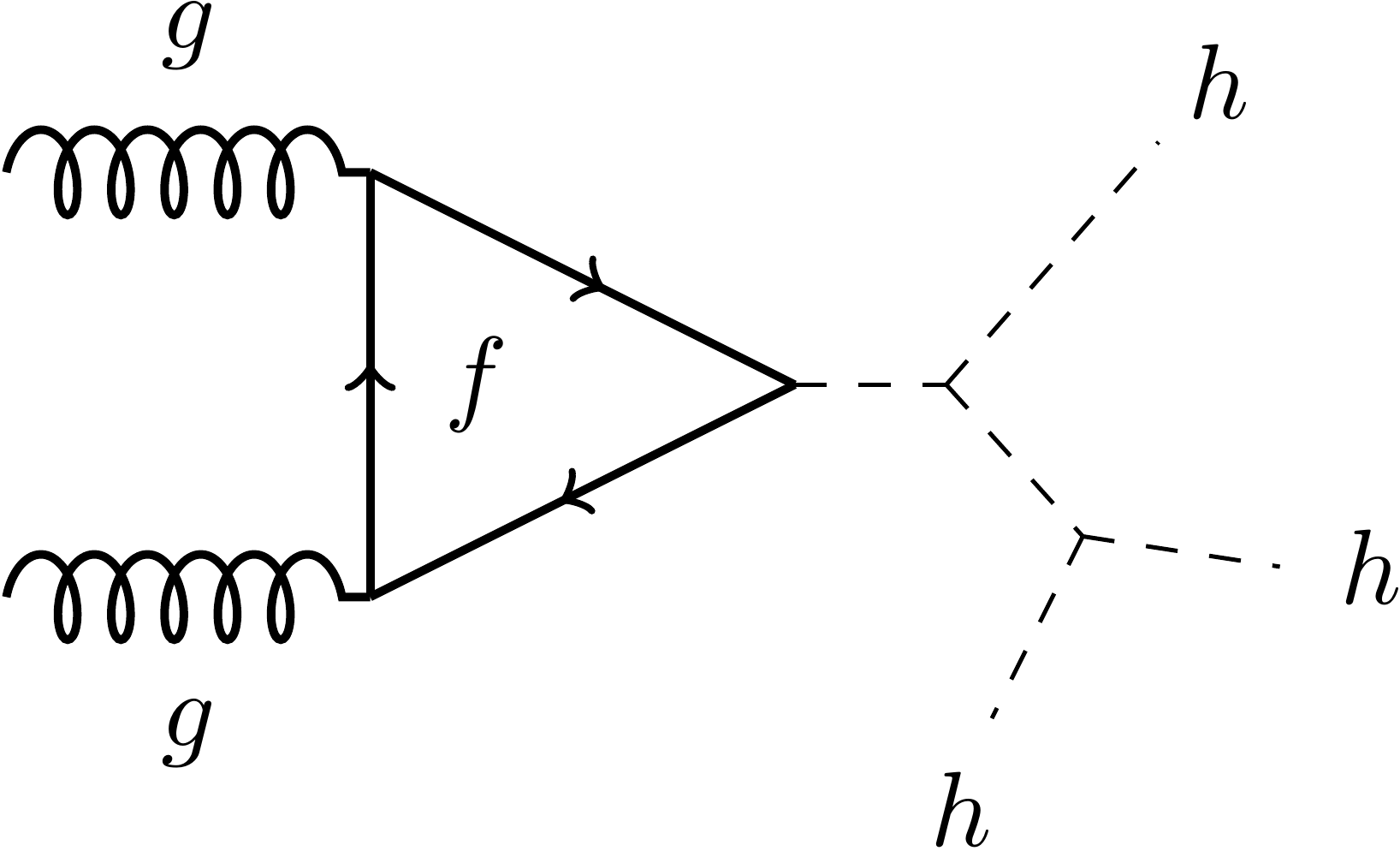}
  \includegraphics[width=.23\textwidth]{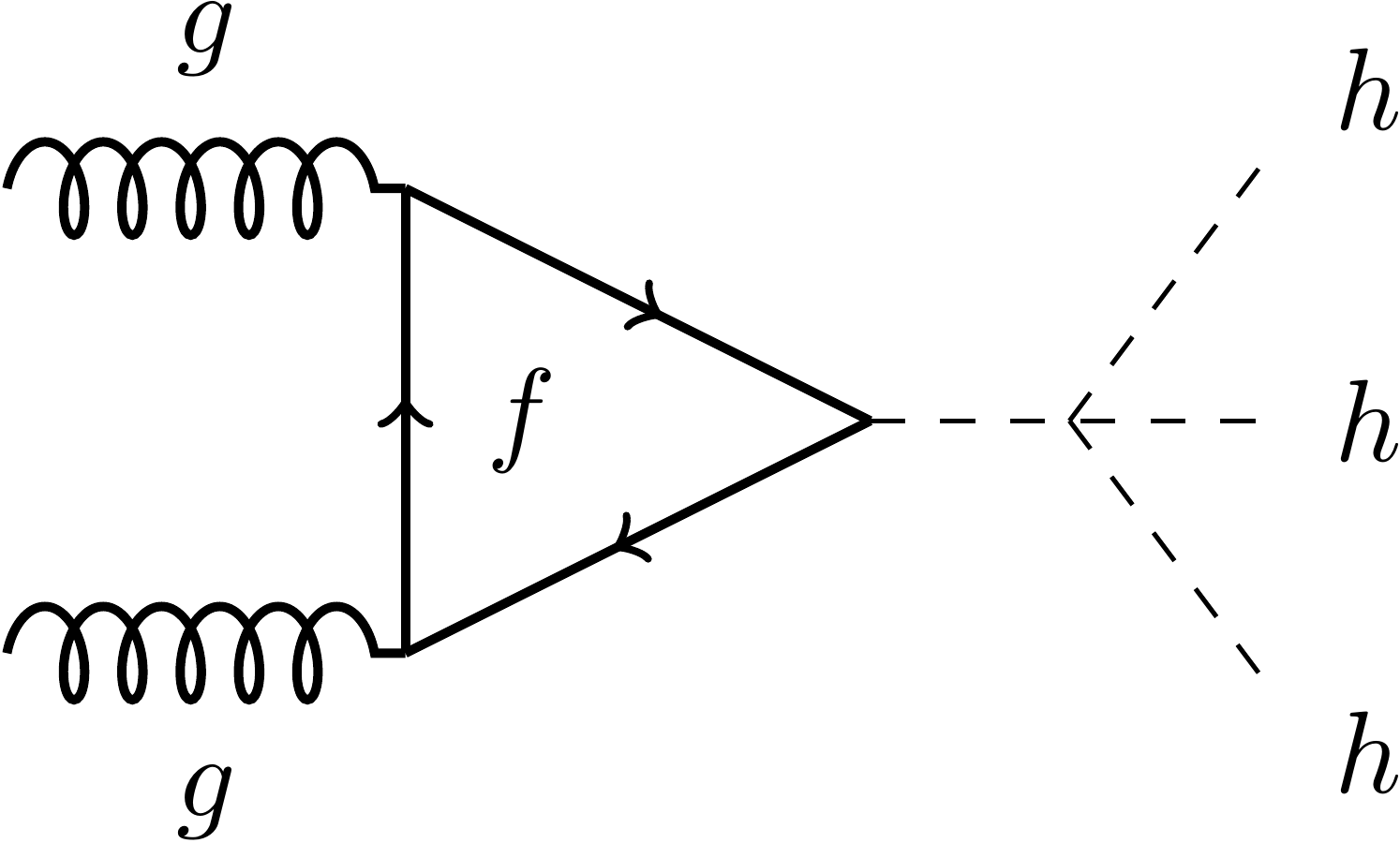}
  \caption{Representative leading-order Feynman diagrams for triple-Higgs production in proton-proton collisions.\label{fig:hhhdiags}}
\end{figure}

Similarly, the quartic Higgs self-coupling, which represents the second key factor in determining the shape of the Higgs potential, can be directly examined through triple-Higgs production and indirectly through loop-corrections in di-Higgs production. In the Standard Model, triple-Higgs production suffers from extremely low cross sections because of large destructive interference between the representative diagrams shown in figure~\ref{fig:hhhdiags}, rendering any expectation at the LHC unrealistic. The total rate at a centre-of-mass energy $\sqrt{s}=\SI{14}{\tera\electronvolt}$ is indeed as low as $\sigma_{hhh} \simeq \SI{0.05}{\femto\barn}^{+31\%}_{-22\%}$, thus exhibiting additionally a large theory uncertainty~\cite{deFlorian:2019app}. However, the prospects for a future proton-proton collider operating at $\sqrt{s}=\SI{100}{\tera\electronvolt}$ are much more promising, particularly in scenarios involving new physics where the cross section could be substantially enhanced. {This requires controlling the associated SM background, a task still lying at the frontier of the state of the art for theoretical predictions, not only because of the problematic of jet flavour tagging in an infrared-safe way and the treatment of the mass of the $b$ quark, but also because of the final-state multiplicity of relevant SM background processes challenging our computing capabilities of achieving precise predictions. These issues are addressed in detail in section~\ref{sec:qcd}. On the other hand, probing the quartic Higgs coupling with di-Higgs and triple-Higgs probes also poses various experimental challenges. Notably, this concerns the choice of the best final state and kinematic configuration to be studied, that both impact our abilities to decipher the relevance of the different diagrams shown in figure~\ref{fig:hhhdiags} as a function of phase space. This is further discussed in section~\ref{sec:chall}. Moreover, heavy-flavour tagging is also instrumental to maximise the potential experimental outcome. In this respect, huge progresses have been realised in the last decade, as detailed in section~\ref{sec:flavotag}.}

\begin{figure}
  \centering
  \includegraphics[width=0.48\textwidth]{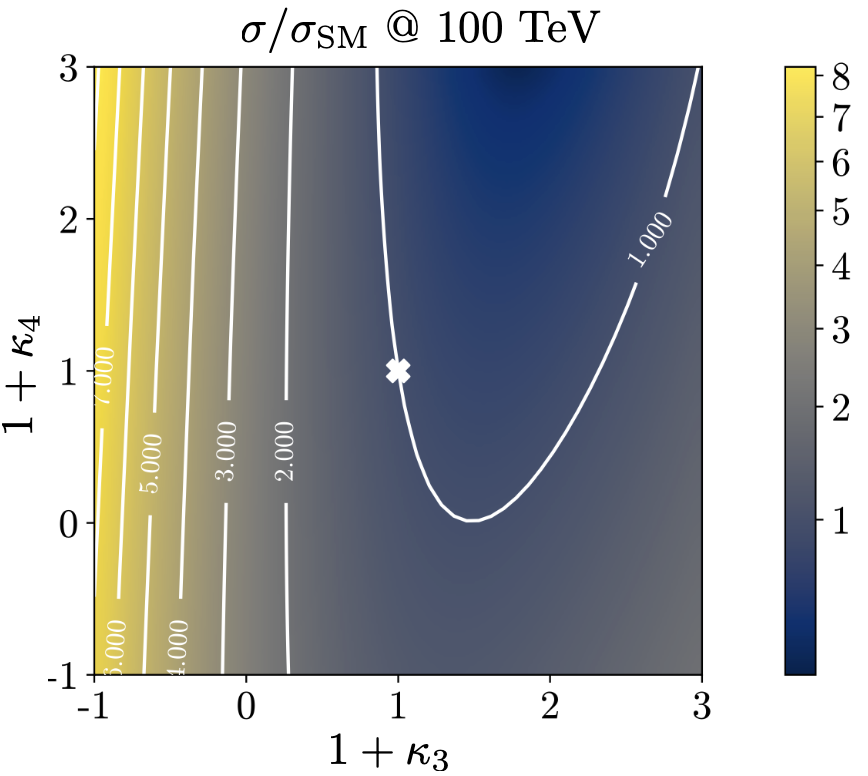}\hfill
  \includegraphics[width=0.40\textwidth]{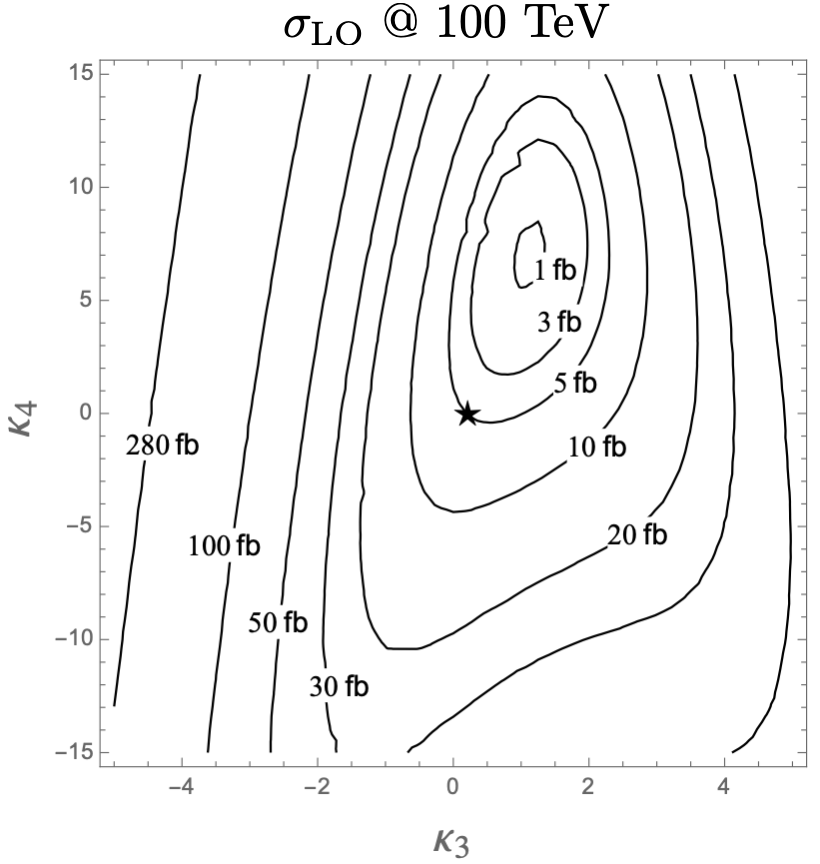}
  \caption{Triple-Higgs production cross section at a proton-proton collider operating at a centre-of-mass energy of $\SI{100}{\tera\electronvolt}$. We present predictions normalised with respect to the Standard Model cross section and with next-to-next-to-leading order corrections in the heavy-top limit included (left, figure adapted from~\cite{deFlorian:2019app}), as well as at leading order without any approximation (right, taken from \cite{Fuks:2015hna}). The star and cross represent the SM scenario.\label{fig:hhhrates}}
\end{figure}

{From now on, we specifically discuss existing studies aiming to assess the prospects of triple Higgs production at present and future colliders. W}ithin the $\kappa$-framework, \textcolor{Maroon}{$pp\to hhh$} production rates could potentially be several times larger. This is illustrated in figure~\ref{fig:hhhrates} where the left panel depicts the ratio between the triple-Higgs production cross section with non-zero $\kappa_3$ and $\kappa_4$ values and the Standard Model predictions (with $\kappa_3=\kappa_4=0$). Theory calculations are state-of-the-art, and incorporate corrections at the next-to-next-to-leading order modelled through form factors expressed in the heavy-top limit so that theory uncertainties are reduced to $5-10\%$~\cite{deFlorian:2019app}. As $\kappa_3$ values are negative and decrease, new physics contributions to the total rate become increasingly dominant, leading to enhancement of 1 to 5 for $-5\lesssim\kappa_3\lesssim-1$. The right panel of the figure presents instead exact leading-order predictions for a wider range of $\kappa_3$ values~\cite{Fuks:2015hna}, demonstrating that the cross section can increase by 1 or 2 orders of magnitude for moderately sized $\kappa_3$ values well below those acceptable by perturbative unitarity~\cite{Stylianou:2023xit}. While these perspectives are promising for observing a triple-Higgs signal at a future collider operating at $\SI{100}{\tera\electronvolt}$, the dependence of the cross section on modifications of the quartic Higgs coupling (through a non-zero $\kappa_4$ parameter) is less pronounced. Moreover, in the unlucky situation in which both $\kappa_3$ and $\kappa_4$ parameters are positive, the cross section suffers for even more destructive interference as in the Standard Model, rendering the situation even more challenging.

Consequently, associated measurements could offer additional insight into $\kappa_3$, which could then be used in combination with the aforementioned di-Higgs searches to refine its determination. However, obtaining the first constraints on the $\kappa_4$ coupling modifier is not straightforward and will require comprehensive phenomenological studies going beyond simple analyses of the total production rates, and where $\kappa_3$ effects must be correlated with $\kappa_4$ effects. This will then have to be confronted to a precise examination of di-Higgs production, where $\kappa_4$ impacts higher-order virtual corrections (similar to $\kappa_3$ for single Higgs production). Such an approach is expected to yield complementary constraints, enabling a more precise determination of $\kappa_4$~\cite{Bizon:2018syu, Borowka:2018pxx}. For instance, for 30~ab$^{-1}$ of $pp$ collisions at centre-of-mass energy of 100~TeV, the parameter $\kappa_4$ can be constrained to a range of $[-3, 13]$ by profiling over $\kappa_3$. On the other hand, studies in the $\kappa$-framework are not the whole story; investigations in the context of well-defined ultaviolet (UV)-complete models are also necessary as they could involve resonant contributions that significantly alter rates and distributions.

Once Higgs-boson decays are taken into account, triple-Higgs production can give rise to a wide variety of final-state signatures. However, due to the diverse magnitude of the different Higgs branching ratios and the expected background levels, only a select few final states have been studied thus far in light of their potentially significant signal-to-background ratios and feasibility for detection. They include cases where all three Higgs bosons decay into bottom quarks~\cite{Papaefstathiou:2019ofh} ($hhh\to bbbbbb$ with a triple-Higgs branching ratio of approximately 19.5\%), topologies in which two Higgs bosons decay into bottom quarks and the third decays into either a pair of photons~\cite{Papaefstathiou:2015paa, Fuks:2015hna, Chen:2015gva} ($hhh\to bbbb\gamma\gamma$ with a triple-Higgs branching ratio of about 0.23\%) or a pair of hadronically-decaying tau leptons~\cite{Fuks:2017zkg} ($hhh\to bbbb\tau\tau$ with a triple-Higgs branching ratio of approximately 6.5\%), and a configuration in which two Higgs bosons decay into a pair of $W$-bosons and the third into bottom quarks~\cite{Kilian:2017nio} ($hhh\to WWWWbb$ with a triple-Higgs branching ratio of around 0.9\%).

All past studies on triple-Higgs production in proton-proton collisions at a centre-of-mass energy of $\SI{100}{\tera\electronvolt}$ have significantly influenced the design requirements for future detectors at such colliders. It has been consistently emphasised, irrespective of the considered $hhh$ decay channel, that excellent $b$-tagging performance is indispensable. This entails achieving a low mistagging rate, even at the expense of a lower tagging efficiency, and ensuring good coverage of the forward region of the detector given that any produced systems tend to be more forward when they originate from collisions at higher centre-of-mass energies. Furthermore, the exploitation of the $hhh\to4b2\gamma$ mode necessitates a high photon resolution to enable the possible selection of a narrow mass window around the true Higgs mass, minimising hence background contamination. Similarly, the $hhh\to4b2\tau$ mode should leverage excellent double-tau-tagging performance, as currently achieved in di-Higgs searches at the LHC. Additionally, efficient reconstruction of boosted-Higgs systems, where the Higgs boson decays into a pair of collimated bottom quarks, is crucial for several signatures. This is essential for disentangling the signal from the overwhelming QCD background featuring light jets. Finally, the incorporation of high-level variables in the analysis, such as the $m_{T2}$ variable~\cite{Lester:1999tx, Barr:2003rg} or the $m_{\tau\tau}^{\rm Higgs-bound}$ and $m_T^{\rm True}$ variables~\cite{Barr:2011he, Barr:2009mx, Barr:2013tda}, could provide excellent handles to discriminate signal and backgrounds.

\begin{figure}
  \centering
  \includegraphics[width=0.37\textwidth, trim={0 0 0 33pt},clip]{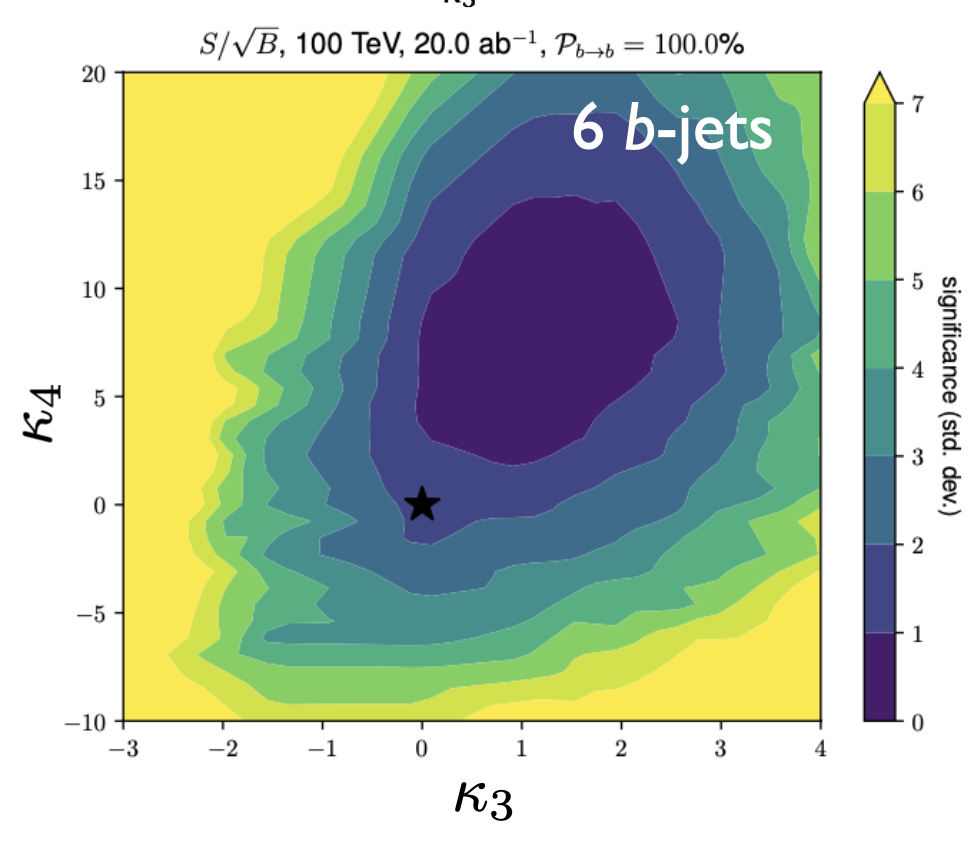}
  \includegraphics[width=0.305\textwidth]{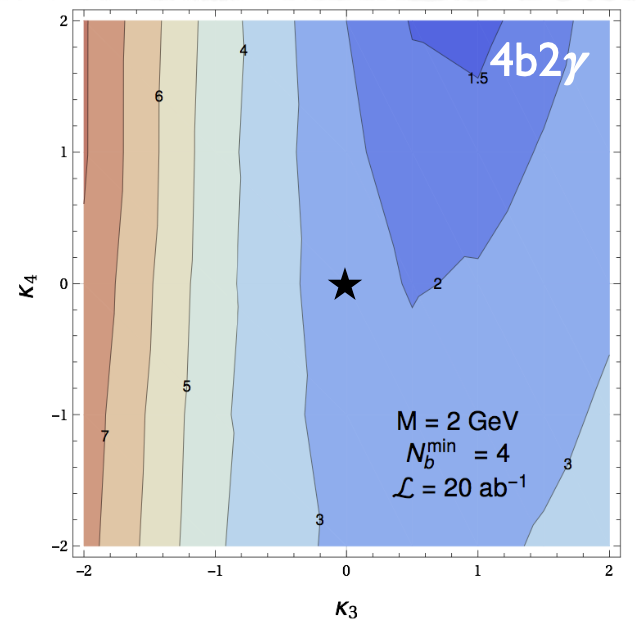}
  \includegraphics[width=0.305\textwidth]{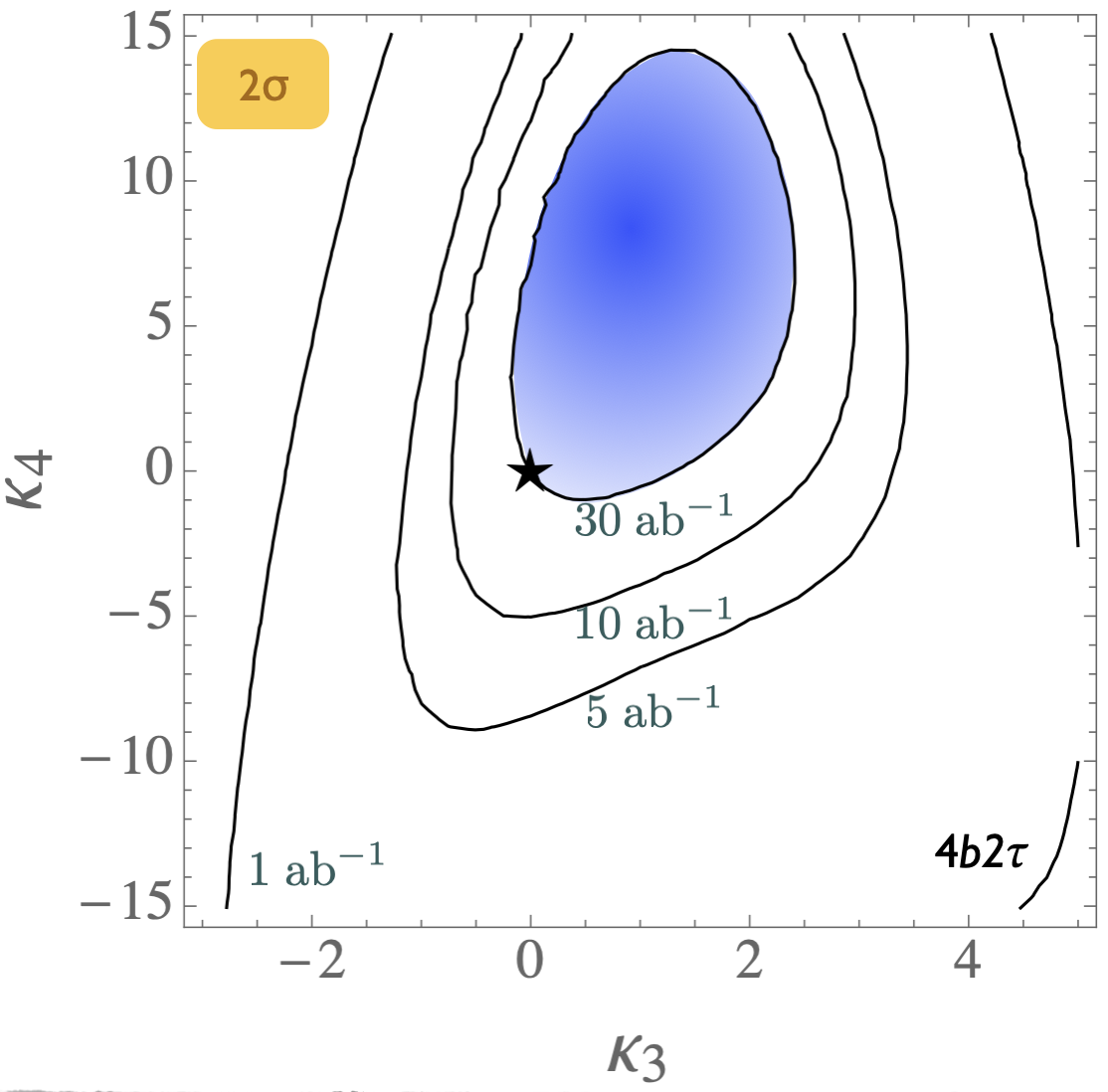}
  \caption{Sensitivity to a triple-Higgs signal at a proton-proton collider operating at a centre-of-mass energy of $\SI{100}{\tera\electronvolt}$. The figure presents the sensitivity in terms of standard deviations for the $hhh\to 6b$ final state (left, figure adapted from~\cite{Papaefstathiou:2019ofh}) and for the $hhh\to4b2\gamma$ final state (centre, figure adapted from~\cite{Fuks:2015hna}), as well as in terms of the luminosity required to achieve a $2\sigma$ sensitivity for the $hhh\to4b2\tau$ mode (right, figure adapted from~\cite{Fuks:2017zkg}). The star represents the SM scenario.\label{fig:hhhreach}}
\end{figure}

In figure~\ref{fig:hhhreach}, we evaluate the capability of detecting a triple-Higgs signal in proton-proton collisions at $\sqrt{s}=\SI{100}{\tera\electronvolt}$ for the three most promising final states. The results, obtained from state-of-the-art Monte Carlo simulations, are presented in the $\kappa$-framework. Technical details and analysis description can be found in~\cite{Papaefstathiou:2019ofh, Fuks:2015hna, Fuks:2017zkg}. The left panel of the figure showcases the sensitivity to an $hhh\to 6b$ signal in terms of standard deviations, and illustrates its dependence on the two $\kappa$-parameters across a wide range of values. Similarly, the central panel focuses on the $hhh\to 4b2\gamma$ mode. Despite potentially aggressive and not always conservative assumptions on detector parametrisation, both analyses demonstrate similar sensitivity. Notably, these pioneering studies indicate that the Standard Model configuration, defined by $\kappa_3=\kappa_4=0$, is theoretically attainable at a $2\sigma$ level. Furthermore, the right panel considers the $hhh\to 4b2\tau$ channel. However, the results are this time displayed in terms of the luminosity required to achieve a $2\sigma$ exclusion for each point in the $(\kappa_3, \kappa_4)$ parameter space. Specifically, we can note that a target luminosity of $\SI{30}{\atto\barn^{-1}}$ ensures a $2\sigma$ exclusion for the Standard Model point. These results underscore the potential of combining all modes, mirroring current practice for single Higgs and di-Higgs experimental studies at the LHC. {Finally, we recall that we can leverage the same $hhh$ studies to get additional handles of non-standard couplings of the Higgs boson to the top quark, as depicted in section~\ref{sec:anomalouscoupl} when the $(\kappa_3 ,\kappa_4)$ parameter space is generalised to include new physics modifiers to the coupling of a top-antitop pair to one, two and even three Higgs bosons.}

Beyond the $\kappa$-framework, triple-Higgs production can be also enhanced through extra diagrams incorporating new physics contributions{. Prime examples include models featuring multiple scalars, such as those explored in section~\ref{sec:thst_models}}. In these scenarios, the enhancement arises from Higgs-to-Higgs cascade decays~\cite{King:2014xwa, Costa:2015llh, Ellwanger:2017skc, Baum:2018zhf, Baum:2019uzg, Robens:2019kga, Papaefstathiou:2020lyp, Englert:2020ntw}. For instance, one or two heavier Higgs bosons could be initially produced and subsequently decay into a set of Standard-Model-like Higgs bosons, potentially leading to abundant production of triple-Higgs systems beyond the Standard Model. {This configuration is realised easily in a model with three Higgs-like particles $(h, h_2, h_3)$, where the heavier $h_2$ and $h_3$ correspond to new physics Higgs states. Well-studied frameworks exhibiting such states is the so-called `Two Real Singlet Model' (TRSM)~\cite{Robens:2019kga, Robens:2022nnw} further explored in sections~\ref{sec:thstud} and \ref{sec:tr_other}, as well as the complex two-Higgs Doublet Model~\cite{Lee:1973iz, Branco:1985aq, Weinberg:1990me, Ginzburg:2002wt, Fontes:2017zfn}, Next-to-Two-Higgs-Doublet Model~\cite{Chen:2013jvg, Muhlleitner:2016mzt, Engeln:2018mbg} and the Non-Minimal Supersymmetric Standard Model~\cite{Barbieri:1982eh, Dine:1981rt, Ellis:1988er, Drees:1988fc, Ellwanger:1993xa, Ellwanger:1995ru, Ellwanger:1996gw, Elliott:1994ht, King:1995vk, Franke:1995tc, Maniatis:2009re, Ellwanger:2009dp} discussed in section~\ref{subsec:introduction}.} A triple-Higgs system can then be produced through the production and decay sequence of sub-processes
\be
  p p \to h_3 \to h_2\, h \to h\,h\,h\,.
\ee
These decay processes here occur due to multi-Higgs interactions included in the scalar potential.

This phenomenon is particularly relevant at the LHC, not only for the planned high-luminosity operations but also for the much closer upcoming Run~3. However, in models featuring additional scalars, the parameter space is often vast and contains numerous free parameters relevant to the Higgs sector. {Nonetheless, studies in the TRSM~\cite{Papaefstathiou:2020lyp}, also detailed in section~\ref{sec:thstud},} have demonstrated that typical scenarios consistent with current constraints on extended scalar sectors, including additional Higgs bosons with masses in the 200-500 GeV range, could yield observable signals at the LHC Run~3 with significance ranging from $2\sigma$ over to $5\sigma$. Furthermore, with an expected accumulated luminosity of $\SI{3}{\atto\barn^{-1}}$ at the high-luminosity LHC, any representative benchmark scenario exhibits a significance exceeding 5 standard deviations. These findings leverage the presence of intermediate resonance effects in triple-Higgs production, and the ability to fully reconstruct the resonant states through kinematic fits of the final state. Consequently, undertaking triple-Higgs searches at the LHC presents promising avenues and there is no need to wait for a future collider that could operate in a few decades from now. {In addition, such analyses involving intermediate scalar resonances and Higgs-to-Higgs cascades could be facilitated by utilising a simplified-model approach, such as the one proposed in section~\ref{sec:simp}.}

\begin{figure}
  \centering
  \includegraphics[width=0.45\textwidth]{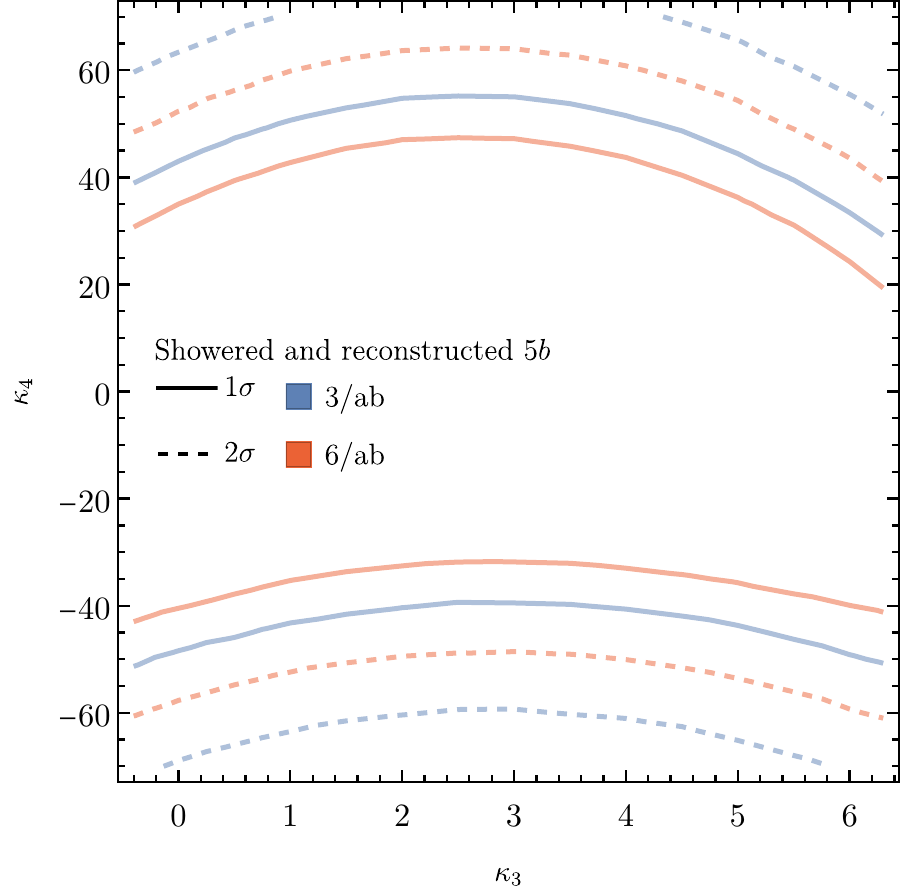}\hfill
  \includegraphics[width=0.45\textwidth]{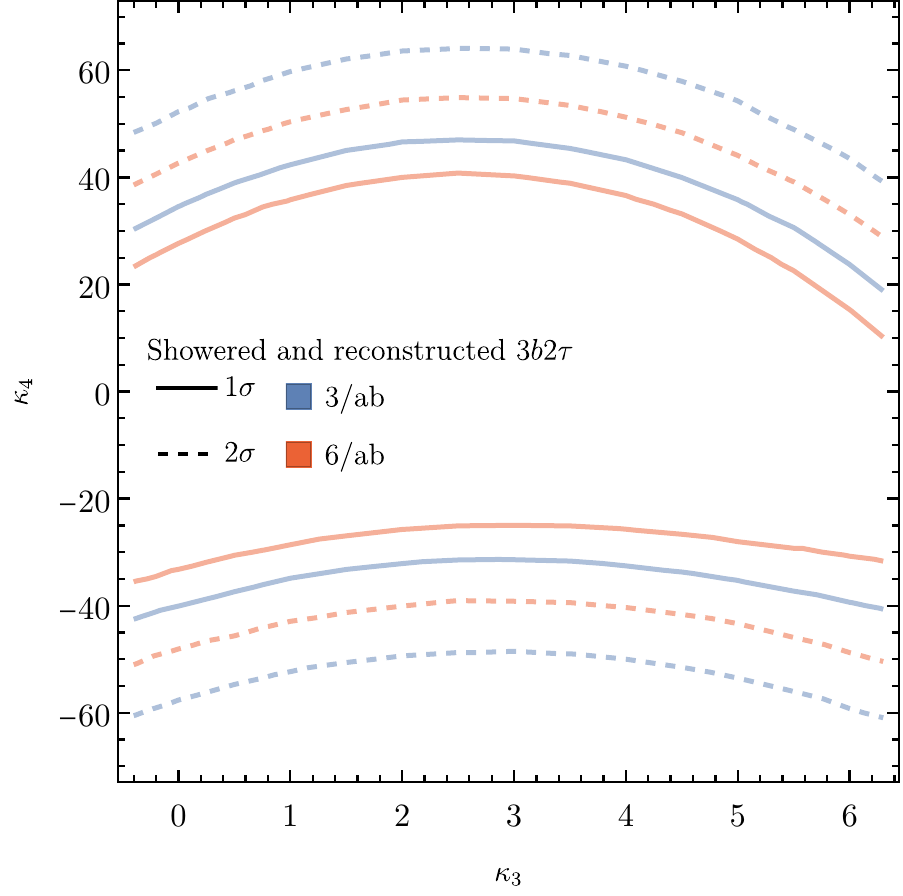}
  \caption{Sensitivity to a triple-Higgs signal at the high-luminosity LHC, after the analysis of signal and backgrounds for the $hhh\to 6b$ final state (left, figure taken from \cite{Stylianou:2023xit}) and for the $hhh\to4b2\tau$ mode (right, figure taken from~\cite{Stylianou:2023xit}).\label{fig:hhh_lhckappa}}
\end{figure}

These promising results should prompt a reevaluation of triple-Higgs phenomenology within the $\kappa$-framework at the LHC, particularly considering that perturbative unitarity allows for $\kappa_3$ and $\kappa_4$ values much larger than those considered in pioneering studies at future colliders, with acceptable values of $|\kappa_3|<10$ and $|\kappa_4| < 65$ using partial wave expansion at the tree level and the optical theorem~\cite{Stylianou:2023xit}. However, despite the larger signal cross sections for more extreme $\kappa$ parameter values, they remain insufficient to ensure potential observations across wide parts of the parameter space. Leveraging advanced machine learning techniques and assuming excellent detector performance for the high-luminosity LHC, along with an aggressive choice for the systematics, it is however possible to show that certain regions of the $(\kappa_3, \kappa_4)$ parameter space are excluded at 95\% confidence level with a luminosity of $\SI{3}{\atto\barn^{-1}}$, or even of $\SI{6}{\atto\barn^{-1}}$ when combinations from both the ATLAS and CMS experiments are considered. This is depicted in figure~\ref{fig:hhh_lhckappa} for the $hhh\to6b$ and $hhh\to 4b2\tau$ channels, the only two modes showing significant potential at the LHC due to their large-enough production cross section (including relevant branching ratio factors){, as well as more comprehensively discussed in section~\ref{sec:pertur}}. Consequently, scenarios with extreme values for the $\kappa_4$ parameter can be possibly excluded, providing further motivation for investigating triple-Higgs production at the LHC.

Throughout our discussion, we delved into the significance of triple-Higgs production in the context of high-energy colliders, particularly focusing on its implications for understanding the Higgs potential and probing physics beyond the Standard Model. We emphasised the importance of the $\kappa$-framework as a mean to both probe the Standard-Model nature of the Higgs self-couplings and provide insights into new physics scenarios. While studies at future colliders indicate promising prospects for observing triple-Higgs events, we highlighted the potential for reevaluating triple-Higgs phenomenology at the LHC both within the $\kappa$-framework and in new physics models with additional scalars. As also further detailed in the next chapters of this work, despite challenges posed, advanced machine learning techniques, high-level variables and excellent detector performance could offer avenues for excluding certain regions of the parameter spaces. {Collider studies should however always try to exploit best any complementary source of information. For instance, as detailed in section~\ref{sec:GW}, gravitational wave properties that are now measured sufficiently well could be interpreted as constraints on the Higgs potential.} In conclusion, undertaking such searches at the LHC could hold the promise of shedding light on fundamental aspects of particle physics, advancing our understanding of the Higgs mechanism and its implications for physics beyond the Standard Model.

\clearpage
\section{{\bfseries QCD overview and possible challenges } \label{sec:qcd}}
{\sl G. Soyez, G. Zanderighi}

{
The investigation of the triple Higgs final state in the SM as well as possible new physics scenarios highly relies on an accurate understanding of the corresponding theoretical predictions for both signal and dominant SM background, related higher-order corrections and scale uncertainties, as well as available tools to describe the corresponding differential contributions as accurately as possible. In addition, it is important to understand the differences in predictions depending on the chosen flavour scheme. We briefly address these issues in this section by describing current state of the art as well as open questions and possible challenges that appear in the investigation of triple Higgs final states at current and future hadron colliders.

}

\subsection{Jet flavour}

As extensively discussed in Sec.~\ref{sec:thover}, the investigation
of triple-Higgs production and the endeavor to extract the quartic
coupling are extremely challenging due to the tiny cross sections for
the production of three Higgs bosons. These cross sections are strongly
suppressed not only because of the large invariant mass of the final
state but also due to the destructive interference between diagrams
involving the triple and the quartic Higgs coupling. Such destructive
interference may persist in models of physics beyond the SM or could
be alleviated, potentially making the signals accessible. However,
even if the signal involving the quartic Higgs coupling were to be
significantly amplified, precisely determining the quartic Higgs
coupling would remain exceedingly challenging due to the overwhelming
background processes to this signal.

As already noted in Sec.~\ref{sec:thover}, the tiny cross-section for
the signal process necessitates a focus on decay channels with the
largest branching ratio of the Higgs boson, notably final states
involving three pairs of $b\bar b$ quarks, two $b \bar b$-pairs and
one $2\tau$, or $4b + 2 \gamma$.
All these decay channels feature at least four $b$ quarks in the final
state. However, $b$ quarks are abundantly produced at the LHC in
numerous processes unrelated to Higgs bosons, such as gluon splitting
or the decay of top quarks, $Z$-bosons, or $W$-bosons. At high
energies, $b$ quarks typically result in $b$-jets, making the study of
the quartic Higgs coupling inseparable from the challenge of
understanding and optimizing $b$-tagging and assigning bottom-flavor
to jets.

While the development of infrared (IR) safe jet algorithms is a solved
problem for unflavored jets, incorporating flavor information into jet
definitions poses challenges. Traditionally, a flavored-jet is
identified by the presence of at least one flavor tag, such as a $B$
or $D$ meson, above a specified transverse momentum
threshold. However, due to collinear or soft wide-angle $g\to Q\bar Q$
splittings, where $Q$ represents a quark with the flavor of interest,
this definition lacks collinear and infrared safety whenever the quarks are treated as massless.
In a calculation which keeps the finite mass of the heavy flavour, 
even though infrared-and-collinear safety is technically restored, 
the infrared sensitivity still manifests itself as large logarithms 
in the ratio of the small mass of the flavoured quark over the hard 
scale of the process. 
As extensively
discussed in Ref.\cite{Banfi:2006hf}, defining jet flavor in
perturbation theory is extremely delicate. Notably, defining a $b$-jet
as a jet containing at least a $b$-quark yields non-infrared finite
cross-sections in the case of calculations performed in the massless limit, and results logarithmically sensitive to the quark mass, when this is kept finite in the calculations. The formulation of a $k_t$-like algorithm ensuring
infrared safety to all orders was attempted in
Ref.\cite{Banfi:2006hf}, predating the anti-$k_t$
algorithm~\cite{Cacciari:2008gp}. Key elements of this definition
include a mechanism preventing soft flavored quarks from contaminating
the flavor of hard flavorless jets and labeling jets containing more
than one $b$-quark as flavorless jets.

This first flavour-algorithm was formulated to address a discrepancy
between data and theory in the context of heavy-flavor production at
the Tevatron~\cite{Banfi:2006hf,Banfi:2007gu}.  However, the proposed
jet-algorithm based on the $k_t$ algorithm was impractical for
experimental implementations and its use was primarily limited to the
development of perturbative predictions involving heavy-flavor.

Recent years have witnessed renewed interest in providing an infrared
safe and practical definition of flavored jets. Given the widespread
use of the anti-$k_t$ algorithm in experimental studies, recent
endeavors have focused on formulating algorithms maintaining the
anti-$k_t$ kinematics of jets while ensuring infrared safety, at least 
to some high order in the perturbation theory, and enabling flavor assignment~\cite{Czakon:2022wam,Gauld:2022lem,Caletti:2022hnc}.

Nevertheless, addressing this problem has proven more complex than
anticipated. A recent breakthrough was achieved with the development
of infrared-safe anti-$k_t$-like jets, accomplished through the
introduction of an interleaved flavor neutralization
procedure~\cite{Caola:2023wpj}. However, experimental challenges related to the identification and separation of two B hadrons which are very close to each other remain. Furthermore, an unfolding procedure will
be indispensable to convert experimental measurements of flavor-$k_t$
jets into a format directly comparable with theoretical
predictions. Further research in this direction is undoubtedly
needed to accurately describe the signals and backgrounds involving
multiple $b$-jets, which is needed to study signal events with two or three Higgs bosons, and their irreducible backgrounds. 
It is interesting to point out that the approach of 
Ref.~\cite{Caola:2023wpj} is also suited for use with the 
Cambridge/Aachen algorithm. This helps jet flavour tagging for a 
large family of jet substructure tools which could also be relevant 
for multi-Higgs tagging (see below).

\subsection{Perturbative challenges}

In addition to the challenges posed by $b$-tagging and flavored jets,
the complexity of the high multiplicity final states resulting from
the production of two or three Higgs bosons presents other significant
challenges.

Advancements in perturbative calculations over the past two decades
have enabled the development of publicly available
codes~\cite{Buccioni:2019sur,Alwall:2014hca,Actis:2016mpe}, which
allow the automatic computation of one-loop amplitudes for final states with a high particle
multiplicity. For a long time the availability of one-loop amplitudes constituted the bottleneck to obtain next-to-leading order (NLO) accurate predictions for these processes. Nowadays, the primary obstacles in obtaining NLO
predictions lie in issues of numerical stability and computational
time rather than theoretical limitations.
Processes featuring six particles in the final state, such as the 
production of three $b\bar b$-pairs, while feasible, still present 
numerical challenges for NLO calculations. These calculations can be
further refined by matching them with all-order parton shower effects
using methods like {\tt POWHEG}~\cite{Nason:2004rx} or {\tt MC@NLO}~\cite{Frixione:2002ik}.

Despite the progress made, the precision of NLO calculations remains
limited, especially for pure QCD processes involving a high particle
multiplicity. For instance, in the QCD production of three pairs of
bottom quarks, the leading-order contributions involve a high power
(6$^{\rm th}$\,power) of the strong coupling constant. In such a case, determining
a preferred renormalization and factorization scale is not
straightforward. Consequently, uncertainties due to missing higher
orders below 10-20\% are not reachable based solely on pure NLO
predictions (see e.g. ref.~\cite{Alwall:2014hca}).
The frontier of next-to-next-to-leading order (NNLO) calculations now
extends, for selected processes, to cross sections with three
particles in the final state. Processes known today at NNLO include 
three photon production~\cite{Chawdhry:2019bji,Kallweit:2020gcp}, two photons and one jet~\cite{Chawdhry:2021hkp}, two
jets and one photon~\cite{Badger:2023mgf}, three-jets~\cite{Czakon:2021mjy}, $Wb\overline{b}$
production~\cite{Hartanto:2022qhh,Buonocore:2022pqq}, $ttH$~\cite{Catani:2022mfv} and $t\bar t W$~\cite{Buonocore:2023ljm}.

However, it is currently unrealistic to
expect NNLO calculations for processes with six particles in the final
state in the near future, which is the typical multiplicity of
backgrounds relevant to triple-Higgs production.

Various approaches are routinely employed to address this issue. One
widely used experimental-driven approach involves extracting precise
estimates of background processes directly from experimental data
using regions which are devoid of signal to normalize the background, and subsequently
extrapolating these backgrounds to the signal region of
interest. These techniques, and extensions thereof, have been
highly successful in searches for new physics, particularly in
excluding regions of parameter space for new physics models. However,
their application to precision measurements is more challenging due to
the difficulty in estimating the uncertainty associated with the
extrapolation from the signal-free region to the region of interest. 
This, coupled with the challenges related to flavor
assignment discussed earlier, makes it particularly challenging to
assign solid theory uncertainties to theory predictions of 
high multiplicity processes such as the production of 4 $b$-jets, 4 $b$-jets and two photons, or 6 $b$-jets.

Several theory-based approaches exist to improve upon NLO
calculations. One widely used and generic approach is the multi-jet
merging of NLO calculations involving different multiplicities~\cite{Hoeche:2012yf,Gehrmann:2012yg,Frederix:2012ps,Hamilton:2012rf}. This
approach is known to work well in practice, particularly concerning
the shapes of distributions.
Alternatively, it is sometimes feasible to include a well-defined
subset of NNLO corrections, such as form factor corrections. Another 
approximation is to work in the leading-color approximation,
which typically captures the bulk of the NNLO corrections.
In some cases, such as the production of top-quarks decaying
to $W$ and bottom quarks or the production of other resonances, it is
possible to consider only factorizing corrections~\cite{Fadin:1993dz}, i.e.\ to separate
the corrections to production and decay, thereby simplifying the structure
of higher-order corrections. This simplification is justified by the
observation that non-factorizable corrections are typically suppressed 
by the small width over the heavy mass of the resonant particles.
Other interesting approximations include, for instance, employing the soft Higgs approximation in the two-loop virtual corrections.
This method bears resemblance to the soft-gluon approximation widely used in perturbative QCD, albeit tailored specifically to the Higgs boson.
Recently, it has been employed to provide an accurate estimate of the NNLO cross-section for $ttH$ production~\cite{Catani:2022mfv} and $t\bar t W$~\cite{Buonocore:2023ljm}.
In these cases, it is possible to validate the soft-boson
approximation at one-loop. Since the predicted two-loop hard
coefficient is found to be very small, even when assigning a very
conservative error to it, the resulting theory uncertainty remains
small.
Another approach to obtaining massive amplitudes involves starting
from massless ones and then incorporating masses through a massification
procedure~\cite{Penin:2005eh,Mitov:2006xs,Becher:2007cu,Engel:2018fsb,Wang:2023qbf}.
It is worth noting that in the case of $t\bar t W$, the massification
procedure of the quarks, or the soft approximation of the $W$, yield
approximate two-loop results that are consistent with one another. 
This observation is particularly intriguing because
both approximations are, in principle, utilized beyond their region of
validity, and the two approaches are conceptually very different.
Yet another standard approximation for the two-loop virtual is to use
 Pad\'e approximants~\cite{Samuel:1995jc,Ellis:1996zn}, which
essentially determines a best estimate of the missing higher-orders based on previous orders. 
To name a few examples,  Pad\'e approximants were used in ref.~\cite{Elias:2000iw} to estimate higher-order effects in the decays of Higgs to $b\bar b$ and Higgs to two gluons, in ref.~\cite{Grober:2019kuf} Pad\`e approximants are constructed from the expansions of the amplitude for large top mass and around the top threshold to estimate the top-quark mass effects in the Higgs-interference contribution to Z-boson pair production in gluon fusion and in ref.~\cite{Davies:2019nhm} the approximation is used to estimate the three-loop corrections to the Higgs boson-gluon form factor, incorporating the top quark mass dependence. 
In general, these approximations and their
practical 
efficiency can only be assessed on a case-by-case basis.

Overall, these and other approximate higher-order results are likely
to drive the progress of theory predictions to achieve the desired
precision for the dominant background processes relevant to the study
of triple Higgs production in different decay channels, while full
NNLO corrections are likely to remain unavailable in the foreseeable
future.

\subsection{Four- versus five-flavour scheme}

When dealing with processes involving bottom quarks,\footnote{Similar
arguments apply to charm quarks.} two commonly used approaches are
the four-flavor scheme (4FS) and the five-flavor scheme (5FS). Each
scheme offers distinct advantages and drawbacks. For a discussion of these see e.g. ref.~\cite{Harlander:2011aa}. 
In the 4FS, the $b$-quark is treated as a massive object at the level of
short-distance matrix elements, and never explicitly appears in the
initial state. Cross-sections in the 4FS typically contain large
logarithms of the ratio of the bottom mass to the hard scale of the
scattering process.
Conversely, in the 5FS, $b$-quarks are treated as light partons in
short-distance matrix elements. They are generated at a scale $\mu\sim
m_b$ in the Dokshitzer-Gribov-Lipatov-Altarelli-Parisi (DGLAP)
evolution of initial state PDFs, and resummation of large logarithms
is achieved through the DGLAP evolution equations of the bottom PDF.

While resummation of large logarithms is not possible in the 4FS, and
large logarithms are included only at fixed order.  This resummation,
included in the 5FS, typically translates into a better perturbative 
convergence for the latter scheme. Computing higher-order effects is 
also more challenging in the 4FS due to the larger multiplicity and 
inclusion of massive quarks in the Born process.  On the other hand 
in the 4FS scheme, mass effects are included exactly, at the order at
which the calculation is carried out. Implementing 4FS calculations in
a Monte Carlo framework is straightforward, whereas in the 5FS 
particular care is needed when dealing with gluon splittings to bottom 
quarks.

When mass effects are significant and the resummation of collinear
logarithms is important, a combination of both schemes is
necessary. The FONLL (Fixed Order plus Next-to-Leading Logarithms)
approach~\cite{Cacciari:1998it} successfully combines the strengths of
both schemes to obtain a best estimate of total cross sections. Essentially, this involves adding the cross-sections
computed in the 4FS and 5FS and subtracting the double-counting at
fixed order. The only subtlety is that, in order to consistently
remove the double-counting, one needs to express both 4FS and 5FS
cross-sections in terms of the same coupling (i.e. involving the same
number of flavours) and the same PDF. Although technically cumbersome,
this procedure is well-understood and has been widely applied in
various contexts.

Having FONLL-matched predictions available for all ranges of signals
and backgrounds relevant to double and triple Higgs production at the
LHC would be highly desirable for more accurate theoretical
predictions and comparisons with experimental data. This would enable
a better understanding of the underlying physics and aid in the
measurements or constraints of triple and quartic Higgs coupling.

\subsection{Monte Carlo predictions}

While perturbative fixed-order calculations provide the best estimates
for inclusive measurements, Monte Carlo (MC) tools are essential for the description of more exclusive observables and for a full
interpretation of LHC data. The sophistication of Monte Carlo tools
has improved over the years, and it is not uncommon to find examples
where, for instance, Pythia outperforms full matrix element generators
even in regions dominated by hard radiation, which should
theoretically be described less accurately by Monte Carlo
generators. However, since Monte Carlos rely on several
approximations, particularly in the generation of the parton shower in
soft and collinear limits, one issue in comparing data to Monte Carlo
predictions is the lack of clarity in assigning a theory uncertainty
to MC predictions.

Over the past few years, a significant effort has been directed towards 
improving generic-purpose Monte Carlo event generators. In particular, 
several new parton shower algorithms have been introduced. 
In this context, considerable progress has been made to formally validate 
the (logarithmic) accuracy of parton showers by comparing their output
to analytic resummation for specific classes of observables. 
Concretely, several groups~(see e.g.~\cite{Dasgupta:2020fwr,vanBeekveld:2022ukn,Nagy:2014mqa,Forshaw:2020wrq,Herren:2022jej}) 
have reported next-to-leading (NLL) logarithmic accuracy for broad 
classes of observables, or even higher accuracy for non-global
observables~\cite{FerrarioRavasio:2023kyg}.
Additionally a substantial progress has been made to include 
subleading-colour contributions in dipole-based parton showers 
(see, for example, Refs.~\cite{Nagy:2019pjp,Hamilton:2020rcu,Forshaw:2021mtj})
We refer to Ref.~\cite{Campbell:2022qmc}, and references therein, 
for a broader overview of recent improvements.

Such progress in Monte-Carlo generators (together with steady progress 
in analytic resummations) can be viewed as complementary to the 
fixed-order perturbative considerations highlighted in the last two 
sections. 
In the context of multi-Higgs production, combining improvements in
fixed-order perturbation theory, all-order resummations (analytically 
or by means of parton shower algorithms), and non-perturbative 
corrections, would largely help  the study of both signals and 
backgrounds. It could, in particular, impact the modelling of 
backgrounds in experimental context.

\subsection{Boosted versus non-boosted}

As a final set of remarks, we wish to comment on possible scenarios 
where one or more Higgs bosons are produced with a transverse 
momentum much larger than its mass. 
This could for example happen in situations where a more massive 
intermediate new particle decays into a pair of Higgs bosons. 

In such a boosted-Higgs case, the angle between the $b$ and $\bar b$ 
quarks becomes small and the Higgs is reconstructed as a single fat jet. 
The event reconstruction therefore has to rely on jet substructure techniques.
While the boosted regime often comes with low, kinematically-suppressed, 
cross-sections, it can offer several advantages that we briefly discuss 
here.

First of all, jet substructure techniques have seen a large amount 
of development over the past decade, establishing themselves as a 
powerful approach to study complex final-states. A wealth of techniques 
have been proposed and can be used to enhance specific aspects of the signal.
The recent years have also seen the rise of Deep-Learning-based 
tools which excel at separating signals from backgrounds in boosted jets.
This is particularly relevant in a discovery context where boosted 
Higgses would appear in a BSM scenario. 

From an event reconstruction perspective, situations with one or more 
boosted Higgs(es) would suffer less from combinatorial issues than 
non-boosted cases. 

It is beyond the scope of this document to dive into specific jet 
substructure tools. We can however redirect the reader to review articles, 
and references therein, for a generic overview of theoretical and 
machine-learning aspects~\cite{Larkoski:2017jix}, for experimental 
aspects~\cite{Kogler:2018hem}, and for a generic introduction with 
emphasis on analytic aspects in QCD~\cite{ Marzani:2019hun}.

We also note that several jet substructure methods of broad interest 
have been introduced since these reports have been written. 
This includes, for example, techniques based on the Lund Jet 
Plane~\cite{Dreyer:2018nbf}, or on energy correlators 
(see e.g.~\cite{Moult:2016cvt}). 
When it comes to using Machine learning algorithms to tag boosted objects, 
techniques such as the ones from Ref.~\cite{Komiske:2018cqr,Qu:2019gqs,Dreyer:2020brq}
have shown good overall performance in different physics scenarios.

A final set of remark concerns the relation between the boosted regime 
and the perturbative QCD aspects discussed in the previous sections. 
Some substructure techniques are amenable to precision calculations. 
This could lead to  situations  where analytic predictions, obtained
through a combination of (approximate) NNLO, analytic resummations 
and parton shower developments allow for better, simplified, theoretical
control over QCD backgrounds.
A word of caution is however needed when relying on machine-learning 
techniques. These would typically involve training a neural network on
Monte Carlo samples. In such a case, aspects of the physics which are 
not accurately described by the Monte Carlo generator would be "learned" 
by the neural network, resulting in potentially spurious discriminating 
power. Besides being aware of this fact when using Deep learning 
techniques, this again points towards pursuing the effort of improving 
the theoretical description of both the multi-Higgs signals and the 
associated backgrounds.

\clearpage
\section{{\bfseries Experimental lessons from HH }}
{\sl T. du Pree, M. Stamenkovic}

\label{sec:exp-hh}
The self-interactions of the Higgs boson are determined by the shape of the Higgs field potential, which can be written as a polynomial function of the Higgs field $h$:
\begin{equation}
V(h) = \frac{1}{2} m^{2}_{H} h^2 + \lambda_3 v h^{3} + \frac{1}{4} \lambda_4 h^4,
\end{equation}
where $m_H$ is the Higgs boson mass, $v$ is the vacuum expectation value of the Higgs field, and $\lambda_3$ and $\lambda_4$ are the coefficients of the cubic and quartic terms, respectively. These coefficients are also known as the trilinear and quartic couplings for the Higgs boson, and they encode the strength of the interactions among three and four Higgs bosons, respectively. In the Standard Model, these couplings are fixed by the Higgs boson mass and the electroweak parameters, and their values are $\lambda_3 = \lambda_4 = m_H^2/(2v^2) \approx 0.13$. The shape of the Higgs potential is a crucial ingredient of the theory that describes the origin and nature of the Higgs boson and its interactions. However, this shape is not predicted by the theory, but rather assumed as an input. It is essential to test this assumption experimentally and measure the shape of the Higgs potential.

\begin{figure}[ht]
\begin{center}
 \includegraphics[width=1.0\columnwidth]{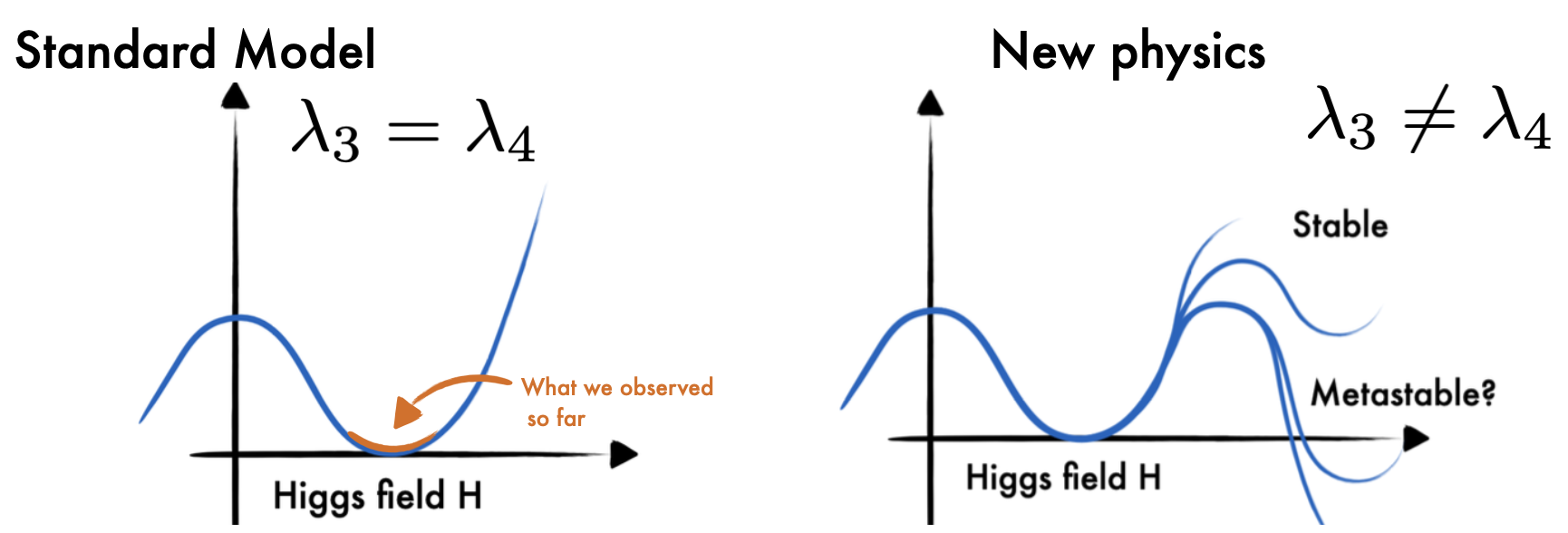}
\caption{Illustration of the shape of the Higgs field potential for the Standard Model ($\lambda_3=\lambda_4$) and for new physics scenarios where the trilinear and quartic self-coupling are not equal ($\lambda_3\neq\lambda_4$).}
\label{fig:sketch_higgs_potential}
\end{center}
\end{figure} 

Figure~\ref{fig:sketch_higgs_potential} illustrates how the shape of the Higgs potential depends on the values of the trilinear and quartic couplings of the Higgs boson, denoted by $\lambda_3$ and $\lambda_4$, respectively. Deviations of these couplings from their expected values in the SM would indicate the presence of new physics beyond the SM. Therefore, measuring these couplings precisely is a powerful way to search for new physics phenomena and to understand the fundamental nature of the Higgs boson and its role in the universe.

The Higgs boson is a key element of the SM of particle physics, responsible for the mass generation of elementary particles. The ATLAS and CMS experiments at the Large Hadron Collider (LHC) have confirmed the existence of the Higgs boson and measured its interactions with gauge bosons and the third-generation fermions. They have also found evidence for its interactions with the second-generation charged leptons~\cite{CMS:2022dwd,ATLAS:2022vkf}. However, the self-interactions of the Higgs boson, which are related to the shape of the Higgs potential, remain untested. The ATLAS and CMS experiments have searched for the production of two Higgs bosons ($HH$), but no significant signal has been observed yet. No results have been reported so far on the $HHH$ production at the LHC.

\begin{figure}[ht]
\begin{center}
 \includegraphics[width=1.0\columnwidth]{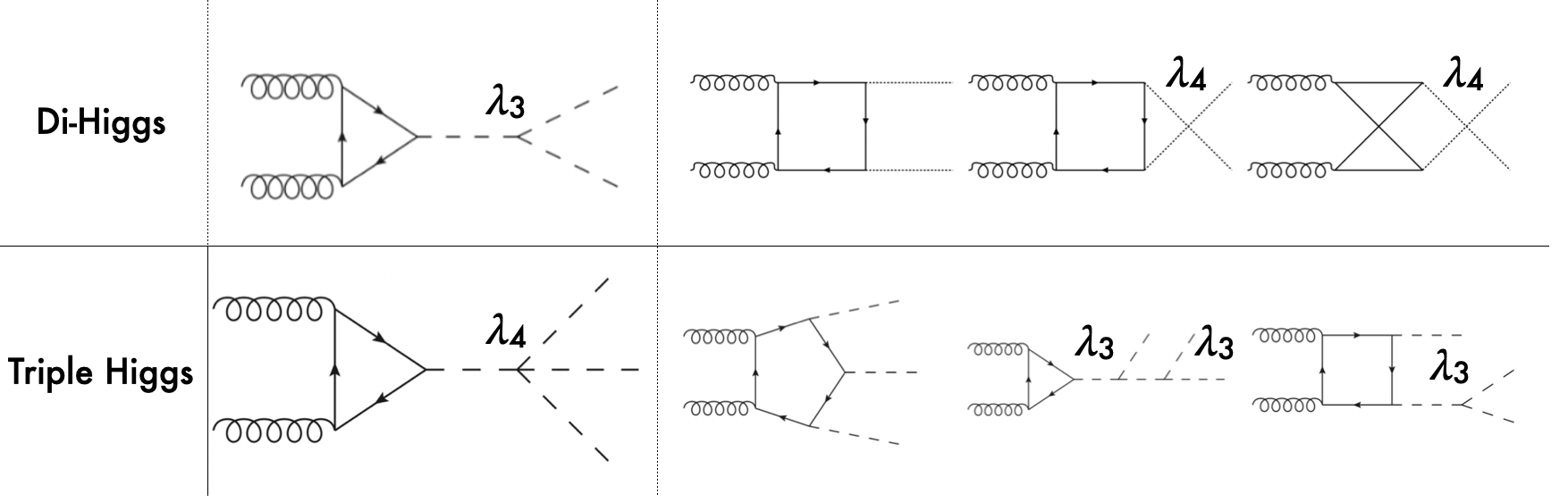}
\caption{Feynman diagram for the gluon-gluon fusion di-Higgs $HH$ and triple Higgs $HHH$ productions at  hadron colliders.}
\label{fig:feynman_hh_hhh}
\end{center}
\end{figure}

The Feynman diagrams for both the $HH$ and $HHH$ production at hadron colliders are shown in Figure~\ref{fig:feynman_hh_hhh}. While the $HH$ production is mostly sensitive to the trilinear coupling $\lambda_3$, the quartic coupling $\lambda_4$ contributes at the next-to-leading order. The $HHH$ production, however, is dominated by both the trilinear and quartic couplings at leading order. 

\textbf{From an experimental point of view, the measurement of the Higgs self-coupling as well as the shape of the potential can only be fully determined from a combined measurement of the $HH$ and $HHH$ processes.}

\subsection{Cross-sections and branching ratios}

At proton-proton colliders, the dominant production mode for the $HH$ and $HHH$ processes is the gluon-gluon fusion production mode. The theoretical and experimental status of the $HH$ production searches, and of the direct and indirect constraints on the Higgs boson self-coupling is extensively discussed in ~\cite{DiMicco:2019ngk}. The cross-sections for both the $HH$ and $HHH$ gluon-gluon fusion production mode, calculated at a center-of-mass $\sqrt{s}=14$ TeV at NNLO, are shown in Table~\ref{tab:xsec_hh_hhh}. The cross-section of the $HH$ production is approximatively 300 times larger than the cross-section of the $HHH$ production. 

\begin{table}[ht]
\begin{center}
\begin{tabular}{c|c|c}
 & $HH$ & $HHH$ \\ 
 \hline
 $\sigma_{\text{NNLO}}$ at $\sqrt{s} = 14$ TeV [fb] & $36.69^{+2.1\%}_{-4.9\%}\pm 3.0\%$ & $0.103^{+5\%}_{-8\%}\pm 15\%$  \\
\end{tabular}
\caption{Cross-section of the gluon-gluon fusion production mode for $HH$~\cite{Shao:2013bz,deFlorian:2015moa,LHCHiggsCrossSectionWorkingGroup:2016ypw} and $HHH$~\cite{deFlorian:2019app} production at NNLO at a center-of-mass $\sqrt{s}=14$ TeV. The uncertainties include the available QCD corrections, as well as the renormalisation and factorisation scales set to $m_{HH} / 2$ and $m_{HHH} / 2$.} 
 \label{tab:xsec_hh_hhh}
\end{center}
\end{table}

\begin{figure}[ht]
\begin{center}
 \includegraphics[width=0.48\columnwidth]{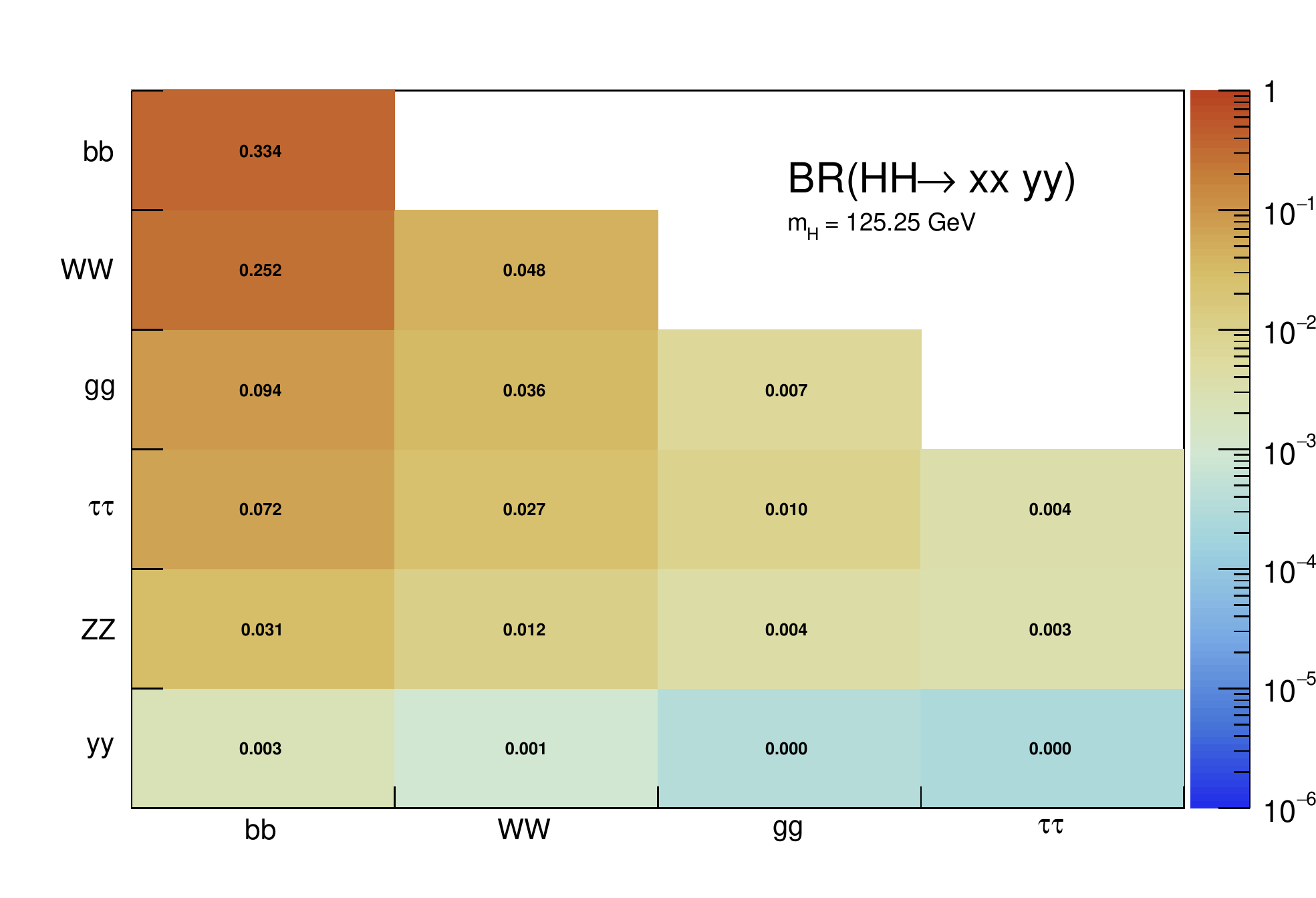}
 \includegraphics[width=0.48\columnwidth]{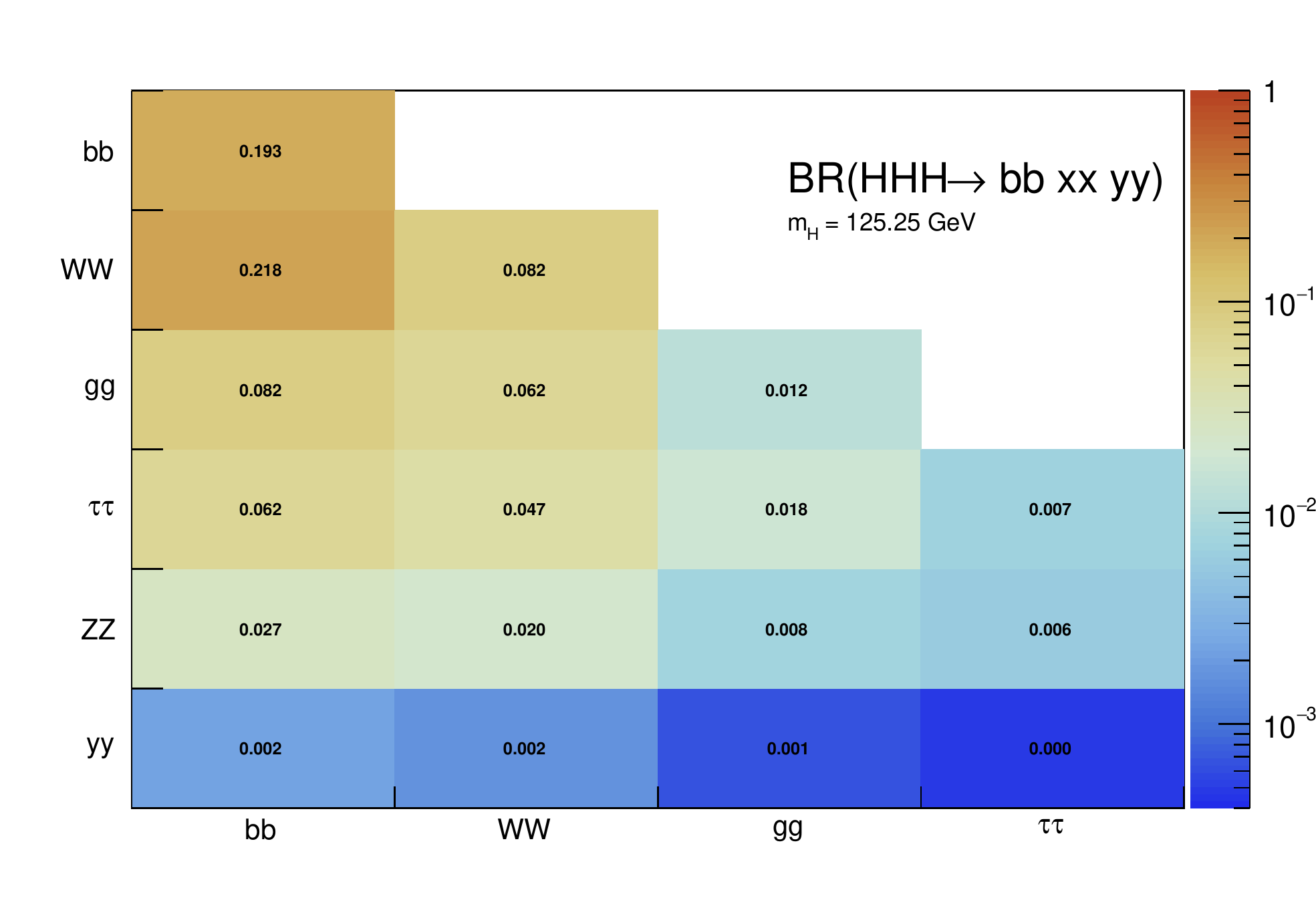}
\caption{Branching ratios for the largest decay mode of the $HH\rightarrow XX~YY$ and $HHH\rightarrow b\bar{b}~XX~YY$ final states assuming a Higgs boson with a mass of 125.25 GeV, rounded to the third decimal.}
\label{fig:br_hh_hhh}
\end{center}
\end{figure}

Under the SM hypothesis, the dominant branching ratios the $HH$ and $HHH$ decay modes are shown in Figure~\ref{fig:br_hh_hhh} for a mass $m_H=125.25$ GeV. Due to the largest branching fraction of the $H\rightarrow b\bar{b}$ decay mode, the largest branching ratio for the $HH$ process is the $HH\rightarrow b\bar{b}~b\bar{b}$ decay mode. In the case of the $HHH$ process, the largest branching ratios are the $HHH\rightarrow b\bar{b}~b\bar{b}~WW$ and $HHH\rightarrow b\bar{b}~b\bar{b}~b\bar{b}$. Furthermore, in the case of $HHH$, about 60\% of the total cross-section is accessible via the $HHH\rightarrow b\bar{b}~b\bar{b}~YY$ decay modes, where $YY=b\bar{b},WW,gg,\tau\tau,ZZ,yy$. The $HH$ and $HHH$ processes have similar decay modes, kinematics and backgrounds. Therefore, the experimental techniques and results obtained from the $HH$ searches can provide useful guidance and input for the $HHH$ searches.

\subsection{Sensitivities to SM $HH$}
From an experimental point of view, the three $HH$ channels with the highest sensitivity are:
\begin{itemize}
    \item $HH\rightarrow b\bar{b}~b\bar{b}$: largest branching ratio (33.4\%) but large contamination from QCD multi-jet background,
    \item $HH\rightarrow b\bar{b}~\tau\tau$: sizable branching ratio (7.2\%) with lower background contamination,
    \item $HH\rightarrow b\bar{b}~yy$: small branching ratio (0.3\%) but low background contamination and better energy resolution on photons.
\end{itemize}

The  $HH\rightarrow b\bar{b}~b\bar{b}$ final state is the most probable decay mode for the $HH$ production, but it also poses several experimental challenges. One of them is the identification of $b$-jets, which requires efficient and precise tagging algorithms to discriminate them from light-flavor jets. Another challenge is the reliable modelling of the dominant background, which is the QCD multi-jet production. This background has a large cross section and is computationally costly to simulate for the ATLAS and CMS experiments. Therefore, data-driven methods are often employed to estimate the QCD multi-jet background from control regions in data and extrapolate it to the signal region. 

A further complication arises from the jet pairing problem, which refers to the ambiguity in assigning the $b$-jets to the Higgs boson candidates. To resolve this problem, a pairing algorithm based on the minimal distance between the invariant masses of the $b$-jet pairs, where the signal uniquely converges to the same mass. This algorithm does not shape the QCD multi-jet background around the Higgs boson mass peak, however the probability to correctly reconstruct the pairs is often lower than in the non-ambiguous decay modes. The jet pairing algorithm is even more important for the $HHH\rightarrow b\bar{b}~b\bar{b}~b\bar{b}$ process, where the additional jets increase the number of possible combinations and therefore degrades the reconstruction efficiency. The usage of modern machine learning methods, such as attention networks~\cite{Shmakov:2021qdz}, or algorithms based on the minimal distance between the jets will be necessary to improve the sensitivity to the $HHH$ processes.

The loss of performance arising from the jet pairing can be mitigated with the usage of a boosted category where the two Higgs boson candidates, recoiling against each other, are reconstructed within a large-radius jets with a transverse momentum of 300 GeV. By exploiting from the recent improvement in boosted Higgs boson tagging, such as ParticleNet~\cite{Qu:2019gqs}, the QCD multi-jet background can be reduced and the sensitivity largely improved. Boosted reconstruction techniques can play a large role in the search for $HHH$. 

The $HH\rightarrow b\bar{b}~\tau\tau$ final state requires both flavour tagging and $\tau$-identification algorithms. While the branching ratio is lower than in the $HH\rightarrow b\bar{b}~b\bar{b}$ final state, the presence of 2 $\tau$-leptons allows to efficiently reduce the background contamination from the QCD multi-jet process. The dominant background is therefore the $t\bar{t}$ process, for which the Monte-Carlo simulation can be used to describe the data accurately. The sensitivity of the analysis is further improved by splitting the signal region in categories depending on the decays of the $\tau$-leptons: $e\tau_{hadronic}$, $\mu\tau_{hadronic}$ and $\tau_{hadronic}\tau_{hadronic}$. The $\tau_{hadronic}\tau_{hadronic}$ channel has the advantage of having a lower contamination from jets from the QCD background misidentified as a $\tau$-lepton, which in turns improves the sensitivity. It is interesting to note that the $HHH\rightarrow b\bar{b}~b\bar{b}~\tau\tau$ final state will benefit from the same advantages as the $HH\rightarrow b\bar{b}~\tau\tau$. In this case, the branching ratio difference with respect to the final state with 6 $b$-quarks is lower than the difference in $HH$, a hint that this channel will play a crucial role in the search for $HHH$.

The $HH\rightarrow b\bar{b}~yy$ final state has a lower branching ratio but benefits from the energy resolution of the ATLAS and CMS experiments, which is of the order of $\mathcal{O}(1)$ GeV with respect to the jets energy resolution of $\mathcal{O}(10)$ GeV. The analysis is designed to measure a narrow resonance in the invariant mass distribution $m_{yy}$, where the dominant background $yy$+jets is estimated from a parametric fit to the sideband. Due to the more precise resolution of the invariant mass of the Higgs candidate, this final state benefits the most from the increased statistics obtained over the years. Regarding $HHH\rightarrow b\bar{b}~b\bar{b}~yy$, the branching ratio is $0.228\%$, resulting in about 1 event produced by the end of the High-Luminosity LHC. This channel therefore constitutes an interesting probe for new physics phenomena. 

\begin{table}[ht]
\begin{center}
\begin{tabular}{c|c|c}
 Final state & ATLAS & CMS \\ 
 \hline
Resolved  $HH\rightarrow b\bar{b}~b\bar{b}$ & $\mu_{HH} < 5.4~(8.1)$~\cite{ATLAS:2023qzf} & $\mu_{HH} < 3.9~(7.8)$~\cite{CMS:2022cpr}   \\
Boosted  $HH\rightarrow b\bar{b}~b\bar{b}$ & - & $\mu_{HH} < 9.9~(5.1)$~\cite{CMS:2022gjd}   \\ 
Combined $HH\rightarrow b\bar{b}~b\bar{b}$ & - & $\mu_{HH} < 6.4~(4.0)$~\cite{CMS:2022dwd}   \\ \hline

$HH\rightarrow b\bar{b}~\tau\tau$ & $\mu_{HH} < 5.9~(3.1)$~\cite{ATLAS:2023vjn}  & $\mu_{HH} < 3.3~(5.2)$~\cite{CMS:2022hgz}   \\ \hline
$HH\rightarrow b\bar{b}~yy$ & $\mu_{HH} < 4.0~(5.0)$~\cite{ATLAS:2023gzn}  & $\mu_{HH} < 7.7~(5.2)$~\cite{CMS:2020tkr}  \\ 
\end{tabular}
\caption{Observed (expected) limit on the signal strength $\mu=\frac{\sigma\times~br}{\sigma_{SM}\times br_{SM}}$ to the SM $HH$ process from the ATLAS and CMS experiments, under the background only hypothesis $\mu_{HH}=0$.} 
 \label{tab:sensitivity_hh}
\end{center}
\end{table}

The limits at 95\% confidence level on the signal strength $\mu=\frac{\sigma\times~br}{\sigma_{SM}\times br_{SM}}$, under the assumption that there is no SM Higgs self-coupling $\mu=0$, are shown in Table~\ref{tab:sensitivity_hh}. In CMS, the combined measurement of the $HH\rightarrow b\bar{b}~b\bar{b}$ analyses results in the highest expected sensitivity. This mostly relies on the inclusion of a category where both the Higgs bosons are reconstructed in a large-radius jet with a transverse momentum of $p_{\text{T}} > 300$ GeV and exploits the ParticleNet machine learning algorithm to select Higgs-like jets and remove the background arising from QCD multijets. This unique signature, where two Higgs bosons recoil again each other, measured in a decay channel with the highest branching ratio, drives the sensitivity to the $HH$ process. The other channels exhibit a similar sensitivity to this boosted category. 

In ATLAS, the decay channel $HH\rightarrow b\bar{b}~\tau\tau$ results in the best sensitivity and drives the search for the $HH$ process. In particular, the category where the two $\tau$-leptons decay hadronically shows the best performance within the analysis. This result outperforms the other leading channels, taken separately, in both ATLAS and CMS by 60\%-70\% and is therefore one of the most promising channel for $HHH$ as well. The gain in signal acceptance outperforms the increase in the dominant $t\bar{t}$ background relevant for this channel. The difference with respect to the CMS result is partly due to the trigger requirements, where the ATLAS experiment recorded signal events more efficiently during Run 2. The Run 3 analyses, which will benefit from optimised strategies in terms of trigger as well as improved machine learning tools for the identification of $b$-jets and $\tau$-lepton, will lead to even better constraints on the $HH$ search and the Higgs self-coupling. 

\begin{table}[ht]
\begin{center}
\begin{tabular}{c|c|c}
 Final state & ATLAS & CMS \\ 
 \hline
Resolved  $HH\rightarrow b\bar{b}~b\bar{b}$ & $-3.5(-5.4)< \kappa_{\lambda} < 11.3(11.4)$ & $-2.3(-5.0)< \kappa_{\lambda} < 9.4(12)$  \\
Boosted  $HH\rightarrow b\bar{b}~b\bar{b}$ & - & $-9.9(-5.1)< \kappa_{\lambda} < 16.9(12.2)$   \\ \hline
$HH\rightarrow b\bar{b}~\tau\tau$ & $-3.2(-2.5)< \kappa_{\lambda} < 9.1(9.2)$  & $-1.7(-2.9)< \kappa_{\lambda} < 8.7(9.8)$   \\ \hline
$HH\rightarrow b\bar{b}~yy$ & $-1.4(-2.8)< \kappa_{\lambda} < 7.8(6.9)$  & $-3.3(-2.5)< \kappa_{\lambda} < 8.5(8.2)$ \\ 
\end{tabular}
\caption{Observed (expected) limit on coupling modifier of $\kappa_{\lambda}$ from the ATLAS and CMS experiments. References for each measurements can be found in Table~\ref{tab:sensitivity_hh}.} 
 \label{tab:sensitivity_kappa_lambda_hh}
\end{center}
\end{table}

These results are interpreted in terms of Higgs self-coupling modifications $\kappa_{\lambda}$ and reported in Table~\ref{tab:sensitivity_kappa_lambda_hh}, where $\kappa_{\lambda}=1$ corresponds to the SM self-coupling. In terms of constraints on the self-coupling, it is interesting to note that the $HH\rightarrow b\bar{b}~yy$ channel drives the sensitivity. This is due to the trigger requirement, which selects events with two photons and allows to record events in the low part of the invariant mass $m_{HH} < 450$ GeV, where the large modifications of the $\kappa_{\lambda}$ coupling are dominant. Under the current assumptions, only coupling modifications to the trilinear $\kappa_{\lambda_3}$ coupling are considered and the modifications to the quartic coupling are currently neglected. In order to relax these assumptions, the combined measurement of $HH$ and $HHH$ will provide complementary constraints. 

\begin{figure}[ht]
\begin{center}
 \includegraphics[width=0.48\columnwidth]{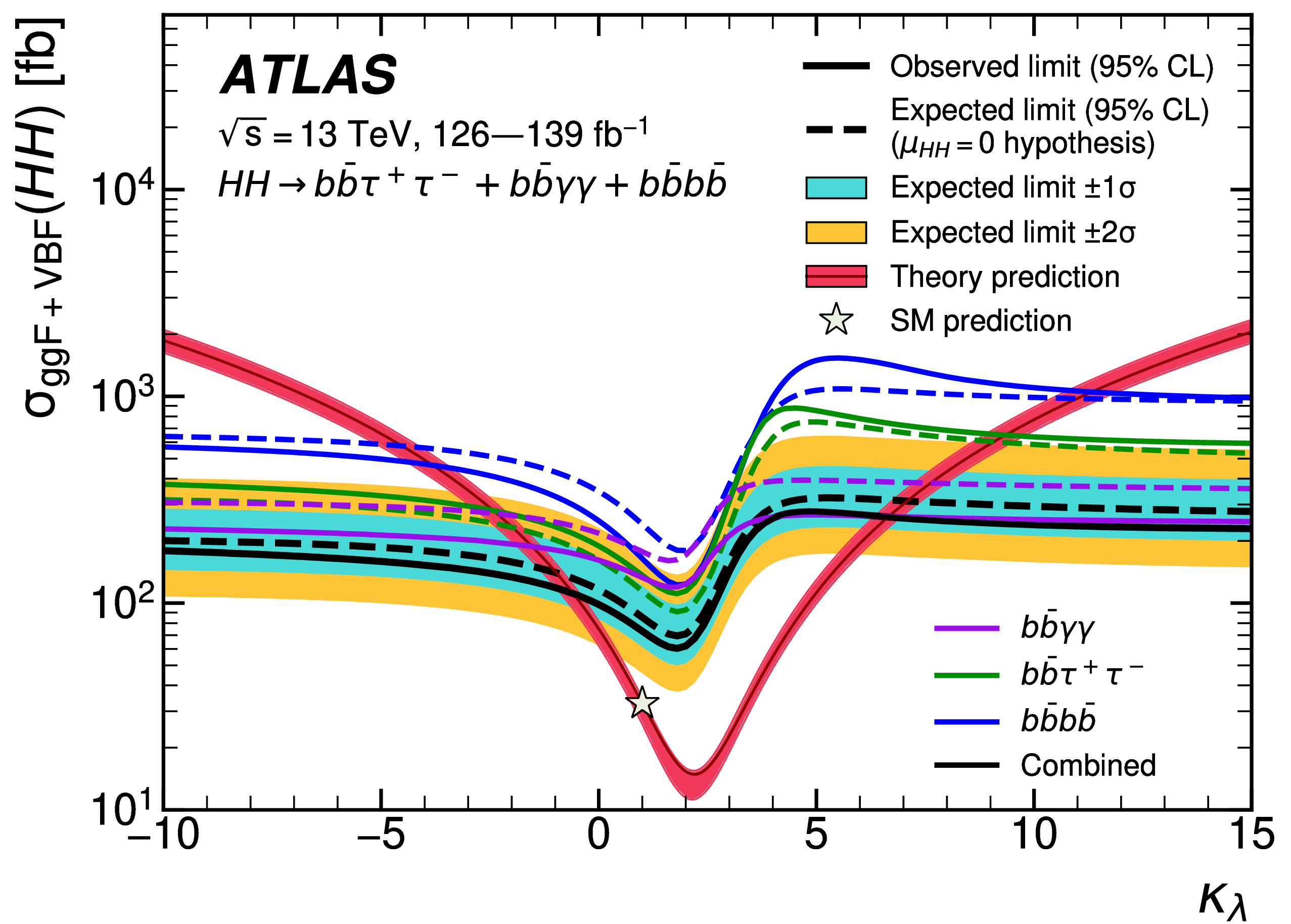}
 \includegraphics[width=0.48\columnwidth]{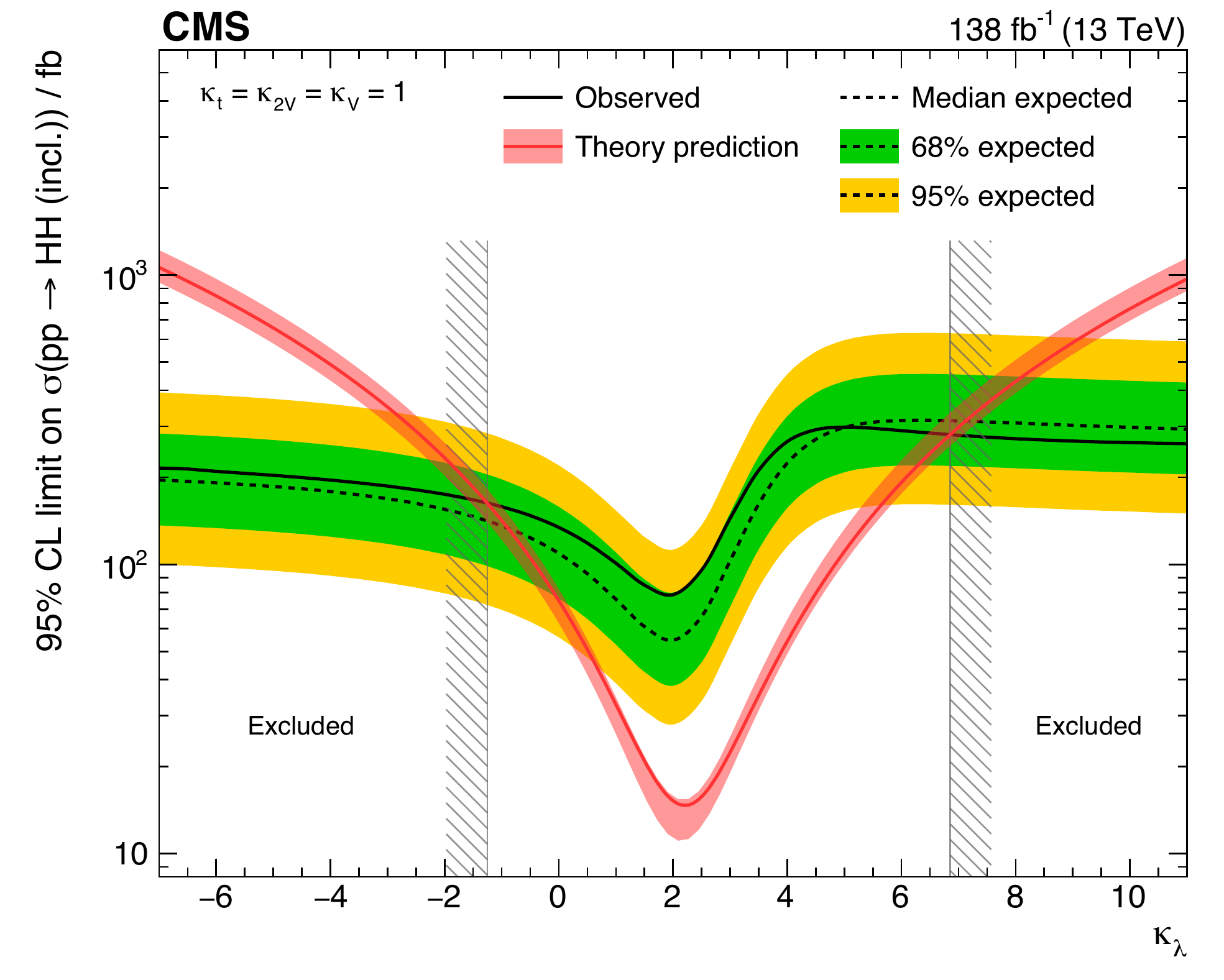}
\caption{Combined measurements of $HH\rightarrow b\bar{b}b\bar{b}$, $HH\rightarrow b\bar{b}\tau\tau$ and $HH\rightarrow b\bar{b}yy$ interpreted in terms of constraints on the coupling modifier $\kappa_{\lambda}$ for ATLAS and CMS~\cite{CMS:2022dwd,ATLAS:2022vkf}.}
\label{fig:combinatin_kappa_lambda}
\end{center}
\end{figure}

Finally, the combination of the main $HH$ analyses allows to set the most stringent constraint on the $\kappa_{\lambda}$ coupling modification, as reported by both the ATLAS and CMS experiments in Figure~\ref{fig:combinatin_kappa_lambda}. A similar combination for the dominant $HHH$ channels is expected to yield in the most stringent constraint on both $\lambda_{3}$ and $\lambda_{4}$ and probe further the potential of the Higgs field. 

\textbf{In summary, while the cross-section of the $HHH$ process is $\approx$300 times smaller than the cross-section of the $HH$ process, this unexplored process at the LHC will allow to test the shape of the Higgs field potential. As both processes depend on the trilinear $\lambda_3$ and quartic $\lambda_4$ couplings, the most promising probe of the self-coupling will be obtained from a combined measurement. From an experimental point of view, the lessons learned during the $HH$ search are the importance of boosted reconstruction techniques to select $H\rightarrow b\bar{b}$ and $H\rightarrow \tau\tau$ signatures in large-radius jets. In addition, signatures including $\tau$-leptons provide a high signal acceptance for a lower background contamination, which in turns result in a large sensitivity. Finally, decay channels including photons $y$, while subject to a small branching ratio, provide excellent probes to test anomalous self-couplings of the Higgs boson, in both $HH$ and $HHH$.}

\clearpage
\section{{\bfseries Experimental prospects and challenges }\label{sec:chall}} 
{\sl H. Arnold, G. Landsberg, B. Moser, M. Stamenkovic}

\subsection{Experimental thoughts}

In this section, we offer a few thoughts on the best ways of tackling various experimental challenges in a search for $HHH$ production, with the focus on LHC and HL~LHC.

\subsubsection{Diagrammatics}
At leading order, there are exactly 100 Feynman diagrams contributing to the standard model like $pp \to HHH$ production: 50 involving the top quark mediated loops and another 50 involving the $b$ quark mediated loops. Ignoring the latter as subdominant contributions, we could focus on the former 50 diagrams. Here by SM-like production, we mean production with SM-like diagrams, i.e., the ones that do not involve new particles, but not necessarily with the SM value of Higgs self-couplings. This is non-resonant $HHH$ production, which results in generally falling $HHH$ invariant mass spectrum.

\begin{figure}[htb]
\centering
\includegraphics[width=0.9\textwidth]{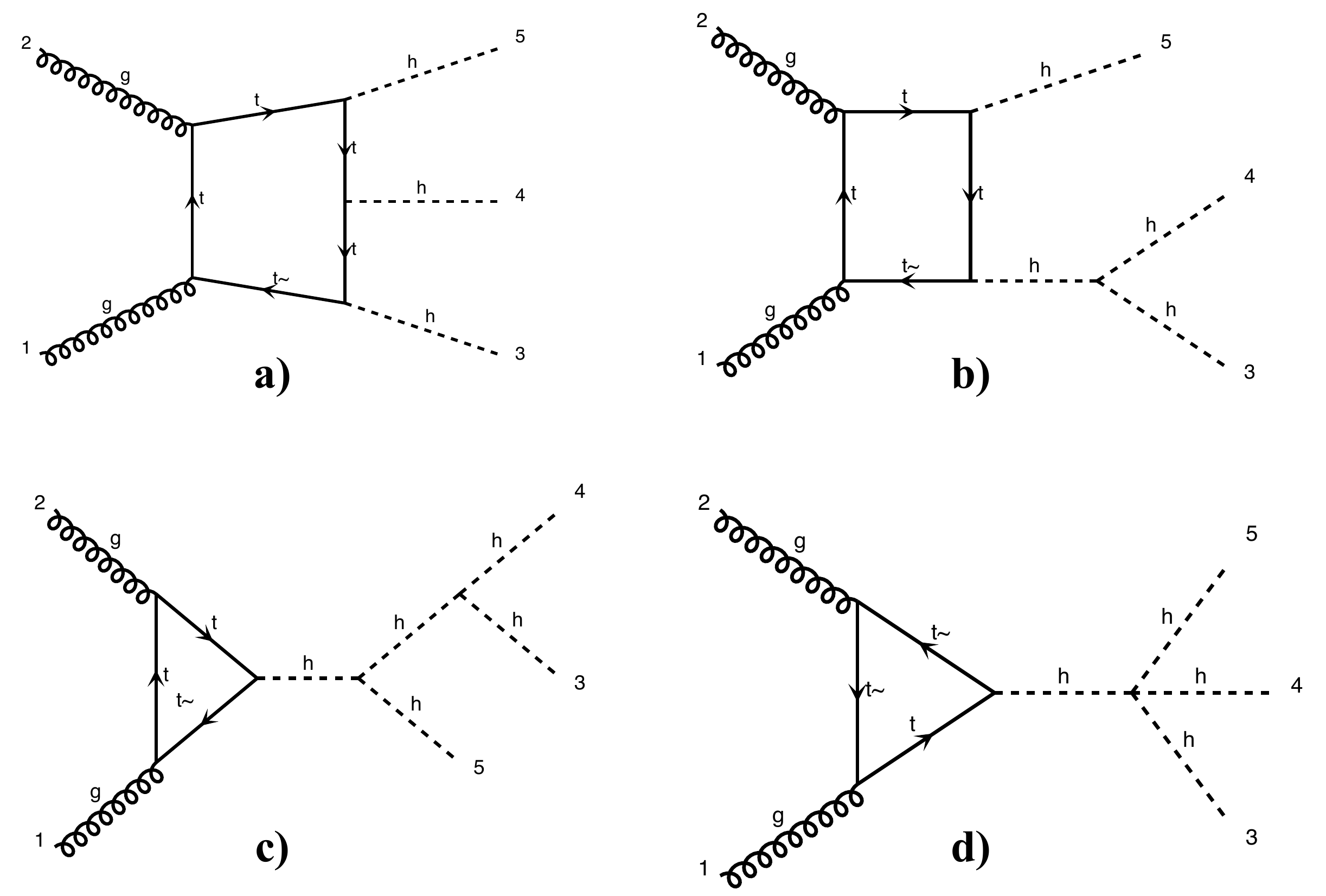}
\caption{Examples of four classes of leading-order diagrams contributing to the $pp \to HHH$ production: a) pentagon; b) box; c) triangle; and d) quartic.}
\label{fig:diagrams}
\end{figure}

These 50 diagrams can be arranged in four broad classes, as shown in Fig.~\ref{fig:diagrams}: 24 pentagon, 18 box, 6 triangle, and 2 quartic diagrams, which generally destructively interfere with each other. The pentagon diagrams constitute the irreducible SM background, as they do not involve either trilinear ($\lambda_3$) or quartic ($\lambda_4$) Higgs self-coupling. In contrast, we will refer to the diagrams that are sensitive to either trilinear or quartic Higgs self-coupling as signal diagrams. The matrix elements of these diagrams are ${\cal M} \sim y_t^3$, where $y_t$ is the top quark Yukawa coupling. The box (triangle) diagram matrix elements are ${\cal M} \sim y_t^2\lambda_3$ (${\cal M} \sim y_t\lambda_3^2$), while the quartic diagrams matrix elements are ${\cal M} \sim y_t\lambda_4$. The box and triangle diagrams interfere destructively with the SM background diagrams, while the quartic diagrams to first order do not interfere with the other three classes. Given that in the SM $\lambda_3 = \lambda_4 \approx 0.13$, the pentagon background diagrams dominate, but this is not necessarily the case when $\lambda_3$ and/or $\lambda_4$ are large. We note that while there are only 2 diagrams involving $\lambda_4$, there are 24 diagrams involving $\lambda_3$, which makes $HHH$ production an excellent laboratory to study trilinear Higgs self-coupling.

{\bf An experimental challenge is to identify the region of phase space where box and triangular diagram contributions dominate, which could improve sensitivity to Higgs self-couplings by not only suppressing the irreducible SM background but also removing the unwanted negative interference with it.} 

\subsection{Branching fractions}
\label{sec:chall_br}
An obvious experimental question is which channels of the $HHH$ system decay are most promising to explore at the (HL-) LHC.

Here we will use the following values of branching fractions  for the major Higgs boson decay modes~\cite{yr4page}, assuming the Higgs boson mass of 125.25~GeV~\cite{Workman:2022ynf}:
\begin{itemize}
    \item ${\cal B}_1 = {\cal B}(H \to b\bar b) = 57.8\%$;
    \item ${\cal B}_2 = {\cal B}(H \to WW) = 21.8\%$;
    \item ${\cal B}_3 = (H \to gg) = 8.17\%$;
    \item ${\cal B}_4 = {\cal B}(H \to \tau^+\tau^-) = 6.23\%$;
    \item ${\cal B}_5 = {\cal B}(H \to ZZ) = 2.68\%$; and
    \item ${\cal B}_6 = {\cal B}(H \to \gamma\gamma) = 0.227\%$.
\end{itemize}

First, we focus on the existing LHC data from Run 2 and assume that we are aiming at probing the $HHH$ cross section at $\sim 100$ times the SM value. The next-to-next-to-leading order (NNLO) cross section for triple Higgs production was evaluated at 14 TeV~\cite{deFlorian:2019app} as $\approx 100$~ab; within the precision we are interested in here, we will assume that this value also applies to the 13~TeV Run 2 center-of-mass energy. That implies that in Run 2, one would expect to produce $\sim 1000$ $HHH$ events per experiment at 100 times the SM cross section, which would correspond to about 100 events after the acceptance and reconstruction efficiency in a typical decay channel (based on a typical efficiency of the $HH$ analyses~\cite{CMS:2022cpr,CMS:2022hgz}). Even if one manages to completely suppress the background, in order to set a 95\% confidence level limit on the $HHH$ cross section, one needs an expectation of at least three observed events. That implies that any decay channel with a branching fraction of less than $\sim 3\%$ is not useful in setting such a limit with the present data set. While these channels may play an important role at the HL-LHC with a full 3~ab$^{-1}$ data set, for practical purposes, we will ignore such channels for now.

Table~\ref{table:channels} lists leading branching fractions of various experimentally feasible $HHH$ decays. We will use the following branching fractions for the dominant decays of the $\tau$ leptons, and $W$ and $Z$ bosons: ${\cal B}(\tau_{\rm h}) = {\cal B}(\tau \to$ hadrons)$ = 64.8\%$, ${\cal B}(W_{\rm h}) = {\cal B}(W \to q\bar q') = 67.4\%$, ${\cal B}(W_\ell) = {\cal B}(W \to e^+e^- + \mu^+\mu^-) = 21.6\%$, and ${\cal B}(Z_{\rm h}) = {\cal B}(Z \to q\bar q) = 69.9\%$.

\begin{table}[hbt]
    \centering
    \caption{Leading branching fraction of the $HHH$ system decay modes.}
    \label{table:channels}
    \begin{tabular}{lcc}
    $HHH \to 3(b\bar b)$ & ${\cal B}_1^3$ & 19.3\% \\
    \hline
    $HHH \to 2(b\bar b)\tau^+\tau^-$ & $3 {\cal B}_1^2 {\cal B}_4$ & 6.24\% \\
    $HHH \to 2(b\bar b)\tau_{\rm h}\tau_{\rm h}$ & $3 {\cal B}_1^2 {\cal B}_4 {\cal B}(\tau_{\rm h})^2$ & 2.62\% \\
    \hline
    $HHH \to 2(b\bar b)W^+W^-$ & $3 {\cal B}_1^2 {\cal B}_2$ & 21.8\% \\
    $HHH \to 2(b\bar b)W_{\rm h}W_{\rm h}$ & $3 {\cal B}_1^2 {\cal B}_2 {\cal B}(W_{\rm h})^2$ & 9.93\% \\
    $HHH \to 2(b\bar b)W_{\rm h}W_\ell$ & $6 {\cal B}_1^2 {\cal B}_2 {\cal B}(W_{\rm h}){\cal B}(W_\ell)$ & 6.36\% \\
    \hline
    $HHH \to 2(b\bar b)gg$ & $3 {\cal B}_1^2 {\cal B}_4$ & 8.19\% \\
    \hline
    $HHH \to b\bar b W^+W^-\tau^+\tau^-$ & $3! {\cal B}_1 {\cal B}_2 {\cal B}_4$ & 4.7\% \\
    $HHH \to b\bar b W_{\rm h}W_{\rm h}\tau_{\rm h}\tau_{\rm h}$ & $3! {\cal B}_1 {\cal B}_2 {\cal B}_4 {\cal B}(W_{\rm h})^2 {\cal B}(\tau_{\rm h})^2$ & 0.898\% \\
    \hline
    $HHH \to b\bar b gg\tau^+\tau^-$ & $3! {\cal B}_1 {\cal B}_3 {\cal B}_4$ & 1.77\% \\
    $HHH \to b\bar b gg\tau_{\rm h}\tau_{\rm h}$ & $3! {\cal B}_1 {\cal B}_3 {\cal B}_4 {\cal B}(\tau_{\rm h})^2$ & 0.741\% \\
    \hline
    $HHH \to 2(b\bar b) ZZ$ & $3 {\cal B}_1^2 {\cal B}_5$ & 2.69\% \\
    $HHH \to 2(b\bar b) Z_{\rm h}Z_{\rm h}$ & $3 {\cal B}_1^2 {\cal B}_5 {\cal B}(Z_{\rm h})^2$ & 1.31\% \\
    \hline
    $HHH \to b\bar b gggg$ & $3 {\cal B}_1 {\cal B}_3^2$ & 1.16\% \\
    \hline
    $HHH \to b\bar b 2(\tau^+\tau^-)$ & $3 {\cal B}_1 {\cal B}_4^2$ & 0.673\% \\
    \hline
    $HHH \to 2(b\bar b) \gamma\gamma$ & $3 {\cal B}_1^2 {\cal B}_6$ & 0.228\% \\
    \end{tabular}
\end{table}

It is quite obvious from this table that the decay modes with two photons, two Z bosons, or four $\tau$ leptons are hopeless with the currently available data. It is further clear that one should instead focus on the all-hadronic channels, as those are the only ones that have sufficiently high branching fraction. The only exception is the $HHH \to 2(b\bar b)W_{\rm h}W_\ell$ channel that has a branching fraction of 6.36\%, but unfortunately this channel does not have a mass peak in the invariant mass distribution of the visible part of the $W^+W^-$ system decay, so it would be quite challenging (but perhaps worth a second look!). Focusing only on the all-hadronic channels, one can see that it is completely dominated by the $4b+$ jets decays, which comprise  40\% of all $HHH$ decays. This is a great news, as we recover 40\% of possible decays in the channel that has been experimentally proven to be feasible through the $pp \to HH \to bb\bar b \bar b$ searches. Requiring at least two extra jets (and further splitting into categories with extra jets being $b$- or $\tau_{\rm h}$-tagged) is certainly a less challenging signature with lower backgrounds than the inclusive $bb\bar b \bar b$ channel, so one could use the background suppression and evaluation techniques developed in the $H(b\bar b)H(b\bar b)$ analyses to search for triple Higgs boson production with high efficiency and acceptance.

{\bf Thus, the all-hadronic $bb\bar b \bar b +$ jets channels is most promising to establish first limits on the $HHH$ production with Run 2 and Run 3 data.}

\subsubsection{Boost or bust!}

We now focus on the $HHH \to 3(b\bar b)$ channel, which comprises about half of the inclusive $bb\bar b \bar b +$ jets branching fraction. In this case, the combinatorics related to pairing of 6 $b$-tagged jets to match the three Higgs boson candidates becomes quite tedious. The number of possible pairings of 6 $b$-tagged jets is equal to $C_6^2 C_4^2 C_2^2 /3! = 15 \times 6 \times 1/6 = 15$ combinations, making it hard to reconstruct individual Higgs bosons reliably.

\begin{figure}[htb]
\centering
\includegraphics[width=0.6\textwidth]{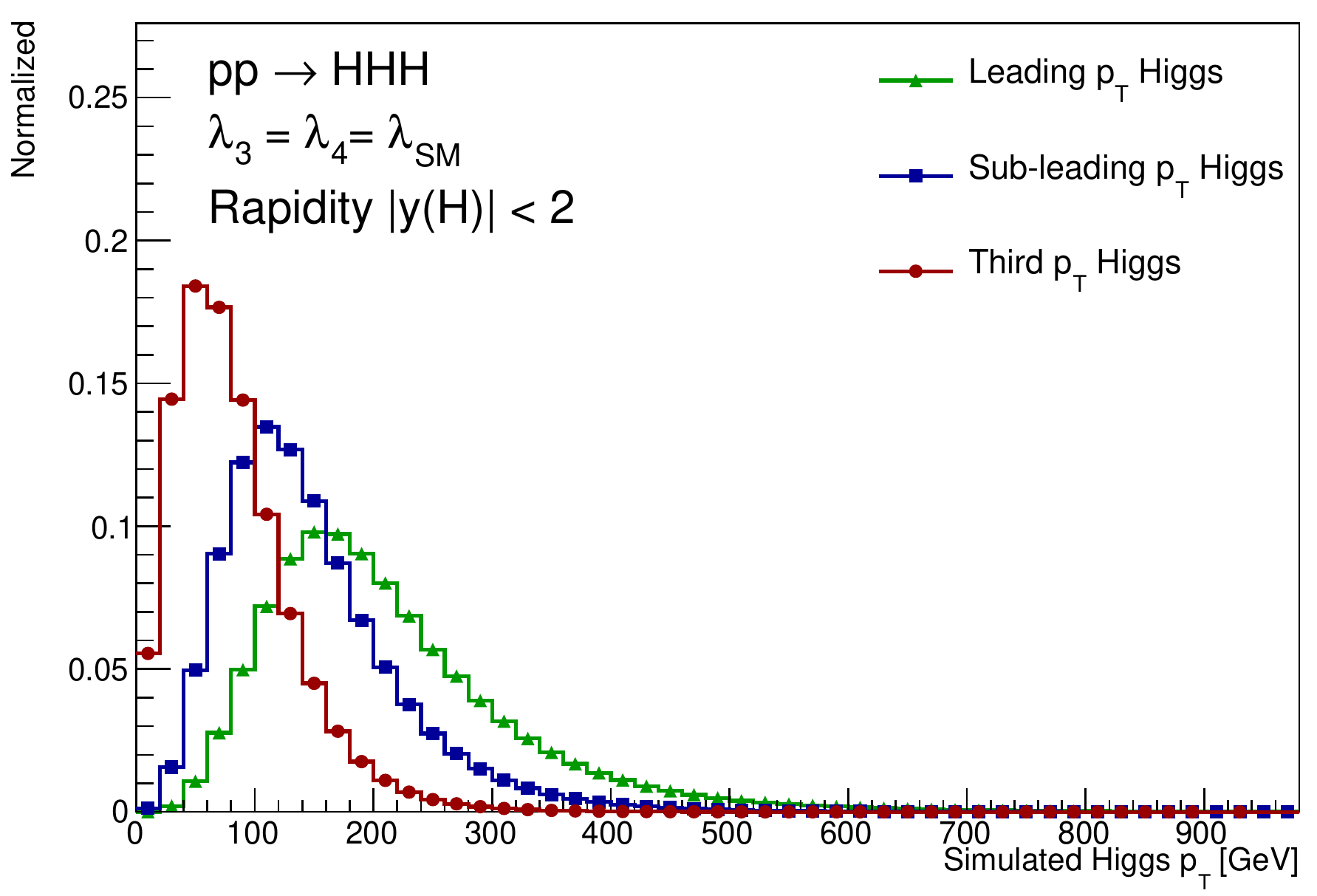}
\caption{Transverse momentum $p_{\rm T}^H$ spectrum of the Higgs bosons in SM triple Higgs boson production.}
\label{fig:HHH-pT}
\end{figure}

This is where the jet merging comes to rescue! It turns out that the Higgs bosons in the $HHH$ production are produced with quite significant transverse momentum $p_{\rm T}$, as shown in Fig.~\ref{fig:HHH-pT}. The distributions for the two leading Higgs boson peak well above 100~GeV, and even for the trailing Higgs boson the median is about 100 GeV. (This is not surprising, as the signal diagrams with trilinear coupling are $t$-channel-like with either the Higgs boson or the top quark as a $t$-channel propagator, so the characteristic $p_{\rm T}$ of the leading Higgs boson or the recoiling di-Higgs system on the other side is of order of the mass of the propagator, i.e., $\sim 150$~GeV.) This implies that it is very likely that at least one of the Higgs bosons within the $HHH$ system has a significant Lorentz boost, resulting in its decay products (a $b$ quark-antiquark pair) to be reconstructed as a single, merged jet, $J$. Indeed, on average, the opening angle between the two decay products of a Lorentz-boosted resonance is given by $\theta = 2/\gamma$, where $\gamma$ is the Lorentz boost. For a Higgs boson with a $p_{\rm T}^H = 250$~GeV, the $\gamma$ factor is 2, so the opening angle is 1 radian. This is similar to a radius parameter of the jet reconstruction used for merged jet analyses (between 0.8 and 1.5).

\begin{figure}[htb]
\centering
\includegraphics[width=0.6\textwidth]{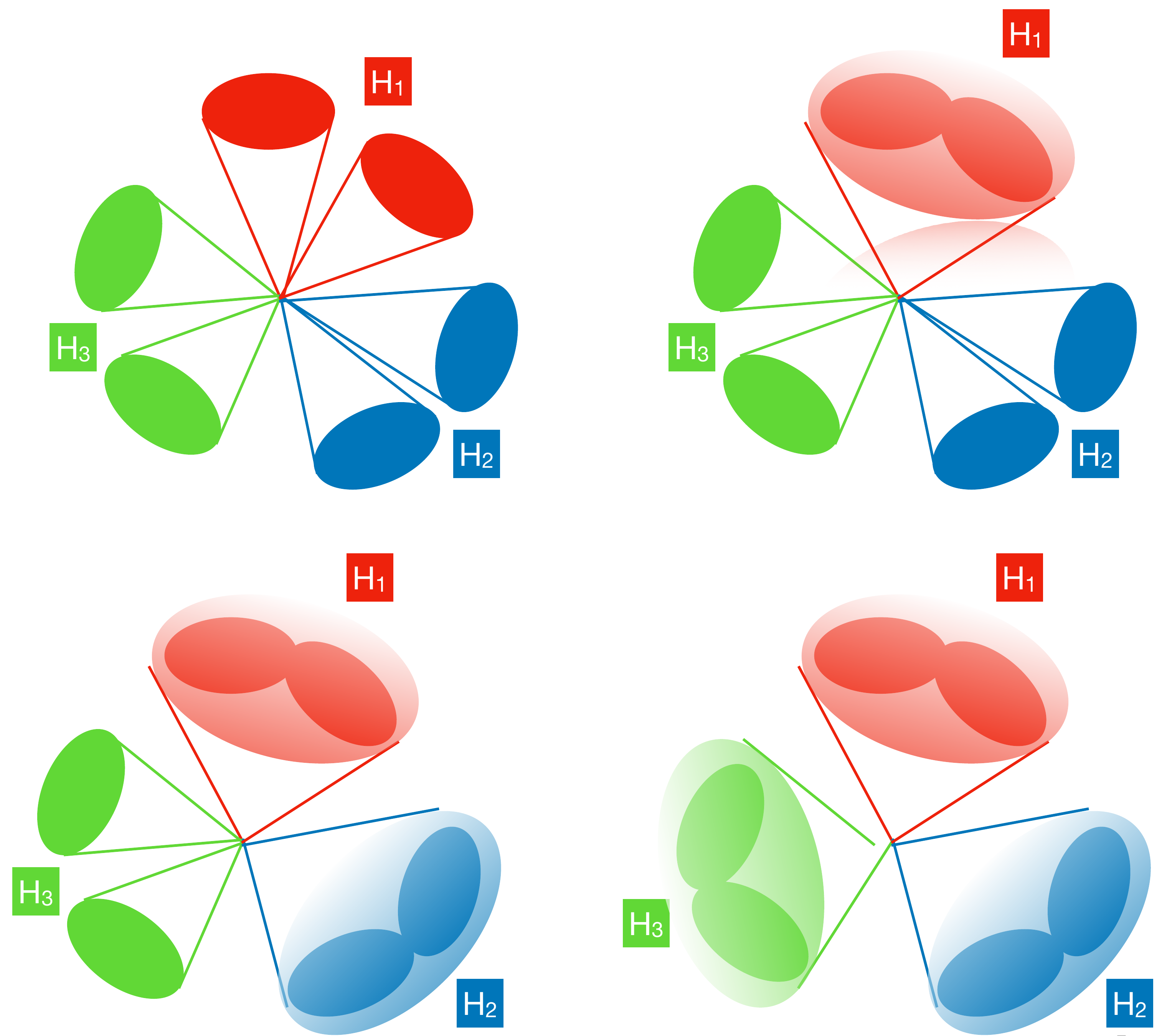}
\caption{Schematics of the reconstruction of the $HHH \to 3(b\bar b)$ system with (upper left to lower right) 0, 1, 2, and 3 Higgs boson decaying into a merged jet topology.}
\label{fig:HHH-merging}
\end{figure}

In the last decade or so, a number of powerful techniques to distinguish such merged jets with a distinct two-prong substructure from regular QCD jets have been developed, which allow for an effective reduction of backgrounds in a boosted topology. (Indeed, the boosted topology is shown to be the most sensitive in the $HH \to bb\bar b \bar b$ searches~\cite{CMS:2022dwd}.) In addition to a powerful background suppression, the boosted topology in the $HHH$ case carries additional benefits: if just one of the Higgs bosons decays into a merged jet, the number of possible jet permutations decreases from 15 to $C_4^2 C_2^2 /2! = 6 \times 1/2 = 3$, and if at least two Higgs bosons are reconstructed as merged jets, there is only one possible permutation, as illustrated in Fig.~\ref{fig:HHH-merging} (as long as we do not distinguish the individual Higgs bosons)! 


The situation becomes even more advantageous for the beyond-the-SM scenarios where the $HHH$ system is produced via resonance decays. For example, in a two real singlet extension of the SM~\cite{Papaefstathiou:2020lyp}, the following process results in a triple Higgs boson production: $pp \to h_3 \to h_2 h_1 \to h_1 h_1 h_1$, where $h_1$ is the SM Higgs boson ($h_1 = H$) and $h_{2,3}$ are the extra scalars. For a typical benchmark with the $h_3$ mass of 500~GeV and $h_2$ mass of 300~GeV, the $pp \to HHH$ production cross section is enhanced by 2.5 orders of magnitude to $\sim 40$~fb, while the relatively high mass of $h_3$ guarantees a large Lorentz boost of the produced Higgs bosons!

As a side remark, generally this and related extensions of the SM should result in resonant production of $HHH$, $VHH$, and $VVH$ systems, with $V = W$ or $Z$ boson. At the LHC, the program of searches for triple-boson resonances is still in its infancy, so it would be very advantageous to mount a broad search for resonant decays into $VHH$ and $VHH$ topologies, in addition to the $HHH$ studies, which are the focus of this paper.

{\bf Requiring one or two of the Higgs bosons to be reconstructed as merged jets with two-prong $b$ jet substructure by employing a large-radius jet algorithm with the radius parameter of about 1.0 offers a powerful way to deal with combinatorics in the $HHH \to 3(b\bar b)$ decays.}

\subsection{$HHH$ estimated sensitivities at the LHC}

The current consensus in the ATLAS and CMS collaborations is that a measurement of the quartic coupling is out of reach. As a consequence, there is currently no estimate of the sensitivity to the triple Higgs production at the LHC. However, from various studies performed by theorists for future colliders, one can estimate the sensitivity range for $HHH$. The predictions at future colliders assume a center of mass energy of $\sqrt{s}=100$ TeV and each prediction focuses on a specific decay mode such as $HHH\rightarrow b\bar{b}b\bar{b}b\bar{b}$~\cite{bjets}, $HHH\rightarrow b\bar{b}b\bar{b}\gamma\gamma$~\cite{gamma} and $HHH\rightarrow b\bar{b}b\bar{b}\tau^+\tau^-$\cite{taus}. A basic event selection is applied, usually similar to the ones used in experimental measurements. 

In order to obtain an estimated result at the LHC, the significance is scaled with respect to the luminosity ratio and the difference in the predictions of the cross-sections. The difference in the cross-section of the signal is a factor $\sigma(HHH)_{13 \mathrm{TeV}}/\sigma(HHH)_{100 \mathrm{TeV}} = 1/60$~\cite{Maltoni:2014eza}. As the background processes for these different modes can vary, two scenarios are investigated: an optimistic scaling using the same reduction as the signal ($1/60$) and a pessimistic scaling assuming a reduction factor of $1/10$ for the background processes only, which corresponds to the ratio of cross-sections for the QCD multi-jet production with 6 $b$-quarks in the final state. This assumption is not optimal for the $HHH\rightarrow b\bar{b}b\bar{b}\gamma\gamma$ and $HHH\rightarrow b\bar{b}b\bar{b}\tau^+\tau^-$ decay modes but it captures the general trend that the background production should be lower at $\sqrt{s}=13$ TeV. 

\begin{table}[ht]
    \begin{center}
    \begin{tabular}{|c|c|c||c|c|c|}
        \hline
       Channel & $\mathcal{L}$ at 100 TeV & Significance & $\mathcal{L}$ at 13 TeV & Pessimistic & Optimistic \\ \hline
       $HHH\rightarrow b\bar{b}b\bar{b}b\bar{b}$~\cite{bjets} & 20 ab$^{-1}$ & $1.6\sigma$ & 139 fb$^{-1}$ & $285\times$ SM & $120\times$ SM \\
       $HHH\rightarrow b\bar{b}b\bar{b}\gamma\gamma$~\cite{gamma} & 20 ab$^{-1}$ & $2.1\sigma$ & 139 fb$^{-1}$ & $220\times$ SM & $90\times$ SM \\
       $HHH\rightarrow b\bar{b}b\bar{b}\tau^+\tau^-$~\cite{taus} & 30 ab$^{-1}$ & $2.0\sigma$ & 139 fb$^{-1}$ & $280\times$ SM & $115\times$ SM \\ \hline
       Combination & 20 ab$^{-1}$ & $2.9\sigma$ & 139 fb$^{-1}$ & $150\times$ SM & $64\times$ SM \\ \hline
    \end{tabular}
\caption{Extrapolation of the main triple Higgs decay modes to the Large Hadron Collider. The results are presented in terms of the limit on the signal strength at 95\% confidence level. The pessimistic scaling assumes a reduction of a factor 10 in the background similar to the reduction of the cross-section of the multijets process with 6 $b$-quarks. The optimistic scaling assumes a reduction of 60 similar to the signal. \label{tab:projected-results}}
\end{center}
\end{table}

A sensitivity estimate at the LHC is presented in the Table~\ref{tab:projected-results} for the main decay modes as well as a potential combination. The combination leads to a sensitivity of 60-150 times the SM prediction. In order to obtain this result, several challenges will have to be resolved. In particular the choice of the trigger, the control and reduction of the background processes as well as the estimation of the systematic uncertainties will need to be studied in details. 

\begin{table}[ht]
    \begin{center}
    \begin{tabular}{|c|c|c|}
        \hline
        $\mathcal{L}$ at 13 TeV & Pessimistic & Optimistic \\ \hline
        139 fb$^{-1}$ & $150\times$ SM & $64\times$ SM \\ \hline
        300 fb$^{-1}$ & $100\times$ SM & $40\times$ SM \\ \hline
        500 fb$^{-1}$ & $80\times$ SM & $35\times$ SM \\ \hline
        3000 fb$^{-1}$ & $30\times$ SM & $15\times$ SM \\ \hline
    \end{tabular}
\caption{Estimated limit on the triple Higgs production from a combination of the $HHH\rightarrow b\bar{b}b\bar{b}b\bar{b}$, $HHH\rightarrow b\bar{b}b\bar{b}\tau^+\tau^-$ and $HHH\rightarrow b\bar{b}b\bar{b}\gamma\gamma$ at 95\% confidence level for different luminosities at a center-of-mass energy of $\sqrt{s} = 13$ TeV.\label{tab:projected-results-lumi}}
\end{center}
\end{table}

While the result of the combination indicates that the evidence for the $HHH$ production might be achieved at a future collider, this result can be improved with more sophisticated analyses techniques than the simple selections applied in the theory studies. These measurements could strongly benefit continuous improvement in $b$-jets and $\tau$-leptons identification as well as analyses design relying on modern machine learning developments. The projections assuming a scaling with the luminosity expected to be achieved in Run 2, Run 3 and the High-Luminosity LHC is shown in Table~\ref{tab:projected-results-lumi}. \textbf{The ATLAS and CMS experiments at the LHC are the only detectors in the world capable of probing electro-weak symmetry breaking through searches for the $HHH$ process.}

\subsection{Complementary between ongoing $HH$ searches and future $HHH$ searches}

As shown in Section \ref{sec:exp-hh}, for multi Higgs boson production, the connection between Higgs boson multiplicity and contributing coupling modifiers is non-trivial: $HH$ and $HHH$ production are both affected by the trilinear coupling modifier $\kappa_3$ and the quartic coupling modifier $\kappa_4$. A combined experimental picture is therefore desirable.

Through a combination of multiple search channels, the ATLAS experiment limits the signal strength of $HH$ production $\mu_{HH}$ to be $< 2.4$ times the SM prediction at the 95\% confidence level, where $2.9 \times$SM is expected~\cite{ATLAS:2022jtk}. The CMS experiment reaches similar sensitivity with an observed limit of $\mu_{HH} < 3.4 \times$SM where $2.5 \times$SM is expected in the absence of any signal~\cite{CMS:2022dwd}. 

In this section we present expected limits on $\kappa_3$ and $\kappa_4$ based on extrapolations of the expected ATLAS $HH$ results, scaled to an integrated luminosity of 450$\,\text{fb}^{-1}$. For $HHH$ production, limits have been estimated extrapolating existing phenomenological studies \cite{Chen:2015gva, Fuks:2015hna, Papaefstathiou:2019ofh} to LHC energies, similar to the previous section. The $\kappa$ limits presented in this section are purely based on re-interpretations of the signal strength limits and neglect any change in the event kinematics induced by anomalous $\kappa_3$ and $\kappa_4$ values. In the case of $\kappa_3$ and $HH$ production for example, this assumption has its limitations as large values of $\kappa_3$ make the $m_{HH}$ spectrum softer and the signal-to-background ratio is lower at low $m_{HH}$~\cite{DiMicco:2019ngk, ATLAS:2024yuv}. Therefore the results in this section are to be seen as qualitative statements. The purpose of these studies is to highlight the complementary between the two channels and to advocate for a more thorough study within the experiments, taking the kinematic changes fully into account.

To calculate likelihood values, the $HH$ and $HHH$ signal strengths are parameterised as a function of $\Delta \kappa_3 = (\kappa_3 - 1)$ and $\Delta \kappa_4 = (\kappa_4 -1)$ based on~\cite{Bizon:2018syu, Gillis:2024cqi}:

\begin{align*}
    \mu_{HH}^{14\, \rm TeV} &= 
        1 - 0.867(\Delta\kappa_3) + 1.48\cdot 10^{-3}(\Delta\kappa_4)
        + 0.329(\Delta\kappa_3)^2 \nonumber \\ &+ 7.80\cdot 10^{-4}(\Delta\kappa_3\Delta\kappa_4) + 2.73\cdot 10^{-5}(\Delta\kappa_4)^2 
         -1.57\cdot 10^{-3}(\Delta\kappa_3)^2(\Delta\kappa_4) \nonumber \\ & -1.90\cdot 10^{-5}(\Delta\kappa_3)(\Delta\kappa_4)^2 
        + 9.74\cdot 10^{-6}(\Delta\kappa_3)^2(\Delta\kappa_4)^2
 \end{align*}

\begin{align*}
\mu_{HHH}^{14\, \rm TeV} &= 1 - 0.921(\Delta \kappa_3) + 0.091 (\Delta \kappa_4) +0.860 (\Delta \kappa_3)^2\\
&- 0.168 (\Delta \kappa_3 \Delta \kappa_4) + 1.71 \times 10^{-2} (\Delta \kappa_4)^2  -0.258 (\Delta \kappa_3)^3 \\
& +4.91 \times 10^{-2}(\Delta \kappa_3)^2 \Delta \kappa_4 + 4.13 \times 10^{-2}(\Delta \kappa_3)^4 \quad .
\end{align*}

\begin{figure}[htb]
\centering
\includegraphics[width=0.6\textwidth]{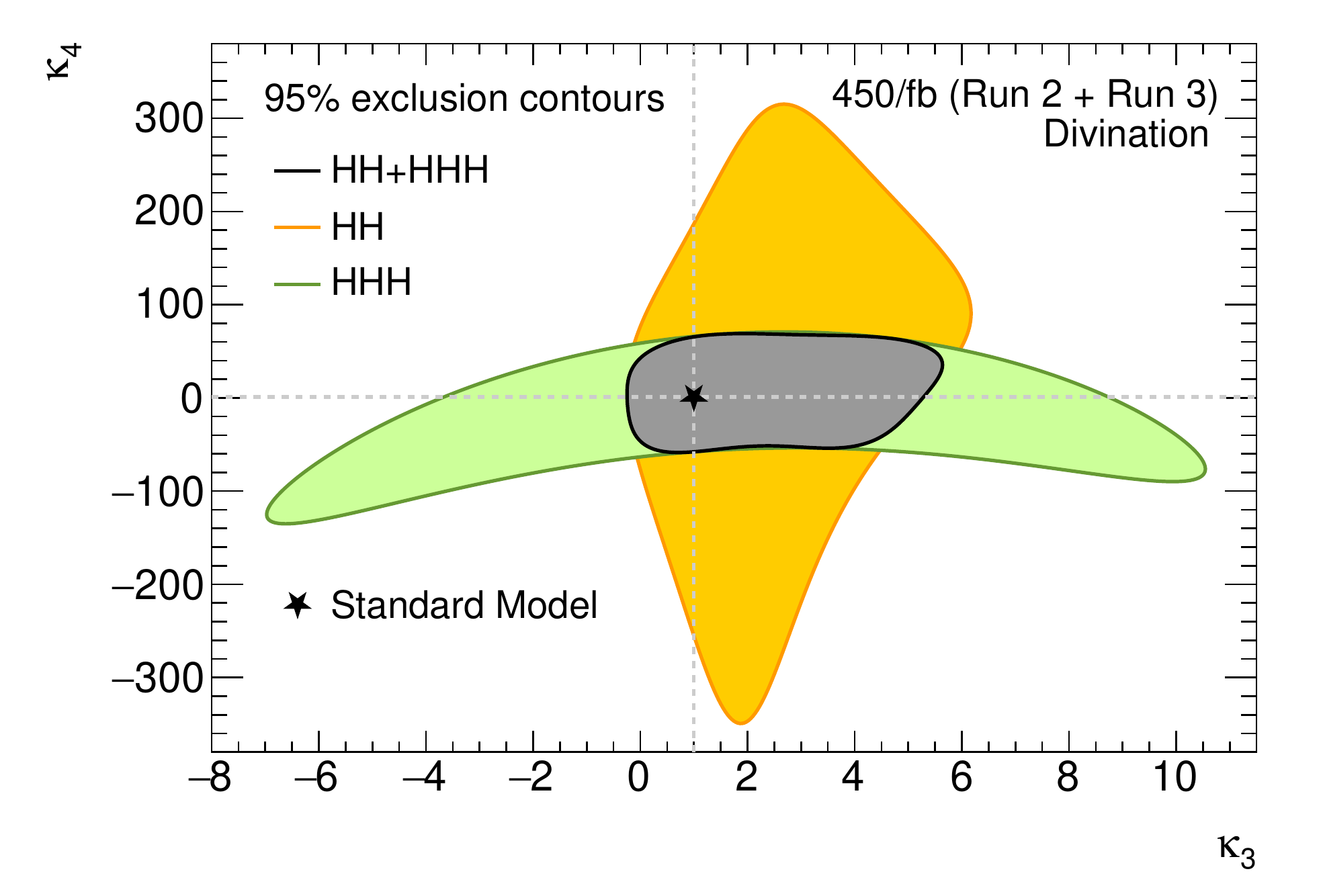}
\caption{Estimated likelihood contours at the 95\% confidence level in the $\kappa_3$ and $\kappa_4$ plane from searches for $HH$, $HHH$, and a combination.}
\label{fig:k3_k4_scan}
\end{figure}

As can be seen, the $HH$ signal strength, for example, depends only weakly on the quartic coupling as it intervenes at a two-loop level. While the absolute cross-section values are $\sqrt{s}$ dependent, the signal strength parameterisations show little dependence on the assumed $\sqrt{s}$. The estimated constraints are shown in Figure \ref{fig:k3_k4_scan} in the two-dimensional $\kappa_3$-$\kappa_4$ plane. The plot highlights the complementary between the two searches. 

\begin{figure}[htb]
\centering
\includegraphics[width=0.6\textwidth]{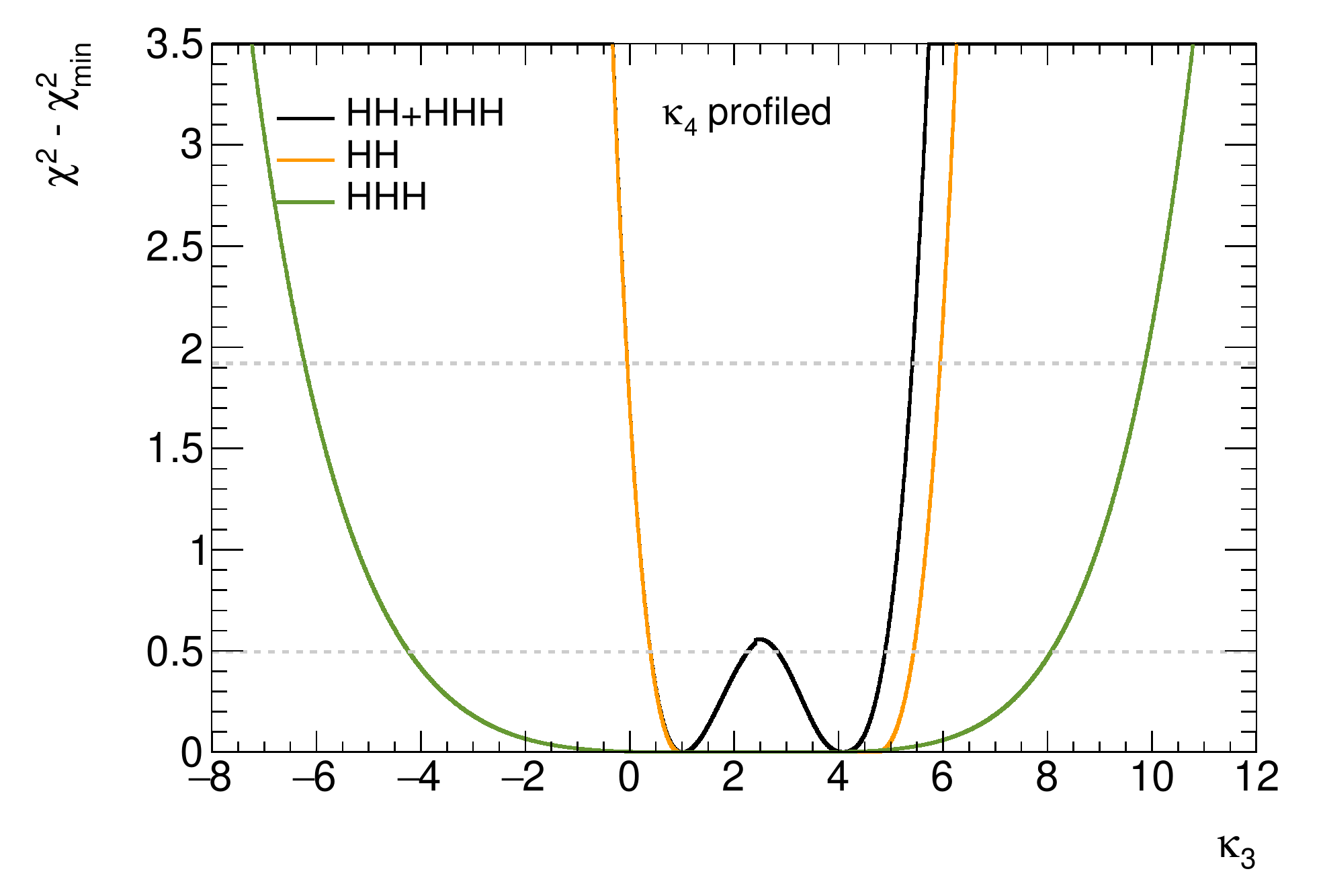}
\caption{Projected constraints on $\kappa_3$ without assumptions on $\kappa_4$ from searches for $HH$, $HHH$, and a combination of both searches. The estimates are based on a total integrated luminosity of $450\,\text{fb}^{-1}$ at $\sqrt{s} = 14\,\text{TeV}$.}
\label{fig:k3_scan}
\end{figure}

\begin{figure}[htb]
\centering
\includegraphics[width=0.6\textwidth]{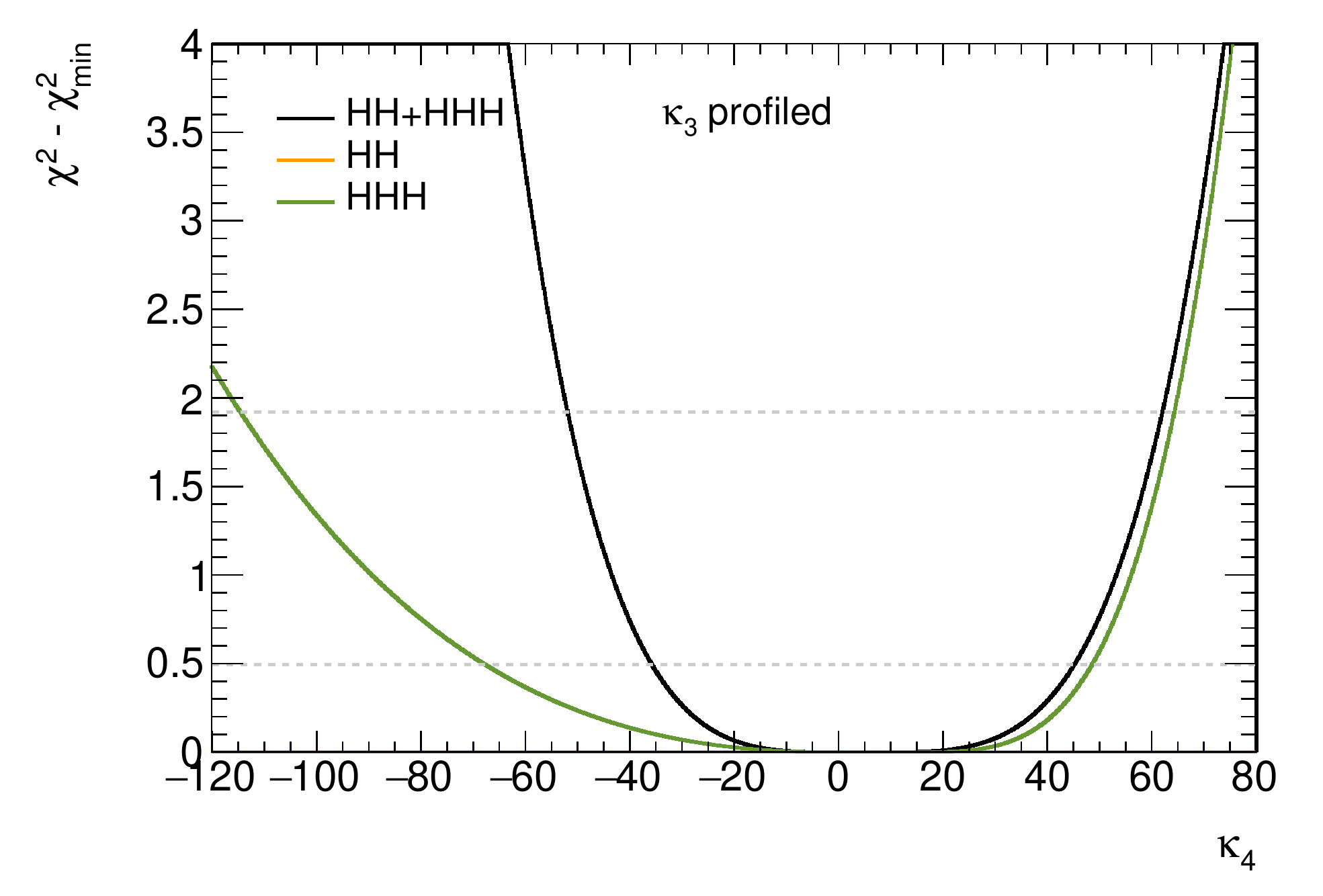}
\caption{Projected constraints on $\kappa_4$ without assumptions on $\kappa_3$ from searches for $HH$, $HHH$, and a combination of both searches. The estimates are based on a total integrated luminosity of $450\,\text{fb}^{-1}$ at $\sqrt{s} = 14\,\text{TeV}$.}
\label{fig:k4_scan}
\end{figure}

One dimensional likelihood contours are shown for $\kappa_3$ in Figure \ref{fig:k3_scan} and for $\kappa_4$ in Figure \ref{fig:k4_scan}. For each of those contours the coupling modifier that is not shown is profiled over. By taking into account also the effect of $\kappa_4$ on $HH$ production, one can derive limits on $\kappa_3$ that do not rely on any assumption on the relationship between $\kappa_3$ and $\kappa_4$ and therefore gain model independence. Furthermore, a $HH$ + $HHH$ combination adds information to the constraint on $\kappa_3$. This is even more so the case for $\kappa_4$, where the combination significantly improves over the constraints from $HHH$ production alone when $\kappa_3$ is profiled over.

\begin{figure}[htb]
\centering
\includegraphics[width=0.6\textwidth]{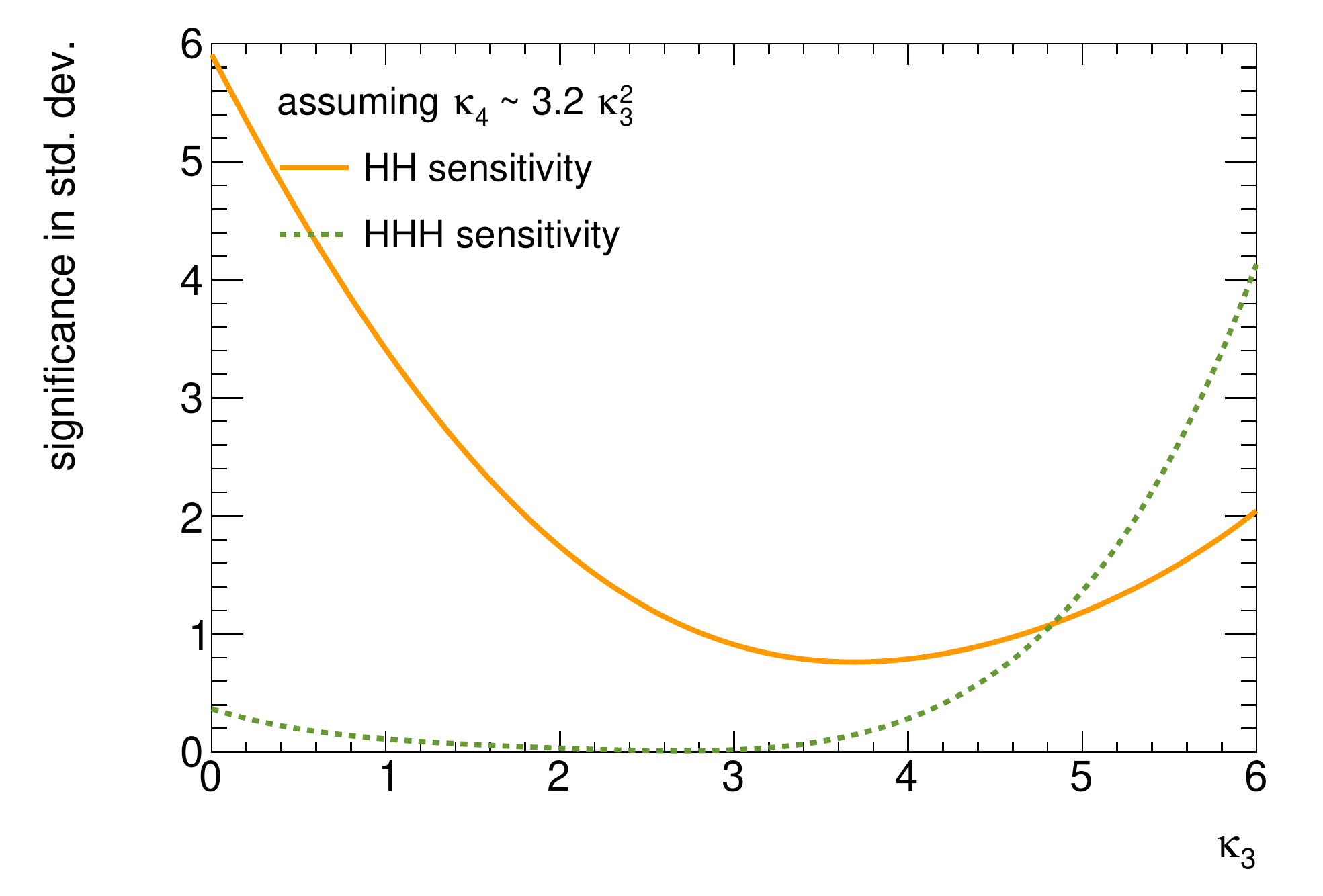}
\caption{Estimated sensitivity of $HH$ and $HHH$ searches to a model where $\kappa_4 = 3.2 \kappa_3^2$ as a function of $\kappa_3$. The estimates are based on a total integrated luminosity of $450\,\text{fb}^{-1}$ at $\sqrt{s} = 14\,\text{TeV}$.}
\label{fig:k3k4_significance}
\end{figure}

The complementary between $HH$ and $HHH$ searches is further illustrated by looking at scenarios where $\kappa_3$ and $\kappa_4$ follow a specific relation, in this case assuming $\kappa_4 = 3.2~\kappa_3^2$. Such a relation would not violate vacuum stability conditions requiring $\kappa_4 \leq \nicefrac{9}{8}~\kappa_3^2$~\cite{Agrawal:2019bpm}. Figure \ref{fig:k3k4_significance} shows the estimated significance with which such a model would show up in $HH$ and $HHH$ searches, respectively. In this scenario, a search for $HHH$ would be equally sensitive as a $HH$ search for larger values of $\kappa_3$ that are currently not yet excluded by experiment.

In conclusion, the complementarity between $HH$ and $HHH$ searches in constraining the trilinear and quartic Higgs boson self-coupling calls for a combination that will allow to determine the shape of the Higgs potential more precisely and with less assumptions. With these studies we hope to trigger more realistic sensitivity estimations, taking into account also signal kinematic changes and refined background contamination estimates. The dependence of $HH$ on $\kappa_3$ and $\kappa_4$, for example, can be simulated with publicly available POWHEG code from Ref.~\cite{Gillis:2024cqi}.

\clearpage
\section{{\bfseries Machine Learning Prospects in Di-Higgs Events } \label{sec:hh_ml}}

{\sl D. Diaz, J. Duarte, S. Ganguly,  B. Sheldon}
\subsection{Introduction}
The di-Higgs production via vector boson fusion (VBF) processes at hadron colliders has been broadly studied in the theory literature~\cite{Dolan:2013rja,Ling:2014sne,Dolan:2015zja,Bishara:2016kjn,Arganda:2018ftn,Cepeda:2019klc}, and only recently investigated experimentally~\cite{CMS:2022gjd,Aad:2020kub}.
Current projections~\cite{Dainese:2703572} achieve an expected significance of approximately $4.0\,\sigma$ from CMS and ATLAS combined for the full HL-LHC data set. 
Measurements of Higgs boson pair production face the difficulty of the small expected event yields even for the mode with the largest branching fraction ($\bbbar\bbbar$) as well as the presence of similar reconstructed QCD multijet events, which occur far more often. 
However, these projections do not include dedicated analyses of highly boosted hadronic final states, which may be especially sensitive to the SM and anomalous Higgs couplings~\cite{Kling:2016lay}. 

If the Higgs boson is highly Lorentz boosted, its hadronic decay products can be reconstructed as one single jet and the jet can be tagged using jet substructure techniques~\cite{Butterworth:2008iy, Abdesselam:2010pt, Kogler:2018hem, Larkoski:2017jix}.
Moreover, several machine learning methods have also been demonstrated to be extremely efficient in jet tagging and jet reconstruction~\cite{Kasieczka:2019dbj}. 
In the present work, we adopt ML algorithms to analyze boosted di-Higgs production in the four-bottom-quark final state at the FCC-hh, which is expected to produce hadron-hadron collisions at $\sqrt{s} = 100\TeV$ and to deliver an ultimate integrated luminosity of 30\,ab$^{-1}$.
We compare our ML-based event selection to a reference cut-based selection~\cite{L-Borgonovi} to demonstrate potential gains in sensitivity.
The rest of the section is organized as follows.
In Section~\ref{sec:boosted} and \ref{sec:ml}, we illustrate the potential of boosted Higgs channels based on the expected yields and introduce the ML methods.
In Section~\ref{sec:cut}, we describe the reference cut-based analysis and in Section~\ref{sec:gnn}, we explain our ML-based analysis.
Finally, we provide a summary and outlook in Section~\ref{sec:outlook}.

\subsection{Boosted Higgs}
\label{sec:boosted}
The hadronic final states of the Higgs boson are attractive because of their large branching fractions relative to other channels. 
While the $\bbbar\gamma\gamma$ ``golden channel'' has a 0.26\% branching fraction, the $\bbbar\bbbar$ and $\bbbar\PW\PW$ channels have a combined 58.8\% branching fraction, which often produce a fully hadronic final state.
At low transverse momentum ($\pt$), these final states are difficult to disentangle from the background, but at high $\pt$, the decay products merge into a single jet, which new ML methods can identify with exceptionally high accuracy.
Even with a requirement on the $\pt$ of the Higgs boson, the hadronic final states are still appealing in terms of signal acceptance.
The efficiency of the $\pt>400\GeV$ requirement on both Higgs bosons is about 4\% at the LHC.
Thus, the boosted $\bbbar\bbbar$ ($\bbbar\PW\PW$) channel with $\pt>400\GeV$ has $5.2$ times ($4.3$ times) more signal events than the ``golden'' $\bbbar\gamma\gamma$ channel at the LHC.
Given the higher center-of-mass energy of the FCC-hh, the boosted fraction would increase.

Based on our preliminary investigations and existing LHC Run 2 results, these boosted channels are competitive with the $\bbbar\gamma\gamma$ channel, which corresponds to an expected significance of 2.7 standard deviations ($\sigma$) with the full ATLAS and CMS HL-LHC data set.
As such, exploring these additional final states with new methods will be crucial to achieving the best possible sensitivity to the Higgs self-coupling.

\subsection{Machine Learning for Di-Higgs Searches}
\label{sec:ml}
Emerging ML techniques, including convolutional neural networks (CNNs) and graph neural networks (GNNs)~\cite{gnn,Battaglia:2016jem,DGCNN}, have enabled better identification of these boosted Higgs boson jets while reducing the backgrounds~\cite{Lin:2018cin,Qu:2019gqs,Moreno:2019bmu,Moreno:2019neq,Bernreuther:2020vhm,Sirunyan:2020lcu}.
CNNs treat the jet input data as either a list of particle properties or as an image. 
In the image representation case, CNNs leverage the symmetries of an image, namely translation invariance, in their structure.
Deeper CNNs are able to learn more abstract features of the input image in order to classify them correctly.
GNNs are also well-suited to these tasks because of their structure, and have enjoyed widespread success in particle physics~\cite{Shlomi:2020gdn,Duarte:2020ngm,Thais:2022iok}.
GNNs treat the jet as an unordered graph of interconnected constituents (nodes) and learn relationships between pairs of these connected nodes.
These relationships then update the features of the nodes in a \emph{message-passing}~\cite{DBLP:journals/corr/GilmerSRVD17} or \emph{edge convolution}~\cite{DGCNN} step.
Afterward, the collective updated information of the graph nodes can be used to infer properties of the graph, such as whether it constitutes a Higgs boson jet.
In this way, GNNs learn pairwise relationships among particles and use this information to predict properties of the jet.

Significantly, it has been shown that these ML methods can identify several classes of boosted jets better than previous methods.
For instance these methods have been used to search for highly boosted $\PH(\bbbar)$~\cite{Sirunyan:2020hwz} and $\PV\PH(\ccbar)$~\cite{Sirunyan:2019qia} in CMS.
Most recently, they have also been shown to enable the best sensitivity to the SM $\PH\PH$ production cross section and to the quartic $\PV\PV\PH\PH$ coupling in CMS using the LHC Run 2 data set~\cite{CMS-PAS-B2G-22-003}.
In this work, we study the impact of the use of these ML algorithms in future colliders like the FCC-hh. 

\subsection{Reference Cut-based Event Selection}
\label{sec:cut}
For the cut-based reference selection, we follow Refs.~\cite{L-Borgonovi,Banerjee:2018yxy}.
In particular, we study the configuration in which the Higgs boson pair recoils against one or more jets.
We use the \textsc{Delphes}-based~\cite{deFavereau:2013fsa} signal and background samples from Ref.~\cite{database}.
The signal sample of $\PH\PH$+jet is generated taking into account the full top quark mass dependence at leading order (LO) with the jet \pt greater than 200\GeV. 
Higher-order QCD corrections are accounted for with a $K$-factor
$K=1.95$ applied to the signal samples~\cite{Banerjee:2018yxy}, leading to $\sigma_{\PH\PH j} = 38\,\text{fb}$ for jet $\pt > 200\GeV$ and $\kappa_\lambda=1$.
The main background includes at least four b-jets, where the two $\bbbar$ pairs come from QCD multijet production, mainly from gluon splitting $\Pg\to\bbbar$.
The LO background cross section for jet $\pt > 200\GeV$ is given by $\sigma_{\bbbar\bbbar j} \text{ (QCD)} = 443.1\,\text{pb}$.

Jets are reconstructed with the anti-\kt~\cite{Cacciari:2008gp,Cacciari:2011ma} algorithm with a radius parameter $R=0.8$ (AK8) and $R=0.4$ (AK4).
The AK8 jets are formed from calorimeter energy clusters whereas the AK4 jets are formed from track elements.

\begin{figure}[htpbp]
	\centering
		\includegraphics[width=0.49\textwidth,clip=true,viewport = 0 0 504 504]{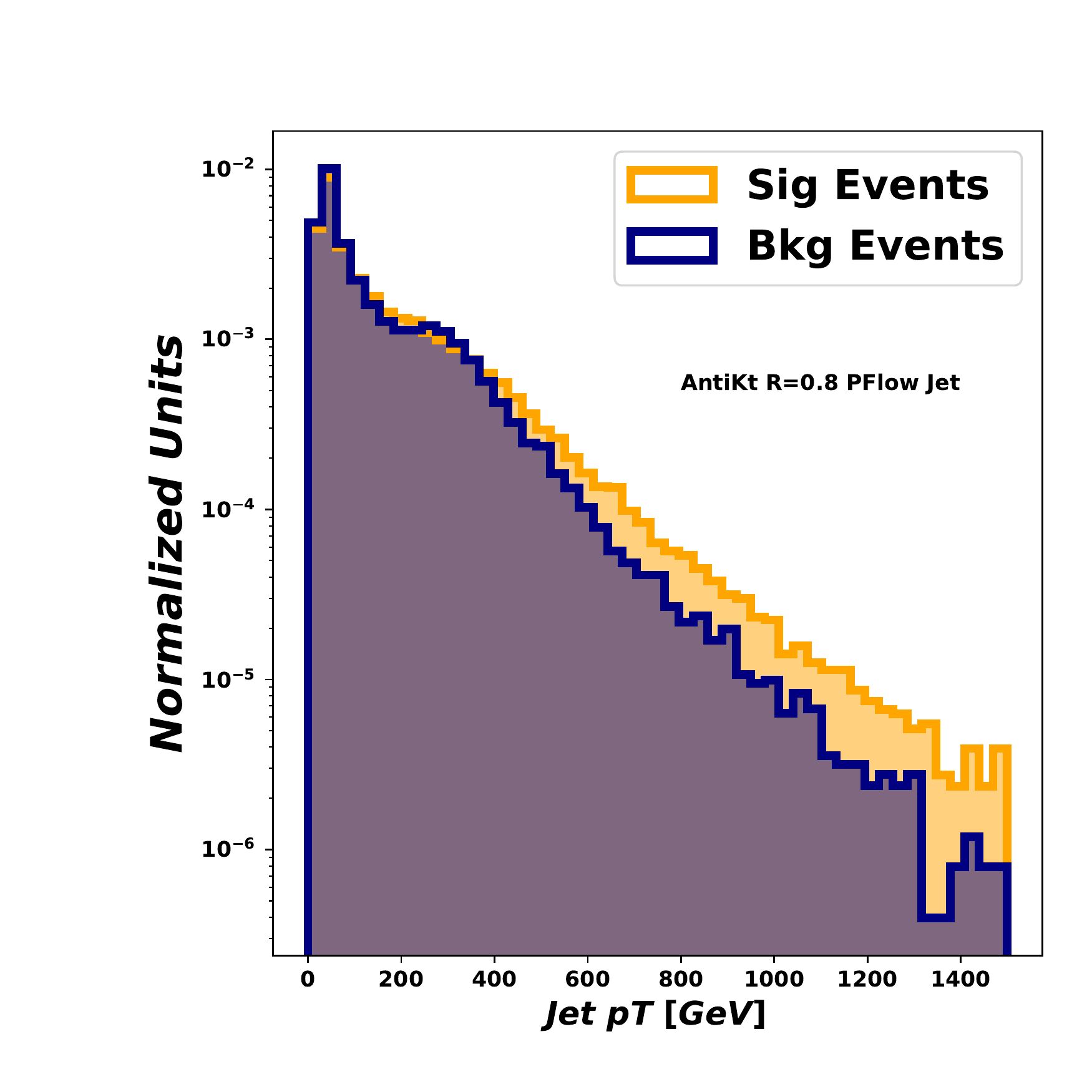}
		\includegraphics[width=0.49\textwidth,clip=true,viewport = 504 0 1008 504]{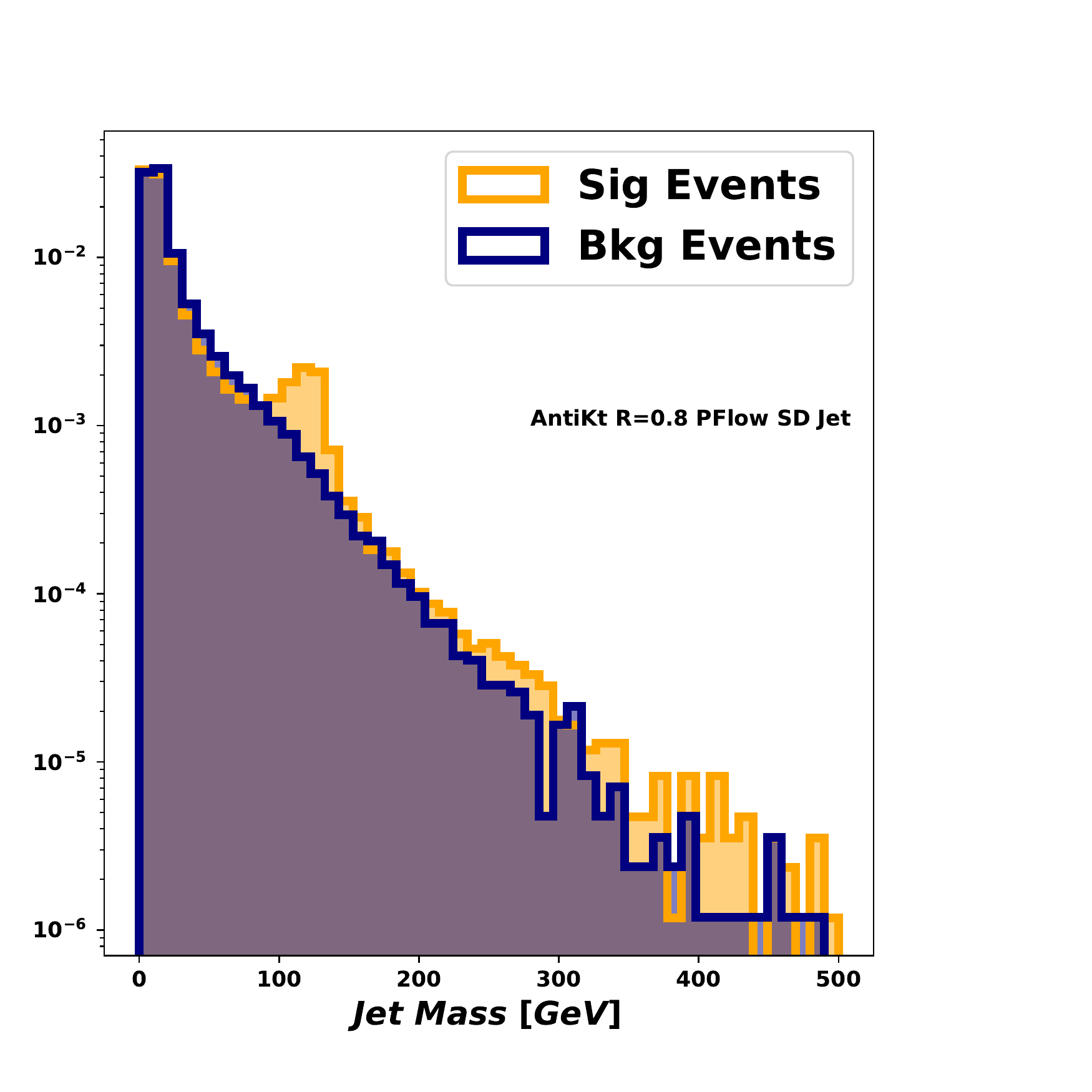}\\
		\includegraphics[width=0.49\textwidth]{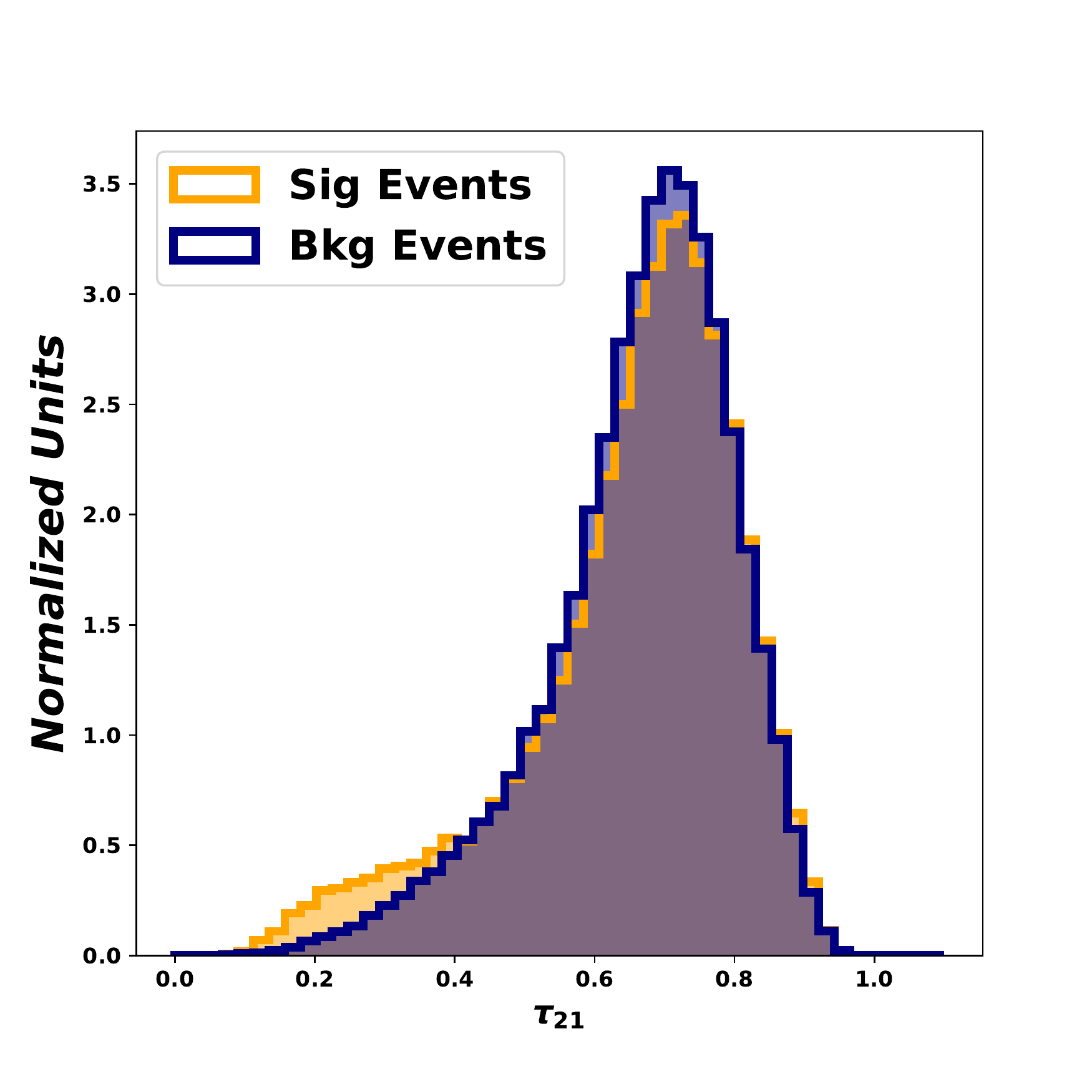}
		\caption{Jet \pt (upper left), soft-drop mass (upper right), $\tau_{21}$ (bottom) and  distribution of anti-$\kt$ $R = 0.8$ PF jets for signal and background events. 
		The shape difference plays a crucial role in identifying signal events over background.}
		\label{fig:ptmtau_dist}
\end{figure}

We require two AK8 jets with $\pt > 300\GeV$ and $|\eta| < 2.5$.
The AK8 jets are considered double b-tagged if they contain two b-tagged AK4 subjets.
This AK4 b-tagging emulation corresponds to a conservative signal efficiency of 70\%.
The two highest \pt double b-tagged AK8 jets constitute the Higgs boson candidates.
We further require the AK8 dijet system to be sufficiently boosted, $\pt^{jj} > 250\GeV$, and the leading jet to have a $\pt > 400\GeV$. 
The jet \pt and soft-drop mass \mSD~\cite{Larkoski:2014wba} distributions are shown in Figure~\ref{fig:ptmtau_dist}, along with the N-subjettiness ratio $\tau_{21} = \tau_2/\tau_1$~\cite{Thaler:2010tr}.
The two Higgs boson candidate AK8 jets are tagged by selecting jets with $\tau_{21} < 0.35$ and $100 < \mSD < 130\GeV$. 

After the selections, the expected signal ($S$) and background ($B$) yields for 30\,ab$^{-1}$ are 12\,700 and 49\,900\,000
events, yielding an approximate significance $S/\sqrt{B} = 1.8$.
The signal and background efficiencies of the cut-based selection are 1.7\% and 0.53\%, respectively. 

\subsection{Graph Neural Network Event Selection}
\label{sec:gnn}
We build a boosted $\PH\PH\to\bbbb$ event classifier using a GNN based on the features of all the AK4 and AK8 jet constituents (tracks and calorimeter clusters, respectively) in the event, as well as additional jet features.
We stress that this approach, an event-level classifier using the information provided in the FCC-hh samples, is conservative as we expect the largest gains in signal-to-background discrimination to arise from including lower-level detector information, including tracking and vertexing information.
Nonetheless, we can still compare this approach with a cut-based selection with access to similar information.

To define the input graph data structure or \emph{point cloud}, each of the jet constituents is treated as a node with its associated pseudorapidity ($\eta$) and azimuthal angle ($\phi$) as coordinates.
The event can then be thought of as a two-dimensional point cloud.
A graph is then formed using the k-nearest neighbor (kNN) algorithm in the $\eta$-$\phi$ plane. 
Each node has four features, namely the four components of the energy-momentum Lorentz vector.
We augment this node representation with three additional variables related to the jet as a whole.
In particular, we include the two- and one-subjettiness ($\tau_2$ and $\tau_1$) as well as $\PQb$-tagging probability. 
Hence this construction associates a feature vector of size 7 to each node. 

For the GNN architecture operation, we use the dynamic edge convolution.
The original idea was proposed for shape classification~\cite{DGCNN}, and was also used for jet classification~\cite{Qu:2019gqs}.
The message-passing (MP) operation, referred to as  
\emph{EdgeConv}, from layer $\ell$ to layer $\ell+1$ consists of the following operations
\begin{align}
 x_{i}^{\ell+1} &= \max_{j \in \mathcal{N}(i)} \left(  \Theta_{x} ( x_{j}^\ell - x_{i}^\ell ) \right) + \Phi_{x} ( x_{i}^\ell ), \\
 e_{i}^{\ell+1} &= \frac{1}{|\mathcal{N}(i)|}\left(\sum_{j \in \mathcal{N}(i)} \Theta_{e} ( e_{j}^\ell - e_{i}^\ell )  \right)  + \Phi_{e}  (e_{i}^\ell),
  \label{eqn:GraphConv}
\end{align}
where $\mathcal{N}(i)$ is the neighborhood of objects connected to object $i$, $|\mathcal{N}(i)|$ is the number of neighboring objects, $x_{i}^\ell$ are the features of node $i$ at layer $\ell$, and $e_{i}^\ell$ are the features of edge $i$ at layer $\ell$.

The implemented model has four such MP layers. 
The output dimensions of the $x$ coordinate after each layers are 3, 5, 4, and 2, respectively, whereas the dimensions of the variable $e$ are chosen to be 4, 5, 6, and 8, respectively.
The energy outputs of each layers are concatenated and passed though a MLP block to predict the output probability of the given event. 
The model is trained using a binary crossentropy loss for the classification task.
The signal events correspond to simulated $\PH\PH(\bbbb)$ events, while background events are from simulated QCD multijet production with four bottom quarks.
The optimizer used is Adam~\cite{DBLP:journals/corr/KingmaB14} with a fixed learning rate of $10^{-3}$. 
For training purposes we have used 50\,k events for training data and 10\,k events for validation data with batch size of five.

\begin{figure}[!htb]
	\begin{center}	
		\includegraphics[width=0.75\textwidth]{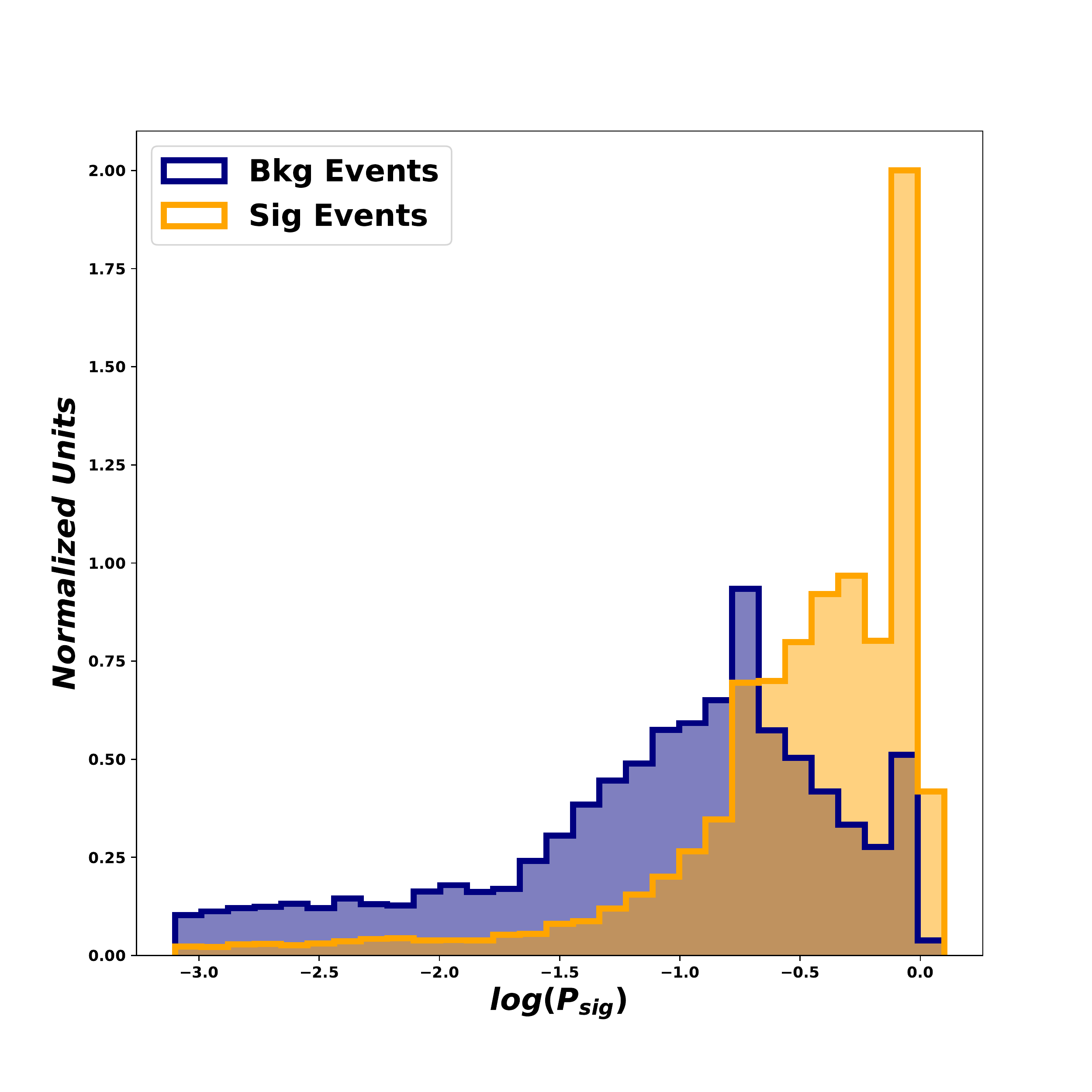}
		\caption{The distribution of natural logarithm of the events of being signal like, evaluated on the signal and background samples, respectively.}
		\label{fig:score_val}
	\end{center}
\end{figure}

\begin{figure}[!htb]
	\begin{center}	
		\includegraphics[width=0.75\textwidth]{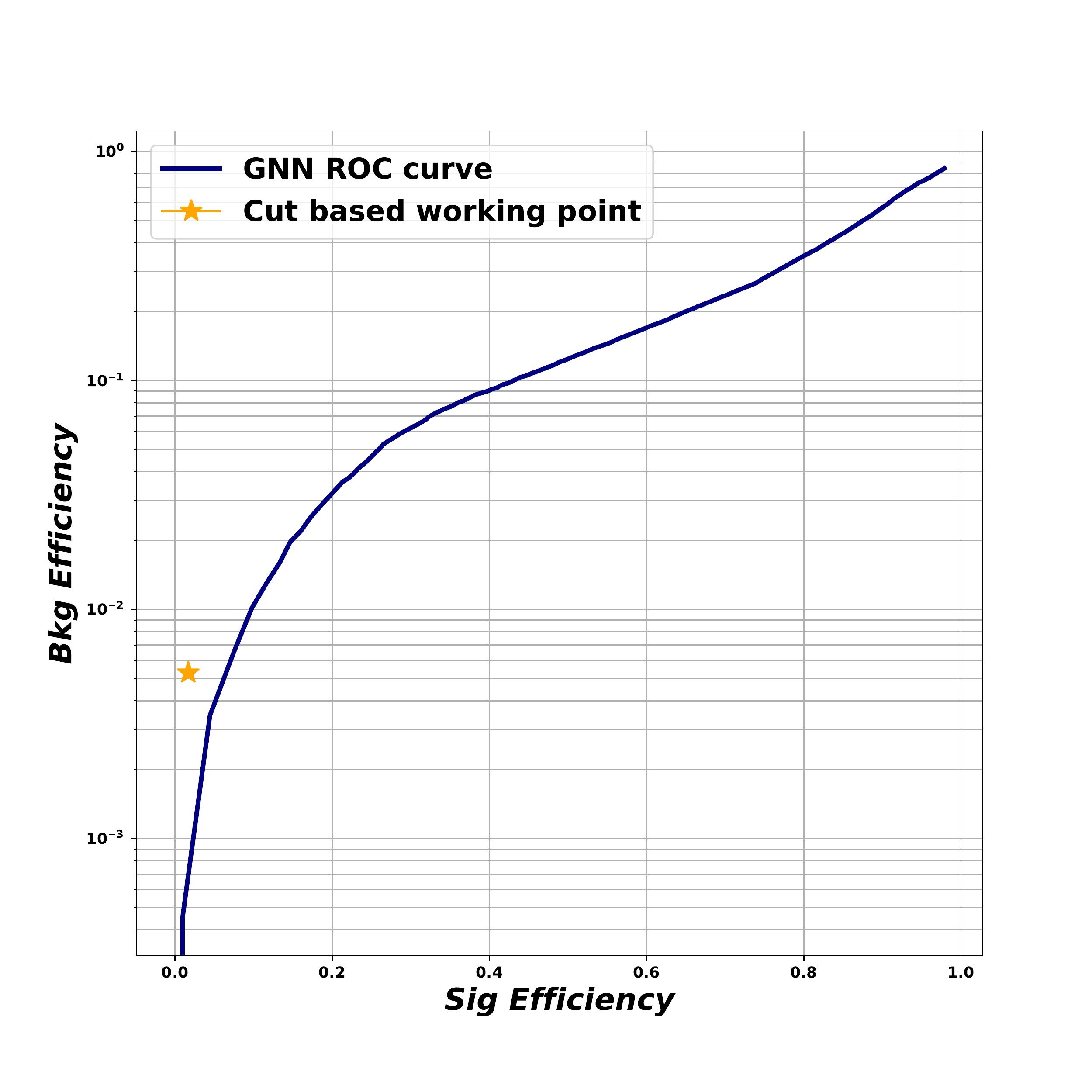}
		\caption{The ROC curve showing the ability of the NN to discriminate between signal and background. 
		With the preliminary study we achieve a signal efficiency of around 40\% for a background efficiency of 9\%.}
		\label{fig:roc_curve}
	\end{center}
\end{figure}

The output of the trained network is evaluated on an independent test sample of signal and background and the logarithm of the signal-like event probability is shown in Figure~\ref{fig:score_val}. 
The distribution demonstrates that a trained network can separate the signal from background.
The level of discrimination is quantified by the receiver operating characteristic (ROC) curve shown in Figure~\ref{fig:roc_curve}. 
This preliminary training can identify signal with 40\% efficiency at the background efficiency level of 9\%.
Compared to the cut-based selection, the event-level GNN can identify $\PH\PH(\bbbb)$ signal events with an efficiency of about 6.1\% for the same background efficiency of 0.53\%, corresponding to a factor of 3 improvement. This leads to a signal efficiency $S/\sqrt{B} = 8.3$ .

\subsection{Summary and Outlook}
\label{sec:outlook}

In summary, we have investigated the feasibility of observing the production of a pair of boosted Higgs bosons in hadronic final states at the Future Circular Collider in hadron mode (FCC-hh) and improving the sensitivity with ML techniques.
The data sets simulated with \textsc{Delphes} corresponds to a center-of-mass energy $\sqrt{s}=100\TeV$ and an integrated luminosity of 30\,ab$^{-1}$
We focused on the four-bottom-quark final state, in which the each $\bbbar$ pair is reconstructed as a large-radius jet. 
We have studied the sensitivity using a traditional cut-based analysis as well as a selection based on an event classifier built using a graph neural network (GNN), as described in Sec.~\ref{sec:gnn}. 
For the cut-based analysis, we leveraged the jet kinematics, substructure variables, and b-tagging for the two leading jets in the event.
For the GNN, we used lower level information, such as the jet constituents' four-momenta, as well as high-level jet substructure and b-tagging variables.
We established that a better sensitivity by a factor of 3 is achievable using the GNN as shown in Fig.~\ref{fig:roc_curve}.

Higgs boson pair production is a crucial process to characterize and measure precisely at future colliders.
In order to do so with the best precision possible, it is important to exploit all possible production and decay modes.
This includes the high-$\pt$ hadronic final states, such as $\bbbar\bbbar$, $\PW(\qqbar)\PW(\qqbar)\bbbar$, $\PW(\qqbar)\PW(\ell\nu)\bbbar$, and $\bbbar\gamma\gamma$, whose sensitivity can be improved with ML methods. 
Beyond $\PH$ jet classification, particle reconstruction~\cite{DiBello:2020bas,Pata:2021oez,Pata:2022wam}, and jet reconstruction~\cite{Guo:2020vvt}, and jet mass regression~\cite{CMS-DP-2021-017} algorithms can also be improved with ML.

Fully quantifying the impact of ML for these final states on the ultimate sensitivity achievable for the $\PH\PH$ cross section, \PH self-coupling, trilinear $\PV\PV\PH$ coupling, and quartic $\PV\PV\PH\PH$ coupling are important goals of future work.
Another important future deliverable is to consider how these ML methods may impact optimal detector design.
In this context, explainable AI methods~\cite{Mokhtar:2021bkf} can be developed to understand the physics learned by the networks, and fully exploit this in future detector design. 
Future work can also explore the impact of using symmetry-equivariant networks~\cite{Bogatskiy:2020tje,Bogatskiy:2022hub} for Higgs boson property measurements at future colliders.

{Future colliders will be able to probe the quartic interaction vertex of SM Higgs by producing three on-shell Higgs boson in the final state. In presence of dominant background, reconstruction of three individual Higgs boson candidates is a complex combinatorial task. The GNN methods, discussed in this section, can be of particular usage for this case. As in the case of HH, the events from HHH decay can be represented as a graph as well. Henceforth, the methods we demonstrated in this section are equally applicable for the reconstruction of HHH events. There are ample opportunities to venture hetero-graph methods in the context of HHH event reconstructions, where each H can decay into a particular mode and different nodes of the heterograph may represent different physics objects (like jets, $\gamma$, leptons etc.). In case of semi-boosted or boosted event topologies with partial overlaps of these objects, sub-graph based methods have further potential which needs to be ventured. GNN based methods are already playing a mainstream role in HH searches and bound to take the center-stage in all upcoming HHH searches.}

\clearpage
\section{{\bfseries Flavour tagging }\label{sec:flavotag}} 
{\sl M. Chen, O. Karkout, M. Kolosova, B. Liu}

\subsection{Introduction} 
\label{sec:flav_intro}

As discussed in Section~\ref{sec:chall_br}, channels involving $H\rightarrow b\overline{b}$ decays give rise to the most
promising channels to search for multiple Higgs-like particles, due to the much larger branching ratios. Therefore, the technique to identify jets
containing $b$-hadrons is instrumental for this search programme. 

Compared to lighter hadrons, $b$-hadron decays have very distinguishable
characteristics. A $b$-hadron can travel for up to a few millimeters in the
detector before decaying, because of its longer lifetime and the typical Lorentz
boost expected at the LHC.  It results in tracks and vertices that are away
from the interaction point.  They are referred to as ``displaced tracks'' and
``displaced vertices''. Because of its heavier mass, there is a larger number
of tracks from $b$-hadron decays. In addition, a $b$-hadron can decay to a
$c$-hadron that subsequently travels for an additional distance in the detector
before decaying. Therefore, typically one expects a Secondary Vertex (SV) from
$b$-hadron decay and a tertiary vertex from the $c$-hadron decay. Last but not
the least, the semi-leptonic branching ratio of $b$-hadrons is larger than that of lighter hadrons, leading
to a higher probability of having a lepton in the decay chain.

A flavour tagging algorithm explores the above unique features of $b$-hadron
decays, aiming at identifying jets containing $b$-hadrons, i.e. $b$-jets.  It
therefore depends on the jet reconstruction algorithm. Two major types of
algorithms have been extensively explored in ATLAS and CMS, one concentrated on jets
with a smaller radius (R = 0.4) containing one $b$-hadron, and one considering
jets with a larger radius (R = 1.0) with two $b$-hadrons inside. The former is
referred to as the ``single-$b$ tagging'', while the latter is referred to as
the ``double-$b$ tagging''. In this section, we will summarize the
state-of-the-art flavour tagging algorithms developed in ATLAS and CMS, and discuss
what can be further improved in the context of searching for multiple
Higgs-like particles.

\subsection{Single-$b$ Tagging in ATLAS} 
\label{sec:singleb}

\begin{figure}[ht]
\begin{center}
 \includegraphics[width=0.6\columnwidth]{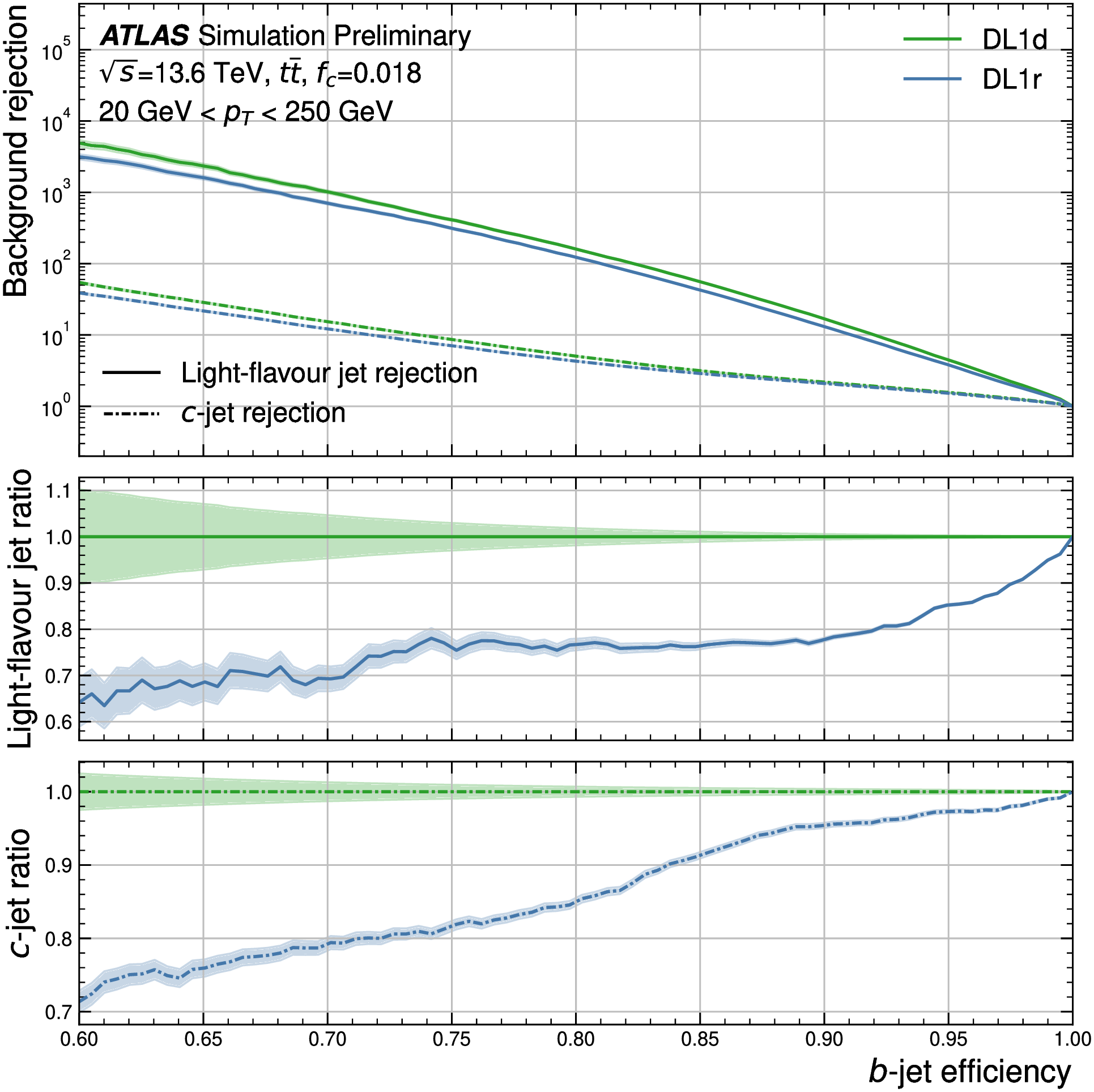}
\caption{The light-flavour jet (solid line) and charm-jet (dashed line) rejection factor for the latest DL1r and DL1d algorithms. The taggers are re-optimised on reprocessed Run 2 simulation. The x-axis corresponds to the $b$-jet efficiency, while the y-axis corresponds to the background rejection in the upper panel. The middle panel shows the ratio of the light-flavour jet rejection while the lower panel shows the ratio of the charm-jet rejection. The uncertainty bands correspond to the statistical uncertainties associated with the test sample~\cite{DL1dplot}.}
\label{fig:dl1d}
\end{center}
\end{figure}

The small-R jets considered in most recent ATLAS analyses are reconstructed by
the particle flow (PFlow) algorithm with the radius set to
0.4~\cite{ATLASPFlow}. A two-step approach is constructed to tag the small-R
PFlow jets. As a first step, various dedicated taggers are designed to explore
the above characteristics of $b$-hadron decays, such as the displaced tracks
and the secondary vertices. The $b$-tagging algorithm developed for small-R jets is suitable for any physics processes as long as the jets only contain one $b$-hadron. Their outputs are then fed into a feedforward
neural network, DL1r. Considering jets with 20 GeV < $p_{\mathrm{T}}$ < 250 GeV
in a $t\overline{t}$ sample, the rejection factor for charm (light) jets is 12
(625), while achieving a 70\% efficiency for $b$-jets~\cite{ATLASDL1r}. Before
DL1r, the MV2 tagger family was widely used in early and partial Run 2 ATLAS
analyses, where the dedicated taggers were fed into a boost decision tree
instead. The DL1d tagger, an updated version of DL1r, replacing the recurrent
neural network impact parameter tagger (RNNIP) with a deep impact parameter set
tagger (DIPS)~\cite{ATLASDIPS}, is used in ATLAS early Run 3
analyses~\cite{ttbarrun3}. The performance of DL1d is increased by 30\% at a $b$-tagging efficiency of 60\%,  
compared to that of DL1r, as shown in Figure~\ref{fig:dl1d}.  The rejection factor of a given type of background jets, which is defined as the multiplicative inverse of its mis-tagging rate, is used as the metric in ATLAS to quantify the performance.

\clearpage

\begin{figure}[ht]
\begin{center}
 \includegraphics[width=1.0\columnwidth]{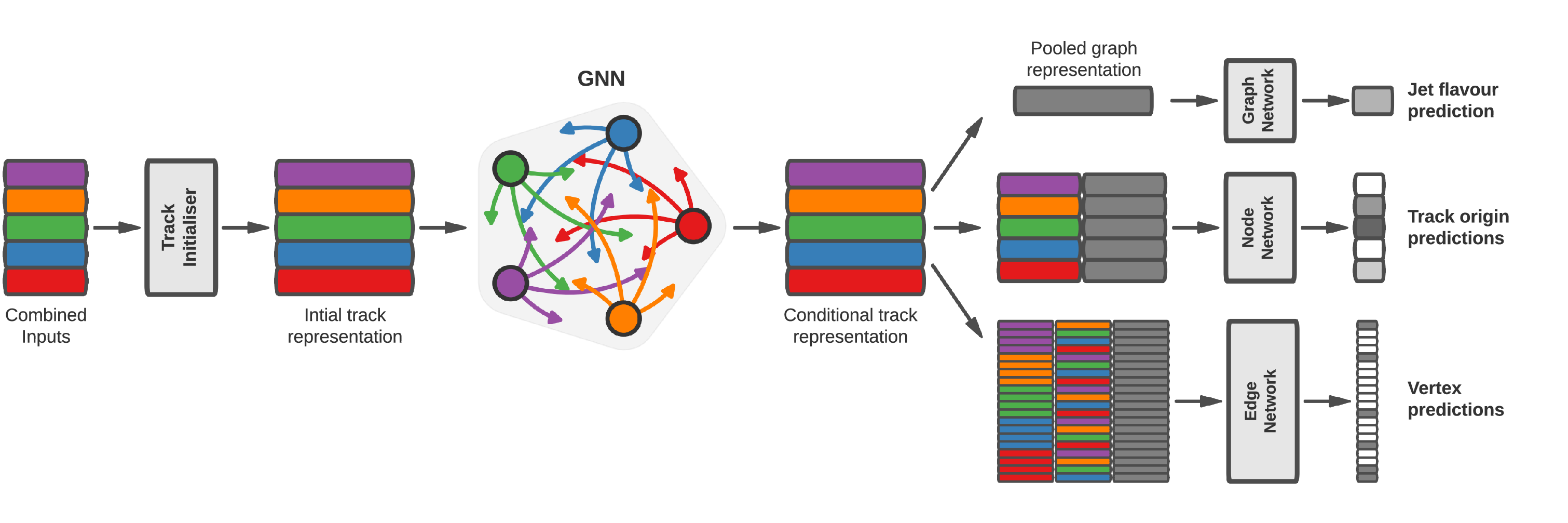}
\caption{Schematic diagram of the GN1 (GN2) tagger~\cite{GN1}.}
\label{fig:GNNArch}
\end{center}
\end{figure}

The most state-of-the-art flavour tagging algorithm in ATLAS is the recently
developed GN2 tagger, a transformer based algorithm~\cite{GN2Plots}. The
previous version, GN1, a graph neural network based algorithm, was also
optimised for the HL-LHC conditions~\cite{GN1HLLHC}. As opposed to the DL1r
architecture, both GN2 and GN1 eliminate the use of intermediate taggers by
utilising tracks as the input directly, as illustrated in
Figure~\ref{fig:GNNArch}. Vertex prediction and track origin prediction are
realised via auxiliary tasks to improve the performance of jet flavour
identification~\cite{GN1}. The flexibility of this architecture makes it
straightforward to be adopted for different tasks as discussed in
Section~\ref{sec:doubleb}.     

\begin{figure}[ht]
\begin{center}
 \includegraphics[width=0.8\columnwidth]{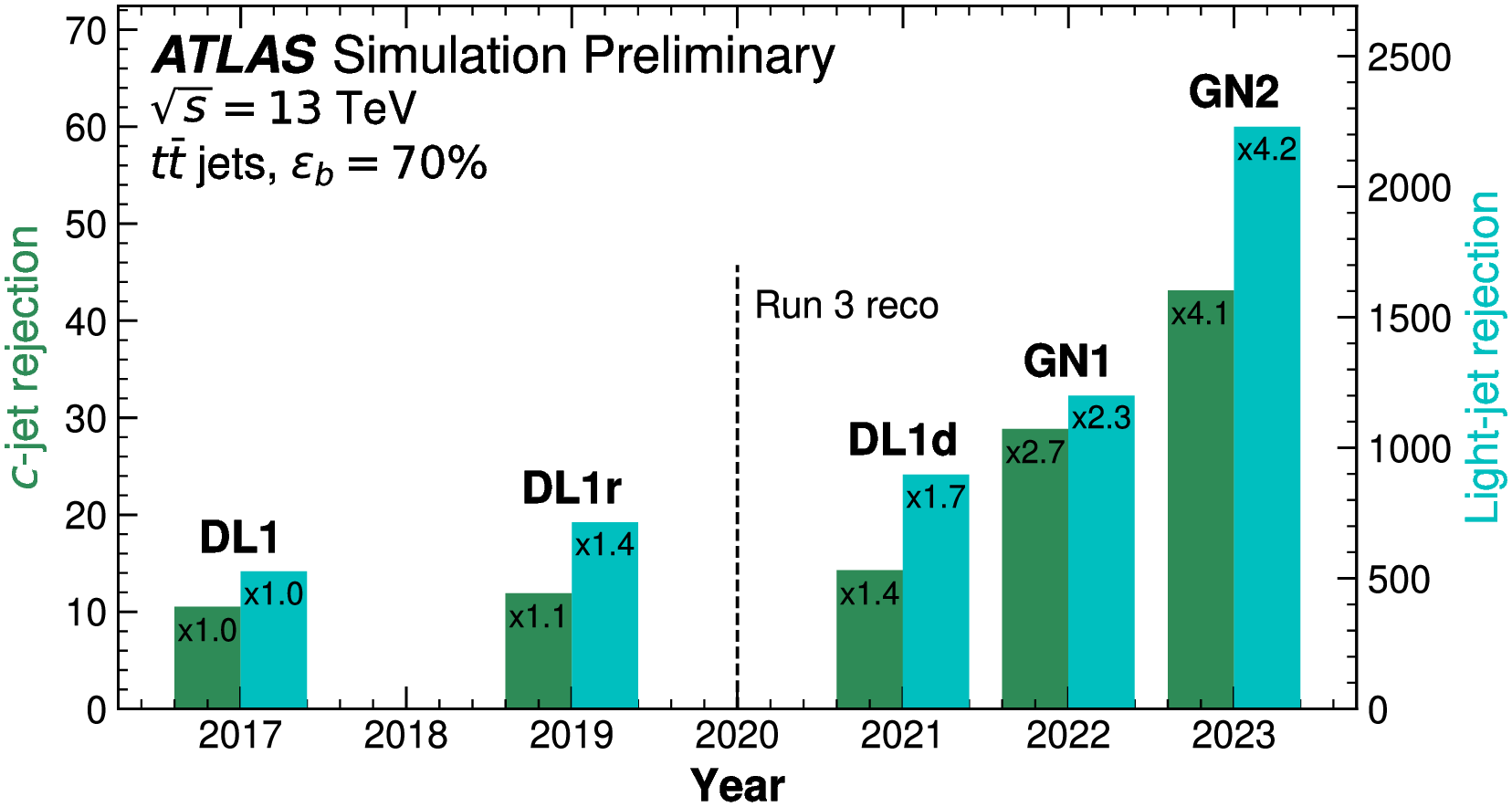}
\caption{Evolution of the ATLAS flavour tagging performance since 2017~\cite{GN2Plots}.}
\label{fig:atlastaggers}
\end{center}
\end{figure}

Figure~\ref{fig:atlastaggers} shows how the flavour tagging performance in
ATLAS has improved in recent years. The charm (light) rejection power is
increased by a factor of 4.1 (4.2) for a fixed 70\% $b$-jet tagging efficiency,
compared to the DL1 tagger. The multi-Higgs related searches will benefit
significantly from the much improved flavour tagging performance.    

\begin{figure}[ht]
\begin{center}
 \includegraphics[width=0.8\columnwidth]{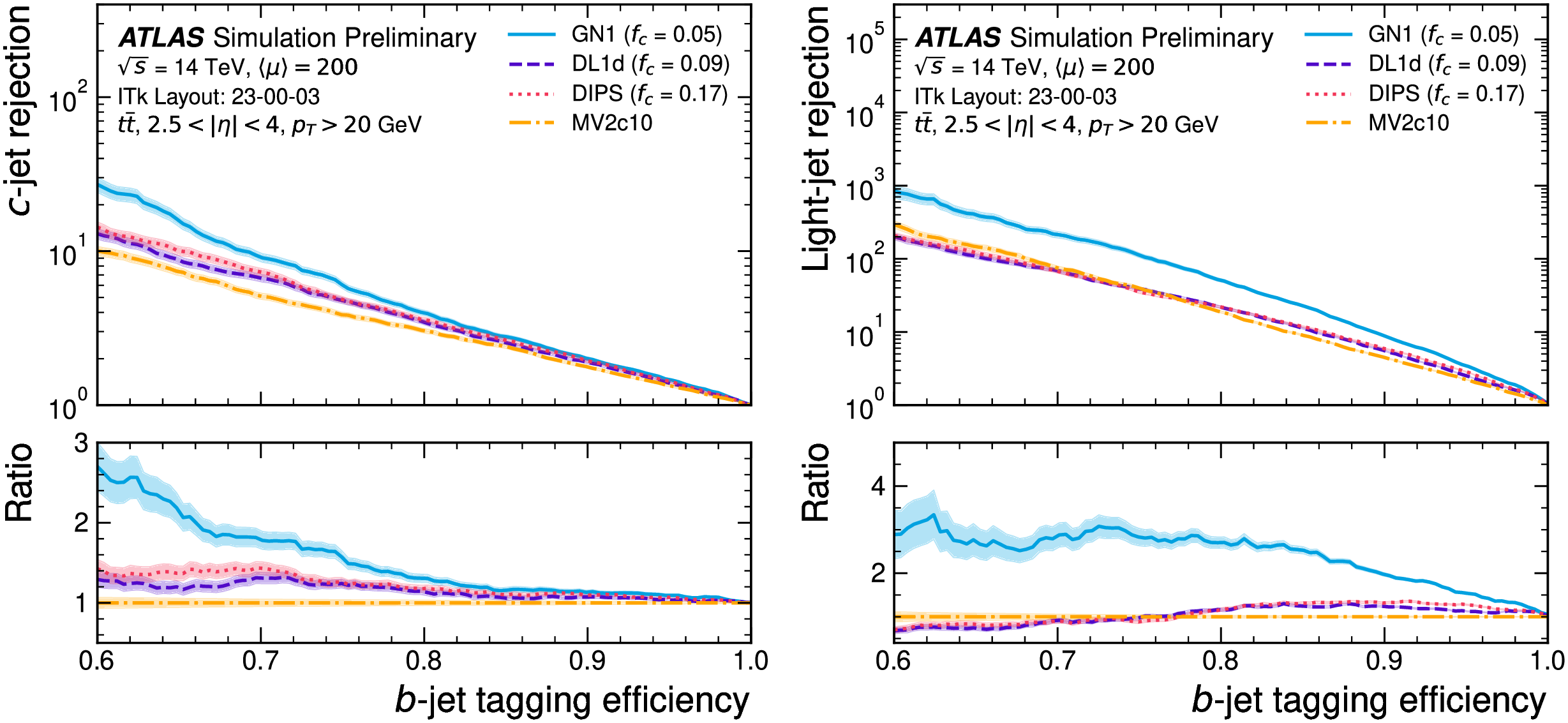}
\caption{The charm-jet (left) and light-flavour jet (right) rejection factors as a function of the $b$-jet tagging efficiency for jets in the $t\overline{t}$ sample with $p_{\mathrm{T}}$ > 20 GeV and 2.5 < |$\eta$| < 4. The uncertainty bands correspond to the statistical uncertainties associated with the test sample~~\cite{GN1HLLHC}.}
\label{fig:GN1Upgrade}
\end{center}
\end{figure}

For the HL-LHC, ATLAS will have a full silicon inner tracker (ITK) to cover the
pseudo-rapidity range up to 4.0. Consequently, the scope of flavour tagging
will be extended to a much larger kinematic region. Figure~\ref{fig:GN1Upgrade}
compares the performance of various taggers in the forward region (2.5 < |$\eta$|
< 4.0).  The charm (light) rejection power of GN1 is a factor of 2 (3)
better~\cite{GN1HLLHC}. It is reasonable to expect the performance to be
further improved with GN2 and larger training samples.     

\subsection{Single-$b$ Tagging in CMS} 
\label{sec:singleb_cms}
\begin{figure}[ht]
\begin{center}
 \includegraphics[width=0.8\columnwidth]{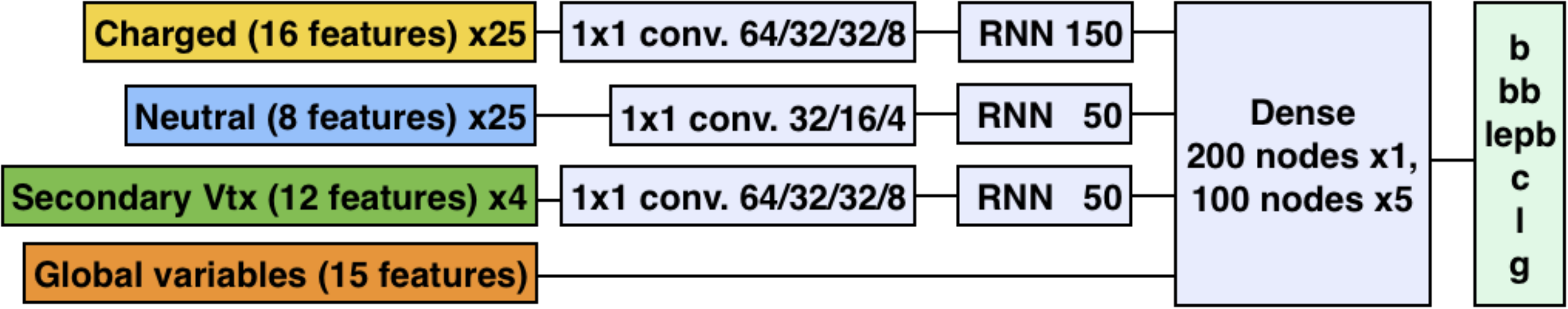}
\caption{Schematic diagram of the DeepJet tagger ~\cite{Bols:2020bkb}.}
\label{fig:DeepJetSchem}
\end{center}
\end{figure}

Within the CMS experiment, heavy flavour tagging on small-R jets has been widely performed during Run 2 using the DNN-based multi-classifier DeepJet~\cite{Bols:2020bkb}. The DeepJet model utilises a total of 650 input variables, including global event variables, charged and neutral particle flow candidate features, and information regarding SVs associated with the small-R jet. DeepJet is a fully connected neural network consisting of 1$\times$1 convolutional layers, which perform some automatic feature preprossesing on each type of jet constituents and SVs. Each of the convolutional layers is followed by a recurrent layer (of LSTM type) which combine the information for each sequence of constituents. The RNN outputs are combined with the global event variables with the use of fully connected layers. DeepJet has 6 output nodes, which can be used for b, c, and quark/gluon tagging. A schematic of the DeepJet architecture is shown in Fig.~\ref{fig:DeepJetSchem}.

\begin{figure}[ht]
\begin{center}
 \includegraphics[width=0.7\columnwidth]{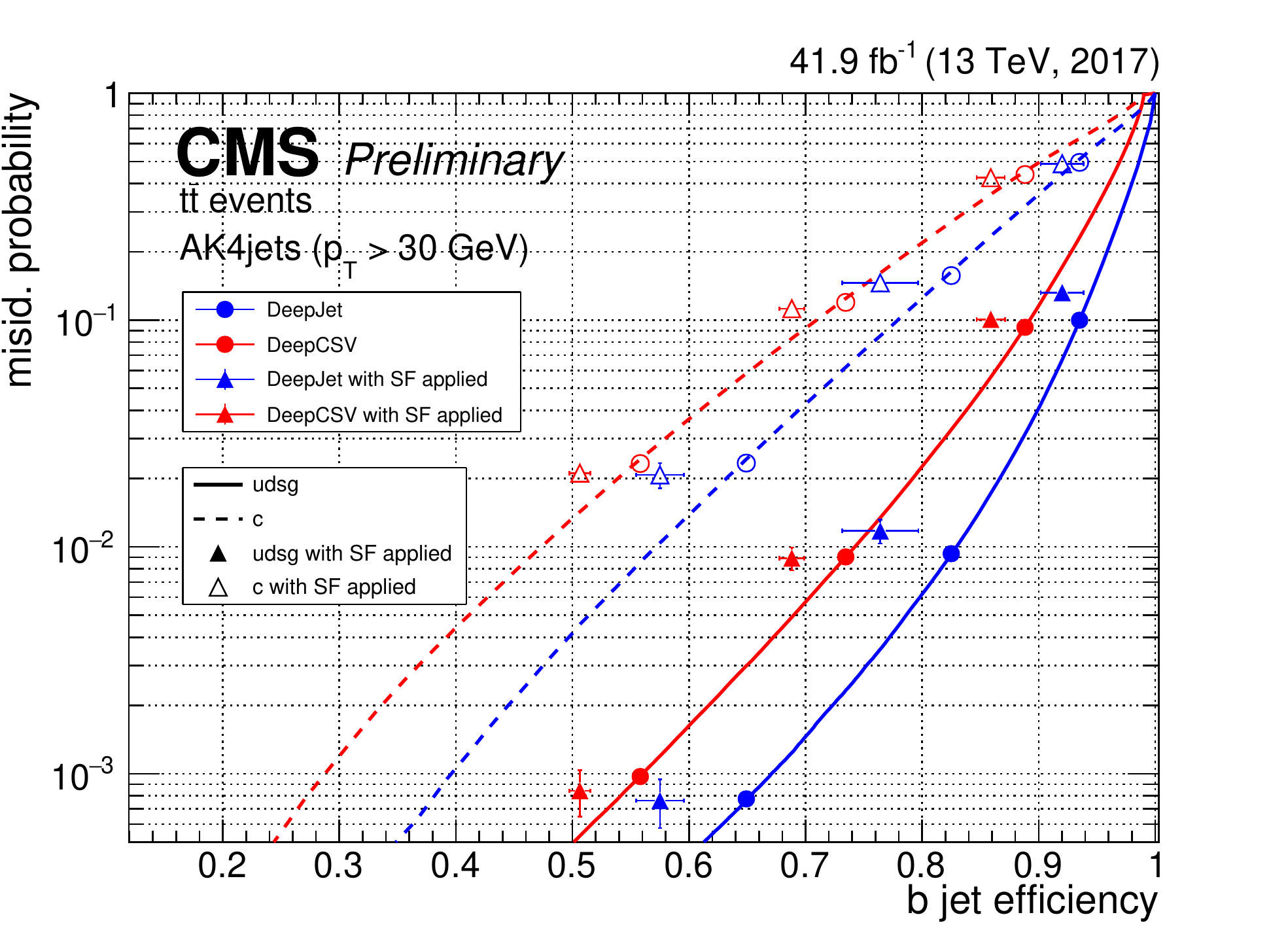}
\caption{The receiver operating characteristic curve (ROC) of the CMS DeepJet (blue) and DeepCSV (red) b-tagging algorithms. The probability of misidentifying non-b jets as b jets is shown with respect to the efficiency of correctly identifying b jets for three working points: loose, medium and tight (represented by the circular markers). A b jet efficiency and a misidentification probability equal to one corresponds to no selection on the b discriminant score. The performance of the algorithms is also shown after the application of the data-to-simulation correction factors (represented by the triangle symbols) ~\cite{CMS-DP-2018-058}.}
\label{fig:DeepJetPerf}
\end{center}
\end{figure}

DeepJet is trained and tested using simulated small-R jets from QCD-multijet and fully hadronic $t\bar{t}$ events.  Figure \ref{fig:DeepJetPerf} shows the performance of DeepJet~\cite{CMS-DP-2018-058} in comparison with its predecessor, the CMS b-tagger DeepCSV~\cite{CMS:2017wtu}, in a simulated $t\bar{t}$ sample containing small-R jets with $p_{\mathrm{T}}$ > 30~GeV and $|\eta|<2.5$. For both algorithms, the b jet identification efficiency is not the same in data and simulation. To account for this difference data-to-simulation correction factors are applied in simulated events.

\begin{figure}[ht]
\begin{center}
 \includegraphics[width=0.6\columnwidth]{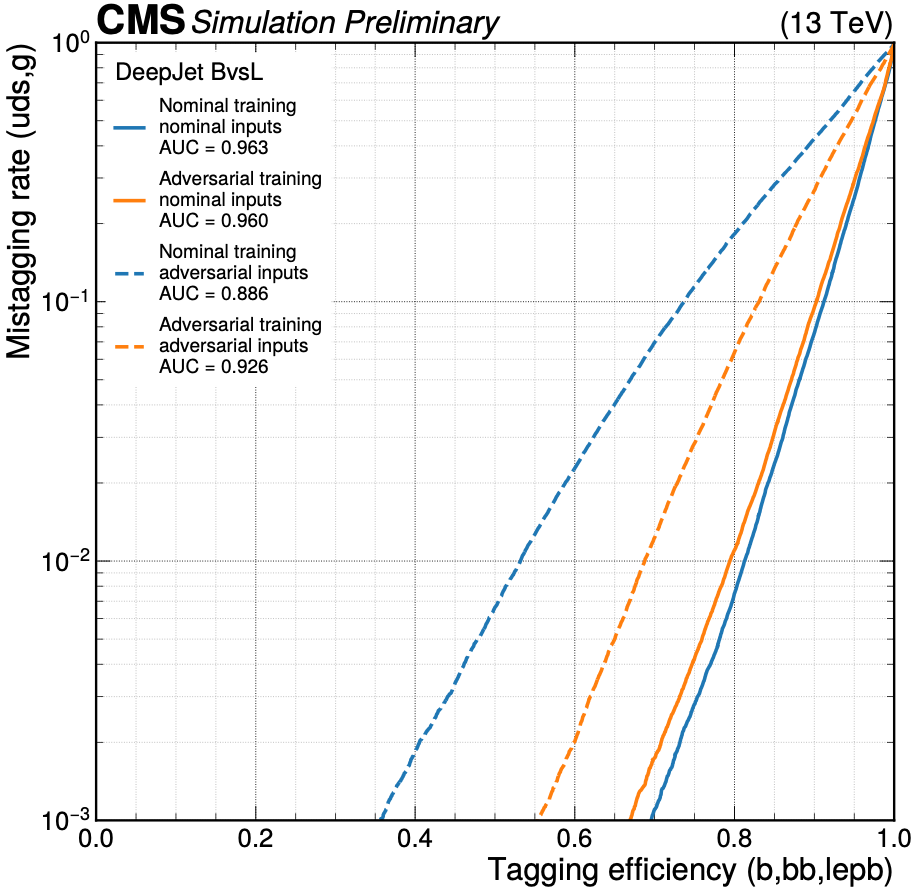}
\caption{Performance of DeepJet b versus light (BvsL) tagger for the nominal and adversarial models as derived from simulated QCD-multijet and $t\bar{t}$ events. The b tagging performance of the nominal (non FSGM-attacked) models are represented with solid lines, while the performance of the FSGM-attacked models are shown with dashed lines. ~\cite{CMS-DP-2022-049}.}
\label{fig:DeepJetAdvAttack}
\end{center}
\end{figure}

\begin{figure}[ht]
\begin{center}
 \includegraphics[width=0.49\columnwidth]{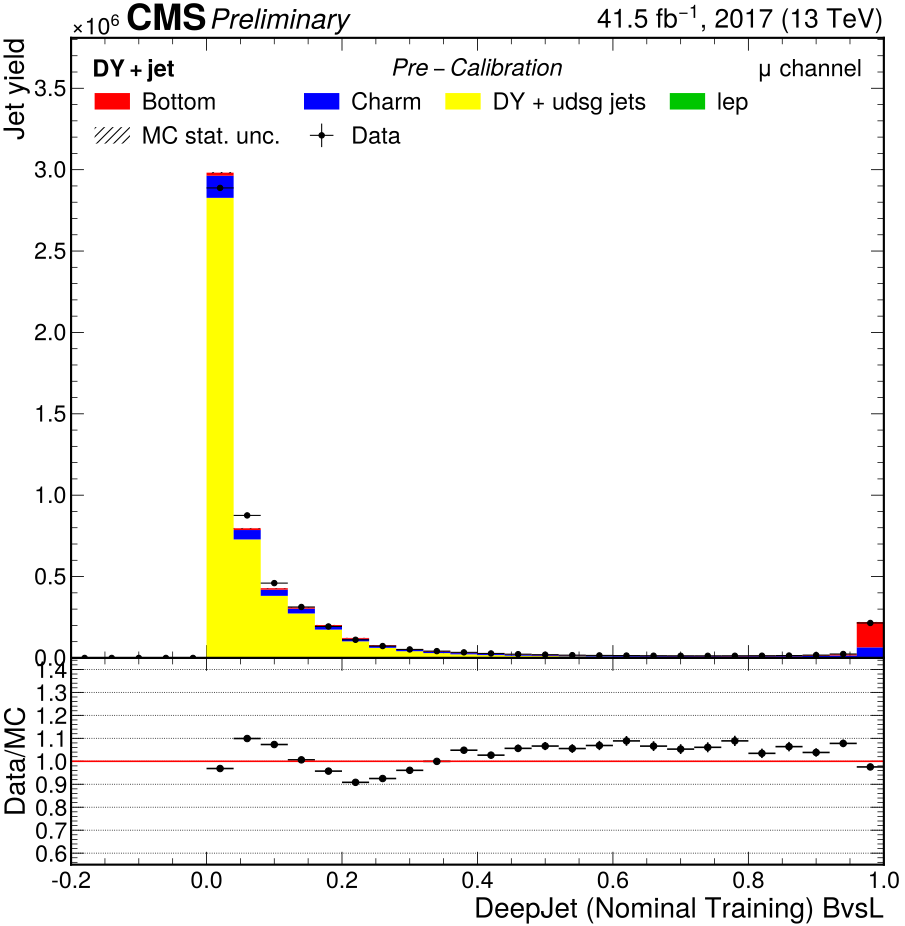}
  \includegraphics[width=0.49\columnwidth]{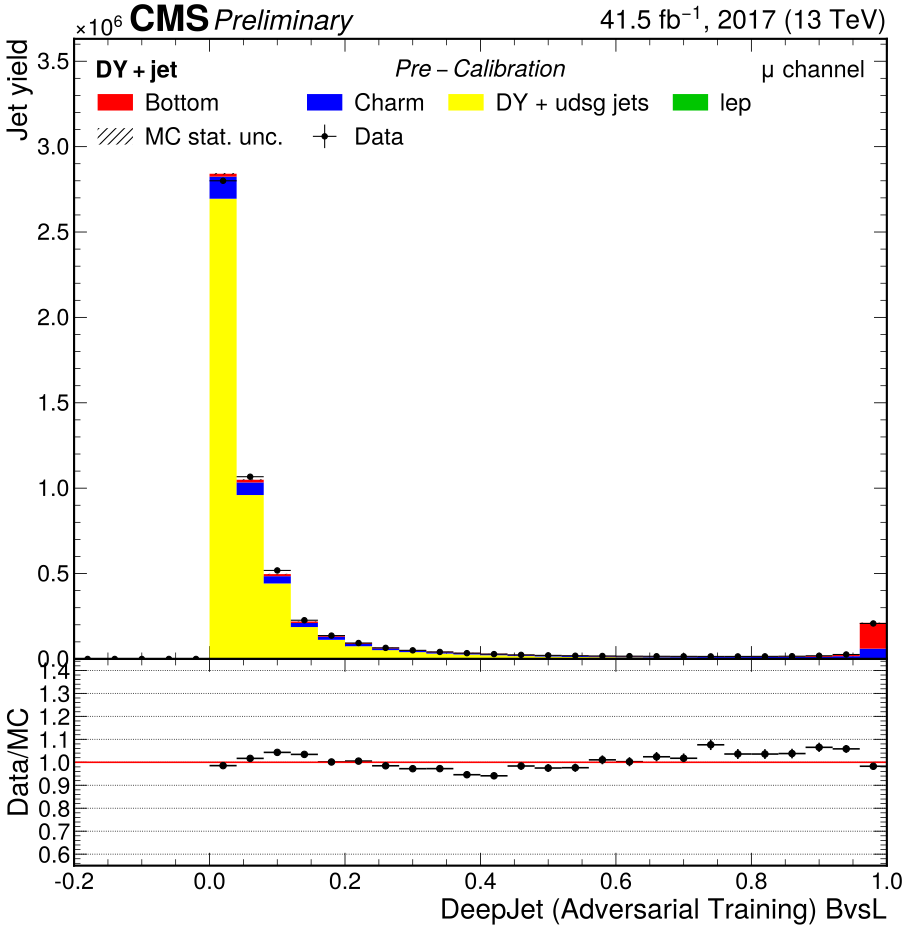}
\caption{Data-to-simulation agreement for the DeepJet b versus light discriminator in a light flavor-enriched selection using the nominal (left) and adversarial (right) trainings ~\cite{CMS-DP-2022-049}.}
\label{fig:DeepJetAdvAttackB}
\end{center}
\end{figure}

Differences in the heavy-flavor tagging performance in data and simulation are observed and must be calibrated against. A recent development~\cite{CMS-DP-2022-049} introduces an adversarial training to the model with the scope of reducing any observed discrepancies between data and simulation before any calibration is applied. This technique improves the robustness of the model, meaning that the model has two tasks to solve simultaneously: optimize classification and hold out against mismodellings that can mimic systematic uncertainties. This is done by applying adversarial attacks (small systematic disturbances) on the input features. Adversarial inputs are generated by the Fast Gradient Sign Method (FSGM) ~\cite{Goodfellow-et-al-2016, goodfellow2015explaining}, which modifies the input features ($x_{\mathrm{raw}}$) in a systematic way in order to increase the loss function, ($J(x_{\mathrm{raw}}, y$):
\begin{equation}
    x_{\mathrm{FGSM}} = x_{\mathrm{raw}} + \varepsilon \cdot sgn(\nabla_{x_{\mathrm{raw}}}J(x_{\mathrm{raw}}, y)),
\end{equation}
where $\varepsilon$ is the (small) distortion parameter, y is the target and $sgn(\alpha)$ stands for the sign of $\alpha$. The FSGM attacks are applied in all DeepJet input features (excluding integer and defaulted values), in every step of the training.  
Figure~\ref{fig:DeepJetAdvAttack} shows the performance in discriminating b from light jets of the nominal and adversarial models with (dashed lines) and without (solid lines) the FSGM-attacks in the input features with a distortion parameter of $\varepsilon=0.01$ and a shift of $x_{\mathrm{raw}}$ not more that 20\% of its original value. The adversarial model not only shows similar performance with the nominal DeepJet training but also provides higher robustness. The hyperparameter optimization is performed based on where the focus is put on: higher performance or higher robustness. Figure~\ref{fig:DeepJetAdvAttackB} shows the data over simulation agreement for the b versus light DeepJet tagger for the nominal (left) and adversarial training (right) in events with at least two well isolated and oppositely charged muons with an invariant mass close to the Z boson mass~\cite{CMS:2021scf}. While data over simulation agreement using the nominal training shows some oscillations, the adversarial training provides a better agreement at a similar performance.  

\begin{figure}[ht]
\begin{center}
 \includegraphics[width=0.65\columnwidth]{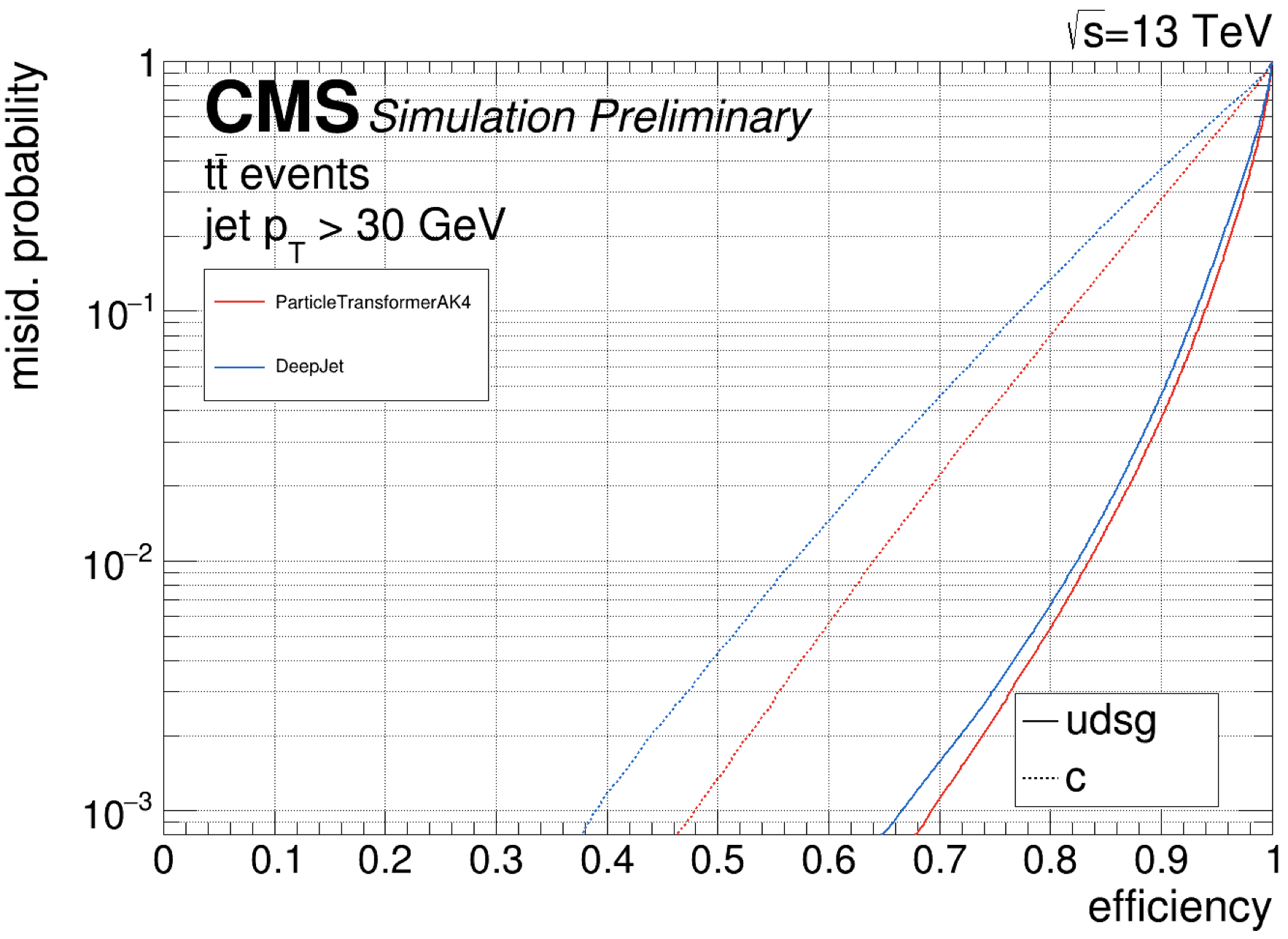}
 \includegraphics[width=0.65\columnwidth]{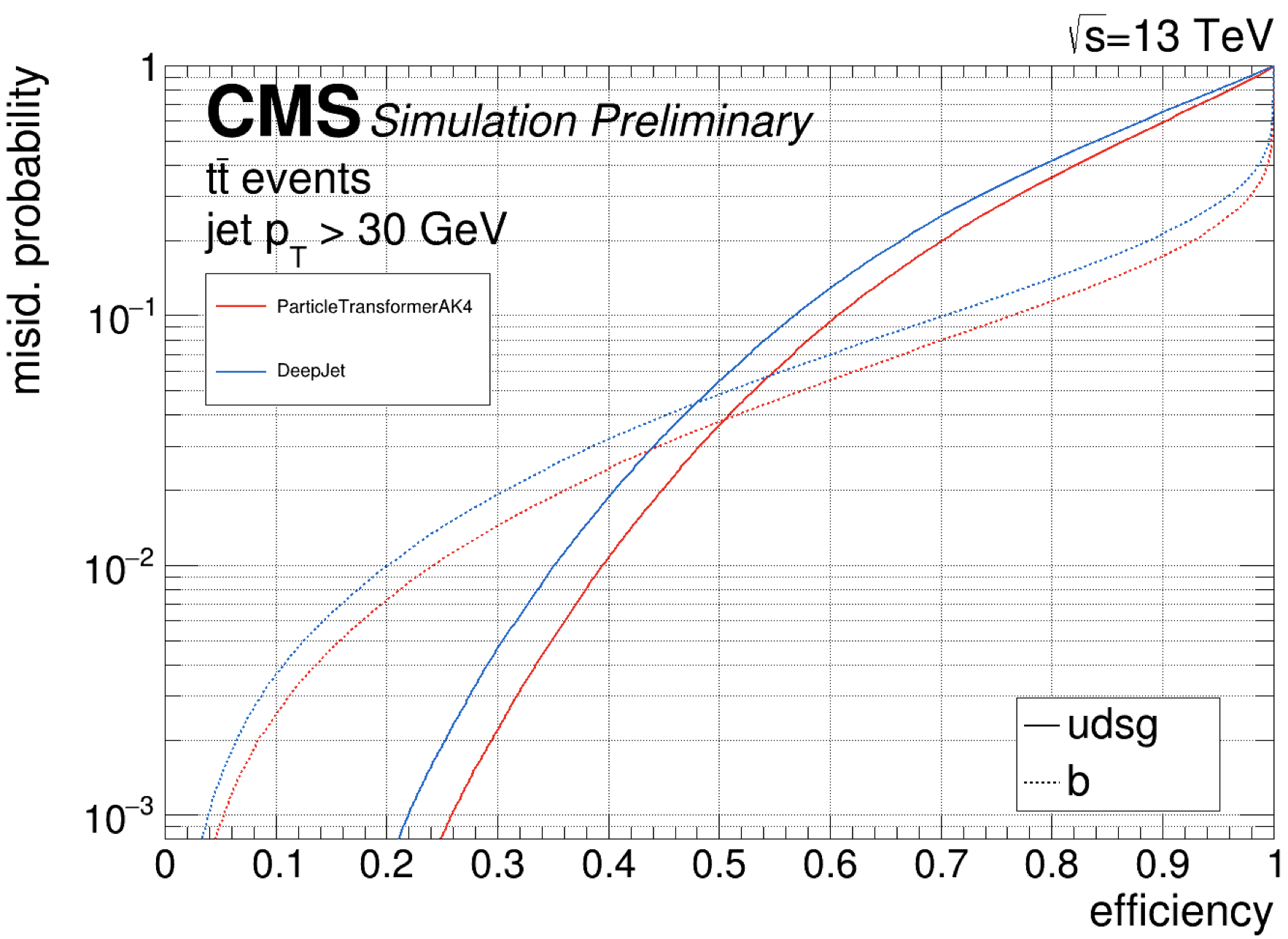}
\caption{Performance of DeepJet (blue) and ParT (red) b-tagging algorithms for identifying small-R jets with $p_{\mathrm{T}}$ > 30~GeV and $|\eta|<2.5$, as measured in simulated $t\bar{t}$ events. Dashed lines correspond to the misidentification rate of udsg jets, while solid lines correspond to the misidentification rate of c (upper) or b (lower) jets~\cite{CMS-DP-2022-050}.}
\label{fig:partvsdeepjet}
\end{center}
\end{figure}

The latest development in heavy flavor tagging at the CMS experiment is the Particle Transformer (ParT) ~\cite{Qu:2022mxj}. ParT has a Transformer-based architecture which incorporates pairwise particle interactions in a tailored attention mechanism~\cite{vaswani2023attention}.
It takes as input information from all jet constituent particles, such as the 4-vector (E, $p_{\mathrm{x}}$, $p_{\mathrm{y}}$, $p_{\mathrm{z}}$), electric charge, particle identity  as determined by the experiment detector and information on the trajectory displacements.
The training is performed on a large dataset containing 100 M jets, called \textsc{JetClass} which includes 10 types of jets: H$\rightarrow b\bar{b}$, H$\rightarrow c\bar{c}$, H$\rightarrow$gg, H$\rightarrow$4q, H$\rightarrow \ell\nu qq\acute{}$, t$\rightarrow bqq\acute{}$, t$\rightarrow b\ell\nu$, W$\rightarrow qq\acute{}$, Z$\rightarrow qq\acute{}$, representing the signal jets and $q/g$, representing the background jets. 
A comparison of the performance of ParT and DeepJet b-tagging algorithms~\cite{CMS-DP-2022-050} is shown on the left (right) plot of 
Figure~\ref{fig:partvsdeepjet}, which shows the probability of misidentifying non-b (non-c) jets as b jets (c jets) with respect to correctly identifying b jets (c jets). ParT shows significant improvements compared to DeepJet and its promising performance makes it a good candidate for becoming the  state-of-the-art heavy flavor tagger for CMS during the Run-3 data-taking period of the LHC.

\subsection{Double-$b$ Tagging in ATLAS} 
\label{sec:doubleb}

\begin{figure}[ht]
\begin{center}
 \includegraphics[width=0.4\columnwidth]{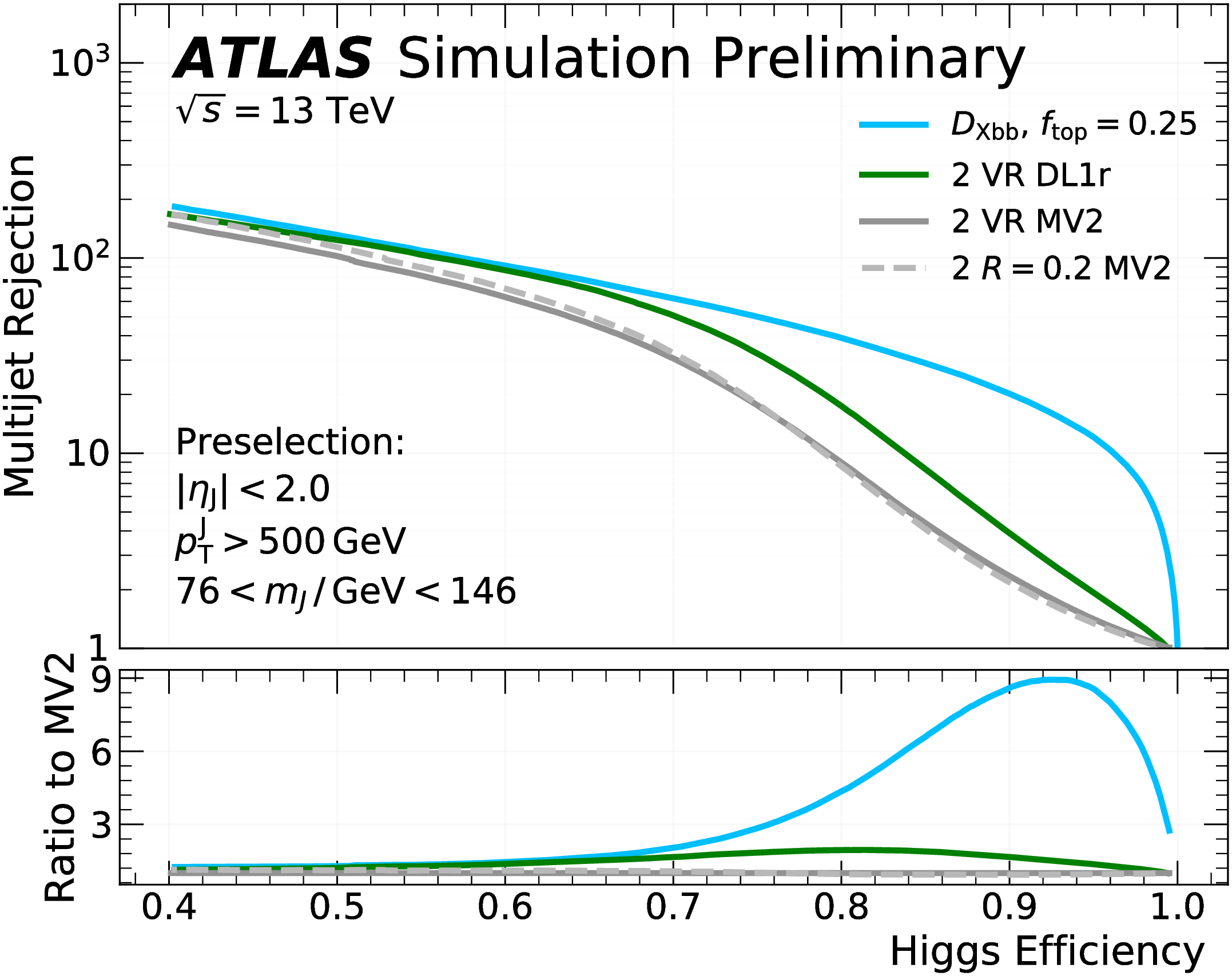}
 \includegraphics[width=0.4\columnwidth]{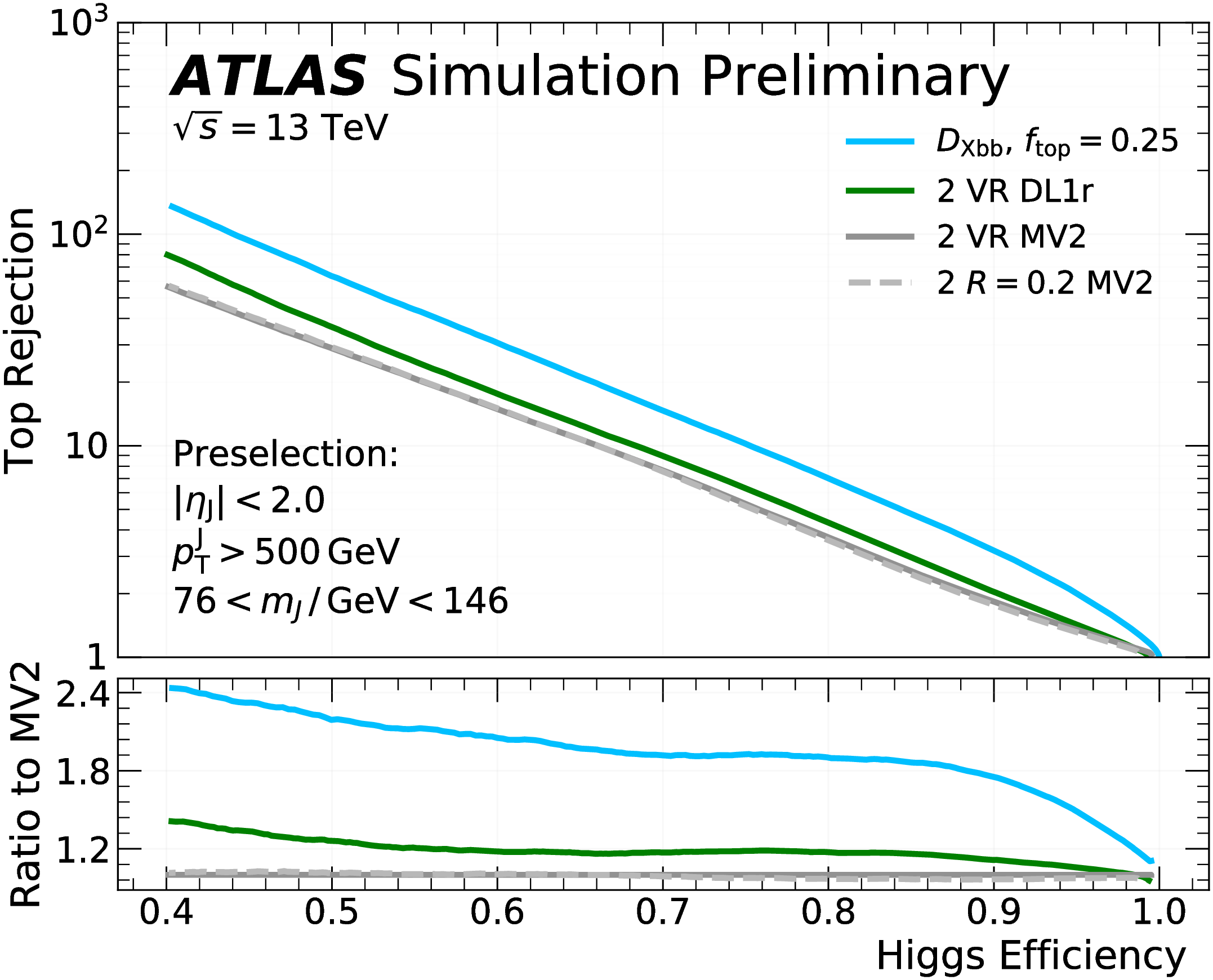}
\caption{Multijet (left) and top jet (right) rejection factors as a function of the $H\rightarrow b\overline{b}$ tagging efficiency, for large-R jet $p_{\mathrm{T}}$ > 500 GeV. Performance of the $D_{Xbb}$ algorithm is compared to 2-VR DL1r and 2-VR MV2. Another variant of the 2-VR MV2 that considers fixed radius (R = 0.2) track jets is also included.}
\label{fig:xbb}
\end{center}
\end{figure}

The double-$b$ tagging algorithm in ATLAS shares the same architecture as that
of the single-$b$ tagging algorithm. The first method developed is to apply the
single-$b$ tagging algorithm directly on the subjets, requiring the two
subjets to be identified as $b$-jets. The subjets are reconstructed using a
variable radius (VR) algorithm considering tracks only~\cite{atlasvr} that can
be associated with the large-R jets reconstructed using local calorimeter
topological clusters (LCTopo). As a consequence, it is referred to as the
``2-VR" method. The first dedicated double-$b$ tagging algorithm,
$D_{Xbb}$~\cite{ATLASDxbb}, is a feed forward neural network using the kinematic
information of the large-R jet and the DL1r outputs evaluated on up to three
subjets as the input. The DL1r tagger was retrained using the VR track jets
in a $t\overline{t}$ sample.  In the boosted regime, the major background
consists of jets from multijet and top events. So the performance is
represented in terms of multijet and top jet rejection factors.
Figure~\ref{fig:xbb} compares the $D_{Xbb}$ tagger and the 2-VR method.
The dedicated algorithm clearly outperforms the latter, especially when the
$H\rightarrow b\overline{b}$ efficiency is high.

\begin{figure}[ht]
\begin{center}
 \includegraphics[width=0.6\columnwidth]{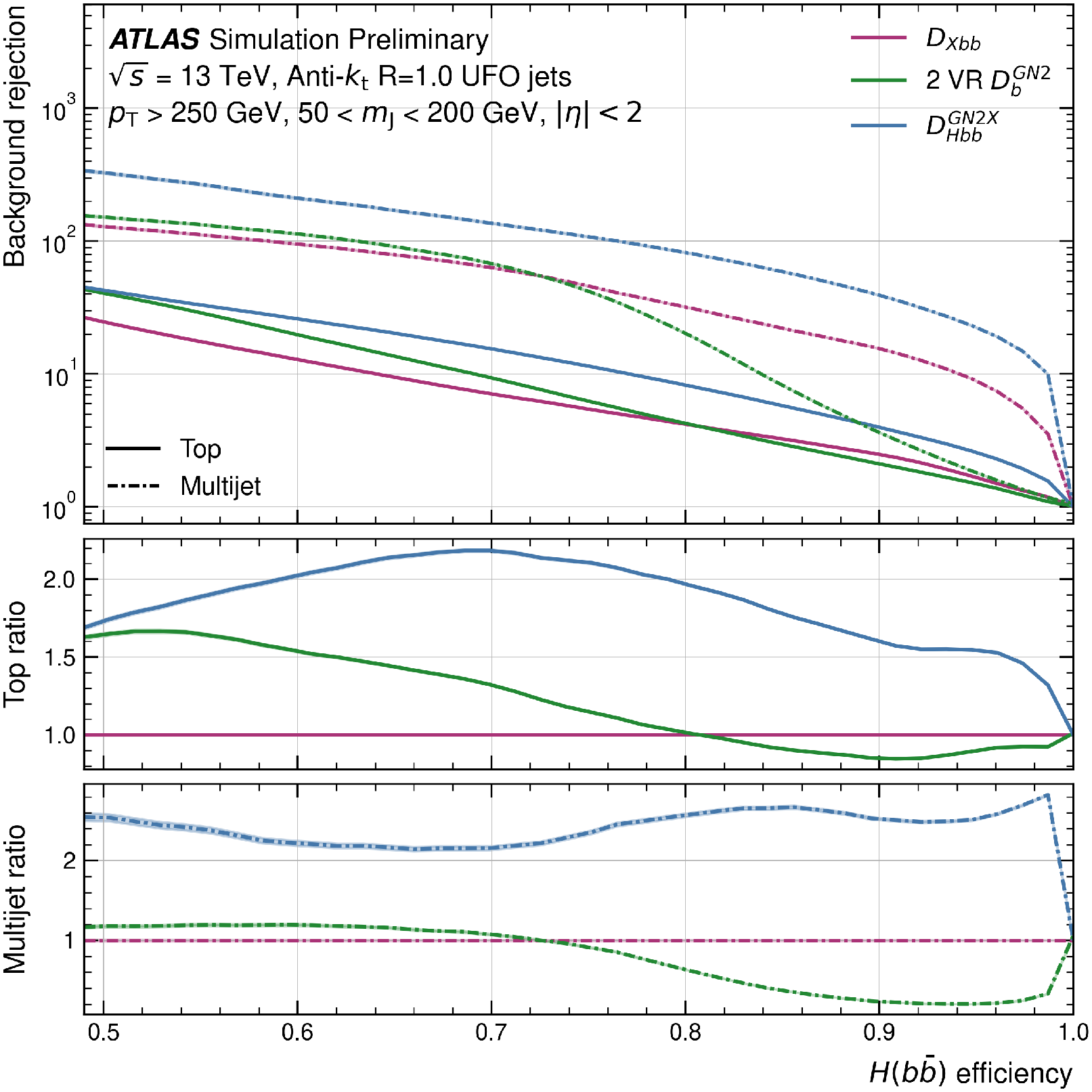}
\caption{Top and multijet rejection factors as a function of the $H\rightarrow b\overline{b}$ tagging efficiency for jets with $p_{\mathrm{T}}$ > 250 GeV and mass ( 50 GeV < $m_{\mathrm{J}}$ < 200 GeV ). Performance of the GN2X algorithm is compared to the $D_{Xbb}$ and VR subjets baselines~\cite{GN2X}.}
\label{fig:GN2XBaseline}
\end{center}
\end{figure}

The flexible architecture of GN2 allows ATLAS to further unify both the
single-$b$ and double-$b$ tagging algorithms. As seen in
Figure~\ref{fig:GNNArch}, the algorithm is agnostic to the jet reconstruction
algorithms. The most state-of-the-art double-$b$ tagging also uses the GN2
architecture but considering the large-R jets reconstructed using the united
flow objects (UFO)~\cite{ATLASUFO}. The baseline model, $D_{Hbb}^{GN2X}$,
similar to the single-$b$ version, only explores the kinematic information of
the large-R jets and the associated tracks. Figure~\ref{fig:GN2XBaseline}
compares the performance of GN2X to that of $D_{Xbb}$. Both the multijet and
top rejection factors are improved by a factor of two when the $H\rightarrow
b\overline{b}$ efficiency is 60\%. In addition, the 2-VR method using GN2 is
added as a reference, showing similar performance as that of
$D_{Xbb}$~\cite{GN2X}.          

\begin{figure}[ht]
\begin{center}
 \includegraphics[width=0.6\columnwidth]{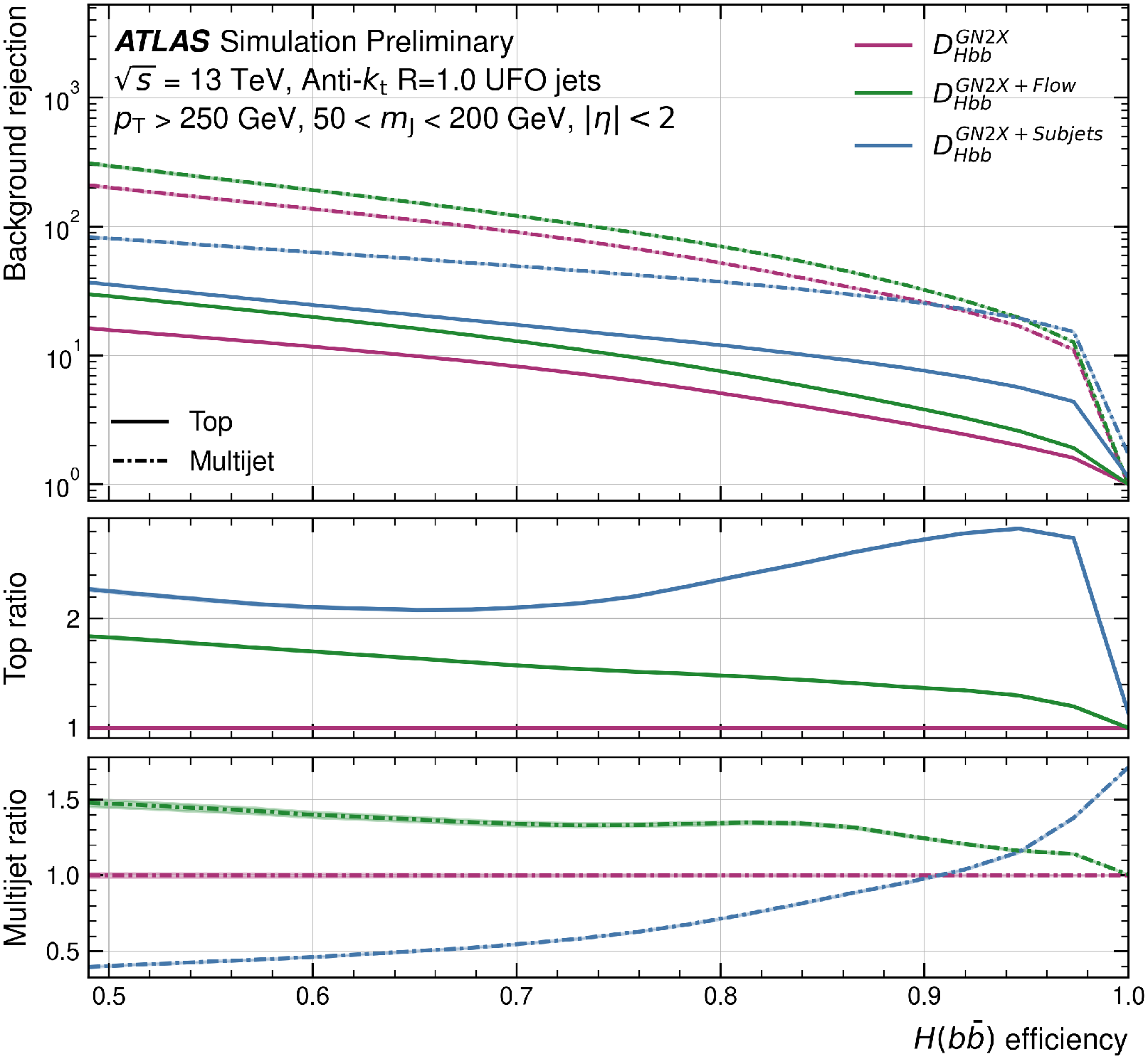}
\caption{The top and multijet background rejection factors as a function of the $H\rightarrow b\overline{b}$ tagging efficiency for the two heterogeneous input type architectures compared to the baseline GN2X model~\cite{GN2X}.}
\label{fig:GN2XVariations}
\end{center}
\end{figure}

Two additional variations are also considered by either adding calorimeter or
subjet information. As seen in Figure~\ref{fig:GN2XVariations}, adding both the
charged and neutral calorimeter information, i.e. the flow objects, improves
the multijet (top) rejection factor by 50\% (80\%). When the kinematic and GN2
output of the VR subjets are included, the top rejection factor is two times
higher while the multijet rejection factor is up to 60\% smaller when the
$H\rightarrow b\overline{b}$ tagging efficiency is below 90\%. The GN2X
performance can be further improved by exploring the above
variations~\cite{GN2X}.        

\subsection{Double-$b$ Tagging in CMS} 
\label{sec:doubleb_cms}

Several machine-learning based algorithms have been developed in CMS to identify highly Lorentz-boosted massive particles. 
These algorithms utilise high level inputs, such as jet substructure observables, and lower level inputs, such as PF candidates or information from secondary vertices associated with the AK8 jets (PF candidates are clustered with the anti-k$_{\mathrm{T}}$ algorithm~\cite{Cacciari:2008gp} with a distance parameter of 0.8). 

\begin{figure}[ht]
\begin{center}
 \includegraphics[width=0.6\columnwidth]{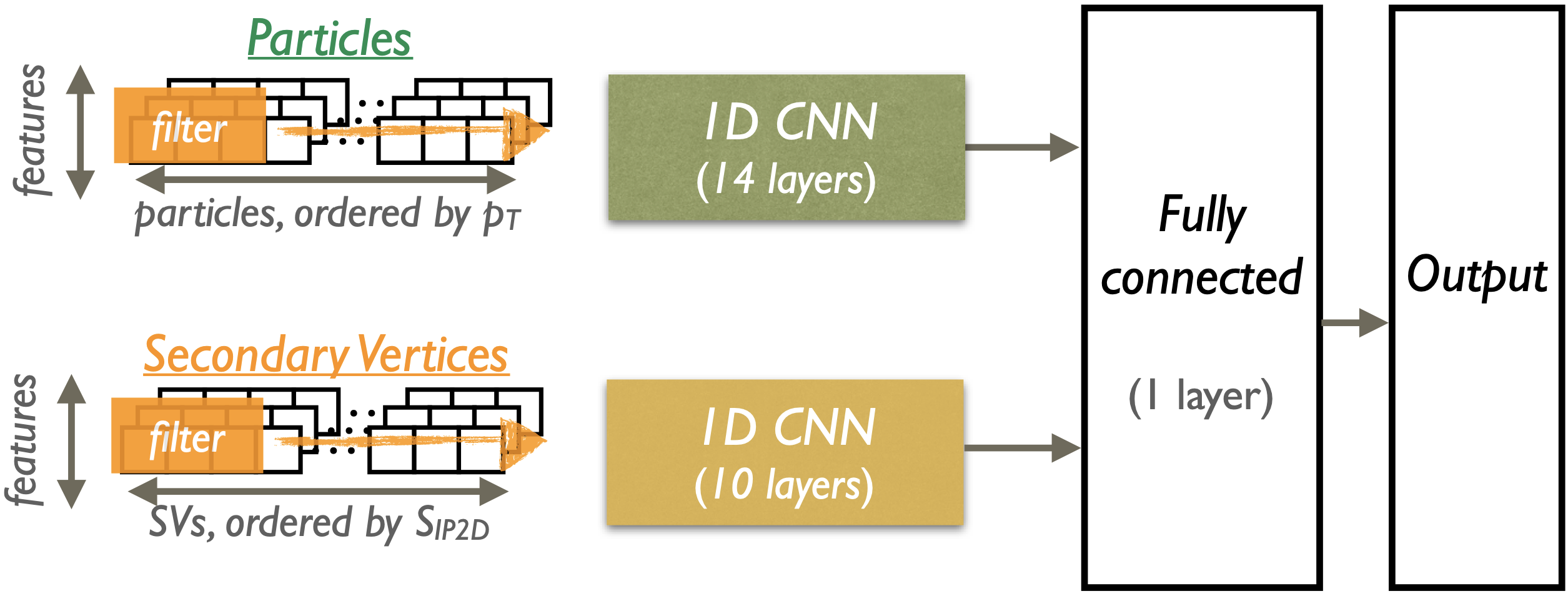}
\caption{The network architecture of the DeepAK8 multi-classifier~\cite{CMS:2020poo}.}
\label{fig:deepak8}
\end{center}
\end{figure}

An example of such algorithm is the DeepAK8~\cite{CMS:2020poo}, a multi-class particle identification algorithm able to discriminate heavy hadronically decaying particles into five categories: W, Z, H, top, or other, and the classes are further subdivided into decay modes (i.e. $b\bar{b}, \ c\bar{c}, \ q\bar{q}$). The DeepAK8 algorithm takes as input up to 100 jet constituent particles, sorted by decreasing $p_{\mathrm{T}}$ and utilises their properties, such as the $p_{\mathrm{T}}$, charge, energy deposit, etc., and information regarding the SVs of the event. The DeepAK8 architecture consists of two parts. The first part consists of two one-dimensional convolutional neural networks (CNNs) which are applied to the particle and SV lists. The CNNs transform the inputs and provide useful features that are in turn processed by the second step, which is a simple fully connected network performing the jet classification. The architecture of DeepAK8 is illustrated in Fig.~\ref{fig:deepak8}.

A jet mass decorrelated version of DeepAK8 is developed, namely DeepAK8-MD. This alternative network is using the same input features as the nominal one and its training samples are reweighted in order to yield a flat transverse momentum and mass distributions. DeepAK8-MD is able to preserve the discrimination power of the original DeepAK8 algorithm using adversarial training~\cite{Louppe:2016ylz}. An additional network, called mass prediction, is added during the training phase and predicts the jet mass from the CNN output. The accuracy of the mass prediction network is subsequently used as a penalty to prevent the tagger from learning specific features correlated with the mass.
A different approach towards jet mass decorrelation is based on the Designing Decorrelated Taggers (DDT) method~\cite{Dolen:2016kst}. In this method, the output of DeepAK8 is transformed as a function of a dimensionless scaling variable $\rho=ln(m_{\mathrm{SD}}^{2}/p_{\mathrm{T}}^{2})$ and the jet $p_{\mathrm{T}}$, where $m_{\mathrm{SD}}$ is the groomed jet mass derived from the soft-drop algorithm~\cite{Larkoski:2014wba} with $\beta=0$ and $z_{\mathrm{cut}} = 0.1$. The resulting output score of DeepAK8-DDT yields into a flat QCD-multijet efficiency across the $m_{\mathrm{SD}}$ and the transverse momentum spectra. Two DeepAK8-DDT models are trained, corresponding to 2\% and 5\% flat background efficiency. 

\begin{figure}[ht]
\begin{center}
 \includegraphics[width=0.3\columnwidth]{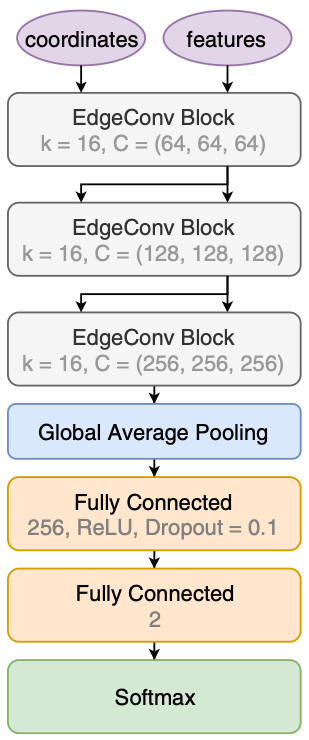}
\caption{The ParticleNet architecture ~\cite{Qu:2019gqs}.}
\label{fig:pnetArch}
\end{center}
\end{figure}

A more recent development is ParticleNet~\cite{Qu:2019gqs}, a multi-classification algorithm that treats jet constituents as a permutation invariant set of particles  (\emph{point cloud}) rather than an ordered structure. ParticleNet is based on customised Dynamic Graph Convolutional Neutral Network (DGCNN)~\cite{wang2019dynamic} and its key building block is the edge convolution (EdgeConv). The EdgeConv represents each point cloud as a graph with each point being the vertex, while the edges of the graph are constructed as connections between each point and its $k-$nearest neighbor. 
The EdgeConv operation can be stacked allowing to form a deep network where local and global structures are learned in a hierarchical way. The architecture of ParticleNet, shown in Fig.~\ref{fig:pnetArch}  consists of three EdgeConv blocks where the first one uses the spatial coordinates in the $\eta-\phi$ plane to compute the distances, while the next two blocks use the learned vectors as the new coordinates. Following the EdgeConv blocks, a global average pooling operation is performed in order to aggregate the output features over all point clouds. Subsequently, there are two fully connected layers of 256 and 2 units, and a softmax function which is used to generate the output of the classifier. As input features, ParticleNet utilises the same inputs as the DeepAK8 algorithm.

A mass-decorrelated (MD) version of ParticleNet that utilises the same inputs and architecture is used to identify highly Lorentz-boosted heavy particles (X) decaying hadronically. The mass decorrelation is achieved by training the network with a simulated signal sample containing Lorentz-boosted spin-0 particles of a flat mass in the range between 15-250~GeV and decay into a quark-antiquark pair. As background, a QCD-multijet sample is used. Both signal and background training samples are subject to reweighting in transverse momentum and mass.

\begin{figure}[ht]
\begin{center}
 \includegraphics[width=0.6\columnwidth]{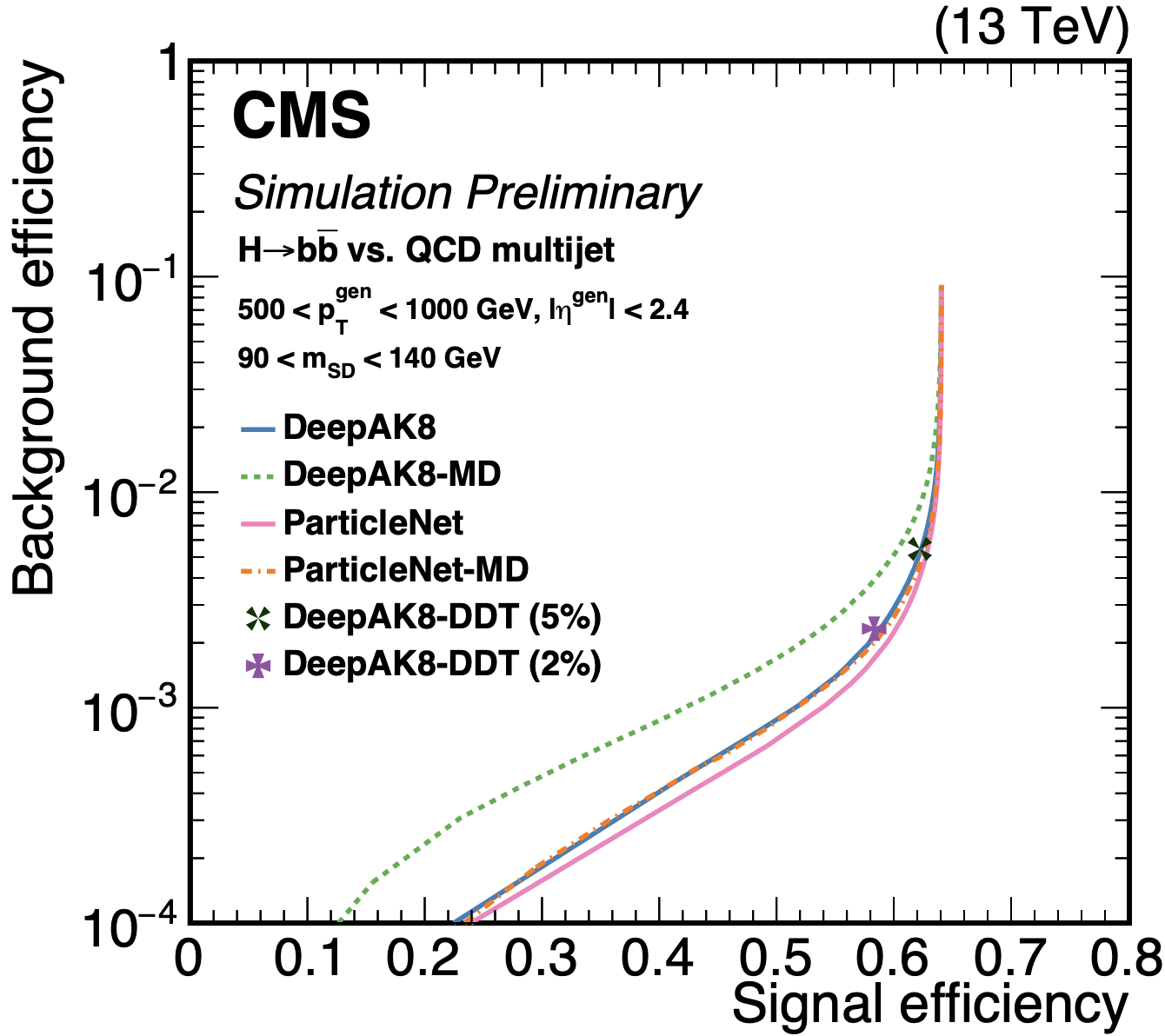}
\caption{Performance of tagging algorithms identifying Higgs bosons decaying into a pair of bottom quarks, as measured in simulated SM H$\rightarrow b\bar{b}$ (signal efficiency) and QCD-multijet (background efficiency) events~\cite{CMS-DP-2020-002}.}
\label{fig:doubleb_cms}
\end{center}
\end{figure}

Figure \ref{fig:doubleb_cms} shows the performance of the aforementioned machine learning algorithms on identifying highly Lorentz-boosted Higgs bosons into a pair of bottom quarks~\cite{CMS-DP-2020-002}. The performance is derived after a selection on the mass of the large-R jets is made, requiring 90 < $m_{\mathrm{SD}}$ < 140~GeV. For simulated SM H$\rightarrow b\bar{b}$ signal (QCD-multijet background) events, the generated Higgs boson (quarks and gluons) candidates are required to have 500 < $p_{\mathrm{T}}$ < 1000~GeV and $|\eta|<2.4$. In the case of background, the efficiency is defined as the ratio of the reconstructed Higgs boson candidates satisfying the selection, over all Higgs boson candidates. ParticleNet shows significant improvements with respect to the previous highly Lorentz-boosted particle taggers. 

\clearpage

\subsection{Flavour Tagging in ATLAS Trigger}
\label{sec:trigger}

\begin{figure}[ht]
\begin{center}
 \includegraphics[width=0.8\columnwidth]{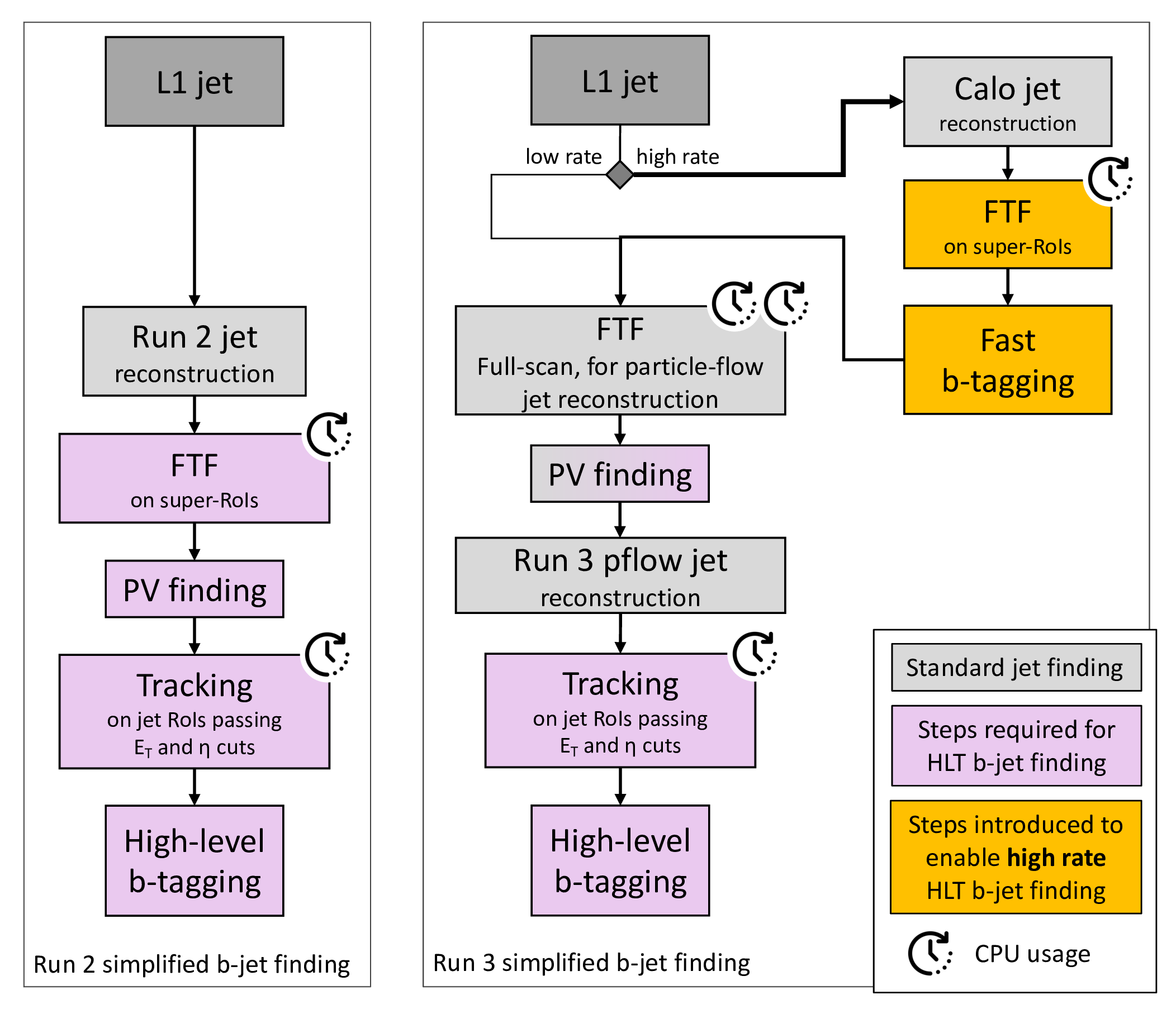}
\caption{Simplified schematic descriptions of the $b$-jet trigger selections in two different ATLAS trigger implementations: the Run 2 implementation on the left, and the Run 3 implementation on the right~\cite{ATLASDIPS}.}
\label{fig:fastdips}
\end{center}
\end{figure}

Trigger performance is vital in the tri-Higgs search programme. In certain low
mass region, applying flavour tagging in the high level trigger (HLT), i.e.
$b$-jet trigger, is the only viable approach. Unlike the offline environment,
the stringent computational requirement for triggers prevents reconstructing
all tracks in the event. Therefore, the overall $b$-jet trigger has the same workflow as the offline reconstruction except the fact that two dedicated track reconstruction
iterations are performed within given regions of interest (ROI). A fast track
finding (FTF) step is employed using the super-ROIs defined by trigger jets
with $p_{\mathrm{T}}$ > 30 GeV, and the tracks are used to reconstruct primary
vertices. Jets passing further kinematic selections define the ROIs for the
precision tracking iteration, and the resulting tracks are used to perform the
flavour tagging algorithms. A simplified flow diagram is shown on the left side
of Figure~\ref{fig:fastdips}.  The MV2 tagger family was adopted in the $b$-jet
triggers during Run 2 data-taking. A detailed documentation can be found in
ref.~\cite{ATLASRun2Trigger}.

\begin{figure}[ht]
\begin{center}
 \includegraphics[width=0.45\columnwidth]{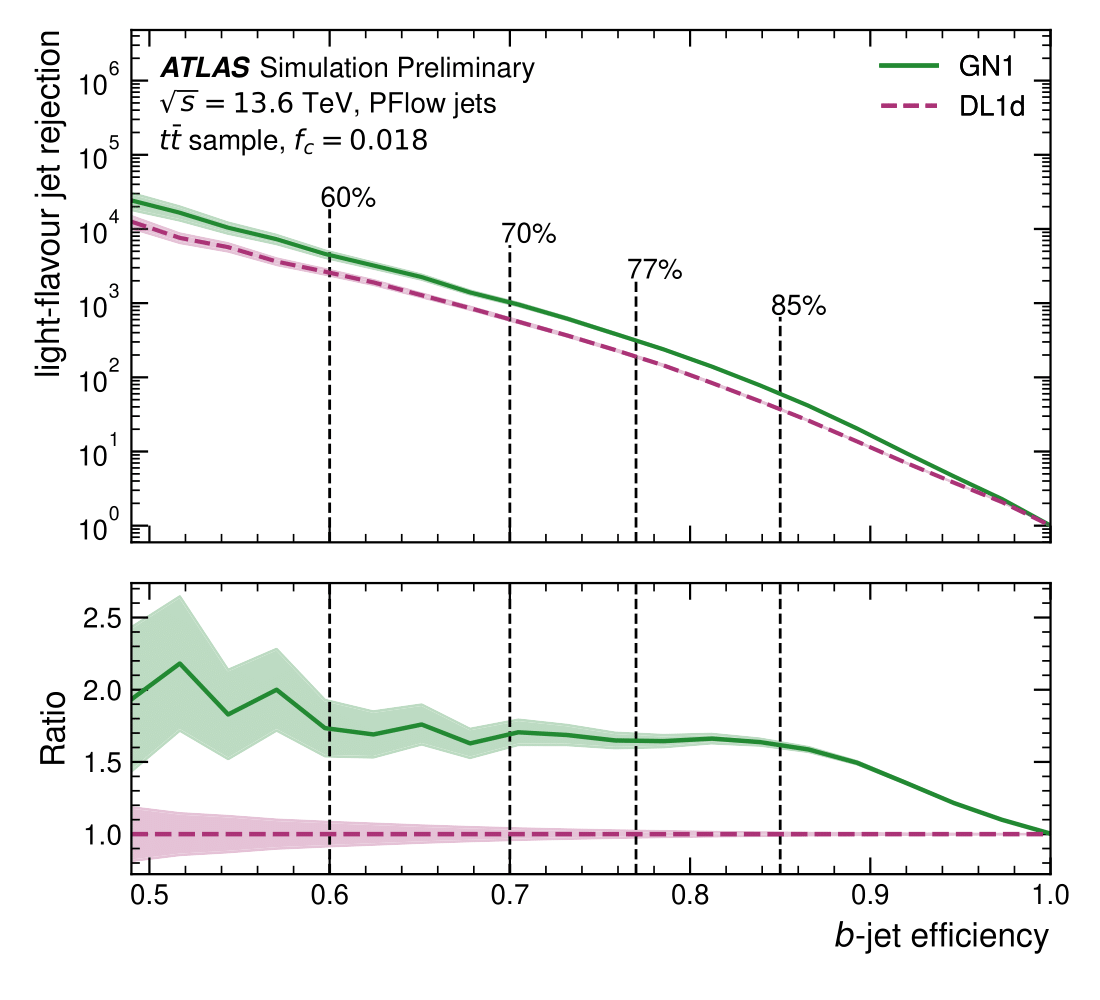}
 \includegraphics[width=0.45\columnwidth]{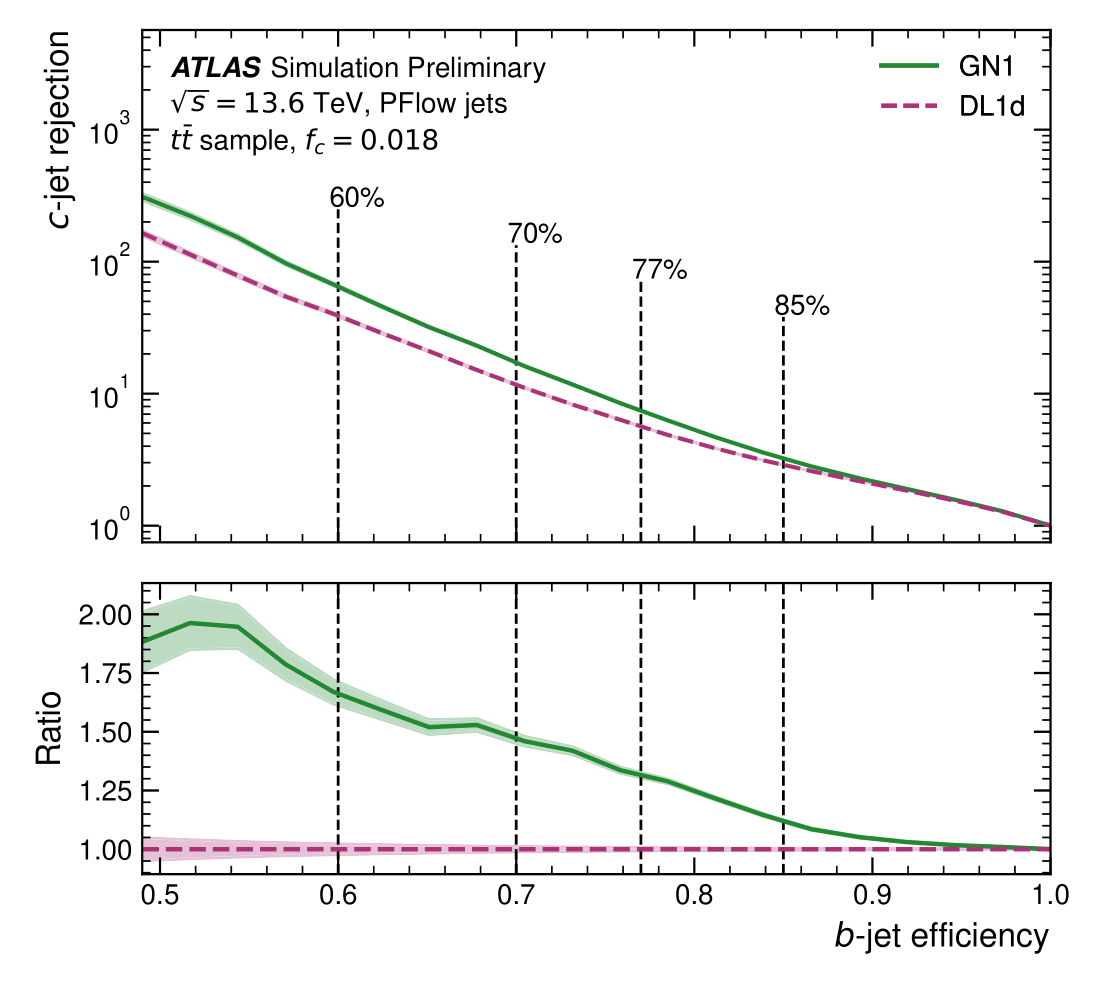}
\caption{Light-flavour jet rejection as a function of $b$-jet efficiency of the GN1 algorithm (green) in comparison to the benchmark DL1d algorithm (purple) which uses the DL1 architecture, evaluated on HLT Particle Flow jets in a $t\overline{t}$ sample. The 60\%, 70\%, 77\% and 85\% $b$-jet efficiency operating points are indicated by vertical black lines~\cite{ATLASRun3BTrigger}.}
\label{fig:Run3TriggerComparison}
\end{center}
\end{figure}

Significant improvements have been introduced to the Run 3 $b$-jet triggers.
New taggers such as DL1d and GN1 were implemented into the trigger, and have
been collecting data efficiently. In addition, a fast $b$-tagging sequence
using the tracks from FTF iteration is introduced to further reduce the rate.
It applies a similar structure as the DIPS tagger so that it is referred to as
the ``fastDIPS'' algorithm. The right side of Figure~\ref{fig:fastdips}
illustrates the new $b$-jet trigger workflow. Nearly all high rate $b$-jet
triggers include the ``fastDIPS'' preselection step~\cite{ATLASDIPS}.
Figure~\ref{fig:Run3TriggerComparison} compares the performance of various
$b$-jet trigger algorithms. The GN1 $b$-jet trigger has a charm (light) jet
rejection larger than 20 (1000) when the $b$-jet tagging efficiency is
70\%~\cite{ATLASRun3BTrigger}.     

\subsection{Flavour Tagging in CMS Trigger}
\label{sec:flavourtagger}

\begin{figure}[ht]
\begin{center}
 \includegraphics[width=0.45\columnwidth]{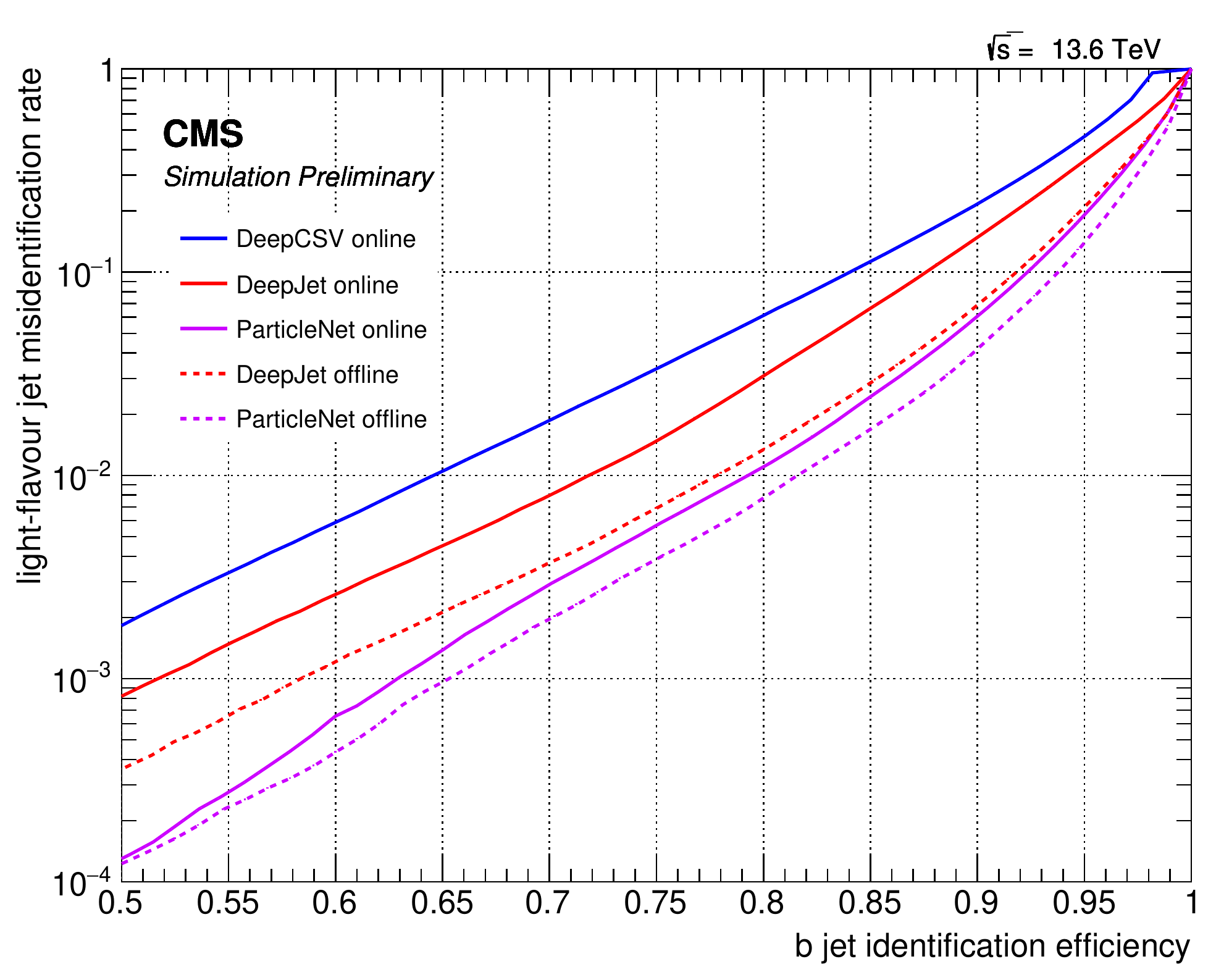}
 \includegraphics[width=0.45\columnwidth]{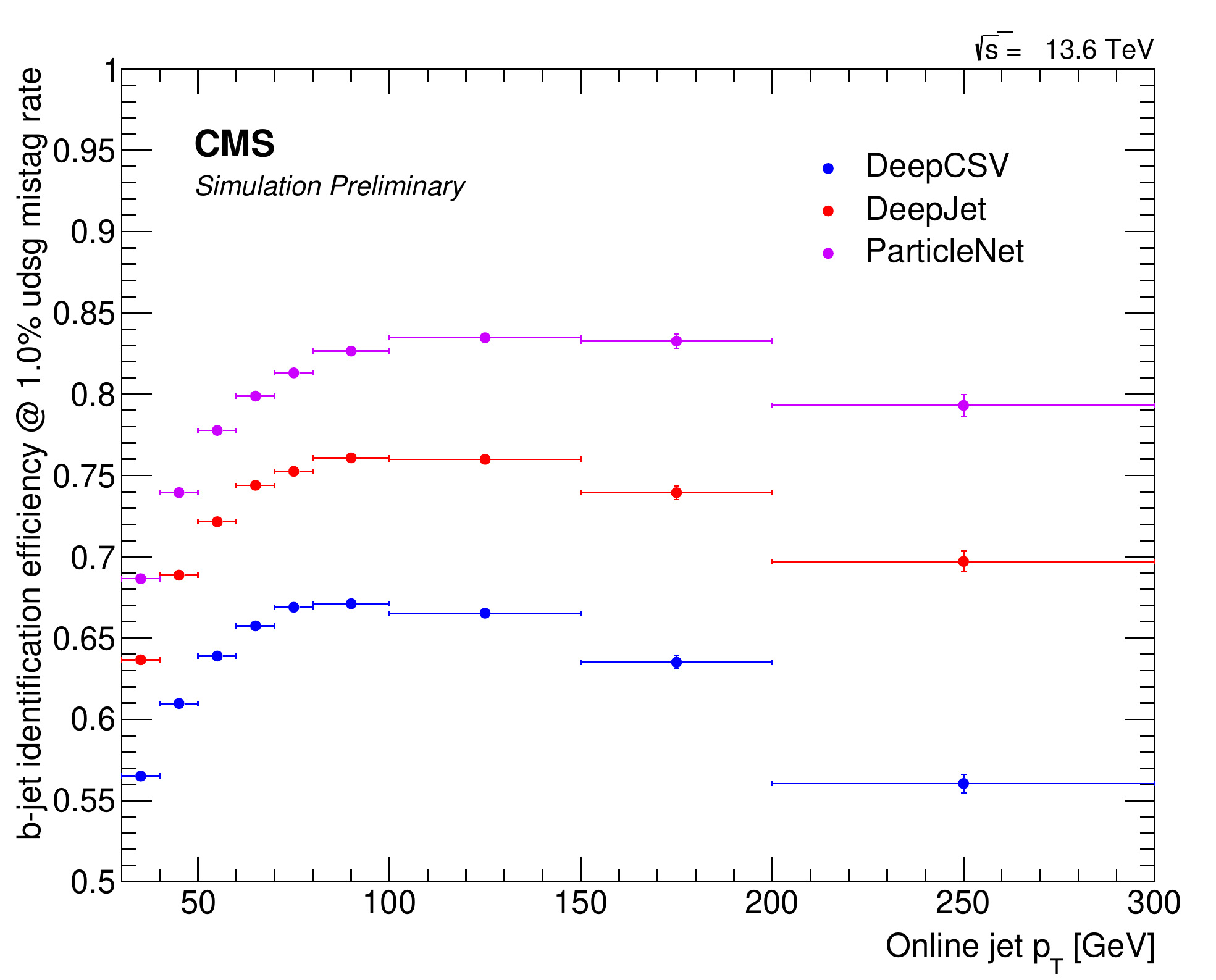}
\caption{Left: Light flavour jet misidentification rate as a function of the efficiency of correctly identifying b jets for the b-taggers DeepCSV (blue), DeepJet (red), and ParticleNet (purple). Solid lines represent the performance for simulated HLT-level jets with $p_{\mathrm{T}}$ > 30~GeV and $|\eta|<2.5$ matched to an offline reconstructed jet with $p_{\mathrm{T}}$ > 25~GeV. Dashed lines correspond to the offline tagging performance on the matched offline jets.
Right: Online b jet identification efficiency at 1\% light flavour jet misidentification rate as a function of the HLT-level jet $p_{\mathrm{T}}$~\cite{CMS-DP-2023-021}.
}
\label{fig:Run3PNet}
\end{center}
\end{figure}

The ParticleNet model was deployed online for the first time in Run 3 2022 with the score of identifying highly Lorentz-boosted heavy particles decaying into a pair of bottom quarks, as well as identifying signatures with b jets (small-R) in the final state.
Figure \ref{fig:Run3PNet} shows the performance of ParticleNet b-tagger~\cite{CMS-DP-2023-021} compared to DeepJet and DeepCSV CMS b-taggers, for HLT-level jets with $p_{\mathrm{T}}$ > 30~GeV and geometrically matched to offline jets. The ParticleNet online b-tagger shows a substantial improvement compared to the previous online DeepJet and DeepCSV b-taggers and its performance approaches that of the offline. Figure \ref{fig:Run3PNet} right, compares the three online b-taggers at a 1\% light-flavour jet misidentification rate, showing that ParticleNet achieves up to 10\% improvement throughtout the jet $p_{\mathrm{T}}$ range.

\begin{figure}[ht]
\begin{center}
 \includegraphics[width=0.68\columnwidth]{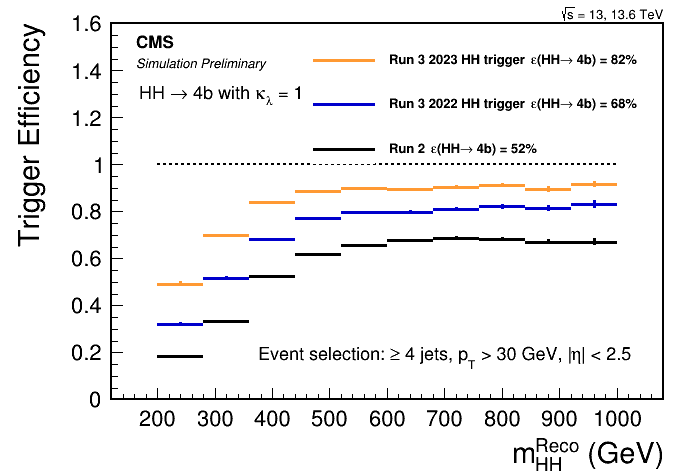}
 \includegraphics[width=0.68\columnwidth]{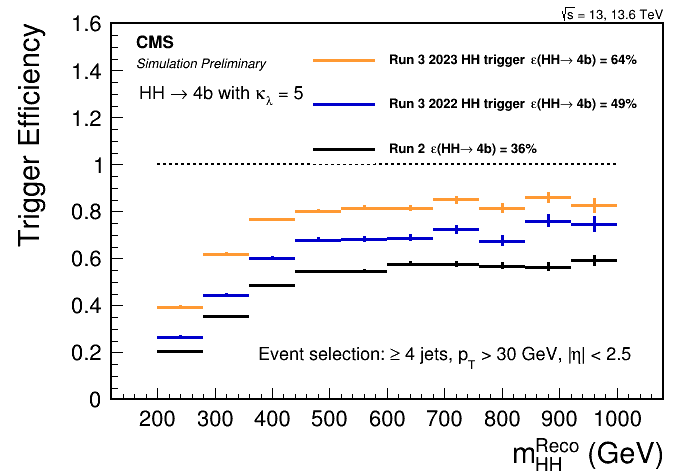}
\caption{Trigger efficiency as a function of the invariant mass of the di-Higgs system ($m_{\mathrm{HH}}^{\mathrm{Reco}}$) for the simulated SM HH$\rightarrow$4b process with $\kappa_{\lambda}$ = 1 (upper) and $\kappa_{\lambda}$=5 (lower), shown for Run 2 (black), Run 3 2022 (blue) and Run 3 2023 (orange) HH triggers~\cite{CMS-DP-2023-050}.}
\label{fig:pnettrga}
\end{center}
\end{figure}

Since the beginning of Run 3 data-taking period, the CMS experiment has exploited the recent improvements in heavy flavour tagging for online HLT-jets~\cite{CMS-DP-2023-021} and deployed online a new trigger strategy~\cite{CMS-DP-2023-050} to record di-Higgs and tri-Higgs production in events with b jets in the final state. In 2022, the trigger targeting HH$\rightarrow$4b production (mentioned below as Run 3 2022 HH trigger) had a rate of around 60~Hz at an instantaneous luminosity of 2$\times 10^{34} cm^{-2}s^{-1}$ and required at least four small-R HLT-jets with $p_{\mathrm{T}}$ > 70, 50, 40, and 35~GeV for the four leading-in-$p_{\mathrm{T}}$ jets and the average score of the two jets with the highest b-tagging score tagged with the ParticleNet online b-tagger to be above 0.65. In 2023, the an updated version of the HH$\rightarrow$4b trigger was deployed in the delayed stream, allowing a higher rate and acceptance at the cost of a delayed event reconstruction. This new trigger (mentioned below as Run 3 2023 HH trigger) recorded events at a maximum rate of 180~Hz at 2$\times 10^{34} cm^{-2}s^{-1}$ and required events to have at least 4 HLT-jets with $p_{\mathrm{T}}$ > 30~GeV, the scalar sum of $p_{\mathrm{T}}$ of all HLT-jets with $p_{\mathrm{T}}$ above 30~GeV ($H_{\mathrm{T}}$) to be above 280~GeV, and the average score of the two leading-in-b-tagging score jets to be at least 0.55. The L1 trigger requirement was also relaxed to allow events with $H_{\mathrm{T}}$ above 280~GeV instead of the 2022 threshold of 360~GeV. Figure \ref{fig:pnettrga} shows the trigger efficiency as a function of the reconstructed invariant mass of the di-Higgs ($m_{\mathrm{HH}}^{\mathrm{Reco}}$) candidate in simulated SM HH$\rightarrow$4b events with $\kappa_{\lambda}=1$ (left) and $\kappa_{\lambda}=5$ (right). The trigger efficiency is defined as:
\begin{equation}
    \varepsilon = \frac{N_{\mathrm{events}}(\text{pass trigger and event selection})}{N_{\mathrm{events}}(\text{pass event selection})},
    \label{eq:trgeff}
\end{equation}
where event selection corresponds to the requirement of at least 4 small-R jets with $p_{\mathrm{T}}$ > 30~GeV and $|\eta|<2.5$. The di-Higgs candidate is reconstructed from the four small-R jets with the highest b-tagging score. The performance of the Run 3 2022 (2023) HH trigger is shown with blue (orange). For comparison, the Run 2 HH trigger~\cite{CMS-DP-2019-042,CMS-DP-2019-026} is also shown with the black line. The aforementioned trigger, which operated at around 8~Hz at 2$\times 10^{34} cm^{-2}s^{-1}$, required an event $H_{\mathrm{T}}$ > 340~GeV and at least four small-R jets with $p_{\mathrm{T}}$ > 75, 60, 45, and 40~GeV, where at least three of those jets were tagged online with the DeepCSV online b-tagger with a working point of 0.24. The overall trigger efficiency achieved by the 2023 trigger strategy for the HH$\rightarrow$4b process with $\kappa_{\lambda}=$1 ($\kappa_{\lambda}=$5) reaches 82\% (64\%), improved by 20\% (30\%) with respect to the 2022 trigger strategy and by 57\% (78\%) with respect to the 2018 one. The Run 3 2023 HH trigger results in higher efficiency on the full $m_{\mathrm{HH}}^{\mathrm{Reco}}$ spectrum.

\begin{figure}[ht]
\begin{center}
 \includegraphics[width=0.68\columnwidth]{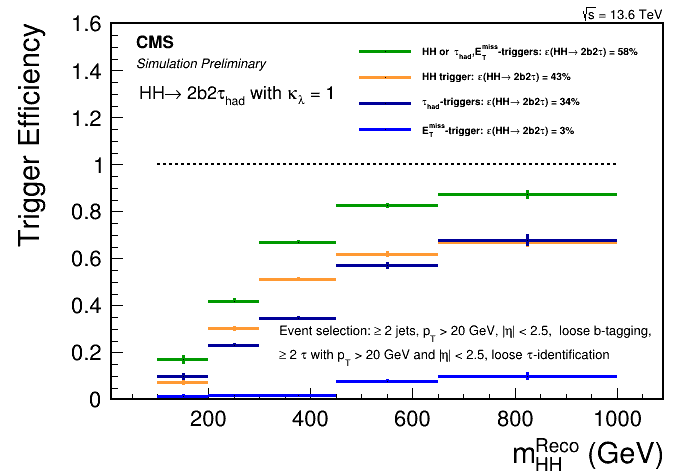}
 \includegraphics[width=0.68\columnwidth]{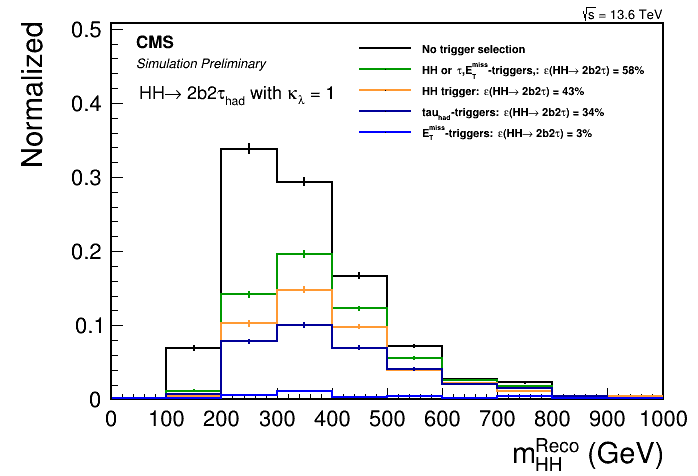}
\caption{Upper: Trigger efficiency as a function of the reconstructed invariant mass of the di-Higgs system ($m_{\mathrm{HH}}^{\mathrm{Reco}}$) for the simulated SM HH$\rightarrow2b2\tau_{\mathrm{had}}$ process with $\kappa_{\lambda}=1$ for the Run 3 hadronic $\tau$ trigger (dark blue), the Run 3 missing transverse momentum trigger (light blue), the Run 3 2023 HH trigger (orange) and the logical OR of all triggers (green). Lower: The $m_{\mathrm{HH}}^{\mathrm{Reco}}$ distribution with and without (black) any trigger requirement applied ~\cite{CMS-DP-2023-050}.}
\label{fig:pnettrgb}
\end{center}
\end{figure}

\newcommand{\tauh}{$\tau_{\mathrm{had}}$\xspace}
The novel Run 3 2023 HH trigger is also used to recover HH$\rightarrow$2b2\tauh-like events that are not recorded by triggers requiring hadronicaly decaying tau leptons (\tauh) or missing transverse energy ($E_{\mathrm{T}}^{\mathrm{miss}}$). The Run 3 \tauh triggers~\cite{CMS-DP-2023-024} have a rate ranging from 17~Hz up to 50~Hz at 2$\times 10^{34} cm^{-2}s^{-1}$ and require the presence of at least two \tauh with $p_{\mathrm{T}}$ > 35~GeV and $|\eta|<2.1$ satisfying the Medium operating point of the DeepTau~\cite{CMS:2022prd} algorithm, or two \tauh with  $p_{\mathrm{T}}$ > 30~GeV  and $|\eta|<2.1$, satisfying the Medium operating point of DeepTau and the presence of an HLT-jet with $p_{\mathrm{T}}$ > 60~GeV, or at least one \tauh with $p_{\mathrm{T}}$ > 180~GeV and $|\eta|<2.1$ and satisfying the loose DeepTau working point. The Run 3 $E_{\mathrm{T}}^{\mathrm{miss}}$ trigger~\cite{CMS-DP-2023-016} operates at around 42~Hz at 2$\times 10^{34} cm^{-2}s^{-1}$ and requires an event $E_{\mathrm{T}}^{\mathrm{miss}}$ of at least 120~GeV. The left plot of Fig. \ref{fig:pnettrgb} shows the trigger efficiency as a function of the $m_{\mathrm{HH}}^{\mathrm{Reco}}$ in simulated SM HH$\rightarrow$2b2\tauh with $\kappa_{\lambda}=1$ for the Run 3 \tauh triggers (dark blue), the Run 3 $E_{\mathrm{T}}^{\mathrm{miss}}$ (light blue), the Run 3 2023 HH trigger (orange) and the logical OR of all trigges (green). The trigger efficiency is defined by Eq. \ref{eq:trgeff} and the event selection requires the presence of at least 2 hadronic small-R jets with $p_{\mathrm{T}}$ > 20~GeV, $|\eta|<2.5$ and identified as b jets with the loose operating point of DeepJet, corresponding to 10\% light-flavor jet misidentification rate, and at least 2 \tauh with $p_{\mathrm{T}} >$20~GeV and $|\eta|<2.5$ with loose identification criteria using the DeepTau algorithm. The di-Higgs candidate is reconstructed using the two b jets and \tauh candidates.  The right plot of Fig. \ref{fig:pnettrgb} shows that the Run 3 2023 HH trigger is able to recover HH$\rightarrow$2b2\tauh events in the full $m_{\mathrm{HH}}^{\mathrm{Reco}}$ spectrum and provides an overall efficiency of 43\%. For values of $m_{\mathrm{HH}}^{\mathrm{Reco}}$ above 650~GeV, the Run 3 2023 HH trigger reaches the one of Run 2 \tauh-triggers. When combined in logical OR with the \tauh- and $E_{\mathrm{T}}^{\mathrm{miss}}$-triggers the overall efficiency reaches 58\%, while the efficiency plateaus at an efficiency of around 85\%.

\begin{figure}[ht]
\begin{center}
 \includegraphics[width=0.68\columnwidth]{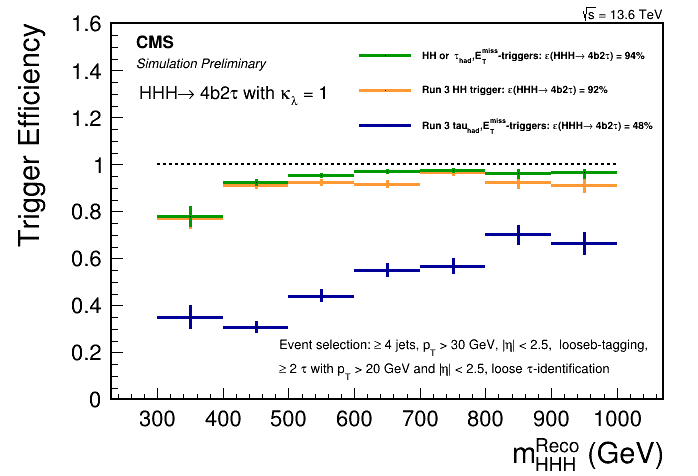}
 \includegraphics[width=0.68\columnwidth]{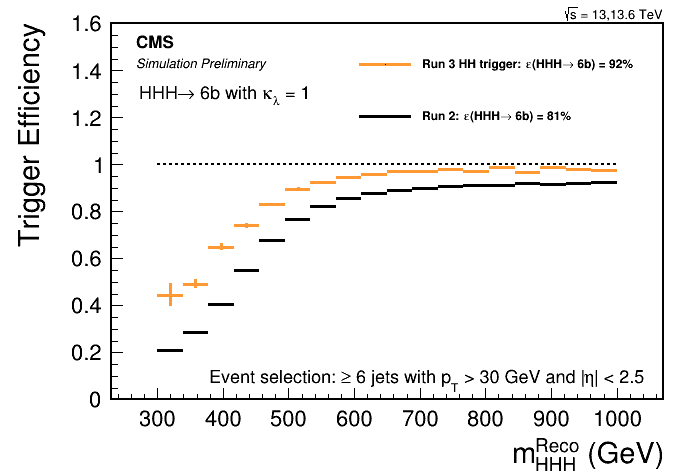}
\caption{Trigger efficiency as a function of the reconstructed invariant mass of the tri-Higgs system ($m_{\mathrm{HHH}}^{\mathrm{Reco}}$) for the simulated SM HHH$\rightarrow$4b2\tauh (upper) and HHH$\rightarrow$6b (lower) processes. ~\cite{CMS-DP-2023-050}.}
\label{fig:pnettrgc}
\end{center}
\end{figure}

The Run 3 2023 HH triggers can also be used to record almost all events from triple Higgs boson production in the 4b2\tauh and 6b final states. Figure \ref{fig:pnettrgc}
 shows the trigger efficiencies as a function of the reconstructed invariant mass of the tri-Higgs system for simulated SM HHH$\rightarrow$4b2\tauh (left) and HHH$\rightarrow$6b (right), both with $\kappa_{\lambda}=1$. 
 In the case of the SM HHH$\rightarrow$4b2\tauh signal, the Run 3 2023 HH trigger can record 92\% of all the events satisfying selection of at least four small-R jets with $p_{\mathrm{T}} >$30~GeV, $|\eta|<2.5$ and satisfying the loose DeepJet working point, and at least 2 \tauh with $p_{\mathrm{T}}$> 20~GeV and $|\eta|<2.5$ and satisfying a loose DeepTau criterion. When combined with the Run 3 \tauh-triggers the effiency reaches 94\%. 
 For the SM HHH$\rightarrow$6b signal, the Run 3 2023 HH trigger can record around 92\% of all events satisfying a basic selection of 6 jets with $p_{\mathrm{T}}$> 30~GeV and $|\eta|<2.5$. Compared to the Run 2 HH trigger, the new trigger strategy improves the acceptance by 14\%. 
 
\clearpage 

\subsection{Flavour Tagging: Outlook}
\label{sec:flavfuture}

Both ATLAS and CMS collaborations have demonstrated the great potential of a unified
end-to-end heavy-flavour tagging architecture. The graph neural network approaches have
been successfully deployed both online and offline. Both the single-$b$ and
double-$b$ tagging performance have been enhanced significantly compared to Run 2 methods. The tri-Higgs
search programme can already greatly benefit from the state-of-the-art taggers. However their final impacts on the physics analyses also depend on the precision of their calibrations in MC. The author would also like to emphasize the importance of such calibration work. Due to the rich phase space of the tri-Higgs models, the search programme will be further extended if the flavour tagging algorithms are
optimised for certain scenarios. In this section, the authors try to offer some
discussion points.

Jets with low momenta play a vital role in certain phase space as seen before.
It is experimentally challenging to identify $b$-jets with low momenta. The
primary reason is that the main characteristics of $b$-hadron decays such as
displaced tracks and secondary vertices diminish when the Lorentz boost is
small. Improving the flavour tagging performance on low-$p_{\mathrm{T}}$ jets
will be appreciated by the tri-Higgs search programme.

Because of the various mass hierarchies and splits, the boosted scenario is
enriched. For instance, decay products of two low mass particles can be
collimated, resulting in jets containing more than two $b$-hadrons. Expanding
the current scope of the double-$b$ tagging algorithms allows the tri-Higgs
search programme to obtain optimal sensitivity.   

It is also important to note that the $H\rightarrow gg$ decay channel should be
investigated. The feasibility depends on the performance of quark-gluon
tagging. So far, the quark-gluon tagging techniques are mainly studied in the
single particle case. A double-gluon tagging algorithm analogous to the
double-$b$ tagging is another possible new avenue.

\clearpage
\section{{\bfseries Theory studies and models, prospects at current and future hadron colliders }\label{sec:thst_models}}
\subsection{The TRSM and triple Higgs production \label{sec:thstud}}

{\sl A. Papaefstathiou, T. Robens, G. Tetlalmatzi-Xolocotzi}

We now turn to studies that investigate triple Higgs production in specific beyond the Standard Model model realizations. As a first example, we consider a model where the SM scalar sector is enhanced by two additional real scalars.  We consider here  the ``Two Real Singlet Model''~\cite{Robens:2019kga,Robens:2022nnw}, where the SM scalar sector is augmented by two additional scalar fields that transform as singlets under the SM gauge group. In addition, two $\mathbb{Z}_2$ symmetries are imposed, leading to a reduction of the available number of degrees of freedom.

The TRSM is characterized by the following scalar potential

\begin{eqnarray}\label{eq:reduced}
V(\Phi, X, S) &=&\mu^2_{\Phi}\Phi^{\dagger}\Phi +\lambda_{\Phi}\Bigl(\Phi^{\dagger}\Phi\Bigl)^2
+\mu^2_S S^2 +\lambda_S S^4 + \mu^2_X X^2  + \lambda_X X^4  \nonumber\\
&&  +\lambda_{\Phi S}\Phi^{\dagger}\Phi S^2 +
\lambda_{\Phi X} \Phi^{\dagger}\Phi X^2 + \lambda_{S X} S^2 X^2\;,
\end{eqnarray}

which contains nine real couplings $\mu_{\Phi}$, $\lambda_{\Phi}$,
$\mu_S$, $\lambda_S$, $\mu_X$, $\lambda_X$, $\lambda_{\Phi S}$,  $\lambda_{\Phi X}$,  $\lambda_{X S}$. All fields are assumed to acquire a vacuum expectation value (VEV). The physical gauge-eigenstates $\phi_{h,S,X}$ then follow from expanding around these according to:

\begin{equation}
    \Phi = \begin{pmatrix} 0\\\frac{\phi_h + v}{\sqrt{2}}\end{pmatrix}\;,
    S = \frac{\phi_S + v_S}{\sqrt{2}}\;, \quad
    X = \frac{\phi_X + v_X}{\sqrt{2}}\;.
    \label{eq:fields}
\end{equation}

The scalars $\phi_h$, $\phi_S$, $\phi_X$ mix into the physical 
states $h_1$, $h_2$ and $h_3$ according to

\begin{equation}
    \begin{pmatrix}
        h_1 \\h_2\\h_3
    \end{pmatrix} = R \begin{pmatrix}
        \phi_h \\\phi_S\\\phi_X
    \end{pmatrix}\;,
\end{equation}

with the rotation matrix $R$ characterized by the angles

\begin{eqnarray}
-\frac{\pi}{2} < \theta_{hS}\;, \theta_{hX}\;, \theta_{SX} < \frac{\pi}{2}\;.
\end{eqnarray}

In our scenario $h_1$ is identified with the SM-like Higgs boson, and $h_2$ and  $h_3$ are two new physical \textit{heavier} scalars obeying the mass hierarchy

\begin{eqnarray}
M_1 \leq M_2\leq M_3\;.
\end{eqnarray}

The identification of $h_1$ as the SM-like scalar fixes 

\begin{center}
\begin{eqnarray}
M_1&\backsimeq& 125\,\rm{GeV},\nonumber\\
v&\backsimeq&246\,\rm{GeV}.
\end{eqnarray}
\end{center}

This leaves us with $7$ independent parameters, which we chose as

\begin{eqnarray}
M_2\;, M_3\;, \theta_{h S}\;, \theta_{h X}\;, \theta_{S X}\;, v_S, v_X\;.	
\label{eq:freeparam}
\end{eqnarray}

As this model contains three CP even neutral scalars, double resonance enhanced production of $h_1\,h_1\,h_1$ is possible and can be realized according to

\begin{equation}\label{eq:chain}
p p \,\rightarrow\, h_3\,\rightarrow\,h_2\,h_1\,\rightarrow\,h_1\,h_1
\,h_1,
\end{equation}
where $h_{1,2,3}$ are the physical scalar states of a model with an extended scalar sector. 
Depending on the values 
that the free parameters of eq.~(\ref{eq:freeparam}) assume, different realisations 
of the TRSM are possible, yielding a rich phenomenology at colliders. Here we concentrate on the ``Benchmark Plane 3'' {\bf{(BP3)}} addressed in~\cite{Robens:2019kga}, which was carefully tailored to allow for a large region in the $(M_2, M_3)$ plane which obeys all current theoretical and experimental constraints, while at the same time allowing for a large $h_1 h_1 h_1$ decay rate. \textbf{BP3} is characterised by the numerical values of the parameters shown in table \ref{tab:BP3}.

\begin{table}
\centering
\begin{tabular}{cc}
Parameter & Value \\	
\hline
$M_1$ & $125.09~\rm{GeV}$\\
$M_2$ & $[125,~500]~\rm{GeV}$\\
$M_3$ & $[255,~650]~\rm{GeV}$\\
$\theta_{hS}$ & $-0.129$\\
$\theta_{hX}$ & $0.226$ \\
$\theta_{SX}$ & $-0.899$\\
$v_S$ & $140~\rm{GeV}$\\
$v_X$ & $100~\rm{GeV}$\\
\end{tabular}	
\caption{\label{tab:BP3} The numerical values for the independent parameter values of eq.~(\ref{eq:freeparam}) that characterise \textbf{BP3}. The Higgs doublet VEV, $v$, is fixed to 246~GeV. The $\kappa_i$ values correspond to the rescaling parameters of the SM-like couplings for the respective scalars and are derived quantities.}
\end{table}

\begin{center}
\begin{figure}
\begin{center}
 \includegraphics[width=0.65\columnwidth]{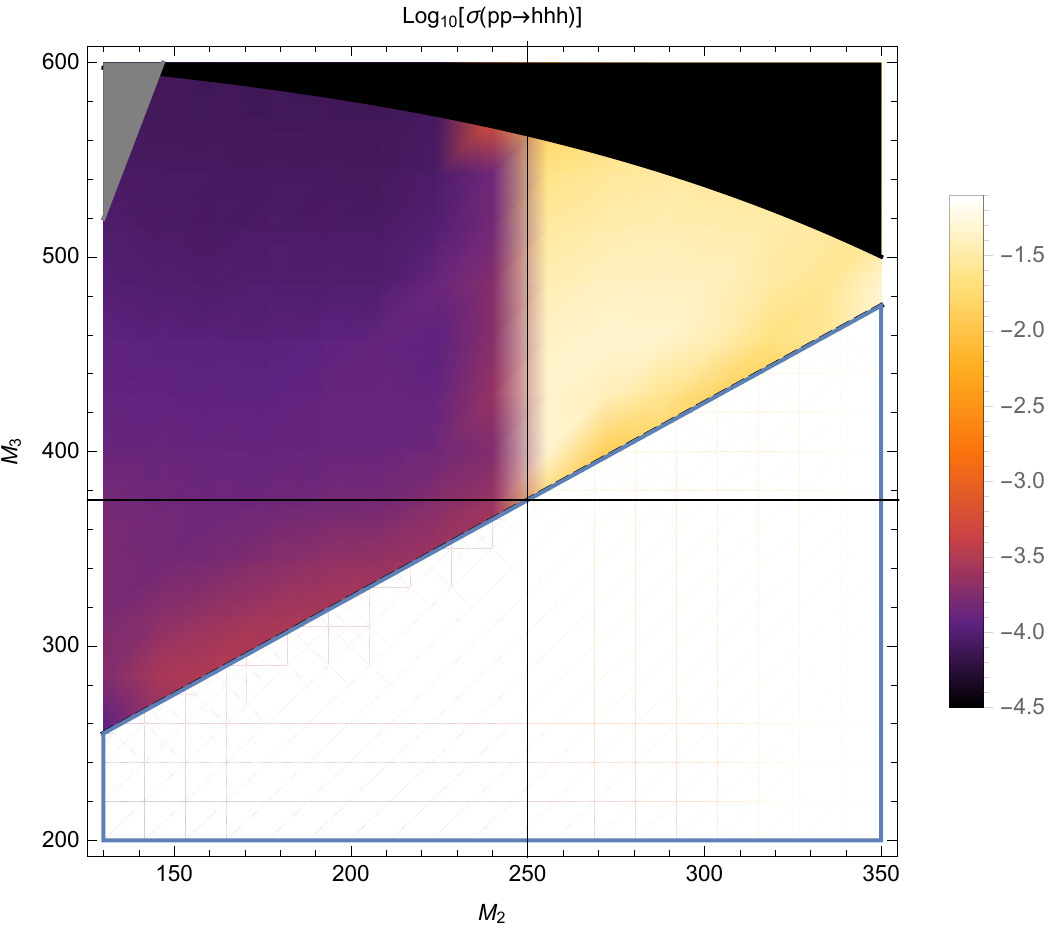}
\caption{The total leading-order gluon-fusion production cross sections for the $p\,p\,\rightarrow\,h_1\,h_1\,h_1$ process at a 14 \TeV~ LHC. No cuts have been imposed. We also show the region excluded by constraints coming from perturbative unitarity in the dark upper part and boundedness from below in the gray wedge. In the allowed region, the leading-order predictions reach cross-section values of up to $\sim 50 \,\fb$. }
\label{fig:crosssections_m2m3}
\end{center}
\end{figure}
\end{center}

The values of the cross sections in the plane $[M_2, M_3]$ are given in Fig.~\ref{fig:crosssections_m2m3}. It can be seen that the regions with maximal values occur when $h_2$ and $h_3$ are produced on-shell.

Note that several regions in that plane are already ruled out by current LHC data, as e.g. $h_{2/3}\,\rightarrow\,h_1\,h_1$ \cite{CMS:2018ipl,ATLAS:2018rnh,CMS:2018qmt,ATLAS:2019qdc}, $h_3\,\rightarrow\,Z\,Z$ \cite{ATLAS:2020tlo}, as well as $h_3\,\rightarrow\,h_1\,h_2$ searches \cite{CMS:2021yci}, see e.g. \cite{Robens:2023pax,Robens:2023oyz}. Concerned are regions for which $M_3\,\lesssim\,350\,-\,450\,\GeV$ or $M_2\,\lesssim\,140\,\GeV$.

The results presented here have been presented in \cite{Papaefstathiou:2019ofh}, to which we refer the reader for more details on the model as well as analysis setup. For reference, we here briefly list the most important details.

An event is analysed if it contains at least $6$ $b$-tagged jets with a transverse
momentum of at least $p_{Tmin, b}=25$~GeV and a pseudo-rapidity no greater than $|\eta_{b, max}|=2.5$. These initial cuts are further optimised for each of our signal samples, which are characterised by different combinations of $M_2$ and $M_3$.

We then select the $6$ $b$-tagged jets with the highest transverse momentum and form pairs in different combinations, with the aim of first reconstructing individual SM-like Higgs bosons, $h_1$, and subsequently the two scalars $h_2$ and $h_3$. Thus, we introduce two observables:

\begin{eqnarray}
\label{eq:chi4}
\chi^{2, (4)}&=&\sum_{q r \in I}\Bigl(M_{q r} - M_1 \Bigl)^2\;,
\end{eqnarray}

\begin{eqnarray}
\label{eq:chi6}
\chi^{2, (6)}&=&\sum_{q r\in J}\Bigl(M_{q r} - M_1 \Bigl)^2\;,
\end{eqnarray}

where we have defined the sets $I=\{i_1 i_2, i_3 i_4\}$ and $J=\{j_1 j_2, j_3 j_4, j_5 j_6\}$, constructed from different pairings of 4 and 6 $b$-tagged jets, respectively.  Moreover, $M_{q r}$ denotes the invariant mass of the respective pairing, $q r$. It should be understood that each jet can appear only in a single arrangement inside $I$ and $J$. We select the combinations of $b$-tagged jets entering in $I$ and $J$ based on the minimisation of the sum
 
 \begin{eqnarray}
 \chi^{2, (6)} + \chi^{2, (4)}\;.
 \end{eqnarray}

The optimisation of the analysis is based on the sequential application of cuts on the different observables
including $p_{Tmin,b}$, $|\eta_b|$,  $\chi^{2, (6)}$, $\chi^{2, (4)}$, $m^{\rm inv}_{6b}$,  $m^{\rm inv}_{4b}$. In addition we consider observables affecting  the pairings of $b$-jets  which define the combinations of six and four elements: (v) $p_{T}(h^i_1)$, (vi) $(\Delta m_{\rm min,~med,~max})$, (vii) $\Delta R(h^i_1, h^j_1)$, (viii) $\Delta R_{bb}(h^i_1)$. We optimize for cuts on the different observables by constructing a grid over each one of them and exploring sequentially combinations of cuts which deliver the maximum rejection of the background while maintaining the highest acceptance for the signal.  The specific values for the cuts depend on the combination of masses for the physical scalars $h_2$ and $h_3$.

We show the results after these selection cuts in table \ref{table:efficiencies}. Note that we show significances with and without taking systematic uncertainties into account. For more details on the actual selection process, we refer the reader to \cite{Papaefstathiou:2019ofh}. 

\begin{center}
\begin{figure}
\begin{center}
\includegraphics[width=0.45\textwidth]{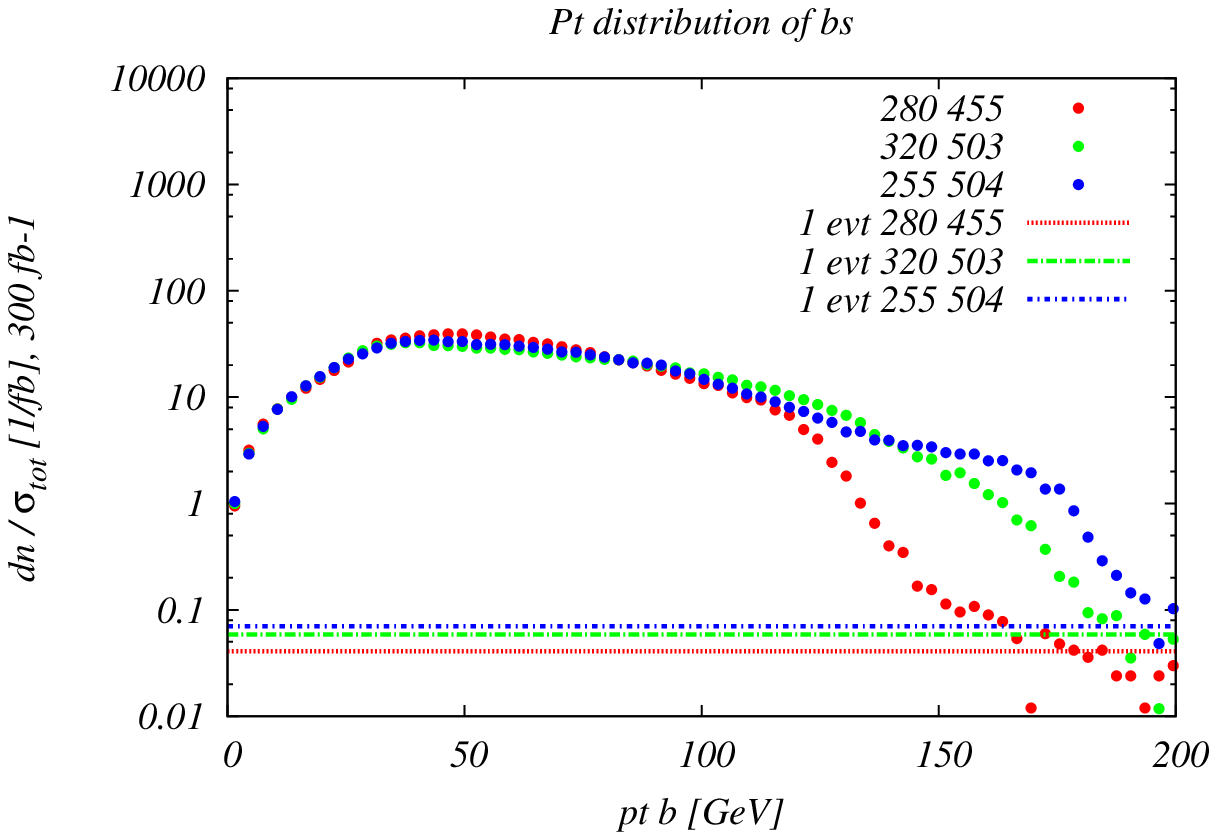}
\includegraphics[width=0.45\textwidth]{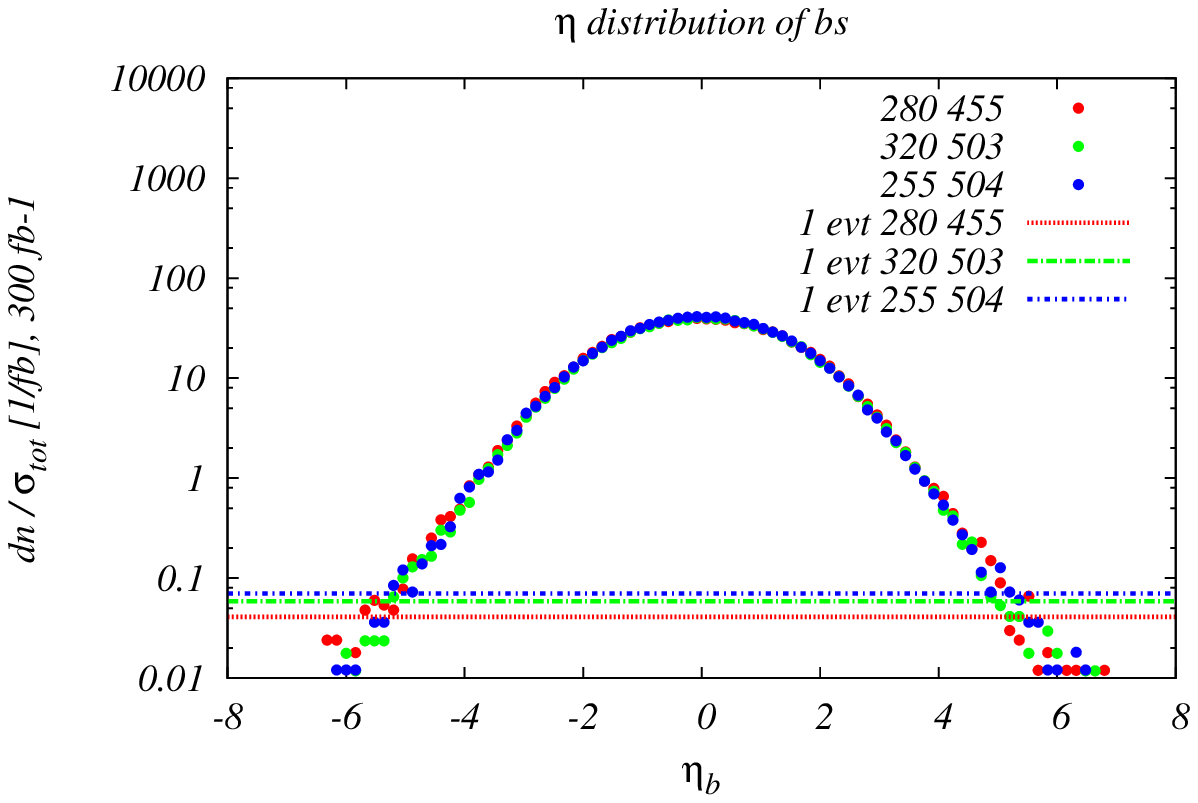}
\caption{\label{fig:bdistis} For various benchmark points from table \ref{table:efficiencies}, we show the $p_\perp$ distribution {\sl (left)} as well as $\eta_b$ distribution {\sl (right)} of the $b$-jets, normalized to the respective production cross sections at 13 \TeV. Shown are points G (280,455) {\sl (red)}, E (320,503) {\sl (green)}, and A (255,504) {\sl (blue)}. Also displayed are lines that would represent 1 event for an integrated luminosity of $300\,\fb^{-1}$. While the rapidity distributions do not display significant differences, the $p_\perp$ distributions show large differences, with a major dependence on the absolute scale $M_3$.}
\end{center}
\end{figure}
\end{center}

We also provide some distributions for the b-jets $p_\perp$ and pseudorapidity in figure \ref{fig:bdistis}. Events have been generated using the TRSM model file available at \cite{gitlabtrsm}, with leading order event generation using Madgraph \cite{Alwall:2011uj}. Distributions have been obtained within the Madanalysis framework\cite{Conte:2012fm}.

\begin{table*}[t!]
\begin{tabular*}{\textwidth}{@{\extracolsep{\fill}}cccccccc@{}}
Label&$(M_2, M_3)$ & $\varepsilon_{\rm Sig.}$& $\rm{S}\bigl|_{300\rm{fb}^{-1}}$ & $\varepsilon_{\rm Bkg.}$ & 
$\rm{B}\bigl|_{300\rm{fb}^{-1}}$ & $\text{sig}|_{300\rm{fb}^{-1}}$ & $\text{sig}|_{3000\rm{fb}^{-1}}$\\
& [GeV] & & & & & (syst.) &(syst.)  
\\
\hline
\textbf{A} &$(255, 504)$ & $0.025$ & $14.12$  & $8.50\times 10^{-4}$ & $19.16$ & $2.92~(2.63)$&$9.23~(5.07)$\\
\textbf{B} & $(263, 455)$ & $0.019$ & $17.03$    & $3.60\times 10^{-5}$ & $ {8.12}$ & $4.78 ~(4.50)$&$15.10~(10.14)$\\
\textbf{C} & $(287, 502)$ & $0.030$ & $20.71$ & $9.13\times 10^{-5}$ & $20.60$  & $4.01~(3.56)$ & $12.68~(6.67)$\\
\textbf{D} & $(290, 454)$ & $0.044$ & $37.32$    & $1.96\times 10^{-4}$ & $44.19$& $5.02~(4.03)$&$15.86~(6.25)$\\
\textbf{E} & $(320, 503)$ & $0.051$ & $ {31.74}$    & $2.73\times 10^{-4}$ & $61.55$& $3.76~( {2.87}) $&$11.88~(4.18)$\\
\textbf{F} & $(264,504)$&$0.028$& $18.18$&$9.13\times 10^{-5}$&$20.60$&$3.56~(3.18) $&$11.27~(5.98)$\\
\textbf{G} & $(280, 455)$&$0.044$& $38.70$ &$1.96\times 10^{-4}$& $44.19$ & $5.18~(4.16)$ &$16.39~(6.45)$\\
\textbf{H} & $(300, 475)$ & $0.054$& $41.27$ & $2.95\times 10^{-4}$& $66.46$ & $4.64~(3.47)$&$ 14.68~( {4.94})$\\
\textbf{I} & $(310, 500)$& $0.063$& $41.43$& $3.97\times 10^{-4}$& $89.59$& $4.09~(2.88) $&$ {12.94~(3.87)}$\\
\textbf{J} & $(280,500)$& $0.029 $& $20.67$&$9.14\times 10^{-5}$& $20.60$&$4.00~(3.56) $&$12.65~(6.66)$\\
\end{tabular*}
\caption{ The resulting selection efficiencies, $\varepsilon_{\rm Sig.}$ and $\varepsilon_{\rm Bkg.}$, number of events, $S$ and $B$ for the signal and background, respectively. A $b$-tagging efficiency of $0.7$ has been assumed. The number of signal and background events are provided at an integrated luminosity of $300~\rm{fb}^{-1}$. Results for $3000~\rm{fb}^{-1}$ are obtained via simple extrapolation. The significance is given at both values of the integrated luminosity excluding (including) systematic errors in the background. Table taken from \cite{Papaefstathiou:2019ofh}.}
\label{table:efficiencies}
\end{table*}

\begin{center}
\begin{figure}[htb]
\begin{center}
\includegraphics[width=0.48\textwidth]{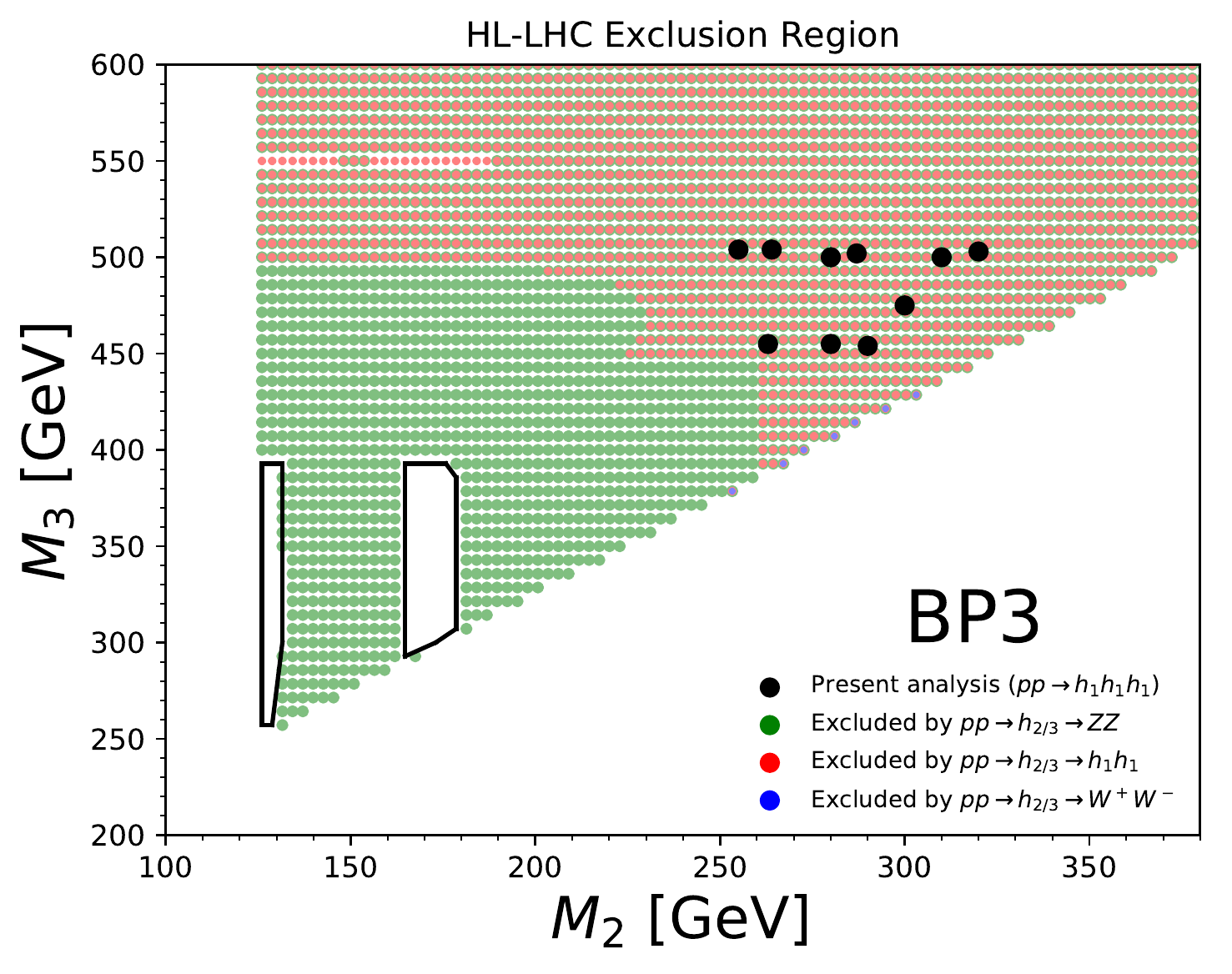}
\end{center}
\caption{Constraints on the $\lb M_2,\,M_3 \rb$ plane from extrapolation of other searches at the HL-LHC from extrapolation (see text for details). Taken from \cite{Papaefstathiou:2020lyp}.}
\label{fig:hlothers}
\end{figure}
\end{center}

Another important question is whether the benchmark points discussed above could already be tested by other channels at the HL-LHC, e.g. via heavy resonance production decaying into a pair of (vector)-bosons. For this, we have extrapolated various analyses assessing the heavy Higgs boson prospects of the HL-LHC in final states originating from $h_i \rightarrow h_1 h_1$ \cite{Sirunyan:2018two,Aad:2019uzh}, $h_i \rightarrow ZZ$ \cite{Sirunyan:2018qlb,Cepeda:2019klc} and $h_i \rightarrow W^+W^-$ \cite{Aaboud:2017gsl,ATL-PHYS-PUB-2018-022}, for $i=2,3$, and combined these with extrapolations of results from 13 TeV where appropriate. We display the results in figure \ref{fig:hlothers}.

In particular $ZZ$ final states can probe nearly all of the available parameter space. However, such searches do investigate different parts of the potential, and therefore can be seen as complementary.

\subsection{Other theory scenarios \label{sec:tr_other}}

{\sl T. Robens}

We here briefly discuss other scenarios that lead to triple scalar final states within the TRSM, as well as other new physics scenarios that can lead to triple scalar final states. TRSM benchmark planes follow the discussion in \cite{Robens:2019kga,Robens:2022nnw}, with more recent updates available in \cite{Robens:2023pax,Robens:2023oyz}.

In addition to the benchmark plane discussed above, two more scenarios can render interesting triple scalar final states. The first one is BP1, where the heaviest scalar is associated with the 125 \GeV~ resonance at the LHC. For this parameter plane, the input parameters are specified in table \ref{tab:BPparams}. 

\begin{table} 
    \centering
    \begin{tabular}{crrrrrr}
        \hline
        Parameter           & \multicolumn{3}{c }{Benchmark scenario}                                                                             \\
                            & \textbf{BP1}                            & \textbf{BP3} &  \textbf{BP6}  \\
        \hline
        $M_1~[ {\GeV}]$ & $[1, 62]$      & $125.09$     & $125.09$      \\
        $M_2~[ {\GeV}]$ & $[1, 124]$                              & $[126, 500]$ & $[126, 500]$  \\
        $M_3~[ {\GeV}]$ & $125.09$                                & $[255, 650]$ & $[255, 1000]$ \\
        $\theta_{hs}$       & $1.435$                              & $-0.129$     & $0.207$       \\
        $\theta_{hx}$       & $-0.908$                             & $0.226$     & $0.146$       \\
        $\theta_{sx}$       & $ -1.456$                            & $-0.899$     & $0.782$       \\
        $v_s~[ {\GeV}]$ & $630$                                  & $140$        & $220$         \\
        $v_x~[ {\GeV}]$ & $700$                                   & $100$        & $150$         \\
        \hline
        $\kappa_1$          & $0.083$                               & $0.966$      & $0.968$       \\
        $\kappa_2$          & $0.007$                             & $0.094$      & $0.045$       \\
        $\kappa_3$          & $-0.997$                           & $0.239$      & $0.246$       \\
        \hline
    \end{tabular}
    \caption{Input parameter values and coupling scale factors, $\kappa_a$
    ($a=1,2,3$), for the three benchmark scenarios discussed here. The doublet VEV is set
    to $v= {246}{\GeV}$ for all scenarios. Table adapted from \cite{Robens:2019kga}.}
    \label{tab:BPparams}
\end{table}

\begin{center}
\begin{figure}
\begin{center}
\includegraphics[width=0.5\textwidth]{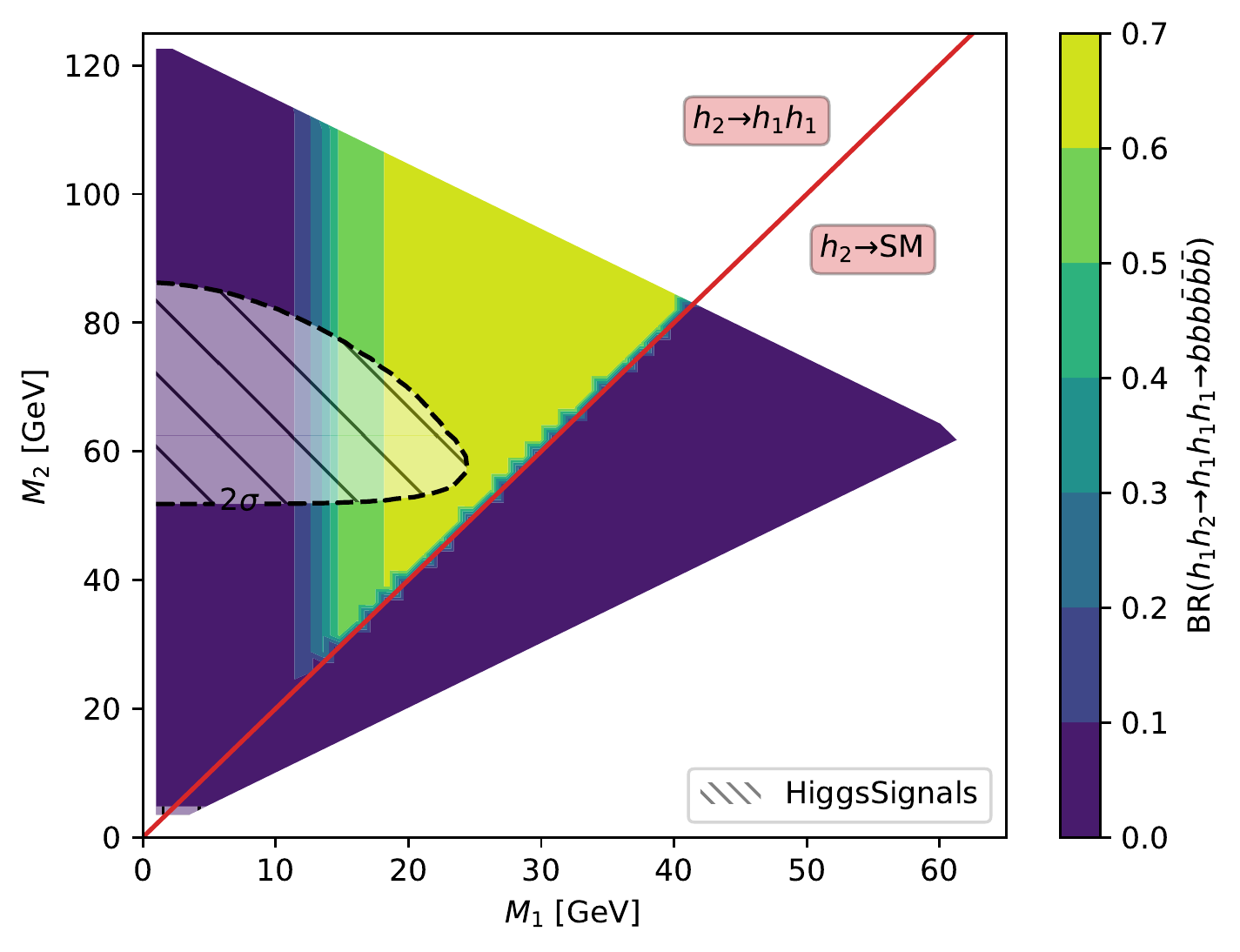}
\caption{\label{fig:bp1_bbbbbb} Branching ratio into $6\,b$ final states for BP1, with triple light scalar productions. Cross sections can reach up to around 2 \pb. Dominant constraints stem from signal strength measurements. Figure taken from \cite{Robens:2019kga}.}
\end{center}
\end{figure}
\end{center}

For the BP1, we show the allowed benchmark plane in figure \ref{fig:bp1_bbbbbb}, where we have already included the branching ratio to $6\,b$ final states that can reach up to $70\%$ depending on the specific mass range. Note that here due to the BP assumptions light scalars have masses $\lesssim\,40\,\GeV$, leading to relatively soft decay products that might be difficult to trigger. Production cross section for the $h_{125}\,\GeV$ scalar is around $48\,\pb$ at 13 \TeV.

\begin{center}
\begin{figure}
\begin{center}
\includegraphics[width=0.45\textwidth]{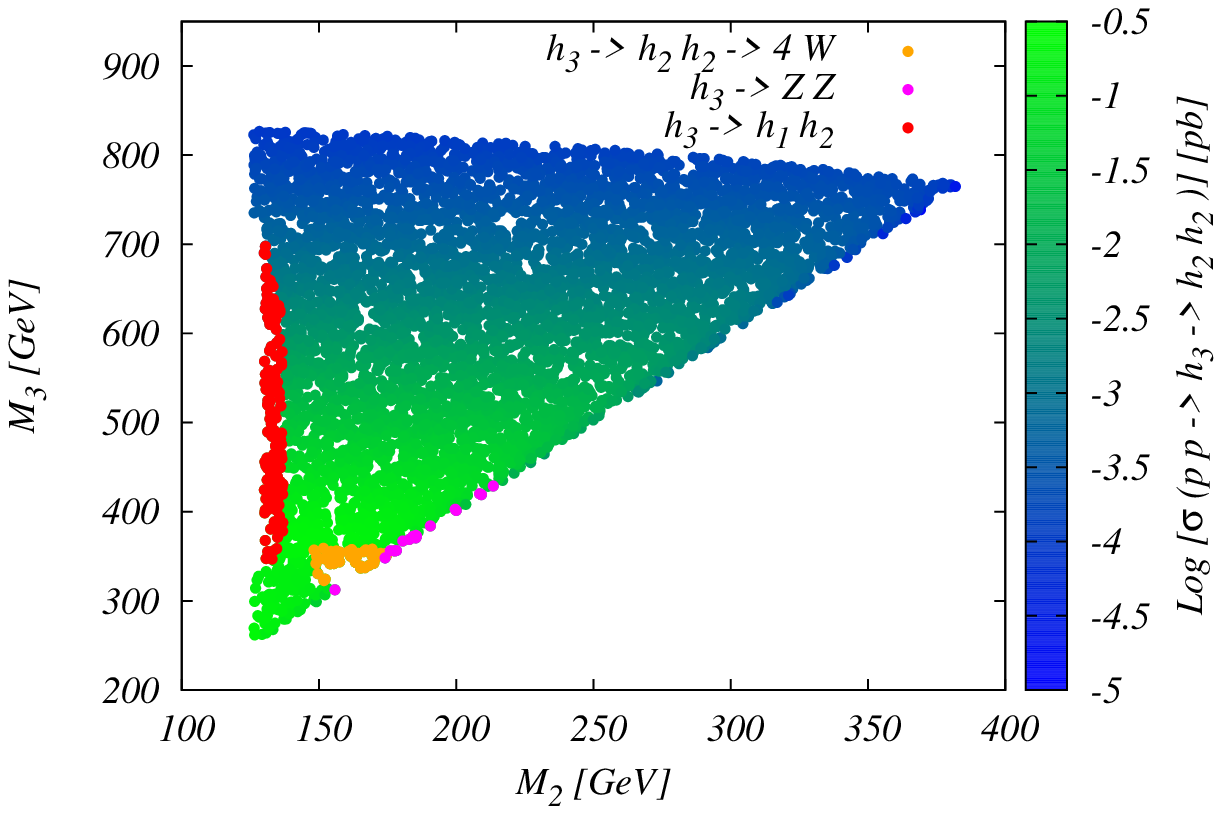}
\includegraphics[width=0.45\textwidth]{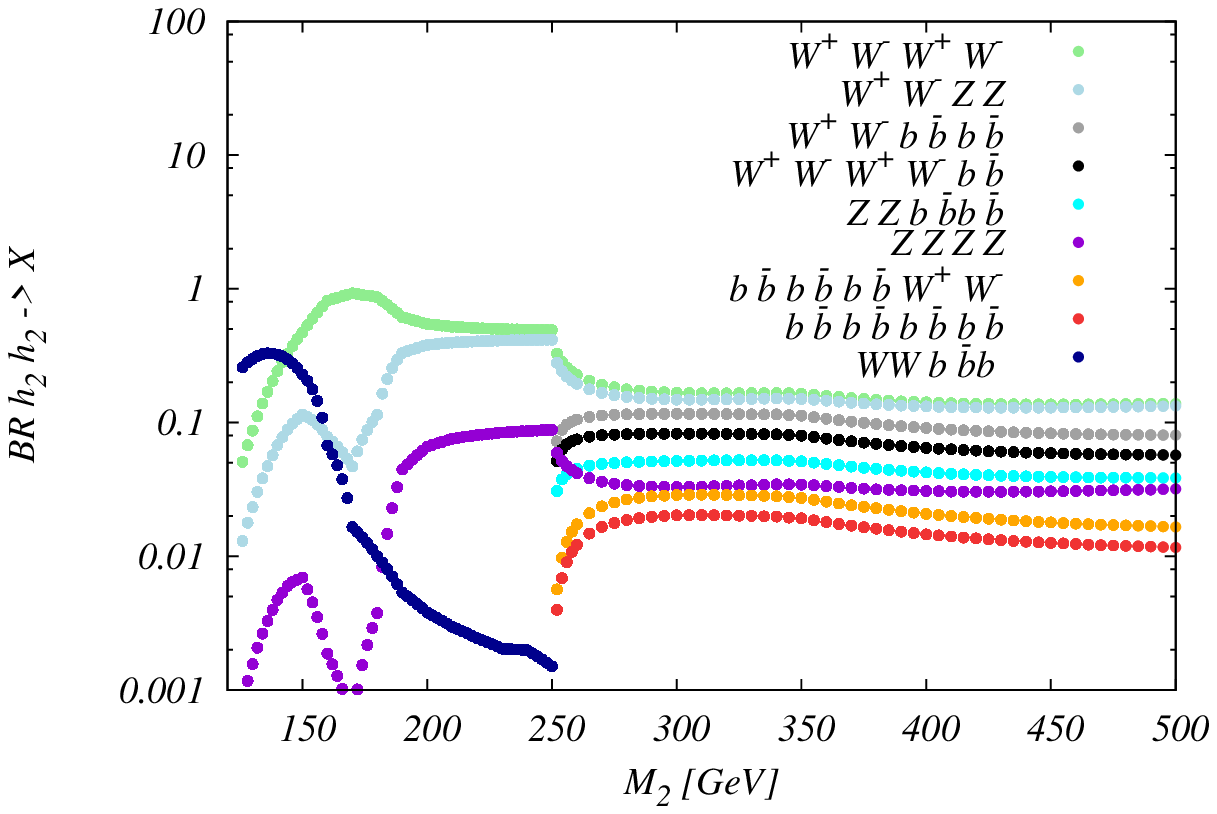}
\caption{\label{fig:bp6} {\sl Left:} Total rate for the $h_2\,h_2$ final state at 13 \TeV in BP6. Exclusions stem from $4\,W$ \cite{ATLAS:2018ili}, $ZZ$ \cite{ATLAS:2020tlo}, and $h_1\,h_2$ \cite{CMS:2021yci} searches. {\sl Right:} Branching ratios of the $h_2\,h_2$ final state as a function of $M_2$.  Figures taken from \cite{Robens:2023pax,Robens:2022nnw}.}
\end{center}
\end{figure}
\end{center}

Another point of interest is BP6, which was targeted for the $p\,p\,\rightarrow\,h_3\,\rightarrow\,h_2\,h_2$ production and subsequent decays, where $M_{2,3}\,\geq\,125\,\GeV$. We display the allowed parameter space in figure \ref{fig:bp6}, and parameters are again defined in table \ref{tab:BPparams}.

For this mass plane, production cross sections can reach up to $0.5\,\pb$ in the low mass region. As soon as the decay $h_2\,\rightarrow\,h_1\,h_1$ is kinematically allowed, interesting novel final states are possible, as e.g. $W^+\,W^-\,b\,\bar{b}\,b\,\bar{b}$ or $W^+\,W^-\,W^+\,W^-\,b\,\bar{b}$. 

\begin{center}
\begin{figure}
\begin{center}
\includegraphics[width=0.45\textwidth]{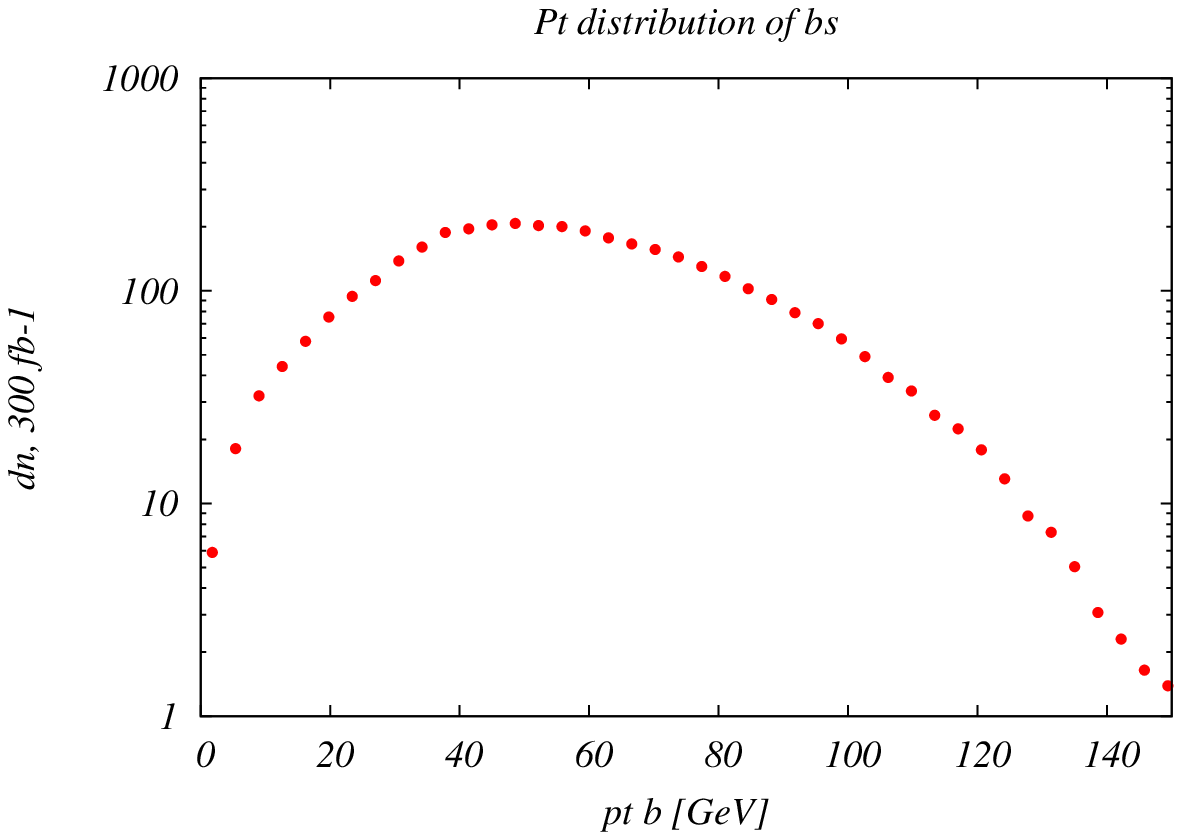}
\includegraphics[width=0.45\textwidth]{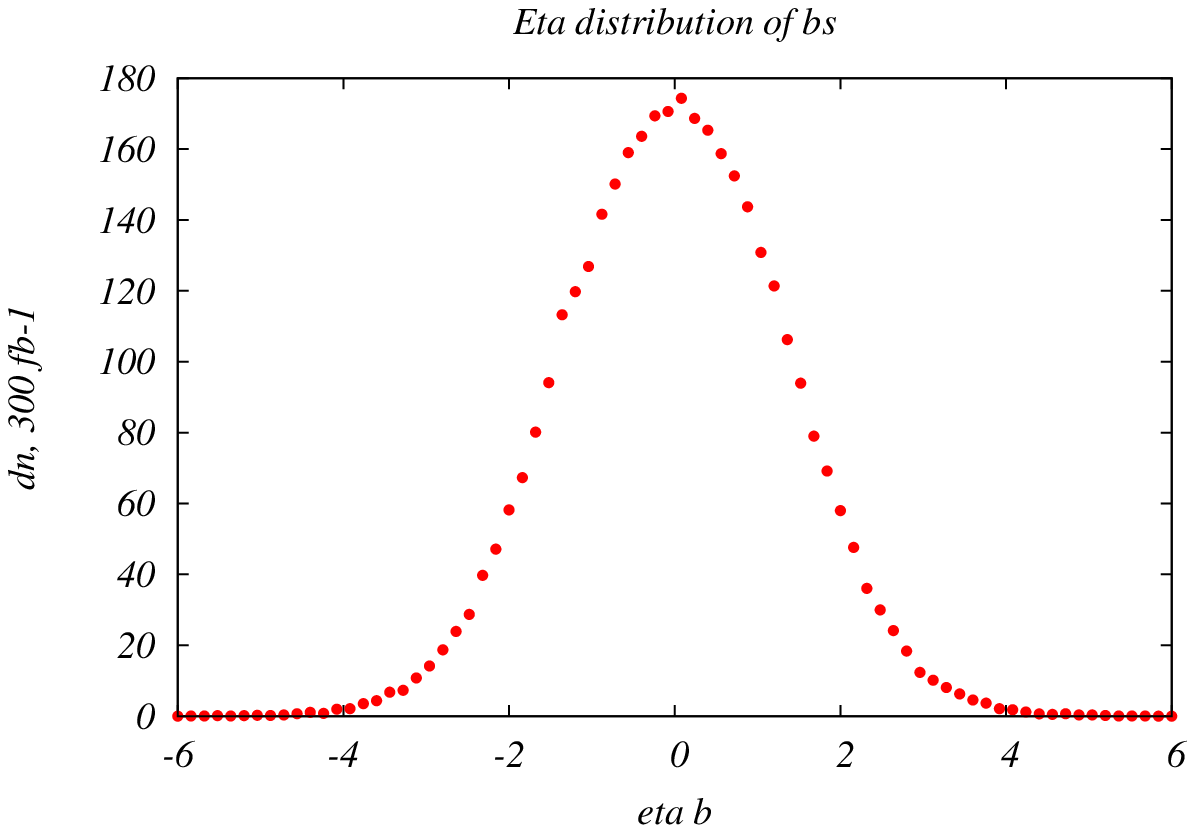}
\caption{\label{fig:distis} Distributions of the b-jets for the process $p\,p\,\rightarrow\,h_3\,\rightarrow\,h_2,h_2\,\rightarrow\,h_2\,h_1\,h_1\,\rightarrow\,W^+\,W^-\,b\,\bar{b}\,b\,\bar{b}$.  }
\end{center}
\end{figure}
\end{center}

In Figure \ref{fig:distis}, we furthermore provide the distributions for the b-jets for a sample point where $M_2\,=\,279\,\GeV,\,M_3\,=\,583\,\GeV$, and $\sigma_{h_1\,h_2}\,=\,185\,\fb$, and where we consider the $W^+\,W^-\,b\,\bar{b}\,b\,\bar{b}$ final state with a total rate $\sim\,21\,\fb$. 

\subsection{Cascade Higgs-to-Higgs decays in the C2HDM, N2HDM and NMSSM} \label{subsec:introduction}

{\sl H. Abouabid, A. Arhrib, D. Azevedo, J. El Falaki, P. Ferreira, M. M\"{u}hlleitner, R. Santos}

In non-minimal Higgs models multi-Higgs final states may arise from cascade Higgs-to-Higgs decays (see for instance a recent study of Ref.~\cite{Ferreira:2022sno}). 
In some models, the cross sections can still be probed during the next LHC run. Moreover, there are scenarios where double Higgs
production becomes more relevant than single Higgs production.
We will discuss three extensions of the Standard Model where these processes are relevant: the complex two-Higgs Doublet Model (C2HDM),
the Next-to-2HDM (N2HDM) and the Non-Minimal Supersymmetric extension
of the SM (NMSSM). A discussion of the models can be found 
in~\cite{Abouabid:2021yvw} where a thorough discussion on double Higgs production in these models is presented. Here we will 
just present very briefly the models and the constraints they are subject to. 

The NMSSM~\cite{Barbieri:1982eh,Dine:1981rt,Ellis:1988er,Drees:1988fc,Ellwanger:1993xa,Ellwanger:1995ru,Ellwanger:1996gw,Elliott:1994ht,King:1995vk,Franke:1995tc,Maniatis:2009re,Ellwanger:2009dp} 
solves the little hierarchy
problem and more easily complies with the discovered SM-like Higgs
mass after inclusion of the higher-order corrections
\cite{Slavich:2020zjv}. The Higgs sector consists of two Higgs
doublets to which a complex singlet superfield is added so that after
electroweak symmetry breaking (EWSB) we have three neutral CP-even,
two neutral CP-odd and two charged Higgs bosons in the spectrum.
Supersymmetric (SUSY) relations constrain the
Higgs potential parameters in a different way than non-SUSY models.
Therefore, we also investigate
non-SUSY Higgs sector extensions where the trilinear couplings are
less constrained from a theoretical point of view. This way we make sure
not to miss some possibly interesting di-Higgs signatures. We start
with one of the most popular extensions complying with $\rho=1$
at tree level, the C2HDM~\cite{Lee:1973iz, Branco:1985aq,Weinberg:1990me, Ginzburg:2002wt, Fontes:2017zfn}  where a second Higgs doublet is added to
the SM sector. Incorporating a minimal set of BSM Higgs bosons (five
in total, three neutral and two charged ones) allows for resonant
di-Higgs enhancement \cite{Grober:2017gut}. 
In this model there are three CP-mixed and two
charged Higgs bosons. In this case the SM-like Higgs couplings can be
diluted by CP admixture, the same happens through
singlet admixture. Thus, light Higgs bosons may not be excluded yet
because they may have escaped discovery through small couplings to the SM
particles. Such a singlet admixture is realized in the next-to-2HDM
(N2HDM) \cite{Chen:2013jvg,Muhlleitner:2016mzt,Engeln:2018mbg} as well as in the complex-singlet extensions of 2HDM with anomaly-free U(1)~\cite{Ordell:2019zws,Ordell:2020yoq}. By adding, for instance, a real singlet field to the 2HDM Higgs sector the Higgs spectrum then consists
of three neutral CP-even Higgs bosons, one neutral CP-odd and two
charged Higgs bosons, allowing for the possibility of Higgs-to-Higgs
cascade decays. This is also possible in the C2HDM and the NMSSM. For simplicity, we will focus on the type I and II versions of the
C2HDM and N2HDM. 

\subsubsection{Models and scans \label{sec:models}}

In this section we just briefly review the models and refer to Ref.~\cite{Abouabid:2021yvw} for details.

\bigskip
\noindent
\textbf{The Real and Complex 2HDM}\\[0.1cm]

The 2HDM was first proposed by Lee in 1973~\cite{Lee:1973iz} to provide an extra source of CP violation via spontaneous symmetry breaking.
The version considered here has a softly broken discrete
$\mathbb{Z}_2$ symmetry under $\Phi_1 \to \Phi_1\;$ and $\Phi_2 \to - \Phi_2\;$. In terms of the two $SU(2)_L$ Higgs
doublets $\Phi_{1,2}$ with hypercharge $Y=+1$, the most general scalar
potential which is $SU(2)_L\times U(1)_Y$ invariant and possesses a softly broken
$\mathbb{Z}_2$ symmetry is given by
\beq
V_{\text{(C)2HDM}} &=& m_{11}^2 |\Phi_1|^2 + m_{22}^2 |\Phi_2|^2 - m_{12}^2 (\Phi_1^\dagger
\Phi_2 + h.c.) + \frac{\lambda_1}{2} (\Phi_1^\dagger \Phi_1)^2 +
\frac{\lambda_2}{2} (\Phi_2^\dagger \Phi_2)^2 \nonumber \\
&& + \lambda_3
(\Phi_1^\dagger \Phi_1) (\Phi_2^\dagger \Phi_2) + \lambda_4
(\Phi_1^\dagger \Phi_2) (\Phi_2^\dagger \Phi_1) + \left[ \frac{\lambda_5}{2}
(\Phi_1^\dagger \Phi_2)^2 + H.c. \right] \;.
\label{eq:2hdmpot}
\eeq
The $\mathbb{Z}_2$ symmetry is introduced in the model in order to avoid tree-level flavour-changing
neutral currents (FCNCs) mediated by the neutral scalar. Since the $\mathbb{Z}_2$ symmetry is extended
to the fermion sector, it will force all families of same-charge fermions to couple to a single
doublet which eliminates tree-level
FCNCs~\cite{Glashow:1976nt,Branco:2011iw}. This implies
four different types of doublet couplings to the fermions listed in
Tab.~\ref{tab:yuycoup}.
\begin{table}
\begin{center}
\begin{tabular}{rccc|ccccc}
\hline
& $u$-type & $d$-type & leptons & $Q$ & $u_R$ & $d_R$ & $L$ & $l_R$ \\
  \hline
type I & $\Phi_2$ & $\Phi_2$ & $\Phi_2$ & + & $-$ & $-$ & + & $-$ \\
type II & $\Phi_2$ & $\Phi_1$ & $\Phi_1$ & + & $-$ & + & + & $-$ \\
flipped (FL) & $\Phi_2$ & $\Phi_1$ & $\Phi_2$ & + & $-$ & $-$ & + & + \\
 lepton-specific (LS) & $\Phi_2$ & $\Phi_2$ & $\Phi_1$ & + & $-$ & + & + & $-$
  \\ \hline
\end{tabular}
\caption{Four left rows: The four Yukawa types of the
  $\mathbb{Z}_2$-symmetric 2HDM, stating which Higgs
  doublet couples to the different fermion types. Five right columns:
  Corresponding $\mathbb{Z}_2$
  assignment for the quark doublet $Q$, the up-type
  quark singlet $u_R$, the down-type quark singlet $d_R$, the lepton
  doublet $L$, and the
  lepton singlet $l_R$. \label{tab:yuycoup}}
\end{center}
\end{table}

In the CP-violating version of the 2HDM, the C2HDM, the parameters $m_{12}^2$ and
$\lambda_5$ can be complex. The two complex doublet fields can be parametrised as
\beq
\Phi_i = \left( \begin{array}{c} \phi_i^+ \\ \frac{1}{\sqrt{2}} (v_i +
    \rho_i + i \eta_i) \end{array} \right) \;,  \ \ i=1,2 \;,
    \label{eq:phidef}
\eeq
with $v_{1,2}$ being the vacuum expectation values (VEVs) of the two
doublets $\Phi_{1,2}$. 
After EWSB three of the eight degrees of freedom initially present in
$\Phi_{1,2}$ are taken by the Goldstone bosons to give masses to the
gauge bosons $W^\pm$ and $Z$, and we are left with five
physical Higgs bosons. In the C2HDM, the three neutral Higgs bosons mix, resulting in three
neutral Higgs mass eigenstates $H_i$ ($i=1,2,3$) with no definite CP
quantum number and which by 
convention are ordered as $m_{H_1} \le m_{H_2} \le m_{H_3}$. The
rotation matrix $R$ diagonalising the neutral Higgs sector can be
parametrised in terms of three mixing angles $\alpha_i$ ($i=1,2,3$) as
\beq
R = \left( \begin{array}{ccc}
c_{1} c_{2} & s_{1} c_{2} & s_{2}\\
-(c_{1} s_{2} s_{3} + s_{1} c_{3})
& c_{1} c_{3} - s_{1} s_{2} s_{3}
& c_{2} s_{3} \\
- c_{1} s_{2} c_{3} + s_{1} s_{3} &
-(c_{1} s_{3} + s_{1} s_{2} c_{3})
& c_{2}  c_{3}
\end{array} \right) \;, \label{eq:rmixmatrixc2hdm}
\eeq
where $s_i \equiv \sin \alpha_i$, $c_i \equiv \cos \alpha_i$, and,
without loss of generality, the angles vary in the range
\beq
- \frac{\pi}{2} \le \alpha_i \le \frac{\pi}{2} \;.
\eeq
We also define 
\beq
\tan\beta = \frac{v_2}{v_1} \; 
\eeq
and  identify
\beq
v = \sqrt{v_1^2 + v_2^2} \;,
\eeq
where $v$ is the SM VEV, $v \approx 246$~GeV.  

In the C2HDM the three neutral Higgs boson masses are not independent. The third
neutral Higgs mass is a dependent quantity and is obtained from the
input parameters, {\it cf.}~\cite{ElKaffas:2007rq}. We choose two of the three
neutral Higgs boson masses as input values and calculate the third one. The
chosen input masses are called $m_{H_i}$ and $m_{H_j}$ with $H_i$ per
default denoting the lighter one, {\it i.e.}~$m_{H_i} <
m_{H_j}$. They denote any two of the three neutral Higgs
bosons among which we take one to be the 125 GeV SM-like scalar. 
We furthermore replace the three mixing
angles $\alpha_{1,2,3}$ by two coupling values of $H_i$ and by a matrix
element of our rotation matrix. These are the squared $H_i$ couplings to the
massive gauge bosons $V$ and to the top quarks $t$, $c^2_{H_i VV}$
and $c^2_{H_i tt}$, respectively, and the neutral mixing matrix entry
$R_{23}$. We furthermore fix the sign
of $R_{13}$, sg($R_{13}$), to either +1 or $-1$ in order to lift the
degeneracy that we introduce by specifying only the squared values of
the $H_i$ couplings. This choice of input parameters complies with
the input parameters of the program code {\tt ScannerS} that we will use
for our parameter scans as explained below. We hence have the 
input parameter set
\beq
v  \;, \quad \tan\beta \;, \quad c_{H_i VV}^2 \;, \quad c_{H_i tt}^2
\;, \quad R_{23} \;, \quad m_{H_i} \;, \quad m_{H_j} \;, \quad m_{H^\pm} \quad
\mbox{and} \quad \mbox{Re}(m_{12}^2) \;.
\label{eq:c2hdminput}
\eeq
One should notice here that in certain multi-Higgs scenarios, featuring flavour symmetries in both the Higgs and fermion sectors, the next-to-lightest scalars may predominantly couple to the light (first- or second-generation) quarks, thus, significantly altering their production and decay observables compared to conventional searches \cite{Camargo-Molina:2017klw}. A comprehensive analysis of the multi-Higgs production channels in such models is a subject of a future work.

\bigskip
\noindent
\textbf{The N2HDM}\\[0.1cm]

We briefly introduce the N2HDM and refer to~\cite{Muhlleitner:2016mzt} for more details.
The scalar potential of the N2HDM can be obtained from the 2HDM potential
by adding a real singlet field $\Phi_S$.

In terms of the two $SU(2)_L$ Higgs doublets $\Phi_1$ and $\Phi_2$,
defined in  Eq.~(\ref{eq:phidef}), and the singlet field, defined as
\beq
\Phi_S=v_S+ \rho_S \;,
\eeq
the N2HDM potential is given by
\beq
V_{\text{N2HDM}} &=&  V_{\text{2HDM}}+ \frac{1}{2} m_S^2 \Phi_S^2 + \frac{\lambda_6}{8} \Phi_S^4 +
\frac{\lambda_7}{2} (\Phi_1^\dagger \Phi_1) \Phi_S^2 +
\frac{\lambda_8}{2} (\Phi_2^\dagger \Phi_2) \Phi_S^2 \;.
\label{eq:n2hdmpot}
\eeq
The above scalar potential is obtained by imposing two $\mathbb{Z}_2$ symmetries,
\begin{eqnarray}
 && \Phi_1 \to \Phi_1\;, \quad \Phi_2 \to - \Phi_2\;, \quad \Phi_S \to
    \Phi_S \quad \mbox{and} \nonumber\\
 && \Phi_1 \to \Phi_1\;, \quad \Phi_2 \to \Phi_2\;, \quad \Phi_S \to -\Phi_S \;.
\label{eq:Z2SYM}
\end{eqnarray}
The first (softly-broken) $\mathbb{Z}_2$ symmetry is the extension of the usual 2HDM
$\mathbb{Z}_2$ symmetry to the N2HDM which, once extended to the Yukawa sector,
will forbid FCNCs at tree level, implying four different N2HDM
versions just like in the 2HDM, {\it cf.}~Tab.~\ref{tab:yuycoup}.
The second $\mathbb{Z}_2$ symmetry is an exact symmetry which will be spontaneously
broken by the singlet VEV and as such does not allow the model to have a
DM candidate. Other versions of the model choose parameters such that $v_S = 0$
yielding very interesting DM phenomenology, but in the current work we
will not consider these possibilities.

After EWSB, we have three neutral CP-even Higgs bosons $H_{1,2,3}$
with masses ranked as 
$m_{H_1}<m_{H_2}<m_{H_3}$, one neutral CP-odd boson $A$ and a pair of charged
Higgs bosons $H^\pm$. The physical states $H_{1,2,3}$ are obtained
from the weak basis $(\rho_1, \rho_2, \rho_S)$ by an orthogonal
transformation $R$ which is defined by 3 mixing angles
$\alpha_{1,2,3}$ that are in the same range as in the C2HDM.
After exploiting the minimisation conditions, we are left with twelve
independent input parameters for the N2HDM. For the scan, we will again replace
the three mixing angles $\alpha_{1,2,3}$ by the squared $H_1$
couplings to massive gauge bosons $V$ and the top quarks $t$,
$c_{H_1VV}^2$ and $c_{H_1tt}^2$, respectively, and the 
neutral mixing matrix element $R_{23}$, so that our input parameters read
\beq
\tan\beta \ , \ c_{H_1 VV}^2 \ , \ c_{H_1 tt}^2 \ , \ R_{23} \ , \
m_{H_1}\ ,\  m_{H_2}\ ,\  m_{H_3}\ ,  \ m_A \ , \  m_{H^\pm}\ , \ v , \ v_s\ ,   \
\text{and} \ m_{12}^2 \;.
\label{eq:n2hdminput}
\eeq
Like in the 2HDM, we fix sg($R_{13}$) to either +1 or $-1$ in order to lift the
introduced degeneracy through the squared values of
the $H_1$ couplings.

\bigskip
\noindent
\textbf{The NMSSM}\\[0.1cm]

As a supersymmetric benchmark model, we consider the Next-to Minimal Supersymmetric SM (NMSSM)
\cite{Ellis:1988er,Drees:1988fc,Ellwanger:1993xa,Ellwanger:1995ru,Ellwanger:1996gw,Elliott:1994ht,King:1995vk,Franke:1995tc,Maniatis:2009re,Ellwanger:2009dp}.
It extends the two doublet fields $\hat{H}_u$ and
$\hat{H}_d$ of the MSSM by a complex superfield $\hat{S}$.
When the singlet field acquires a non-vanishing VEV, this not
only solves the $\mu$ problem \cite{Kim:1983dt} but, compared to the
MSSM, it also relaxes the tension on the stop mass values that
need to be large for the SM-like Higgs boson mass value to be
compatible with the measured 125.09 GeV. Indeed in supersymmetry
the neutral Higgs masses are given in terms of the gauge parameters at
tree level so that there is an upper mass bound on the lightest
neutral scalar which, in the MSSM, is given by the $Z$ boson
mass. Substantial higher-order corrections to the Higgs boson mass are
therefore required to obtain phenomenologically valid mass
values for the SM-like Higgs boson. The additional singlet contribution to the
tree-level mass of the lightest neutral Higgs boson shifts its mass to
larger values compared to the MSSM prediction, thus no longer requiring large radiative
corrections. The scale-invariant NMSSM superpotential that is added to
the MSSM superpotential $W^{\text{MSSM}}$ reads
\beq
W^{\text{NMSSM}}&=&- \lambda \hat{S} \hat{H}_u \cdot\hat{H}_d + \frac{\kappa}{3}
\hat{S}^3+ W^{\text{MSSM}}\,, \quad \mbox{with} \nonumber\\
W^{\text{MSSM}}&=&-  y_t
\widehat{Q}_3\widehat{H}_u\widehat{t}_R^c + y_b \widehat{Q}_3
\widehat{H}_d\widehat{b}_R^c  + y_\tau \widehat{L}_3 \widehat{H}_d
\widehat{\tau}_R^c \; ,
\label{eq:nmssmsuperpot}
\eeq
where for simplicity we only included the third generation fermion
superfields, given by the left-handed doublet quark ($\widehat{Q}_3$)
and lepton ($\widehat{L}_3$) superfields, and the right-handed singlet
quark ($\widehat{t}_R^c,\widehat{b}_R^c$) and lepton
($\widehat{\tau}_R^c$) superfields. The NMSSM-type couplings $\lambda$
and $\kappa$ are dimensionless and taken real since we consider the
CP-conserving NMSSM. The Yukawa couplings $y_t, y_b, y_\tau$ can
always be taken real. The scalar part of $\hat{S}$ will develop a VEV
$v_S/\sqrt{2}$, which dynamically generates the effective $\mu$
parameter $\mu_{\text{eff}}=\lambda v_S/\sqrt{2}$ through the first term in
the superpotential. The second term, cubic in $\hat{S}$, breaks the
Peccei-Quinn symmetry and thus avoids a massless axion, and
$W^{\text{MSSM}}$ contains the Yukawa interactions. The symplectic
product $x \cdot y = \epsilon_{ij} x^i y^j$ ($i,j=1,2$) is built by
the antisymmetric tensor $\epsilon_{12}= \epsilon^{12} = 1$.
The soft SUSY breaking Lagrangian reads
\beq
\label{eq:Lagmass}
 {\cal L}_{\text{soft,NMSSM}} &=&
 - m_{H_u}^2 | H_u |^2 -  m_{H_d}^2 | H_d|^2 - m_{{\widetilde
     Q}_3}^2|{\widetilde Q}_3^2|-  m_{\widetilde t_R}^2 |{\widetilde t}_R^2|
 -  m_{\widetilde b_R}^2|{\widetilde b}_R^2| -
 m_{{\widetilde L}_3}^2|{\widetilde L}_3^2| \nonumber \\
&& - m_{\widetilde  \tau_R}^2|{\widetilde \tau}_R^2|
+ (y_t A_t H_u \cdot \widetilde Q_3 \widetilde t_R^c - y_b A_b H_d \cdot
\widetilde Q_3  \widetilde  b_R^c - y_\tau A_\tau H_d \cdot \widetilde L_3 \widetilde \tau_R^c + \mathrm{H.c.})
\nonumber \\
&& - \frac{1}{2} \bigg( M_1 \widetilde{B}
\widetilde{B} + M_2 \sum_{a=1}^3 \widetilde{W}^a \widetilde{W}_a +
M_3 \sum_{a=1}^8 \widetilde{G}^a \widetilde{G}_a  \ + \ {\rm H.c.}
\bigg) \nonumber \\
&& - m_S^2 |S|^2 + (\lambda A_\lambda S H_d \cdot H_u - \frac{1}{3}
\kappa  A_\kappa S^3 + \mathrm{H.c.}) \;,
\eeq
where again only the third generation of fermions and sfermions have been taken into
account. The tilde over the fields denotes the complex scalar component
of the corresponding superfields. The soft SUSY
breaking gaugino parameters $M_k$ ($k=1,2,3$) of the bino, wino and
gluino fields $\widetilde{B},$ $\widetilde{W}$ and $\widetilde{G}$, as
well as the soft SUSY breaking trilinear couplings $A_x$ ($x=\lambda,
\kappa, t, b, \tau$) are in general complex, whereas the soft SUSY
breaking mass parameters of the scalar fields, $m_X^2$
($X=S,H_d,H_u,\widetilde{Q}, \widetilde{u}_R, \widetilde{b}_R,
\widetilde{L}, \widetilde{\tau}_R$) are real. Since we consider
the CP-conserving NMSSM, they are all taken real.
In what follows, we will use conventions such that $\lambda$ and $\tan\beta$
are positive, whereas $\kappa, A_\lambda, A_\kappa$ and
$\mu_{\text{eff}}$ are allowed to have both signs.

After EWSB, we expand the Higgs fields around their VEVs $v_u$,
$v_d$, and $v_S$, respectively, which are chosen to be real and
positive
\beq
H_d = \left( \begin{array}{c} (v_d + h_d + i a_d)/\sqrt{2} \\
   h_d^- \end{array} \right) \,, \;
H_u = \left( \begin{array}{c} h_u^+ \\ (v_u + h_u + i a_u)/\sqrt{2}
 \end{array} \right) \,, \;
S= \frac{v_s+h_s+ia_s}{\sqrt{2}}.
\label{eq:Higgs-para}
\eeq
This leads to the mass matrices of the three scalars $h_d, h_u,
h_s$, the three pseudoscalars $a_d, a_u, a_s$, and the charged Higgs
states $h_u^\pm,h_d^\mp$, obtained from the second derivatives of the
scalar potential. The mass matrix is diagonalised with orthogonal
rotation matrices, mapping the gauge eigenstates to the mass
eigenstates. These are the three neutral CP-even Higgs bosons $H_1, H_2,
H_3$ that are ordered by ascending mass with $m_{H_1} \le m_{H_2} \le
m_{H_3}$, the two CP-odd mass eigenstates $A_1$ and $A_2$ with
$m_{A_1} \le m_{A_2}$, and a pair of charged Higgs bosons $H^\pm$.

After applying the minimisation conditions, we choose as independent
input parameters for the tree-level NMSSM Higgs sector the following,
\beq
\lambda\ , \ \kappa\ , \ A_{\lambda} \ , \ A_{\kappa}, \
\tan \beta =v_u/ v_d \quad \mathrm{and}
\quad \mu_\text{eff} = \lambda v_s/\sqrt{2}\; .
\eeq
Further parameters
will become relevant upon inclusion of the higher-order corrections to
the Higgs boson mass that are crucial to shift the SM-like Higgs boson
mass to the measured value.

\bigskip
\noindent
\textbf{Scans and Theoretical and Experimental Constraints}\\[0.1cm]

We performed the scans with the help of the program {\tt ScannerS}
\cite{Coimbra:2013qq,ScannerS,Muhlleitner:2020wwk} for all models
except for the NMSSM. There are various scenarios with respect to which neutral Higgs boson takes the role of
the SM-like Higgs which we will denote $H_{\text{SM}}$ from now
on. We distinguish the cases ``light'' where the lightest of the
neutral Higgs bosons is SM-like ($H_1 \equiv H_{\text{SM}}$),
``medium'' with $H_2 \equiv H_{\text{SM}}$, and ``heavy'' with the
heaviest being SM-like ($H_3 \equiv H_{\text{SM}}$). Note also that we restrict ourselves to the type
I and II models. For all these models we apply the same theoretical constraints, 
which have different expressions for each model,
requiring that all potentials are bounded from below,
that perturbative unitarity holds and that the electroweak vacuum is
the global minimum. 
In the C2HDM we use the
discriminant from  \cite{Ivanov:2015nea}.

As for experimental constraints, we impose compatibility with the
electroweak precision data by demanding the computed $S$,
$T$ and $U$ values to be within $2\sigma$ of the SM fit \cite{Baak:2014ora}, taking into account the full
correlation among the three parameters.
We require one of the Higgs bosons to have a mass of \cite{Aad:2015zhl}
\beq
m_{H_{\text{SM}}} = 125.09 \, \mbox{GeV} \,,
\eeq
and to behave SM-like. Compatibility with the Higgs signal data is checked
through {\tt HiggsSignals} version 2.6.1 \cite{Bechtle:2013xfa} which is linked to
{\tt ScannerS}. We furthermore suppress interfering Higgs signals by
forcing any other neutral scalar mass to deviate by more than $\pm 2.5$ GeV
from $m_{H_{\text{SM}}}$. 
Scenarios with neutral Higgs bosons that are close in mass are
particularly interesting for non-resonant di-Higgs production as they
may have discriminating power with respect to the SM case. The
appearance of non-trivial interference effects requires, however,  a
dedicated thorough study that is beyond the focus of this study and
is left for future work.
We require 95\% C.L. exclusion limits on non-observed scalar states by using {\tt
HiggsBounds} version 5.9.0 \cite{Bechtle:2008jh,Bechtle:2011sb,Bechtle:2013wla}.
Additionally, we checked our sample with respect to the recent ATLAS
  analyses in the $ZZ$ \cite{ATLAS:2020tlo} and $\gamma\gamma$
  \cite{ATLAS:2021uiz} final 
  states that were not yet included in {\tt HiggsBounds}.
Consistency with recent flavour constraints is ensured by testing for
the compatibility 
with $\mathcal{R}_b$ \cite{Haber:1999zh,Deschamps:2009rh} and
$B\rightarrow X_s \gamma $
\cite{Deschamps:2009rh,Mahmoudi:2009zx,Hermann:2012fc,Misiak:2015xwa,Misiak:2017bgg, Misiak:2020vlo} in the
$m_{H^{\pm}}-\tan\beta$ plane (the code SuperIso~\cite{Arbey:2018msw} used for flavour physics is interfaced with ScannerS). For the
  non-supersymmetric type II models, we imposed the
latest bound on the charged Higgs mass given in \cite{Misiak:2020vlo},
$m_{H^\pm} \ge 800$~GeV for essentially all values of $\tan\beta$, whereas
in the type I models this bound is much weaker and is strongly correlated
with $\tan\beta$.

Lower values for $m_{H^{\pm}}$ allow, via electroweak precision
constraints, different ranges for the masses of the neutral Higgs bosons, which will
therefore affect our predictions for di-Higgs production.

In the C2HDM, we additionally have to take into account
constraints on CP violation  in the Higgs sector arising from
electric dipole moment (EDM) measurements. Among these, the data from the EDM
of the electron imposes the strongest constraints~\cite{Inoue:2014nva}, with the
current best experimental limit given by the ACME
collaboration~\cite{ACME:2013pal}. We demand compatibility with the
values given in~\cite{ACME:2013pal} at 90\% C.L.

In the NMSSM, we use the program {\tt NMSSMCALC} \cite{Baglio:2013iia,King:2015oxa} and compute
the Higgs mass corrections up to ${\cal
  O}((\alpha_t+\alpha_\lambda+\alpha_\kappa)^2 + \alpha_t \alpha_s)$
\cite{Muhlleitner:2014vsa,Dao:2019qaz,Dao:2021khm}
with on-shell renormalisation in the top/stop sector. We demand the
computed SM-like Higgs boson mass to
lie in the range $122 \mbox{ GeV}... 128 \mbox{GeV}$ which accounts for
the present typically applied theoretical error of 3~GeV
\cite{Slavich:2020zjv}. We use {\tt HiggsBounds} and {\tt HiggSignals}
to check for compatibility with the Higgs constraints. Furthermore, we
omit parameter points with the
following mass configurations for the lightest chargino
$\tilde{\chi}_1^\pm$ and the lightest stop $\tilde{t}_1$,
\beq
m_{\tilde{\chi}_1^\pm} < 94 \mbox{ GeV} \,, \; m_{\tilde{t}_1} <
1\mbox{ TeV} \;,
\eeq
to take into account lower limits on the
lightest chargino and the lightest stop mass.
The experimental limits given by the LHC experiments ATLAS and CMS
rely on assumptions on the mass
spectra and are often based on simplified models. The quotation of a
lower limit therefore necessarily requires a scenario that matches the
assumptions made by the experiments. For our parameter scan we therefore
chose a conservative approach to apply limits that roughly
comply with the recent limits given by ATLAS and CMS \cite{atlaslim,cmslim}.
For further details of the Higgs mass
computation and of the input parameters as well as their scan
ranges, we refer to \cite{Dao:2021khm}.

\subsubsection{Multi-Higgs Final States}
In non-minimal Higgs models like the C2HDM, N2HDM, and NMSSM we can
have multi-Higgs final states from cascade Higgs-to-Higgs decays. In
the production of a SM-like plus non-SM-like Higgs final state,
$H_{\text{SM}} \Phi$, we found that both the Higgs-to-Higgs decay of
the SM-like Higgs or the non-SM-like one can lead to substantial final
state rates. The largest next-to-leading-order (NLO) rates that we found above 10~fb, in the multi-Higgs final
state, are summarised in Tab.~\ref{tab:multihiggs}. In the
C2HDM, we did not find NLO rates above 10~fb. We maintain the ordering of particles with regards to their decay chains, so that it becomes clear which Higgs boson decays into which Higgs pair. We give the rates in the
$(6b)$ final state as they lead to the largest cross sections for all
shown scenarios. In the following, we highlight a few benchmark
scenarios from the table. 

\begin{table}[!ht]
\begin{center}
{\small \begin{tabular}{|c|c|c|c|c|c|}
\hline
Model & Mixed Higgs State & $m_{\Phi_1}$ [GeV] & $m_{\Phi_2}$ [GeV]
  & Rate [fb] & $K$-factor \\ \hline \hline
N2HDM-I
& $H_2 H_3(\equiv H_{\text{SM}}) \to H_1 H_1  (b\bar{b}) \to
  (b\bar{b}) (b\bar{b}) (b\bar{b})$ & 98 & 41 & 15& 1.95
  \\ 
     & $H_2 H_1(\equiv H_{\text{SM}}) \to H_1 H_1 (b\bar{b})\to (b\bar{b})
  (b\bar{b}) (b\bar{b})$ & 282 & - & 40 & 1.96 \\ 
& $H_2 H_1(\equiv H_{\text{SM}}) \to A A (b\bar{b}) \to (b\bar{b})
  (b\bar{b}) (b\bar{b})$ & 157 & 73 & 33 & 2.05\\ 
& $H_1 H_2(\equiv H_{\text{SM}}) \to (b\bar{b}) H_1 H_1 \to (b\bar{b})
  (b\bar{b}) (b\bar{b})$ & 54 & - & 111& 2.09 \\
& $H_3 H_2(\equiv H_{\text{SM}}) \to H_1 H_1 (b\bar{b})\to (b\bar{b})
  (b\bar{b}) (b\bar{b})$ & 212 & 83 & 8 & 1.93  \\
\hline
N2HDM-II
& $H_2 H_1(\equiv H_{\text{SM}}) \to H_1 H_1 (b\bar{b}) \to (b\bar{b})
  (b\bar{b}) (b\bar{b}) $ & 271 & - & 3 & 1.87\\ 
\hline
NMSSM
& $H_2 H_1(\equiv H_{\text{SM}}) \to H_1 H_1 (b\bar{b}) \to
  (b\bar{b}) (b\bar{b}) (b\bar{b})$ & 319 & - & 11 & 1.90\\  
& $H_2 H_1(\equiv H_{\text{SM}}) \to A_1 A_1 (b\bar{b}) \to
  (b\bar{b}) (b \bar{b}) (b\bar{b})$ & 253 & 116 & 26 & 1.92 \\  
\hline
\end{tabular}}

\vspace*{0.4cm}
\begin{tabular}{|c|c|c|c|}
	\hline
	Model & Mixed Higgs State & $m_\text{res.}$ [GeV] & res. rate [fb] 
	 \\ \hline \hline
	N2HDM-I
	& $H_2 H_3(\equiv H_{\text{SM}}) \to H_1 H_1  (b\bar{b}) \to
	(b\bar{b}) (b\bar{b}) (b\bar{b})$ & --- & ---
	\\ 
	& $H_2 H_1(\equiv H_{\text{SM}}) \to H_1 H_1 (b\bar{b})\to (b\bar{b})
	(b\bar{b}) (b\bar{b})$ & 441 & 39  \\ 
	& $H_2 H_1(\equiv H_{\text{SM}}) \to A A (b\bar{b}) \to (b\bar{b})
	(b\bar{b}) (b\bar{b})$ & 294 &  37 \\ 
	& $H_1 H_2(\equiv H_{\text{SM}}) \to (b\bar{b}) H_1 H_1 \to (b\bar{b})
	(b\bar{b}) (b\bar{b})$ & 229  & 119  \\
	& $H_3 H_2(\equiv H_{\text{SM}}) \to H_1 H_1 (b\bar{b})\to (b\bar{b})
	(b\bar{b}) (b\bar{b})$ &---  &---    \\
	\hline
	N2HDM-II
	& $H_2 H_1(\equiv H_{\text{SM}}) \to H_1 H_1 (b\bar{b}) \to (b\bar{b})
	(b\bar{b}) (b\bar{b}) $ & 615  & 2 \\ 
	\hline
	NMSSM
	& $H_2 H_1(\equiv H_{\text{SM}}) \to H_1 H_1 (b\bar{b}) \to
	(b\bar{b}) (b\bar{b}) (b\bar{b})$ & 560  &  11\\  
	& $H_2 H_1(\equiv H_{\text{SM}}) \to A_1 A_1 (b\bar{b}) \to
	(b\bar{b}) (b \bar{b}) (b\bar{b})$ &  518& 26   \\  
	\hline
\end{tabular}

\caption{Upper: Maximum rates for multi-Higgs final
    states given at NLO QCD in the heavy-top mass limit. The
    $K$-factor is given in the last column. In the 
  third and fourth column we also give the mass values $m_{\Phi_1}$ and
  $m_{\Phi_2}$ of the non-SM-like Higgs
  bosons involved in the process, in the order of their
  appearance. Lower: In case of resonantly enhanced production the mass of the
  resonantly produced Higgs boson is given together with the
  next-to-next-to-leading order (NNLO) QCD
  production rate. More details on these points can be provided on request. \label{tab:multihiggs}} 
\end{center}
\end{table}

\bigskip
\noindent
\textbf{Non-SM-like Higgs Search: Di-Higgs beats Single Higgs}\\[0.1cm]

In the following we present N2HDM-I and NMSSM scenarios with three
SM-like Higgs bosons in the final
states with $H_1$ being SM-like and with NLO rates above 10~fb. These benchmark
points are special in the sense
that the production of the non-SM-like Higgs boson $H_2$ from di-Higgs
states beats, or is at least comparable to, its direct
production. This
appears in cases where the non-SM-like Higgs is singlet-like and/or
is more down- than up-type like. The latter suppresses direct production from
gluon fusion. The former suppresses all couplings to SM-like
particles. In these cases the heavy non-SM-like Higgs
boson might rather be discovered in 
the di-Higgs channel than in direct single Higgs production.

The input parameters for the N2HDM-I point are given in Tab.~\ref{tab:3sm-I}.
\begin{table}[!ht]
\begin{center}
\begin{tabular}{|c|c|c|c|c|c|}
\hline
$m_{H_1}$ [GeV] & $m_{H_2}$ [GeV] & $m_{H_3}$ [GeV] & $m_{A}$ [GeV]
  & $m_{H^\pm}$ [GeV] & $\tan\beta$ \\ \hline
125.09 & 281.54 & 441.25 & 386.98 & 421.81 & 1.990
\\ \hline \hline
$\alpha_1$ & $\alpha_2$ & $\alpha_3$ & $v_s$ [GeV] &
$\mbox{Re}(m_{12}^2)$ [GeV$^2$] & \\ \hline
1.153 & 0.159 & 0.989 & 9639 & 29769
& \\ \hline
\end{tabular}
\caption{Di-Higgs beats single Higgs: N2HDM-I input parameters
\label{tab:3sm-I}}
\end{center}
\vspace*{-0.4cm}
\end{table}
With the values for the NLO $H_1 H_2$ cross section and the
branching ratios $\mbox{BR}(H_2\to H_1 H_1)$ and
$\mbox{BR}(H_1 \to b\bar{b})$ we get the following rate in the $6b$
final state,
\beq
\sigma_{H_1 H_2}^{\text{NLO}} \times \mbox{BR}(H_2\to H_1 H_1) \times
\mbox{BR}(H_1 \to b\bar{b})^3 = 509 \cdot 0.37 \cdot
0.60^3 \mbox{ fb} = 40 \mbox{ fb} \;.
\eeq
We can compare this with direct $H_2$ production (we use the NNLO
value calculated with {\tt SusHi}
\cite{Harlander:2012pb,Liebler:2015bka,Harlander:2016hcx}) in either
the $4b$ 
final state from the $H_2 \to H_1 H_1$ decay,
\beq
\sigma^{\text{NNLO}} (H_2) \times \mbox{BR}(H_2 \to H_1 H_1) \times
\mbox{BR}(H_1\to b\bar{b})^2 = 161 \cdot 0.37 \cdot 0.60^2 \mbox{ fb}
= 21 \mbox{ fb} \;, 
\eeq
or direct $H_2$ production in the other dominant decay channel given
by the $WW$ final state,
\beq
\sigma^{\text{NNLO}}  (H_2) \times \mbox{BR}(H_2 \to WW) = 161 \cdot 0.44 \mbox{
  fb} = 71 \mbox{ fb} \;.
\eeq
Note that the $H_2$ branching ratio into $(b\bar{b})$ is tiny.
The second lightest Higgs boson $H_2$
has a significant down-type and large singlet admixture but only a small
up-type admixture so that its production in gluon fusion
is not very large\footnote{The production in association with
  $b$ quarks is very small for the small $\tan\beta$ value of this
  scenario.} and also its decay branching ratios into a lighter Higgs
pair are comparable to the largest decay rates into SM particles. In
this case, the non-SM-like Higgs boson $H_2$ has better chances of being discovered
in di-Higgs when compared to single Higgs channels. Note, that the $W$ bosons
still need to decay into fermionic final
states where additionally the neutrinos are not detectable so that the
$H_2$ mass cannot be reconstructed.

The input parameters for the first NMSSM scenario that we discuss here
are given in Tab.~\ref{tab:beat1}. We also specify in Tab.~\ref{tab:beat2} the
parameters required for the
computation of the Higgs pair production cross sections through {\tt
  HPAIR}.
\begin{table}[!ht]
\begin{center}
\begin{tabular}{|c|c|c|c|c|c|}
\hline
$\lambda$ & $\kappa$ & $A_\lambda$ [GeV] &  $A_\kappa$ [GeV] &
$\mu_{\text{eff}}$ [GeV] & $\tan\beta$
\\ \hline
0.593 & 0.390 & 296 & 5.70 & 200 & 2.815
\\ \hline
$m_{H^\pm}$ [GeV] & $M_1$ [GeV] & $M_2$ [GeV] & $M_3$ [TeV] & $A_t$
[GeV] & $A_b$ [GeV] \\ \hline
505 & 989.204 & 510.544 & 2 &-2064 & -1246\\ \hline
$m_{\tilde{Q}_3}$ [GeV] & $m_{\tilde{t}_R}$ [GeV] & $m_{\tilde{b}_R}$ [GeV] & $A_\tau$
[GeV] & $m_{\tilde{L}_3}$  [GeV] & $m_{\tilde{\tau}_R}$ [GeV] \\ \hline
1377 & 1207 & 3000 & -1575.91 & 3000 & 3000 \\ \hline
\end{tabular}
\caption{Di-Higgs beats single Higgs: NMSSM input parameters required by
  {\tt NMSSMCALC} for the
  computation of the NMSSM spectrum.  \label{tab:beat1}}
\end{center}
\vspace*{-0.8cm}
\end{table}
\begin{table}[!ht]
\begin{center}
\begin{tabular}{|c|c|c|c|c|}
\hline
$m_{H_1}$ [GeV] & $m_{H_2}$ [GeV] & $m_{H_3}$ [GeV] & $m_{A_1}$ [GeV] &
$m_{A_2}$ [GeV] \\ \hline
127.78 & 253 & 518 & 116 & 508
\\ \hline
$\Gamma^{\text{tot}}_{H_1}$ [GeV] & $\Gamma^{\text{tot}}_{H_2}$ [GeV] &
$\Gamma^{\text{tot}}_{H_3}$ [GeV] & $\Gamma^{\text{tot}}_{A_1}$ [GeV] &
$\Gamma^{\text{tot}}_{A_2}$ [GeV]
\\ \hline
  4.264 $10^{-3}$ & 0.466 & 3.145 & $9.9 10^{-7}$ & 4.750
\\ \hline
$h_{11}$ & $h_{12}$ & $h_{13}$ & $h_{21}$ & $h_{22}$ \\
\hline
0.325 & 0.939 & -0.112 & 0.234 & 0.034
\\ \hline
$h_{23}$ & $h_{31}$ & $h_{32}$ & $h_{33}$ & $a_{11}$  \\ \hline
0.971 & 0.916 & -0.321 & -0.209 & -0.0063
\\ \hline
$a_{21}$ & $a_{13}$ & $a_{23}$ & & \\
\hline
-0.0022 & 0.999 & 0.0067 & & \\ \hline
\end{tabular}
\caption{These input parameters and those given in the first line of
  Tab.~\ref{tab:beat1} are required by {\tt HPAIR} for the computation of the
  Higgs pair production cross sections. The total width of the charged
  Higgs boson is not required but given here for completeness
  $\Gamma^{\text{tot}}_{H^\pm}=3.94$ GeV. \label{tab:beat2}}
\end{center}
\vspace*{-0.4cm}
\end{table}

Since $H_2$ is rather singlet-like, its production cross section through
gluon fusion is small and also its decay branching ratios into
SM-final states. The gluon fusion production cross section amounts to
\beq
\sigma^{\text{NNLO}}  (H_2) = 13.54 \mbox{ fb} \;.
\eeq
Its dominant branching ratio is given by the decay into $A_1 A_1$,
reaching
\beq
\mbox{BR} (H_2 \to A_1 A_1) = 0.887 \;.
\eeq
We hence get for direct $H_2$ production in the $A_1 A_1$ final state
the rate
\beq
\sigma^{\text{NNLO}}  (H_2) \times \mbox{BR} (H_2 \to A_1 A_1) = 12.01 \mbox{
  fb} \;.
\eeq
On the other hand, we have for di-Higgs production of $H_1 H_2$ at NLO
QCD where $H_1$ is the SM-like Higgs state,
\beq
\sigma^{\text{NLO}} (H_1 H_2) = 111 \mbox{ fb} \;.
\eeq
With
\beq
\mbox{BR} (H_1 \to b\bar{b}) = 0.539,
\eeq
and the $H_2$ branching ratio into $A_1 A_1$ given above we hence have
\beq
\sigma^{\text{NLO}} (H_1 H_2) \times \mbox{BR} (H_1 \to b\bar{b}) \times \mbox{BR}
(H_2 \to A_1 A_1) = 53 \mbox{ fb} \;.
\eeq
With
\beq
\mbox{BR} (A_1 \to b \bar{b}) = 0.704
\eeq
we then obtain in double Higgs production in the $6b$ final state the rate
\beq
\sigma^{\text{NLO}} (H_1 H_2)_{6b} = 53 \times 0.704^2 \mbox{ fb} = 26 \mbox{ fb} \;.
\eeq
On the other hand, we have in single Higgs production for the $4b$ final state
\beq
\sigma^{\text{NNLO}}  (H_2)_{4b} &=& \sigma^{\text{NNLO}}  (H_2) \times
\mbox{BR}(H_2 \to A_1 A_1) \times \mbox{BR}(A_1 \to b\bar{b})^2
\nonumber \\
&=& 13.54 \times 0.887 \times 0.704^2 \mbox{ fb} = 5.95
\mbox{ fb} \;.
\eeq
Note that direct $H_2$ production with subsequent decay into $W^+W^-$ only reaches a rate of 1~fb. We clearly see that
di-Higgs beats single Higgs production and the 
non-SM-like singlet-dominated state $H_2$ might be first discovered in
di-Higgs production instead directly in single $H_2$ production through gluon fusion.

For the second NMSSM benchmark scenario that we present here the input
parameters for {\tt NMSSMCALC} and {\tt 
  HPAIR} are summarized in Tabs.~\ref{tab:beat3} and \ref{tab:beat4}.
\begin{table}[!ht]
\begin{center}
\begin{tabular}{|c|c|c|c|c|c|}
\hline
$\lambda$ & $\kappa$ & $A_\lambda$ [GeV] &  $A_\kappa$ [GeV] &
$\mu_{\text{eff}}$ [GeV] & $\tan\beta$
\\ \hline
0.545 & 0.598 & 168 & -739 & 258 & 2.255
\\ \hline
$m_{H^\pm}$ [GeV] & $M_1$ [GeV] & $M_2$ [GeV] & $M_3$ [TeV] & $A_t$
[GeV] & $A_b$ [GeV] \\ \hline
548 & 437.872 &  498.548 &2 & -1028& 1083  \\ \hline
$m_{\tilde{Q}_3}$ [GeV] & $m_{\tilde{t}_R}$ [GeV] & $m_{\tilde{b}_R}$ [GeV] & $A_\tau$
[GeV] & $m_{\tilde{L}_3}$  [GeV] & $m_{\tilde{\tau}_R}$ [GeV] \\ \hline
1729 & 1886 & 3000 & -1679.21 & 3000 & 3000 \\ \hline
\end{tabular}
\caption{Di-Higgs beats single Higgs: NMSSM input parameters required by
  {\tt NMSSMCALC} for the
  computation of the NMSSM spectrum.  \label{tab:beat3}}
\end{center}
\vspace*{-0.4cm}
\end{table}
\begin{table}[!ht]
\begin{center}
\begin{tabular}{|c|c|c|c|c|}
\hline
$m_{H_1}$ [GeV] & $m_{H_2}$ [GeV] & $m_{H_3}$ [GeV] & $m_{A_1}$ [GeV] &
$m_{A_2}$ [GeV] \\ \hline
123.20 & 319 & 560 & 545 & 783
\\ \hline
$\Gamma^{\text{tot}}_{H_1}$ [GeV] & $\Gamma^{\text{tot}}_{H_2}$ [GeV] &
$\Gamma^{\text{tot}}_{H_3}$ [GeV] & $\Gamma^{\text{tot}}_{A_1}$ [GeV] &
$\Gamma^{\text{tot}}_{A_2}$ [GeV]
\\ \hline
  3.985 $\times\,10^{-3}$ & 0.010 & 4.207 & 6.399 & 6.913
\\ \hline
$h_{11}$ & $h_{12}$ & $h_{13}$ & $h_{21}$ & $h_{22}$ \\
\hline
0.419 & 0.909 & 0.015 & 0.187 & -0.102
\\ \hline
$h_{23}$ & $h_{31}$ & $h_{32}$ & $h_{33}$ & $a_{11}$  \\ \hline
0.977 & 0.889 & -0.407 & -0.212 & 0.908
\\ \hline
$a_{21}$ & $a_{13}$ & $a_{23}$ & & \\
\hline
-0.104 & 0.114 & 0.994 & & \\ \hline
\end{tabular}
\caption{These input parameters and those given
  in the first line of
  Tab.~\ref{tab:beat1} are required by {\tt HPAIR} for the computation of the
  Higgs pair production cross sections. The total width of the charged
  Higgs boson is not required but given here for completeness
  $\Gamma^{\text{tot}}_{H^\pm}= 5.503$ GeV. \label{tab:beat4}}
\end{center}
\vspace*{-0.4cm}
\end{table}
The singlet-like $H_2$ dominantly decays into an SM-like pair $H_1
H_1$, and  in the $H_1 H_1$ final state we obtain the rate
\beq
\sigma^{\text{NNLO}} (H_2) \times \mbox{BR}(H_2 \to H_1 H_1) =
134.95 \cdot 0.566 \mbox{ fb} = 76.38 \mbox{ fb} \;.
\eeq
With BR$(H_1 \to b\bar{b})= 0.636$ this results in the $4b$ rate
\beq
 \sigma^{\text{NNLO}} (H_2) \times \mbox{BR}(H_2 \to H_1 H_1) \times
 \mbox{BR}^2 (H_1 \to b\bar{b})= 31.00 \mbox{ fb} \;.
\eeq
On the other hand, with BR$(H_2 \to b\bar{b})= 0.103$, we have the $2b$
final state rate
\beq
\sigma^{\text{NNLO}} (H_2) \times \mbox{BR}(H_2 \to b\bar{b}) =
134.95 \cdot 0.104 \mbox{ fb} = 14.03 \mbox{ fb} \;.
\eeq
The rate for direct $H_2$ production in the $4b$ final state via its
decay into $H_1 H_1$ beats the one of direct $H_2$ production in the $2b$ final state
by more than a factor of 2. Note finally that the $6b$ rate for $H_2$
production, through $H_1 H_2$ production and further $H_2$ decay into
Higgs pairs, amounts to
\beq
\sigma^{\text{NLO}} (H_1 H_2) \times \mbox{BR} (H_2 \to H_1 H_1)
\times \mbox{BR}^3 (H_1 \to b\bar{b})= 75 \cdot 0.566 \cdot 0.636^3 \mbox{
fb} = 11 \mbox{ fb} \;,
\eeq
which is not much below the $2b$ final state rate.


\bigskip
\noindent
\textbf{Non-SM-Like Higgs Pair Final States \label{sec:nonsm}}
For non-SM-like Higgs pair production, we can have a large plethora of all
possible Higgs pair combinations inducing final states with multiple
Higgs bosons, two or three Higgs bosons in association with one or two gauge bosons,
or also with a top-quark pair, resulting finally in multi-fermion,
multi-photon or multi-fermion plus multi-photon final states (see e.g.~Ref.~\cite{Ferreira:2022sno}). We
present a few selected interesting signatures from non-SM-like Higgs
pair production in Tab.~\ref{tab:selected}. More signatures and benchmark points can be
provided on request. As we can infer from the table, we can have high
rates in non-SM-like Higgs pair production, {\it e.g.}~up to 9~pb in
the $4b$ final state from non-SM-like $H_1H_1$ production in the
N2HDM-I with $H_2 \equiv H_{\text{SM}}$ (marked by a '*' in Tab.~\ref{tab:selected}).

\begin{table}[!ht]
\begin{center}
\begin{tabular}{|c|c|c|c|c|c|}
\hline
Model & SM-like Higgs & Signature  & $m_\Phi$ [GeV] & Rate [fb] & $K$-factor \\ \hline \hline
 N2HDM-I
& $H_3$ & $H_1 H_1 \to (b\bar{b}) (b\bar{b})$ & 41 & 14538 & 2.18 \\
& $H_3$ & $H_1 H_1 \to (4b) ; (4\gamma)$ & 41 & 4545 ; 700 & 2.24 \\
& $H_1$ & $A A \to (b\bar{b}) (b\bar{b})$ & 75 & 6117 & 2.11 \\
& $H_1$ & $H_2 H_2 \to (b\bar{b}) (b\bar{b})$ & 146 & 73 & 2.01\\
& $H_2$ & $A A \to (b\bar{b}) (b\bar{b})$& 80 & 2875 & 2.13 \\
& $H_2$ & $A H_1 \to (b\bar{b}) (b\bar{b})$ & $m_A:$ 87& 921 & 2.09\\
&            &                                                    & $m_{H_1}:$ 91  & & \\
* & $H_2$ & $H_1 H_1 \to (b\bar{b}) (b\bar{b})$ & 47 & 8968 & 2.17 \\
\hline
N2HDM-II
& $H_2$ & $H_1 H_1 \to (b\bar{b}) (b\bar{b})$ & 44 & 1146 & 2.18 \\
\hline
C2HDM-I
& $H_1$ & $H_2 H_2 \to  (b\bar{b}) (b\bar{b})$ & 128 & 475 & 2.07 \\
& $H_2$ & $H_1 H_1 \to (b\bar{b}) (b\bar{b})$ & 66 & 814 & 2.16 \\
& $H_3$ & $H_1 H_1 \to (b\bar{b}) (b\bar{b})$ & 84 & 31 & 2.09 \\
\hline
NMSSM
& $H_1$ & $A_1 A_1 \to (b\bar{b}) (b\bar{b})$ & 166 & 359 & 1.95 \\
& $H_1$ & $A_1 A_1 \to (\gamma\gamma) (\gamma\gamma)$ & 179 & 34 & 1.96 \\
& $H_2$ & $H_1 H_1 \to (b\bar{b}) (b\bar{b})$ & 48 & 3359 & 2.18 \\
& $H_2$ & $A_1 A_1 \to (b\bar{b}) (b\bar{b})$ & 54 & 1100 & 2.18\\
& $H_1$ & $A_1 A_1 \to (t\bar{t}) (t\bar{t})$ & 350 & 20 & 1.82 \\
\hline
\end{tabular}

\vspace*{0.4cm}
\begin{tabular}{|c|c|c|c|c|c|}
	\hline
	Model& Signature & $m_\text{res.}$ [GeV] & res. rate [fb]& $m_\text{res.}$ 2 [GeV] & res. rate 2 [fb]  \\ \hline \hline
	N2HDM-I
	& $H_1 H_1 \to (b\bar{b}) (b\bar{b})$& 125.09 & 621 &98& 17137\\
	& $H_1 H_1 \to (4b) ; (4\gamma)$ & 125.09&126; 19  & 94& 5445; 839\\
	& $A A \to (b\bar{b}) (b\bar{b})$ &1535& $<$0.1 & 323&482\\
	 & $H_2 H_2 \to (b\bar{b}) (b\bar{b})$&360&76&---&---\\
	& $A A \to (b\bar{b}) (b\bar{b})$&178&3191&---&---\\
	 & $A H_1 \to (b\bar{b}) (b\bar{b})$ &  --- & --- &---&---\\
	 & $H_1 H_1 \to (b\bar{b}) (b\bar{b})$ &588&22&125.09&997 \\
	\hline
	N2HDM-II
	 & $H_1 H_1 \to (b\bar{b}) (b\bar{b})$ &520& $<$ 0.1 & 125.09& 1330\\
	\hline
	C2HDM-I
	 & $H_2 H_2 \to  (b\bar{b}) (b\bar{b})$&266& 497  &---&---\\
	 & $H_1 H_1 \to (b\bar{b}) (b\bar{b})$&151&598&---& ---\\
	 & $H_1 H_1 \to (b\bar{b}) (b\bar{b})$&---&---&---&---\\
	\hline
	NMSSM
	 & $A_1 A_1 \to (b\bar{b}) (b\bar{b})$ &552& 31 &453& 332\\
	 & $A_1 A_1 \to (\gamma\gamma) (\gamma\gamma)$&796&$<0.01$&444& 34\\
	 & $H_1 H_1 \to (b\bar{b}) (b\bar{b})$&882& $<$0.1   &125.59& 4173\\
	 & $A_1 A_1 \to (b\bar{b}) (b\bar{b})$&676& $<0.1$&122.99&1353\\
	 & $A_1 A_1 \to (t\bar{t}) (t\bar{t})$&741& 7  &705 & 14\\
	\hline
\end{tabular}
\caption{Upper: Selected rates for non-SM-like Higgs pair final states at NLO
  QCD. We specify the model, which of the Higgs bosons is the
  SM-like one, the signature and its rate as well as the
  $K$-factor. In the
  fourth column we also give the mass value $m_{\Phi}$ of the non-SM-like Higgs
  boson involved in the process. 
Lower: In case of resonantly enhanced cross sections, the mass of the
resonantly produced Higgs boson is given together with the NNLO QCD
production rate. Some scenarios contain two heavier Higgs bosons that
can contribute to resonant production. 
All benchmark details can 
  be provided on request. \label{tab:selected}}
\end{center}
\vspace*{-0.4cm}
\end{table}

\paragraph{Cascade Decays with Multiple Higgs Final States \label{sec:cascade}}
As already stated, in non-mimimal Higgs extensions, we can have
Higgs-to-Higgs cascade decays that can lead to multiple Higgs final
states. The largest rate at NLO QCD that we found, for a final state with more than
three Higgs bosons, is given in the N2HDM-I, where we have 
\beq
\sigma(pp \to H_2 H_2 \to H_1 H_1 H_1 H_1 \to 4 (b\bar{b}))=
1.4 \mbox{ fb} \;.
\eeq
The SM-like Higgs is $H_1$ and the $K$-factor for the NLO QCD
production of $H_2 H_2$ is 1.82. Also in the NMSSM and C2HDM we can
have multiple Higgs production but the rates are below 10~fb after the
decays of the Higgs bosons. In the N2HDM, we can even produce up to
eight Higgs bosons in the final states but the rates are too small to be measurable.


\clearpage

\clearpage
\subsection{Constraints on the trilinear and quartic Higgs self-couplings at HL-LHC \label{sec:pertur}}

{\sl P. Stylianou, G. Weiglein}

\subsubsection{Introduction}
In the SM, the self-interactions of the Higgs boson depend on the form of the potential,
\begin{equation}
\label{eq:smhiggs}
	V(\Phi) = \lambda (\Phi^\dagger \Phi)^2 - \mu^2 \Phi^\dagger \Phi,
\end{equation}
and are parameterised by $\lambda$ and $\mu$. The potential can be reparameterised in terms of the Higgs mass $M_H$ and VEV $v$, which have been measured experimentally. However, a more complicated potential could be realised in nature, arising from models with a richer scalar sector, and measuring the Higgs self-couplings will thus provide concrete information on the exact shape beyond the SM ansatz. One should notice also that the Higgs self-interactions may, in general, receive large radiative corrections which are strongly sensitive to the existence of New Physics states coupled to the Higgs boson. For a systematic procedure of calculation of such corrections starting from the one-loop effective potential, see Ref.~\cite{Camargo-Molina:2016moz}.

Deviations away from the SM values can be experimentally studied in the $\kappa$-framework, where for the self-interactions we define $\kappa_i = g_i / g^\text{SM}_i$ for $i=3,4$, and $g^\text{SM}_i$ is the SM coupling at leading order. The production of a Higgs pair allows the direct probe of the trilinear coupling $\kappa_3$, and the ATLAS~\cite{ATLAS:2022jtk} and CMS~\cite{CMS:2022dwd} experiments provide limits on by combining the gluon fusion and weak boson fusion (WBF) production channels with different decays of the Higgs. ATLAS additionally includes information from single-Higgs channels, where $\kappa_3$ enters at next-to-leading order, and provides the stringest limit of $\kappa_3 \in \left[ -0.4, 6.3 \right]$. 

Triple Higgs production is sensitive to both the trilinear and quartic self-couplings and could enable establishing the first limit on $\kappa_4$, beyond theoretical constraints. However, it is known to suffer from small cross sections at the LHC. We motivate the possibility of large values of $\kappa_3$, $\kappa_4$ in Sec.~\ref{subsubsec:pertunitarity} and explore the potential constraints at HL-LHC in Sec.~\ref{subsubsec:hllhcreach}.

\subsubsection{Theoretical motivation and perturbative unitarity}
\label{subsubsec:pertunitarity}

\begin{figure}[!ht]
	\begin{center}
	\includegraphics[width=0.45\textwidth]{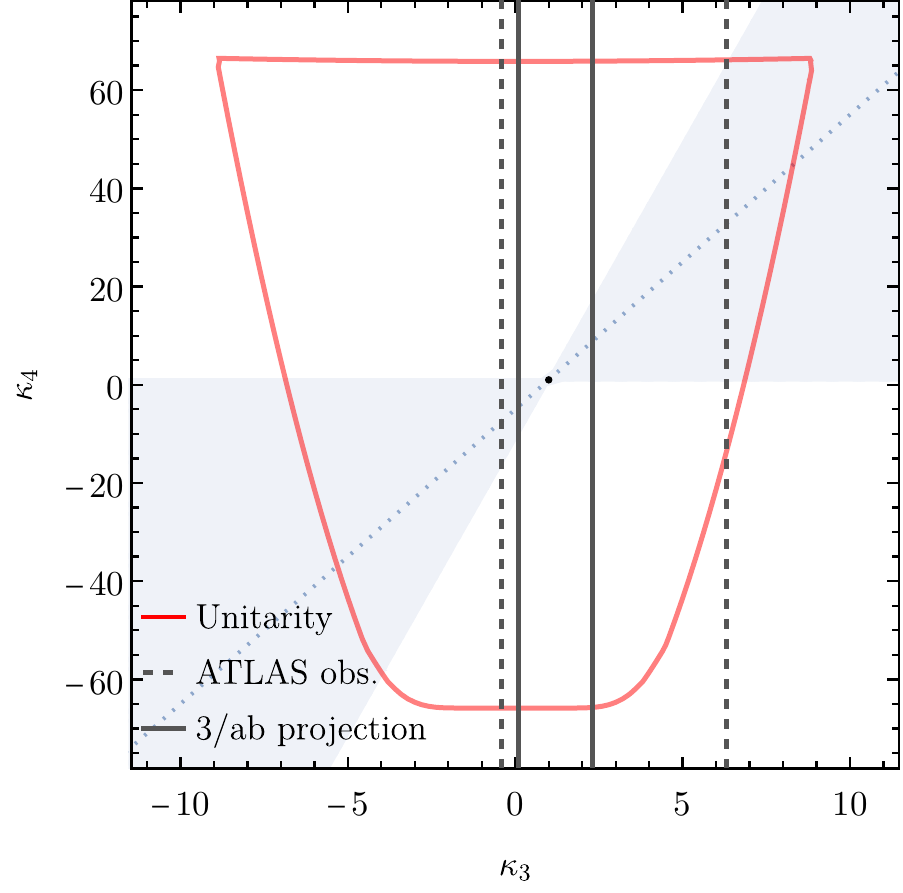}
	\caption{
		Perturbative unitarity bounds for $\kappa_3$ and $\kappa_4$, as well as current experimental bounds (black dashed lines) and HL-LHC projections (black solid lines). The region where dimension-eight contributions to $\kappa_4$ are smaller than dimension-six is shown as a blue region, while the dotted blue line corresponds to 
    $\kappa_4 - 1 \simeq 6 (\kappa_3 - 1)$. \label{fig:unitarity}}
	\end{center}
\end{figure}

Theoretical bounds for $\kappa_3$ and $\kappa_4$ can be established by requiring perturbative unitarity to be satisfied\cite{Liu:2018peg}. Focusing on the $HH \rightarrow HH$ scattering at tree level, which is the relevant channel to extract perturbative unitarity bounds for the self-couplings, the zeroth partial wave is given by
\begin{equation}
	\label{eq:zerothpw}
	\begin{split}
		a^0 =  \frac{ 3 M_H^2 \sqrt{s^2 - 4 M_H^2 s}}{ 32 \pi s (s - M_H^2) v^2} \left[  \kappa_4 (s - M_H^2) - 3 \kappa_3^2 M_H^2 + \frac{6 \kappa_3^2 M_H^2 (s - M_H^2)}{s - 4 M_H^2} \log\left( \frac{s}{M_H^2} - 3\right)\right] ,
	\end{split}
\end{equation}
and the requirement $\lvert{\text{Re}\left(a^0\right)}\rvert \leq 1/2$ gives the region that satisfies perturbative unitarity. 
For large values of $s$, $a^0$ depends only on $\kappa_4$ and yields a lower and upper bound on the quartic coupling, while for lower energies unitarity is violated depending on the value of $\kappa_3$, as shown in Fig.~\ref{fig:unitarity}. The latter occurs at relatively low energies, so in practice we check that a particular value of $\kappa_3$ does not violate unitarity up to $10$~TeV (this upper limit only matters for the corners of the contour at $\kappa_4 \sim 67$ and $\lvert \kappa_3 \rvert \sim 9$). The current ATLAS bounds on $\kappa_3 \in [-0.4, 6.3]$, as well as the $95\%$ combined ATLAS and CMS projection for the HL-LHC $\kappa_3 \in [0.1, 2.3]$~\cite{Cepeda:2019klc} are also shown.\footnote{We note that the current bound is the observed one, with a best-fit value of $\kappa_3 = 3.0^{1.8}_{-1.9}$
. Additionally the negative log-likelihood ratio as a function of $\kappa_3$ obtained by experiments is also asymmetric, implying that the central value does not correspond to the best-fit value~\cite{ATLAS:2022jtk,CMS:2023cee}.} The theoretical constraints on $\kappa_3$ are considerably stronger than on $\kappa_4$, which can be understood in terms of both an effective field theory prescription and concrete UV-models as discussed below.

A generic extension of the SM potential with higher dimensional operators included as a power expansion in inverse powers of a UV-scale $\Lambda$ can be written as~\cite{Boudjema:1995cb,Maltoni:2018ttu}
\begin{equation}
	\label{eq:highpot}
		V_\text{BSM} = \frac{C_6}{\Lambda^2} \left( \Phi^\dagger \Phi - \frac{v^2}{2} \right)^3 + \frac{C_8}{\Lambda^4} \left(\Phi^\dagger \Phi - \frac{v^2}{2} \right)^4 + {\cal{O}}({\frac{1}{\Lambda^6})}\;.
\end{equation}
where the Higgs doublet can be expanded as $\Phi = \left(0, (v + H)/\sqrt{2}\right)$ and $v$, $H$ are the vaccuum expectation value and the $125$~GeV Higgs, respectively. Parameterising the additional terms of the potential in this way ensures that $\kappa_3$ ($\kappa_4$) receives contributions only from dimension-six (dimension-six and -eight) operators but not ones of higher order. The coupling modifiers with this parameterisation are then given by 
\begin{equation}
	\label{eq:k3k4rel}
    \begin{split}
        (\kappa_3 - 1) &= \frac{C_6 v^2}{\lambda \Lambda^2}\;,\\
        (\kappa_4 - 1) &= \frac{6 C_6 v^2}{\lambda \Lambda^2} + \frac{4 C_8 v^4}{\lambda \Lambda^4}. 
    \end{split}
\end{equation}
Requiring that dimension-eight operators vanish yields $(\kappa_4 - 1) \simeq 6 (\kappa_3 - 1)$ (also shown as a line in Fig.~\ref{fig:unitarity}) hinting that deviations on the quartic coupling can be more sizeable than the trilinear coupling, in-line with the weaker constraint of perturbative unitarity for $\kappa_4$. Relaxing the assumption of vanishing dimension-eight operators and requiring that the dimension-eight contribution to $\kappa_4$ is smaller than the dimension-six yields the condition $\lvert (\kappa_4 - 1) - 6 (\kappa_3 - 1) \rvert < 6 \lvert \kappa_3 - 1\rvert$. 

\begin{figure*}[t!]
\subfigure{\includegraphics[width=8.3cm]{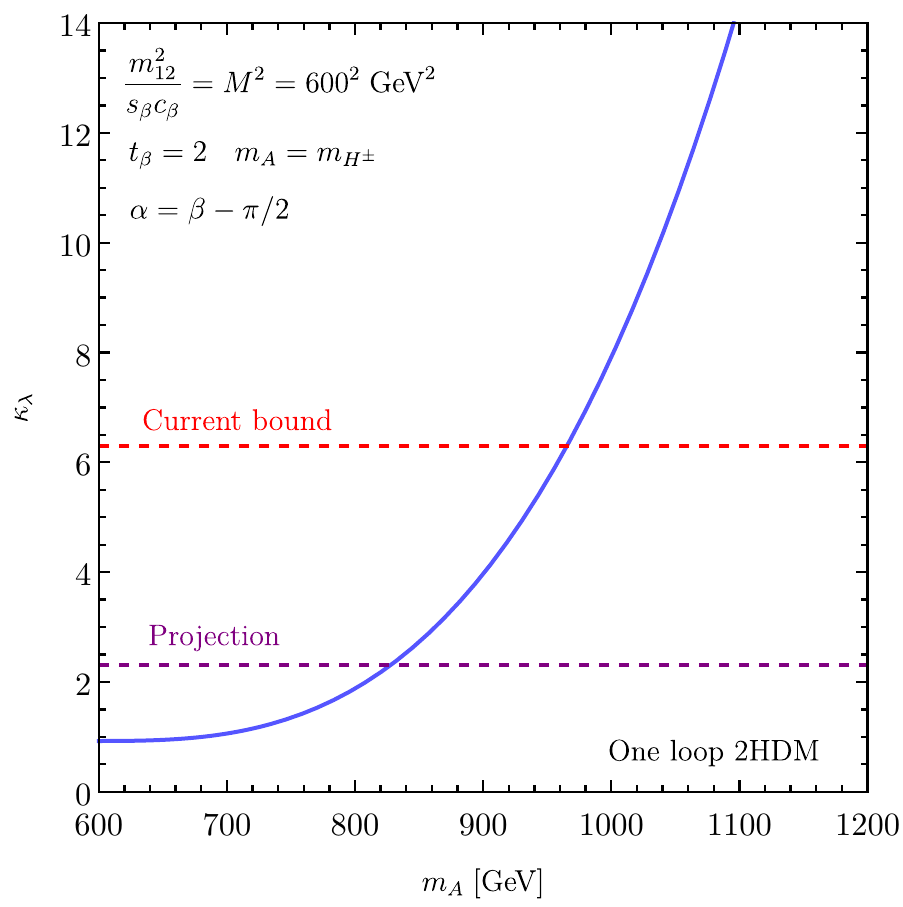}}
\hfill
\subfigure{\includegraphics[width=8.3cm]{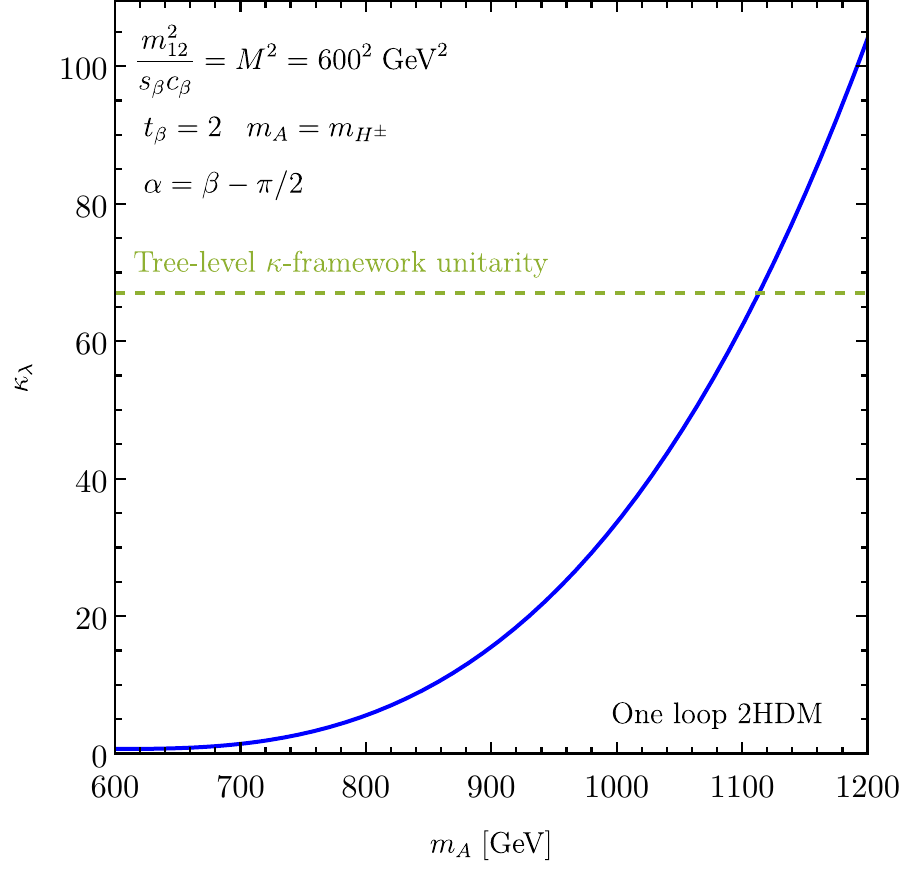}}
	\caption{One loop corrections to the trilinear coupling $\kappa_3$ (left) and to the quartic $\kappa_4$ (right). The scale $M = m_{12}/(c_\beta s_\beta)$ is fixed to $M = m_H = 600$~GeV and
 $m_A = m_{H^\pm}$ 
 is varied. $s_\beta$, $c_\beta$ and $t_\beta$ correspond to $\sin\beta, \cos\beta$ and $\tan\beta$, respectively. 
 \label{fig:loop}}
\end{figure*}

As an example from a specific model, we focus on the Two-Higgs Doublet Model\footnote{For a review, see Ref.~\cite{Branco:2011iw}.} and a specific benchmark point in the alignment limit from Ref.~\cite{Bahl:2022jnx} that is currently not excluded by experiments. The particular benchmark point yields sizeable corrections to $\kappa_3$ at loop-level, and to show this we reproduce the one-loop result for $\kappa_3$ from Ref.~\cite{Bahl:2022jnx} in Fig.~\ref{fig:loop}, showing however also the the one-loop corrections to $\kappa_4$. Consistent with the effective approach and perturbative unitarity, $\kappa_4$ rises to significantly larger values than $\kappa_3$ hinting that if a deviation away from $\kappa_3 \sim 1$ is realised in nature, the deviation on $(\kappa_4 - 1)$ could in fact be much larger. 

\begin{figure*}[t!]
	\begin{center}
	\includegraphics[width=8.3cm]{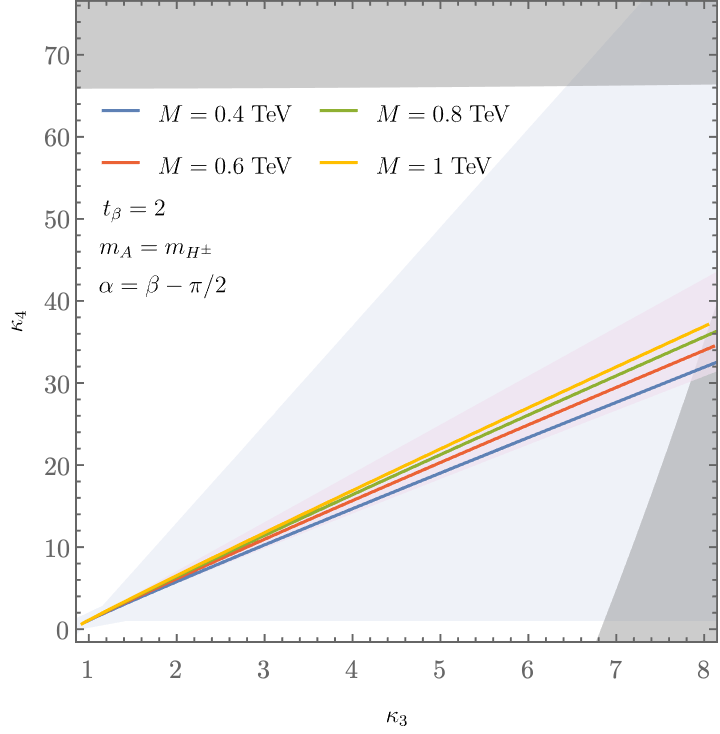}
	\caption{
 Correlation of $\kappa_3$ and $\kappa_4$ when $M = m_H$ and $m_A = m_{H^\pm}$ are varied. The solid lines correspond to particular values of $M$, while the purple area is the range of self-coupling values obtained for $m_A, M \in [0.3, 10]$ TeV. We also overlay blue shaded region of Fig.~\ref{fig:unitarity} and the gray region shows the area excluded by perturbative unitarity. 
	} \label{fig:2hdm}
	\end{center}
\end{figure*}

The correlation between $\kappa_3$ and $\kappa_4$ in the 2HDM (at one-loop) is also shown in Fig.~\ref{fig:2hdm} for different values of the scale $M = m_{12}/ (c_\beta s_\beta)$ and $m_A$. The charged Higgs mass $m_{H^\pm}$ is kept equal to $m_A$ which avoids 2HDM contributions to the oblique parameters S, T and U at one-loop level. A linear relation is maintained between the self-couplings for the particular parameter choices in the 2HDM and these values are within the region of well-behaved EFT framework with higher order operators added to the potential. A non-linear approach (e.g. HEFT) would be required to study regions with small deviations of $\kappa_3$ but large $\kappa_4$ that do not lie in the well-behaved EFT shaded region. For the rest of this work we remain model-agnostic and use the $\kappa$-framework.

\subsubsection{Triple Higgs at the HL-LHC}
\label{subsubsec:hllhcreach}
Setting limits on the self-couplings through triple Higgs production is a challenging task at the HL-LHC due to the small cross section rates and the difficult final states. To counter the former we focus on the dominant production through gluon fusion with on-shell Higgs bosons decaying to $b$-quarks and $\tau$-leptons. In particular, we investigate the $6b$ final state with at least $5$ tagged $b$-quarks and the $4b2\tau$ final state with at least $3$ tagged $b$-quarks and $2$ tagged $\tau$ leptons. The included background contamination for the $6b$ channel consists of multi-jet QCD events, while for the $4 b 2\tau$ we include $WWbbbb$ (including $ttbb$), $Zbbbb$, $ttH$ $ttZ$ and the $tttt$ production of the SM. Analyses with these final states have been previously performed for FCC-energies~\cite{Papaefstathiou:2019ofh,Papaefstathiou:2020lyp,Fuks:2017zkg}. 

Events are generated using {\sc{Madgraph}}~\cite{Alwall:2014hca,Hirschi:2015iia}\footnote{We generate signal events for $p p \rightarrow h h h$ and subsequently decay them on-shell with {\sc{Madspin}}~\cite{Artoisenet:2012st}.}
 and a generation-level cut of $350$~GeV is imposed on the minimum invariant mass of the process. Additionally, relaxed cuts are imposed on the transverse momentum $p_T$ and pseudorapidity $\eta$ of the $b$-quarks and $\tau$-leptons: $p_T(b) > 30$~GeV, $p_T(\tau) > 10$~GeV, $\lvert \eta (b) \rvert < 2.5$ and $\lvert \eta (\tau) \rvert < 2.5$. Furthermore, at least one pair of $b$-jets or $\tau$-leptons should yield an invariant mass close to the SM Higgs mass, $\left[110, 140\right]$~GeV. In order to include higher order effects, K-factors of 1.7 and 2 are applied on cross sections for signal~\cite{deFlorian:2019app} and background, respectively. The tagging efficiencies for both $b$-quarks and $\tau$ leptons is assumed to be 0.8, and for the $4b2\tau$ analysis, at least one $\tau$ is assumed to decay hadronically. 

To identify the appropriate signal region, a Graph Neural Network (GNN) with the EdgeConv~\cite{edgeconv} operation (similar to Refs.~\cite{Atkinson:2021jnj,Anisha:2022ctm}) is trained on simulated signal data with $(\kappa_3, \kappa_4) = (1,1)$ and background events. Nodes are added for $b$-tagged jets (and $\tau$-tagged leptons), as well as pairs of $b$-quarks or $\tau$-leptons that are close to the Higgs mass (for more details, see Ref.~\cite{Stylianou:2023xit}). The features for each node are $[p_T, \eta, \phi, E, m, \text{PDGID}]$. 

In the $6b$ analysis the network discriminates between two classes (signal and background), and the signal region is identified with a background rejection of $\sim 0.4$. We instead use multi-class classification for the $4b2\tau$, maintaining the different background contributions as different classes. The selection region is identified by requiring $P[W W bbbb] < 3\%$, $P[Z bbbb] < 10\%$ and $P[tt(H\rightarrow \tau \tau)] < 30\%$.

\begin{figure*}[t!]
\subfigure{\includegraphics[width=8.3cm]{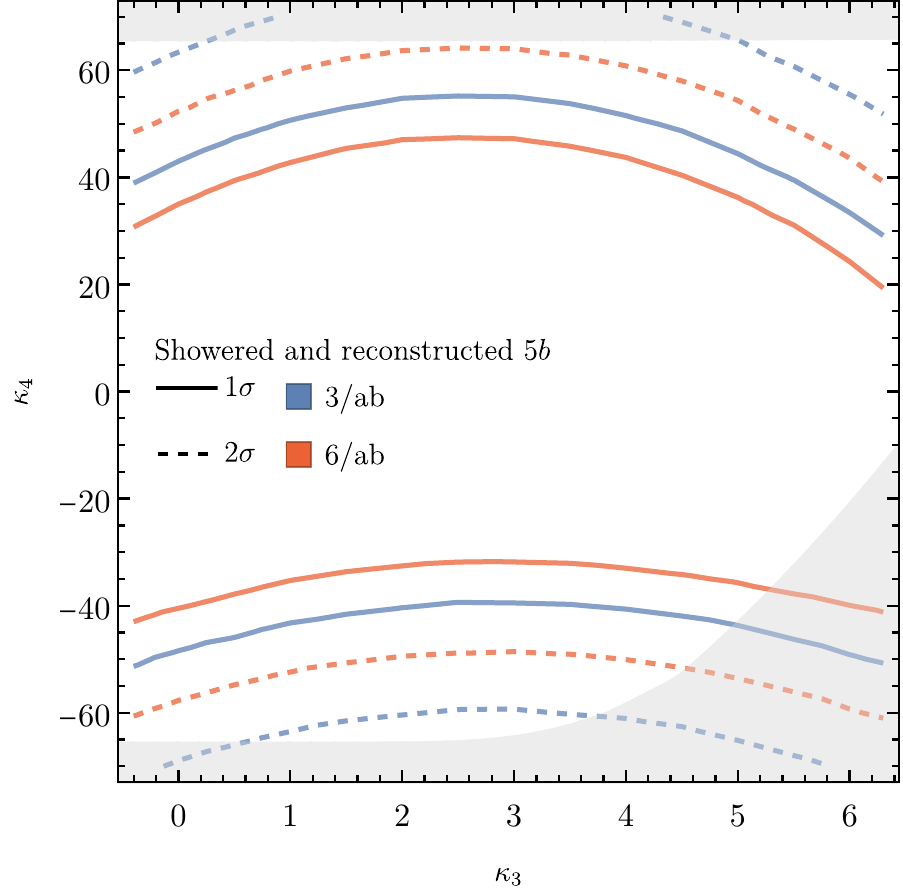}}
\hfill
\subfigure{\includegraphics[width=8.3cm]{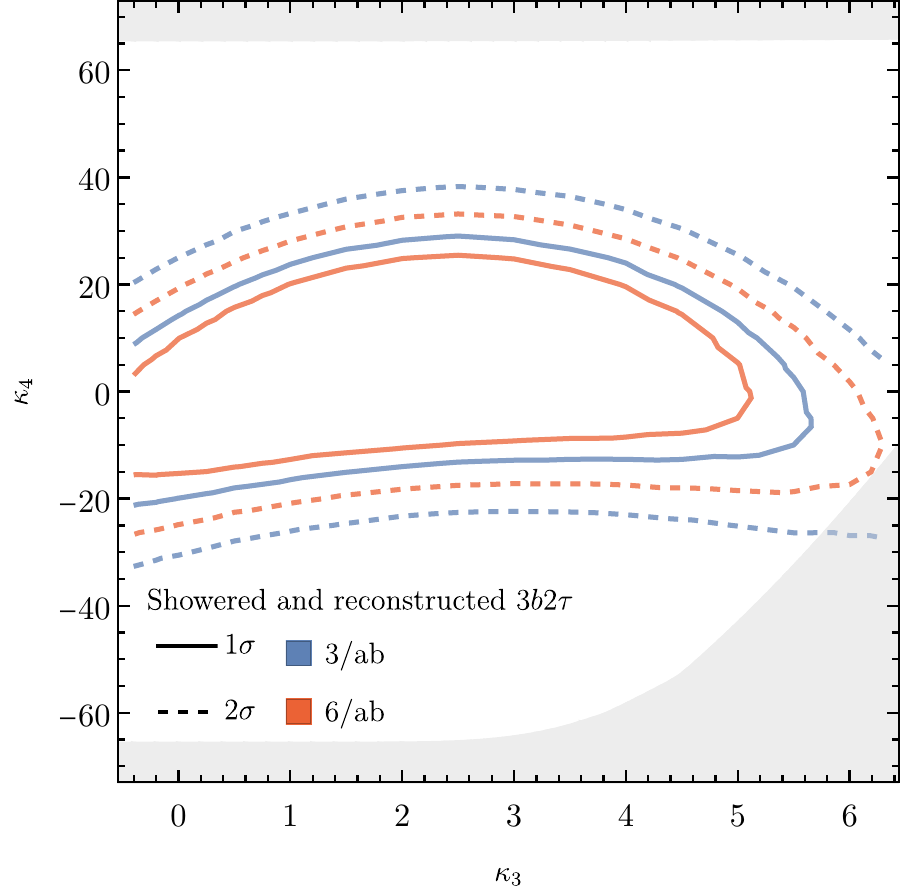}}
	\caption{
		The $1\sigma$ and $2\sigma$ bounds in the $\kappa_3$--$\kappa_4$ plane from the $6b$ analysis are shown on the left, while ones from the $4b2\tau$ analysis are shown on the right. The light gray area corresponds to the region excluded by unitarity.
		\label{fig:contoursreco}}
\end{figure*}

\begin{figure}[!ht]
	\begin{center}
	\includegraphics[width=0.45\textwidth]{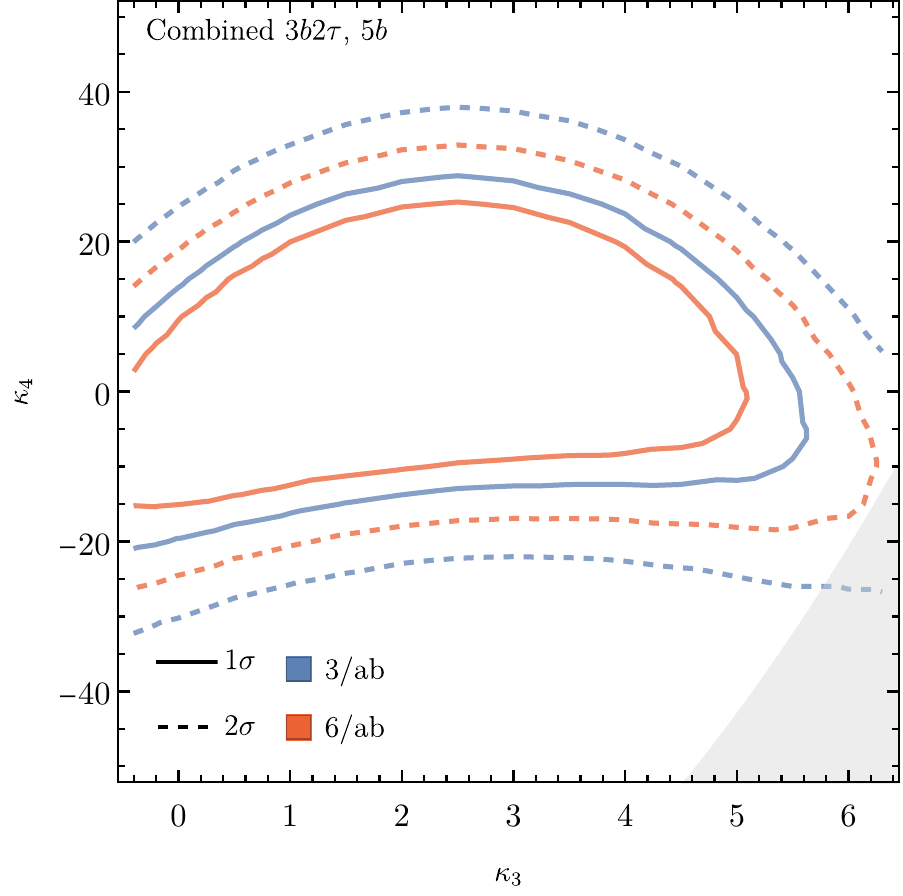}
	\caption{The $1\sigma$ and $2\sigma$ bounds in the $\kappa_3$--$\kappa_4$ plane after a combination of the $6b$ and $4b2\tau$ analyses are shown. Perturbative unitarity is overlaid as a gray shaded area. \label{fig:contourcomb}}
	\end{center}
\end{figure}

Based on the signal $S$ and background events $B$ in the signal region, we calculate the significance
\begin{equation}
    Z = \sqrt{2 \bigg( (S + B) \ln{(1+ \frac{S}{B})} - S\bigg)} \;.
\end{equation}
by assuming a HL-LHC luminosity of $3$/ab and the resulting $1\sigma$ and $2\sigma$ contours for each channel are shown in Fig.~\ref{fig:contoursreco}, as well as the case of a combined ATLAS \& CMS luminosity of $6$/ab. Assuming no correlations, we additionally obtain the combined significance $Z_\text{comb} = \sqrt{Z_{5 b}^2 + Z_{3 b 2 \tau}^2}$, and show the relevant contours in Fig.~\ref{fig:contourcomb}.

\subsubsection{Conclusions}
\label{subsubsec:psgwconc}
Despite its low cross section rates, triple-Higgs production offers valuable insights into the Higgs self-couplings. The correlation between $\kappa_3$ and $\kappa_4$ can enable distinguishing between beyond the SM scenarios. A significant deviation in $\kappa_4$ while maintaining consistency of $\kappa_3$ with the SM may suggest the presence of non-linear effects. Conversely, deviations in both couplings could align with expectations from certain specific models such as the 2HDM. Although sensitivity to $\kappa_3$ at HL-LHC will primarily come from di-Higgs production, incorporating $HHH$ can still remain beneficial for combinations. Our findings suggest that triple-Higgs production at HL-LHC is expected to establish the first experimental constraints on $\kappa_4$ beyond theoretical bounds from perturbative unitarity.

\clearpage
\subsection{Triple Higgs Boson Production with Anomalous
Interactions \label{sec:anomalouscoupl}}

{\sl A. Papaefstathiou, G. Tetlalmatzi-Xolocotzi}

\subsubsection{Introduction}

Novel Higgs boson interactions, to new or to SM particles, may arise at the electro-weak scale, but they may also appear at higher scales, $\mathcal{O}(\mathrm{few~TeV})$. If this is the case, we can parametrize our ignorance using a higher-dimensional effective field theory (EFT), see, e.g.~\cite{Buchmuller:1985jz,Grzadkowski:2010es,Elias-Miro:2013mua}. Neglecting lepton-number violating operators, the lowest-dimensionality EFT that can be written down consists of $D=6$ operators. Upon electro-weak symmetry breaking, the Higgs boson would acquire a VEV and these operators would result in several new interactions, as well modifications of the SM interactions of the Higgs boson. Several of the operators in the physical basis of the Higgs boson scalar ($h$) would then have coefficients that are correlated, according to $D=6$ EFT. These correlations, however, may be broken by even higher-dimensional operators (e.g.\ $D=8$), particularly if the new phenomena are closer to the electro-weak scale. Therefore, it may be beneficial to lean towards a more agnostic, and hence more phenomenological, approach and, while still remaining inspired by $D=6$ EFT, consider fully uncorrelated, ``anomalous'' interactions of the Higgs boson with the SM. This is the approach that was pursued in~\cite{Papaefstathiou:2023uum}.

\subsubsection{Phenomenological Lagrangian for Anomalous Interactions}

The implementation of this study further modifies the $D=6$ EFT Lagrangian relevant to the Higgs boson's interactions, see, e.g.~\cite{Goertz:2014qta}, to allow for uncorrelated, anomalous coefficients in the interactions. In addition, to match more closely the LHC experimental collaboration definitions, we define the following phenomenological Lagrangian \cite{Carvalho:2015ttv, Capozi:2019xsi}: 
\begin{equation}\label{eq:LgghhPhenoExp}
\begin{split}
\mathcal{L}_\mathrm{PhenoExp} = &- \lambda_\mathrm{SM} v \left( 1 + d_3 \right) h^3 - \frac{ \lambda_\mathrm{SM} } { 4 }  \left( 1  + d_4 \right) h^4
\\
&+ \frac{\alpha_s } {12 \pi } \left( c_{g1} \frac{h}{v} - c_{g2} \frac{h^2} {2v^2} \right) G_{\mu\nu}^a G^{\mu\nu}_a
\\
&- \left[ \frac{m_t}{v} \left( 1 + c_{t1}   \right) \bar{t}_L t_R h +  \frac{m_b}{v} \left( 1 + c_{b1}   \right) \bar{b}_L b_R h + \text{h.c.} \right]
\\
&- \left[ \frac{m_t}{v^2} c_{t2} \bar{t}_L t_R h^2 +  \frac{m_b}{v^2} c_{b2} \bar{b}_L b_R h^2 +  \text{h.c.} \right] 
\\
&- \left[ \frac{m_t}{v^3}\left( \frac{c_{t3}}{2} \right) \bar{t}_L t_R h^3 +  \frac{m_b}{v^3}\left( \frac{c_{b3}}{2}  \right) \bar{b}_L b_R h^3 +  \text{h.c.} \right],
\end{split}
\end{equation}
where we have taken $\lambda_\mathrm{SM} \equiv m_h^2 / 2v^2$. 

The CMS parametrization is then obtained by setting: $\kappa_\lambda = (1+d_3)$, $k_t = c_{t1}$, $c_2 = c_{t2}$, $c_{g} = c_{g1}$, $c_{gg} = c_{2g}$ and the ATLAS parametrization by $c_{hhh} = (1+d_3)$, $c_{ggh} = 2 c_{g1}/3$, $c_{gghh} = - c_{g2}/3$ (see, e.g.~\cite{xandatalk}). The Lagrangian of eq.~\ref{eq:LgghhPhenoExp} encapsulates the form of the interactions that we employ for the rest of our phenomenological analysis. 

\subsubsection{Monte Carlo Event Generation}

The Lagrangian of eq.~\ref{eq:LgghhPhenoExp} has been implemented in \texttt{MadGraph5\_aMC@NLO} (\texttt{MG5\_aMC})~\cite{Alwall:2011uj,Hirschi:2015iia}, following closely the instructions for proposed code modifications found in~\cite{loopxtree}.\footnote{Suggested by Valentin Hirschi.} These modifications essentially introduce tree-level diagrams in the form of fake ``UV counter-terms", that are generated along with any loop-level diagrams, therefore allowing the calculation of interference terms between them, which are otherwise not technically possible. The model presented in this section, created through the procedure briefly outlined, has been fully validated by direct comparison to an implementation of Higgs boson pair production in $D=6$ EFT in the \texttt{HERWIG 7} Monte Carlo, and by taking the limit of a heavy scalar boson for those vertices that do not appear in that process. See appendix B of~\cite{Papaefstathiou:2023uum} for further details of the latter effort. The necessary modifications to the \texttt{MG5\_aMC} codebase,\footnote{At present available for versions 2.9.15 and 3.5.0.} as well as the model can be found in the public gitlab repository at~\cite{gitlabrepo}.\footnote{It is interesting to note here that there exists a more comprehensive \texttt{MG5\_aMC} treatment of one-loop computations in the standard-model effective field theory at $D=6$ (dubbed ``smeft@nlo'')~\cite{Degrande:2020evl}, which should directly map to the $D=6$ limit of the present section.}

\subsubsection{Phenomenological Analysis}

To obtain constraints on anomalous triple Higgs boson production at proton colliders, we have performed a hadron-level phenomenological analysis of the 6 $b$-jet final-state originating from the decays of all three Higgs bosons to $b\bar{b}$ quark pairs. We closely follow the analysis of Refs.~\cite{Papaefstathiou:2019ofh,Papaefstathiou:2020lyp}. Parton-level events have been generated using the \texttt{MG5\_aMC} anomalous couplings implementation presented here, with showering, hadronization, and simulation of the underlying event, performed via the general-purpose \texttt{HERWIG 7} Monte Carlo event generator~\cite{Bahr:2008pv, Gieseke:2011na, Arnold:2012fq, Bellm:2013hwb, Bellm:2015jjp, Bellm:2017bvx, Bellm:2019zci,Bewick:2023tfi}. The event analysis was performed via the \texttt{HwSim} framework addon to \texttt{HERWIG 7}~\cite{hwsim}. No smearing due to the detector resolution or identification efficiencies have been applied to the final objects used in the analysis, apart from a $b$-jet identification efficiency, discussed below. 

The branching ratio of $h\rightarrow b\bar{b}$ will be modified primarily due to $c_{t1}$, $c_{g1}$, indirectly through modifications to the $h\rightarrow gg$ and $h \rightarrow \gamma \gamma$ branching ratios, and directly through $c_{b1}$. To take this effect into account, we employed the \texttt{eHDECAY} code~\cite{Contino:2014aaa}. The program eHDECAY includes QCD radiative corrections, and next-to-leading order EW corrections are only applied to the SM contributions. For further details, see Ref.~\cite{Contino:2014aaa}. We have performed a fit of the $\texttt{eHDECAY}$ branching ratio $h\rightarrow b\bar{b}$, and we have subsequently normalized this to the latest branching ratio provided by the Higgs Cross Section Working Group's Yellow Report~\cite{yr4page,LHCHiggsCrossSectionWorkingGroup:2016ypw}, $\mathrm{BR}(h\rightarrow b\bar{b})=0.5824$. The fit is then used to rescale the final cross section of $pp \rightarrow hhh \rightarrow (b\bar{b})(b\bar{b})(b\bar{b})$. The background processes containing Higgs bosons turned out to be subdominant with respect to the dominant QCD 6 $b$-jet and $Z$+jets backgrounds, and therefore we did not modify these when deriving the final cross sections. 

For the generation of the backgrounds involving $b$-quarks \textit{not} originating from either a $Z$  or Higgs boson, we imposed the following generation-level cuts for the 100~TeV proton collider: $p_{T,b} >30~\mathrm{GeV}$, $|\eta _j|< 5.0$, and $\Delta R_{b,b} >0.2$. The transverse momentum cut was lowered to $p_{T,b} >20~\mathrm{GeV}$ for 13.6 TeV, except for the QCD 6 $b$-jet background, for which we produced the events inclusively, without any generation cuts.\footnote{In general, the simulation of the QCD induced process $p p\rightarrow (b\bar{b}) (b\bar{b}) (b\bar{b})$ is one of the most challenging aspects of the phenomenological study. The samples are produced in parallel using OMNI cluster at the University of Siegen using the ``gridpack'' option available in \texttt{MG5\_aMC}.} The selection analysis was optimized considering as a main backgrounds the QCD-induced process $p p\rightarrow  (b\bar{b}) (b\bar{b}) (b\bar{b})$, and the $Z$+jets process (represented by $Z+(b\bar{b}) (b\bar{b})$), which we found to be significant at LHC energies. 

The event selection procedure for our analyses proceeds as follows: as in \cite{Papaefstathiou:2019ofh}, an event is considered if there are at least six $b$-tagged jets, of which only the six ones with the highest $p_{T}$ are taken into account. A universal minimal threshold for the transverse momentum, $p_{T,b}$, of any of the selected $b$-tagged jets is imposed. In addition a universal cut on their maximum pseudo-rapidity, $|\eta_b|$, is also applied. We subsequently make use of the observable:
\begin{eqnarray}
\chi^{2, (6)}=\sum_{qr \in I} (m_{qr}-m_h)^2\;,  
\label{eq:chi2}
\end{eqnarray}
where $I=\{jb_{1}jb_{2},jb_{3}jb_{4},jb_{5}jb_{6}\}$ is the set of all possible 15 pairings of 6-$b$ tagged jets. Out of all the possible combinations we pick the one with the smallest value $\chi_{\rm min}^{2, (6)}$. The pairings of $b$-jets defining $\chi_{\rm min}^{2, (6)}$ constitute our best candidates for the reconstruction of the three Higgs bosons, $h$. Our studies have demonstrated that $\chi_{\rm min}^{2, (6)}$ is one of the most powerful observables to employ in signal versus background discrimination. 

We further refine the discrimination power of the $\chi_{\rm min}^{2, (6)}$ variable by using the individual mass differences $\Delta m=|m_{q r} -m_h|$ in eq.~(\ref{eq:chi2}), sorting them out according to $\Delta m_{\rm min}< \Delta m_{\rm med} < \Delta m_{\rm max}$, and imposing independent cuts on each of them. We also consider the transverse momentum $p_{T}(h^i)$ of each reconstructed Higgs boson candidate. These reconstructed particles are also sorted based on the value of $p_{T}(h^i)$, on which we then impose a cut. Besides the universal minimal threshold on $p_{T,b}$, introduced at the beginning of this section, we impose cuts on the three $b$-jets with the highest transverse momentum $p_{T, b_i}$, for $i=1,2,3$. The set of cuts $p_{T, b_3}<p_{T, b_2}<p_{T, b_1}$ is the second most powerful discriminating observable in our list. Finally, we also considered two additional geometrical observables. The first of them is the distance between $b$-jets in each reconstructed Higgs boson $\Delta R_{bb}(h^i)$. The second one is the distance between the reconstructed Higgs bosons $\Delta R(h^i, h^j)$ themselves. 

Our optimization process then proceeds as in \cite{Papaefstathiou:2019ofh,Papaefstathiou:2020lyp}: we sequentially try different combination of cuts over the observables introduced above on our signal and background samples until we achieve a significance above $2$ or when our number of Monte Carlo events is reduced so drastically that no meaningful statistical conclusions can be derived if this number becomes smaller (this happens for instance when for a given combination of cuts, we are left with less than 10 Monte Carlo events of signal or background).\footnote{Note the Poisson uncertainty on 10 Monte Carlo events is $\sim^{+2.3}_{-4.3}$, resulting in a worst-case scenario uncertainty of $\sim^{+23\%}_{-43\%}$ on our event rates. In practice, this only occurs for certain backgrounds, and given the rest of the uncertainties in the present phenomenological analysis, we have deemed this to be acceptable.} The optimal set of cuts is shown in Table \ref{tab:cuts}, and the resulting signal and background cross sections and number of events are shown in Table \ref{tab:backgrounds}.

\begin{table}
\begin{center}
\begin{tabular}{ |l|l|l| } 
 \hline
 \multicolumn{3}{|c|}{\bf {Optimized cuts}}\\
 \hline
Observable & $13.6$ TeV & $100$ TeV \\ 
  \hline
 $p_{T,b}>$ & $25.95$ GeV &   $35.00$ GeV\\ 
 $|\eta_{b}|<$ & $2.3$ &  $3.3$ \\ 
  $\Delta R_{bb}>$&$0.3$&$0.3$\\
$p_{T,b_i}>$ &$[25.95, 25.95, 25.95]$ GeV $i=1,2,3$&$[170.00, 135.00, 35.00]$ GeV\\
$\chi^{2,(6)} <$ &$27.0$ GeV & $26.0$ GeV\\
$\Delta m_{\rm{ min, med, max}}<$&$[100, 200, 300]$ GeV& $[8, 8, 8]$ GeV\\
$\Delta R_{bb}(h^i)<$ & $[3.5, 3.5, 3.5]$&$[3.5, 3.5, 3.5]$\\
 $\Delta R(h^i,h^j)<$ & $[3.5, 3.5, 3.5]$ & $[3.5, 3.5, 3.5]$\\
 $p_{T}(h^i)>$ & $[0.0, 0.0, 0.0]$ GeV&$[200.0, 190.0, 20.0]$ GeV\\
$p_{T \rm jet}>$ & $25$~GeV &$25$~GeV\\
$|\eta_{\rm jet}|<$ & $4.0$  &  $4.0$ \\
 \hline
\end{tabular}
\caption{Optimized cuts determined for the phenomenological analysis. The indices $i,j$ can take the values $i,j=1,2,3$. For the cut $\Delta R(h^i, h^j)$ the three pairings correspond to $(1,2), (1,3), (2,3)$. The indexed elements should be read from left to right in increasing order. The last two rows refer to cuts over light jets.}
\label{tab:cuts}
\end{center}
\end{table}

\begin{table}
\begin{center}
\begin{tabular}{ |l|l|l|l| } 
 \hline
 \multicolumn{4}{|c|}{\bf{LHC Analysis $(13.6 ~\rm{TeV})$}}\\
 \hline
 &&&\\
 Process & $\sigma_{\rm{NLO}} (6~b\mathrm{-jet})~[\rm{fb}]$  & $\varepsilon_{\rm{analysis}}$&$N^{\rm{cuts}}_{3 \times 10^{3}~ \rm{fb}^{-1}}$\\
 &&&\\
 \hline
 $ h h h (\rm{SM})$& $1.97 \times 10^{-2}$& $0.12$& $2.77$\\
 \hline
 QCD $(b \bar{b}) (b \bar{b}) (b \bar{b})$& 6136.12  & $1.00 \times 10^{-5}$& 69.67 \\ \hline
   $pp \rightarrow Z (b \bar{b}) (b \bar{b})$ & 61.80  & 0.0045 & 318.17 \\\hline
   $pp \rightarrow Z Z (b \bar{b})$ &  2.16   & 0.0059  & 14.3 \\\hline
   $pp \rightarrow h Z (b \bar{b})$ &  0.45   & 0.0159 & 8.1 \\\hline
   $pp \rightarrow h h Z$ &  0.0374   & 0.034  & 1.45 \\\hline
   $pp \rightarrow h h (b \bar{b})$ & 0.0036& 0.028 & 0.11 \\\hline
   LI $gg \rightarrow h Z Z$ & 0.143   &0.022 & 3.62 \\\hline
   LI $gg \rightarrow Z Z Z$ & 0.124   &0.013 & 1.76 \\\hline
   LI $gg \rightarrow h h Z$ & 0.0458  &0.047 & 2.42\\\hline
\multicolumn{4}{c}{}\\
\hline
\multicolumn{4}{|c|}{\bf{FCC-hh Analysis $(100 ~\rm{TeV})$}}\\
 \hline
 &&&\\
 Process & $\sigma_{\rm{NLO}} (6~b\mathrm{-jet})~[\rm{fb}]$ & $\varepsilon_{\rm{analysis}}$&$N^{\rm{cuts}}_{20 ~ \rm{ab}^{-1}}$\\
 &&&\\
 \hline
 $ h h h (\rm{SM})$& $1.14$ & $0.0115$& $98.90$\\
 \hline
 QCD $(b \bar{b}) (b \bar{b}) (b \bar{b})$&  $56.66\times 10^{3}$  & $1.12 \times 10^{-5} $&4777.71\\\hline
    $pp \rightarrow Z (b \bar{b}) (b \bar{b})$ & 1285.37  & $3.04\times 10^{-5}$ & 294.63 \\\hline
   $pp \rightarrow Z Z (b \bar{b})$ & 49.01 & $2.02\times10^{-5}$  & 7.48\\\hline
   $pp \rightarrow h Z (b \bar{b})$ & 9.87  & $3.04\times 10^{-5}$& 2.26 \\\hline
   $pp \rightarrow h h Z$ &  0.601  & $5.95\times 10^{-4}$& 2.70 \\\hline
   $pp \rightarrow h h (b \bar{b})$ & 0.096  & $8.095\times 10^{-5}$& $\ll 1$ \\\hline
   LI $gg \rightarrow h Z Z$ &   8.28  &  $1.62\times 10^{-4}$& $10.12$ \\\hline
   LI $gg \rightarrow Z Z Z$ &  6.63  & $4.05\times 10^{-5}$ & $2.03$ \\\hline
   LI $gg \rightarrow h h Z$ &   2.65   & $2.54\times10^{-4}$& 5.07\\\hline 
 \end{tabular}
 \caption{The lists of processes considered during our phenomenological analysis, along with their respective cross sections to the 6 $b$-jet final state. The efficiencies $\varepsilon_{\rm analysis}$ and number of events $N_{\mathcal{L}}^{\rm cuts}$, correspond to those obtained after applying the set of cuts given in table~\ref{tab:cuts}. A $b$-jet identification efficiency of $0.85$ (for each $b$-jet) has also been applied to obtain the number of events. For the HL-LHC we considered an integrated luminosity of $\mathcal{L}=3000~\hbox{fb}^{-1}$, and for the FCC-hh a luminosity of $\mathcal{L}=20~\hbox{ab}^{-1}$. To approximate higher-order corrections, a $K$-factor $K=2$ has been included for all processes, with respect to the leading-order cross section. The background processes that are shown to be initiated via ``$pp$'' constitute the LO (tree-level) contributions, whereas those marked with ``LI" represent loop-induced contributions that form NLO corrections. Since they do not interfere with the LO processes, they have been generated separately. If a $Z$ boson is not stated in the process definition, the $(b\bar{b})$ has a QCD origin.}
 \label{tab:backgrounds}
 \end{center}
\end{table} 

To employ the results of the SM analysis over the whole of the parameter space we are considering, we have performed a polynomial fit of the efficiency of the analysis on the signal, $\varepsilon_\mathrm{analysis}(hhh)$, at various, randomly-chosen, combinations of anomalous coefficient values. In combination with the fits of the cross section, and the fit of the branching ratio of the Higgs boson to $(b\bar{b})$, we can estimate the number of events at a given luminosity, for a given collider for any parameter-space point within the anomalous coupling picture, which we dub $S(\{c_i\})$. 

\begin{figure}[htp]
\begin{center}
    \includegraphics[width=0.495\columnwidth]{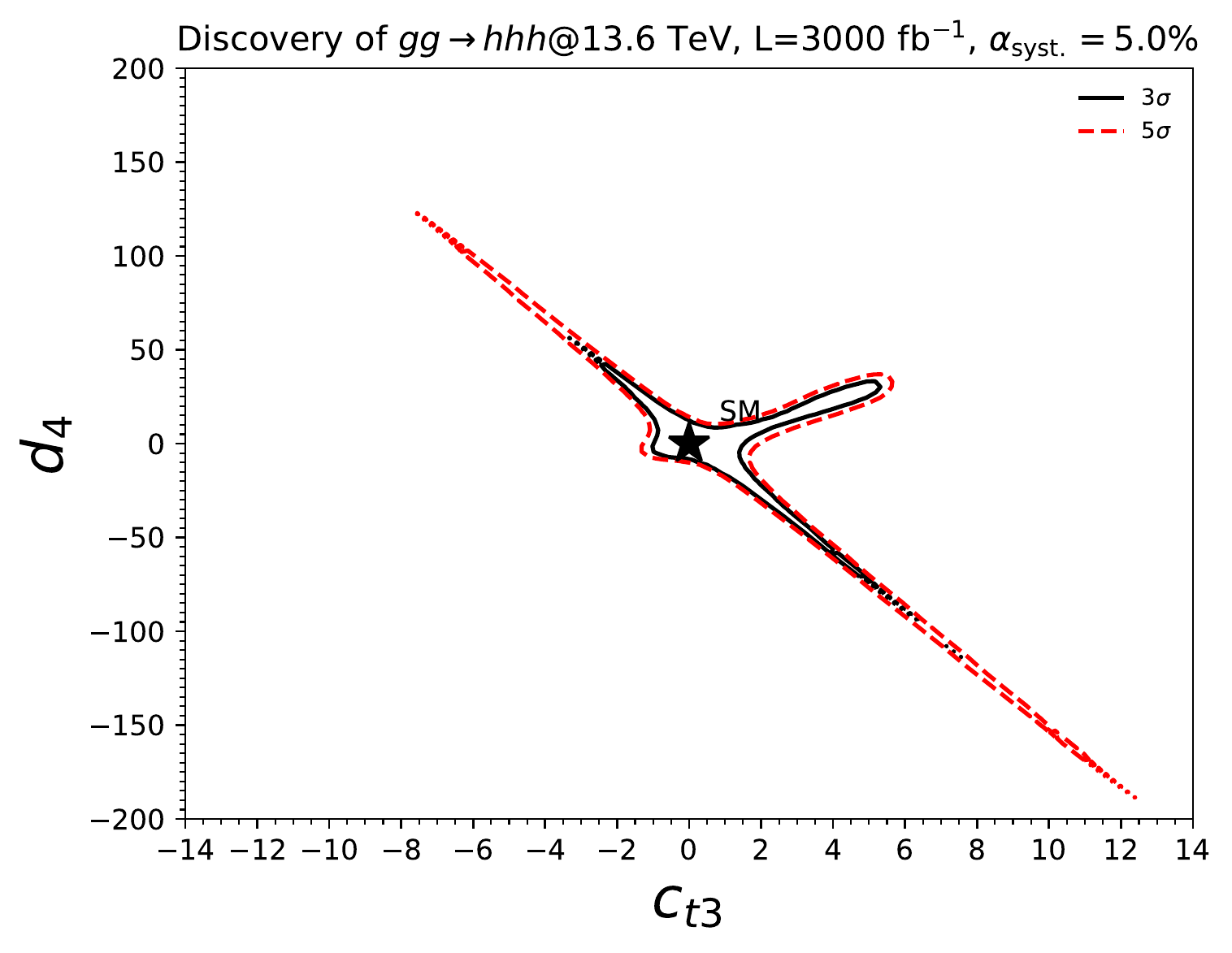}
  \includegraphics[width=0.495\columnwidth]{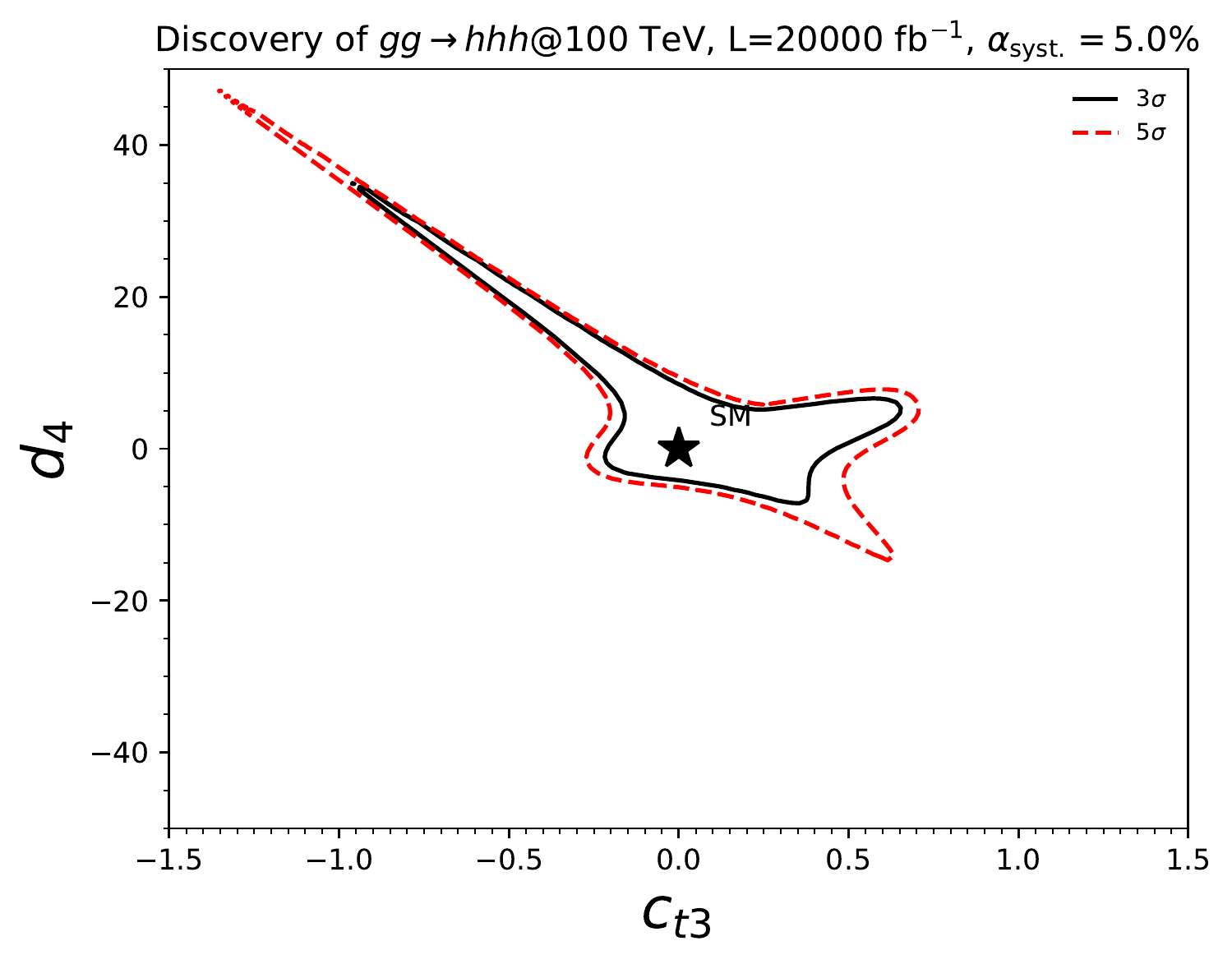}
\caption{\label{fig:2ddiscovery} The 3$\sigma$ evidence (black solid) and 5$\sigma$ discovery (red dashed) curves on the $(c_{t3},d_4)$-plane for triple Higgs boson production at 13~TeV/3000~fb$^{-1}$ (left), and 100~TeV/20~ab$^{-1}$ (right), marginalized over the $c_{t2}$ and $d_3$ anomalous couplings. Note the differences in the axes ranges at each collider.}
\end{center}
\end{figure}

\begin{figure}[htp]
\begin{center}
    \includegraphics[width=0.495\columnwidth]{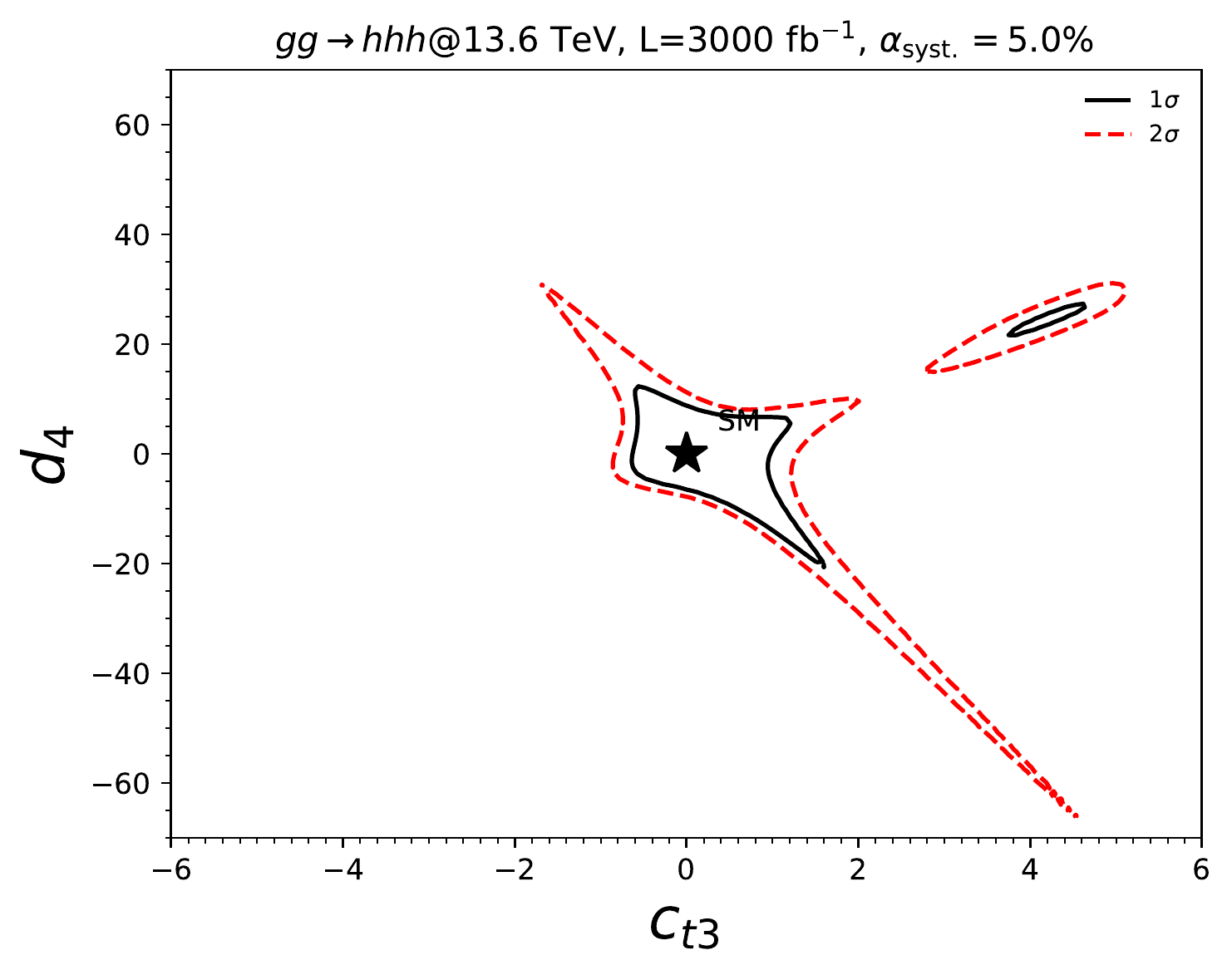}
  \includegraphics[width=0.495\columnwidth]{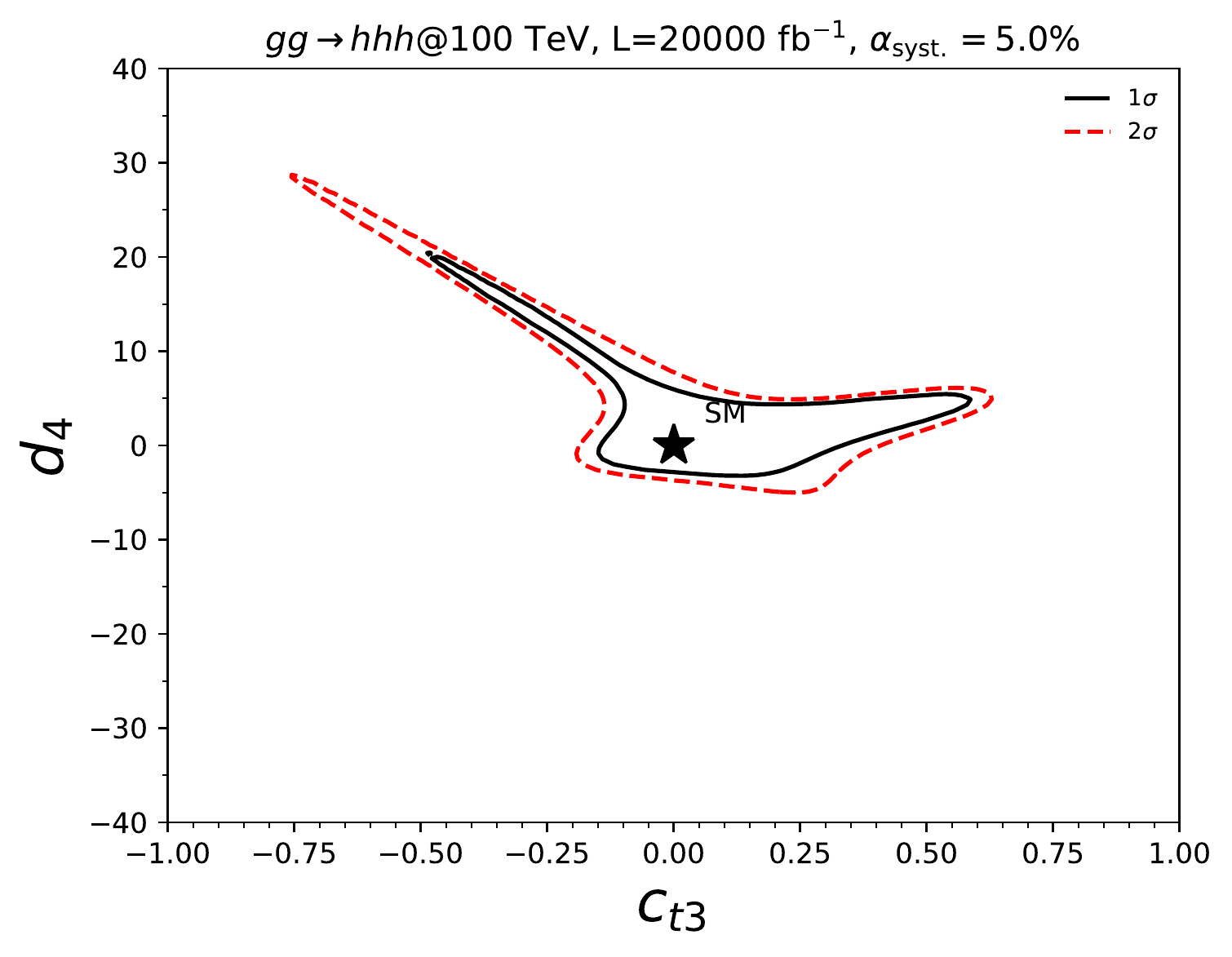}
\caption{\label{fig:2dconstraints} The 68\% C.L. (1$\sigma$, black solid) and 95\% C.L (2$\sigma$, red dashed) limit on the $(c_{t3},d_4)$-plane for triple Higgs boson production at 13~TeV/3000~fb$^{-1}$ (left), and 100~TeV/20~ab$^{-1}$ (right), marginalized over the $c_{t2}$ and $d_3$ anomalous couplings. Note the differences in the axes ranges at each collider. }
\end{center}
\end{figure}

\begin{table}[htp]
    \centering
\begin{tabular}{|c|c|c||c|c|}
\hline
& HL-LHC 3$\sigma$ & HL-LHC 5$\sigma$ & FCC-hh 3$\sigma$ & FCC-hh 5$\sigma$ \\\hline
$d_4$ & $[-28.0, 41.7]$  & $[-99.5, 152.9]$  & $[-24.9, 20.8]$ & $ [-40.8, 23.1]$     \\\hline
$c_{t3}$ & $[-2.1,5.5]$ & $[-7.1, 11.3]$ & $[-0.8,0.6]$ & $[-1.2, 0.7]$  \\\hline
\end{tabular}
   \caption{ The 3$\sigma$ evidence and 5$\sigma$ discovery limits on for triple Higgs boson production, for the $c_{t3}$ and $d_4$ coefficients at 13~TeV/3000~fb$^{-1}$, and 100~TeV/20~ab$^{-1}$, marginalized over $c_{t2}$, $d_3$ and either $d_4$, or $c_{t3}$. }\label{tab:disclimits}
\end{table}

\begin{table}[htp]
    \centering
\begin{tabular}{|c|c|c||c|c|}
\hline
& HL-LHC 68\% & HL-LHC 95\% & FCC-hh 68\% & FCC-hh 95\%   \\
\hline
$d_4$ & $[-6.6, 12.4]$  & $[-10.0, 21.3]$  & $[-3.9, 10.5]$ & $ [-10.6, 18.8]$     \\
\hline
$c_{t3}$ & $[-0.6, 1.1]$ & $[-0.9, 3.6]$ & $[-0.1, 0.3]$ & $[-0.4,  0.6]$  \\
\hline
\end{tabular}
   \caption{ The 68\% C.L. (1$\sigma$) and 95\% C.L (2$\sigma$) limits on $c_{t3}$ and $d_4$ for triple Higgs boson production at 13~TeV/3000~fb$^{-1}$, and 100~TeV/20~ab$^{-1}$, marginalized over $c_{t2}$, $d_3$ and either $d_4$, or $c_{t3}$. }\label{tab:limits}
\end{table}

The main results of our two-dimensional analysis over the $(c_{t3},d_4)$-plane are shown in figs.~\ref{fig:2ddiscovery} and~\ref{fig:2dconstraints}. In particular, fig.~\ref{fig:2ddiscovery} shows the potential ``evidence" and ``discovery'' regions (3$\sigma$ and 5$\sigma$, respectively) for triple Higgs boson production at the high-luminosity LHC on the left (13.6 TeV with 3000~fb$^{-1}$ of integrated luminosity), and at a FCC-hh (100 TeV, with 20~ab$^{-1}$) on the right. Evidently, very large modifications to the quartic self-coupling are necessary for discovery of triple Higgs boson production at the HL-LHC, ranging from $d_4\sim 125$ for $c_{t3}~\sim-8$, to $d_4\sim \pm 40$ for $c_{t3}\sim 0$ and then down to $d_4\sim -200$ for $c_{t3}\sim 12$. The situation is greatly improved, as expected, at the FCC-hh, where the range of $d_4$ is reduced to $d_4~\sim 40$ for $c_{t3}\sim-1.5$, and to $d_4\sim -20$ for $c_{t3}\sim 1.0$. It is interesting to note that the whole of the parameter space with $c_{t3} \gtrsim 1.0$, or with $c_{t3} \lesssim -1.5$ is discoverable, at the FCC-hh at 5$\sigma$. For the potential 68\% (1$\sigma$) and 95\% C.L. (2$\sigma$) constraints of fig.~\ref{fig:2dconstraints}, the situation is slightly more encouraging for the HL-LHC, with the whole region of $d_4 \gtrsim 40$ or $d_4 \lesssim -60$ excluded at 95\% C.L.. The corresponding region at 68\% C.L. is $d_4 \gtrsim 20$ and $d_4 \lesssim -30$. For $c_{t3}$, it is evident that all the region $c_{t3} \lesssim -2$ and $c_{t3} \gtrsim 5$ will be excluded at 95\% C.L. and $c_{t3} \lesssim -1$, $c_{t3} \gtrsim 4$ at 68\% C.L.. On the other hand, the FCC-hh will almost be able to exclude the whole positive region of $d_4$ for any value of $c_{t3}$ at 68\% C.L.. This will potentially be achievable if combined with other Higgs boson triple production final states. For the $c_{t3}$ coupling, both the constraints reach the $\mathcal{O}(\mathrm{few}~10\%)$ level for any value of $d_4$. 

The one-dimensional analysis' results, presented in tables~\ref{tab:disclimits} and~\ref{tab:limits}, for the ``evidence'' and ``discovery'' potential, and exclusion limits, respectively, reflect the above conclusions. For instance, it is clear by examining table~\ref{tab:disclimits}, that the HL-LHC will only see evidence of triple Higgs boson production in the 6 $b$-jet final state only if $d_4$ has modifications of $|d_4| \sim \mathcal{O}(\mathrm{few}~10)$, and will only discover it if $|d_4| \sim \mathcal{O}(100)$. On the other hand, there could be evidence or discovery of Higgs boson triple production if $|c_{t3}|\sim \mathcal{O}(\mathrm{1-10})$. The 1$\sigma$ and 2$\sigma$ exclusion regions are much tighter, as expected, with $|d_4| \sim \mathcal{O}(10)$ at 1$\sigma$ or 2$\sigma$ at the HL-LHC, improving somewhat at the FCC-hh, and $|c_{t3}| \sim \mathcal{O}(0.1-1)$, both at the HL-LHC and FCC-hh.

\subsubsection{Conclusions}

The results of our study demonstrate the importance of including additional contributions, beyond the modifications to the self-couplings, when examining multi-Higgs boson production processes, and in particular triple Higgs boson production. We are looking forward to a more detailed study for the HL-LHC, conducted by the ATLAS and CMS collaborations, including detector simulation effects, and the full correlation between other channels. From the phenomenological point of view, improvements will arise by including additional final states, e.g.\ targetting the process $pp \rightarrow hhh \rightarrow (b\bar{b})(b\bar{b})(\tau^+\tau^-)$, or by performing an analysis that leverages machine learning techniques to maximize significance.\footnote{This approach was taken in~\cite{Stylianou:2023xit} at the HL-LHC for modifications of the Higgs boson's self-couplings.} In summary, we believe that the triple Higgs boson production process should constitute part of a full multi-dimensional fit, within the anomalous couplings picture. 

\subsection{{Probing the Higgs potential through gravitational waves} \label{sec:GW}}
{\sl R. Pasechnik}

\subsubsection{Cosmological phase transitions}

At very early times, the Universe rapidly went through a so-called electroweak (EW) phase transition into a ground state in which almost all elementary particles became massive through interaction with the Higgs field. It is also believed that in the course of this transition the observed matter-antimatter asymmetry has been created through a mechanism widely known as EW baryogenesis. A strongly first-order phase transition manifests, similarly to the boiling of water, as a violent process of bubble nucleation away from thermal equilibrium. Such a process is realised either via quantum tunneling through the potential barrier determined by instanton solutions (at low temperatures)~\cite{Linde:1981zj,Dine:1992wr}, or through thermal jumps over the barrier (at high temperatures). The dynamics of both types of processes can be described as a classical motion in Euclidean space governed by the three-dimensional action:
\begin{equation}
\hat{S}_3(\hat{\phi},T) = 4 \pi \int_0^\infty \mathrm{d}r \, r^2 \left\{ \frac{1}{2} 
\left( \frac{\mathrm{d}\hat{\phi}}{\mathrm{d}r} \right)^2 + V_{\rm eff}(\hat{\phi},T) \right\} \,,
\label{S3}
\end{equation}
where $\hat{\phi}$ is a classical solution of the equation of motion found as the path in the configuration (field) space that minimizes the action \cite{Coleman:1977py,Wainwright:2011kj}, and the thermal effective potential of the underlined theory \cite{Quiros:1999jp,Curtin:2016urg}
\begin{equation}
V_{\rm eff}(\hat{\phi},T) = V_0 + V_{\rm CW} + \Delta V(T) + V_{\rm ct}\,,
\label{eff-potential}
\end{equation}
that consists of the tree-level potential $V_0$ of a given model, the Coleman-Weinberg potential generated by radiative corrections at zero temperature $V_{\rm CW}$, the thermal corrections $\Delta V(T)$, and the counter-term potential $V_{\rm ct}$. 

A phase transition is characterised by the critical temperature $T_c$, at which a local minimum of $V_{\rm eff}(T)$ evolves to become equal to its global one, and by the transition rate \cite{Ellis:2020nnr,Ellis:2022lft}
\begin{equation}
    \Gamma(T) \approx T^4 \left( \frac{\hat{S}_3}{2 \pi T} \right)^{3/2} e^{-\hat{S}_3/T}\,.
    \label{Gamma-nucl_rate}
\end{equation}
At $T<T_c$, the nucleation process becomes effective due to large thermal fluctuations. The nucleation temperature $T_n$ is then found requiring as a point in cosmological history when a single true vacuum bubble nucleates per cosmological horizon satisfying,
\begin{equation}
    \int_{T_n}^{T_c} \frac{d T}{T} \frac{\Gamma(T)}{H^4} = 1\,, \qquad
    H^2 = \frac{g_\ast(T) \pi^2 T^4 }{90 M_\mathrm{Pl}^2} + \frac{\Delta V_{\rm vac}}{3 M_\mathrm{Pl}^2}\,,
    \label{Tn-def}
\end{equation}
in terms of the Hubble rate $H=H(T)$ including both the radiation (first term) and vacuum density $\Delta V_{\rm vac}$ (second term) contributions~\cite{Brdar:2018num}, and the effective number of relativistic d.o.f.~in the considered epoch at a temperature $T$ is denoted as $g_\ast(T)$. Here, $M_\mathrm{Pl}$ is the Planck mass. For practical purposes, it is particularly useful to define a percolation temperature $T_*< T_n < T_c$ at which about $34\%$ of the Universe has transited to the true (stable) vacuum \cite{Ellis:2018mja}, i.e.
\begin{align}
I(T_*) = 0.34\,, \qquad 
I(T) = \frac{4\pi v_b^3}{3} \int_T^{T_c} \frac{\Gamma(T')dT'}{T'^4 H(T')}\left(\int_T^{T'}\frac{d\tilde{T}}{H(\tilde{T})}\right)^3\,,
\label{T_perc}
\end{align}
Eventually, the bubbles of the energetically favoured vacuum state take over giving rise to today's Universe where the EW symmetry is broken.

The energy budget of a first-order phase transition, and hence its effective strength, is typically characterised by difference in the trace anomaly between the two (initial metastable $i$ and final stable $f$) phases or vacua relative to the radiation density of the universe $\rho_\gamma$ at the nucleation epoch $T=T_*$ \cite{Hindmarsh:2015qta,Hindmarsh:2017gnf}
\begin{equation}
\alpha = \frac{1}{\rho_\gamma} \left[ \Delta V - \dfrac{T}{4} \left( \frac{\partial \Delta V}{\partial T} \right) \right] \,, \quad \Delta V = V_i - V_f \,, \quad \rho_\gamma = g_\ast(T_*) \frac{\pi^2}{30} T_*^4\,,
\label{alpha}
\end{equation}
where $V_i\equiv V_{\rm eff}(\phi^i,T_*)$ and $V_f\equiv V_{\rm eff}(\phi^f,T_*)$ are the effective potential values in the two phases, respectively. Besides, $\alpha$ one also considers the inverse time scale of the phase transition defined in units of the cosmological horizon time-scale given by~$H(T)$ as
\begin{equation}
\frac{\beta}{H} = T_\ast  \left. \frac{\partial}{\partial T} \left( \frac{\hat{S}_3}{T}\right) \right|_{T_\ast}\,.
\label{beta_H}
\end{equation}

\subsubsection{Primordial gravitational waves: a window into New Physics}

The bubble nucleation process triggers ripples in space-time through bubble collisions, sound-waves and turbulence in the cosmic plasma contributing to the gravitational-wave background. A large amount of energy released in the phase transition over a very short time scale could be sufficient to produce primordial gravitational waves potentially observable today \cite{Kosowsky:1992rz}. The key observable is the energy-density of the gravitational radiation per logarithmic frequency found as (see for instance Refs.~\cite{Caprini:2001nb,Figueroa:2012kw,Hindmarsh:2016lnk} and references therein)
\begin{equation}
h^2 \Omega_{\rm GW}(f) \equiv \frac{h^2}{\rho_c} \frac{\partial \rho_{\rm GW}}{\partial \log f} = h^2 \Omega_\mathrm{GW}^\mathrm{peak} \left(\dfrac{4}{7}\right)^{-\tfrac{7}{2}} \left(\dfrac{f}{f_\mathrm{peak}}\right)^3 \left[1 + \dfrac{3}{4} \left(\dfrac{f}{f_\mathrm{peak}}\right) \right]^{-\tfrac{7}{2}} \,, 
\label{GW-spectrum}
\end{equation}
in terms of the known critical density of the modern Universe $\rho_c$ as well as the peak-frequency $f_\mathrm{peak}$ and the peak-amplitude $h^2 \Omega_\mathrm{GW}^\mathrm{peak}$ of the gravitational signal. The latter quantities can be parameterised in terms of the basic phase transition characteristics $T_*$, $\alpha$ and $\beta/H$ as well as the bubble wall velocity $v_w$ (see Refs.~\cite{Caprini:2019egz,Caprini:2024hue} for explicit formulas).
Hence, the measurement of primordial gravitational waves is often considered as an indirect way to probe the Higgs potential at temperatures close to the EW phase transition epoch in the early Universe \cite{LISACosmologyWorkingGroup:2022jok,Caprini:2024hue}.

Observable consequences of the EW phase transition in cosmology are strongly connected to the structure of the Higgs potential. A large ongoing effort in the literature is devoted to analysis of an interplay between direct probes of the Higgs potential (such as e.g.~Higgs boson pair and triple-Higgs production etc) at high-energy particle colliders and indirect probes of the EW phase transition in the early Universe using primordial gravitational waves (for such a discussion, see e.g.~Refs.~\cite{Karkout:2024ojx,Caprini:2024hue,Ahriche:2023jdq,Addazi:2023ftv,Biekotter:2023eil,Biekotter:2022kgf,Biekotter:2021ysx,Camargo-Molina:2021zgz,Alves:2020bpi,Eichhorn:2020upj,Alves:2019igs,Morais:2019fnm,Huang:2018aja,Chala:2018opy,Alves:2018oct,Huang:2016cjm,Hashino:2016rvx,Hashino:2016xoj,Kakizaki:2015wua} and references therein). Existing and planned gravitational-wave facilities, hence, offer a plethora of new opportunities to explore many different New Physics scenarios in a way that is complementary to direct observations at particle colliders \cite{Kosowsky:1992rz} (for a recent review, see e.g.~Ref.~\cite{Athron:2023xlk} and references therein). For the recent measurements of the stochastic gravitational waves' background by NANOGrav and its implications for New Physics, see e.g.~Ref.~\cite{NANOGrav:2023hvm} and references therein.

The LHC or its high-luminosity upgrade HL-LHC are expected to provide the data sufficient for a consistent measurement of the Higgs boson self-coupling depending on whether or not the Higgs potential is minimal~\cite{DiMicco:2019ngk,ATLAS:2018rvj}. On the other hand, the next generation, space-based, gravitational-wave experiment LISA~\cite{LISA:2017pwj} will have a large enough sensitivity to probe the stochastic gravitational-wave background of the Universe sourced by the EW phase transition for the first time~\cite{Caprini:2019egz,Caprini:2024hue}. The recent approval of LISA Phase B2 on January 25, 2024, has marked the start of its hardware implementation, with the launch expected in mid-2030s. Thus, we live in a historic moment when the collider data from terrestrial measurements such as LHC can be combined with the astrophysical data from space-based gravitational-wave experiments such as LISA to uncover the true dynamics of the EW phase transition and thereby to probe possible variants of New Physics beyond the Standard Model (BSM).

It is well-known that in the SM featuring the minimal Higgs sector, the EW phase transition is a continuous second-order transformation between the EW-symmetric and EW-broken phases. As result, neither the observed baryon asymmetry nor primordial gravitational waves can be produced in this case. These and other well-known shortcomings of the SM (such as a lack of CP violation required by baryogenesis, the absence of a neutrino mass generation mechanism and of a suitable particle Dark Matter candidate) have historically been the strongest points favouring New Physics still waiting for its direct experimental observation.

Typical New Physics scenarios feature additional states coupled to the Higgs field. These affect the shape of the effective Higgs potential which strongly depends on temperature, such that it is important to probe it at different epochs of cosmic evolution. Indeed, the curvature of the Higgs potential is determined by the measured Higgs mass ($m_h$) and triple-Higgs self-coupling ($\lambda_{hhh}$) whose value as a function of quartic Higgs coupling ($\lambda_{hhhh}$) is known in the SM but may change in BSM theories e.g.~via radiative corrections involving non-SM particles. A systematic and generic way of deriving the effective Higgs self-interactions starting from the one-loop effective potential has been advised in Ref.~\cite{Camargo-Molina:2016moz}. A precision measurement of both $\lambda_{hhh}$ (e.g.~through Higgs pair production) and $\lambda_{hhhh}$ (through triple Higgs production) is instrumental for probing the influence of BSM physics on the Higgs potential and, hence, for the physics of EW phase transition. Such a measurement is one of the key reasons for LHC upgrades as well as for building new collider facilities at the high-energy frontier.

Characteristics of the phase transitions and the corresponding spectrum of primordial gravitational waves in a particular model strongly depend on the thermal effective Higgs potential, which is determined by the particle spectrum of the model and their interactions with the Higgs boson. Many SM extensions feature very complicated, non-minimal Higgs potentials as is the case, for instance, in composite Higgs and supersymmetric theories which also provide an explanation for Dark Matter and yield the Higgs boson mass protected from large radiative corrections. For the purpose of detailed exploration of the EW phase transition, instead of considering the full UV-complete models, it is more consistent to utilise simplified TeV-scale Effective Field Theory (EFT) approximations to them focusing only on a few operators and states relevant below a TeV energy scale such as popular 2HDMs and a scalar singlet-extended SM. 

At high temperatures, additional sub-TeV scalar states and their interactions with the Higgs boson effectively induce a barrier between the initial and final phases in the Higgs vacuum, thus, enabling the EW phase transition to become first-order which, along with the enhanced CP violation, is a vital part of the EW baryogenesis mechanism \cite{Sakharov:1967dj}. Indeed, adding a single EW-singlet field at EW scale is already enough to trigger a first-order phase transition that is sufficiently strong for efficient EW baryogenesis \cite{Espinosa:2007qk}. Knowing the effective Higgs potential at finite temperatures, it is rather straightforward to compute the phases of the theory at a given temperature and such characteristics of transitions between them as the strength of the transition (usually attributed to the released latent heat), the bubble nucleation and percolation temperatures and the duration of a given phase transition. The latter, in turn, determine the spectrum of the produced gravitational waves, although still with significant uncertainties (see e.g.~Ref.~\cite{Freitas:2021yng}).

\subsubsection{Constraining Higgs self-interactions with gravitational waves: the case of Majoron EFT}

One of the simplest SM extensions -- a singlet scalar extended SM -- provides a rich playground for studies of the impact of extra scalar states and interactions on the shape of the effective Higgs potential at finite temperatures, the phase transition dynamics and the related production of primordial gravitational waves\footnote{For earlier works exploring the rich vacuum structure, phase transitions and gravitational wave signatures in extended models featuring more complicated scalar sectors, see e.g.~Refs.~\cite{Vieu:2018nfq,Wang:2019pet,Morais:2019fnm,Bonilla:2023aij} and references therein}. For recent studies of the interplay between the collider signatures and gravitational wave signals in the real singlet scalar extended SM, see e.g.~Refs.~\cite{Beniwal:2017eik,Alves:2018oct,Alves:2019igs,Alves:2020bpi} and references therein. In what follows, we elaborate on such an interplay for a complex singlet scalar extended SM representing another simplest and popular class of BSM scenarios having important implications for both Dark Matter physics and neutrino mass generation (see e.g.~Refs.~\cite{Espinosa:2007qk,Ashoorioon:2009nf,Freitas:2021yng,Goncalves:2021egx}). 

The potential of complex-singlet extended model possesses a global softly-broken U(1) symmetry, and the Higgs weak-doublet $H$ and a complex EW-singlet $\sigma$ can have a non-zero charge under this symmetry. Such a potential provides a basis for the so-called Majoron model \cite{Chikashige:1980ui,Schechter:1981cv,Gonzalez-Garcia:1988okv} where the global U(1) is extended to the lepton sector (thus, considered to be the lepton number symmetry) with additional Majorana neutrinos giving rise to the neutrino mass generation via an inverse seesaw mechanism. 

The simplest renormalisable (dimension-4) variant of the Majoron model that assumes no UV completion does not yield primordial gravitational waves detectable by LISA simultaneously with featuring a suitable Dark Matter candidate \cite{Addazi:2019dqt}. Unknown UV physics at a large seesaw scale $\Lambda$, however, can generate higher-dimensional operators in the low-energy EFT that may enhance the potential barrier between false and true vacua. The emergence of a first-order EW phase transition induced by a dimension-six $\propto (HH^\dagger)^3$ operator in the effective Higgs potential has been studied in Refs.~\cite{Huang:2016odd,Grojean:2004xa,Delaunay:2007wb}. In the case of Majoron EFT, this occurs without immediately conflicting with the invisible Higgs decay constraints for light Majorons \cite{Addazi:2023ftv}. 

Considering the contribution from the U(1)-preserving dimension-six operators, the potential of such a Majoron EFT valid at energies below the cutoff scale $\Lambda$ reads:
\begin{equation}
    V_{0} (H,\sigma) = V_{\mathrm{SM}}(H) + V_{{4\mathrm{D}}}(H,\sigma) + V_{{6\mathrm{D}}}(H,\sigma) + V_{\mathrm{soft}} (\sigma)\,,
    \label{eq:V-EFT}
\end{equation}
where
\begin{equation}
    \begin{aligned}
    V_{\mathrm{SM}}(H) &= \mu_h^2 H^{\dagger}H + \lambda_h(H^{\dagger}H)^2\,,
    \\
    V_{{4\mathrm{D}}}(H,\sigma) &= \mu_{\sigma}^2 \sigma^{\dagger}\sigma + \lambda_{\sigma}(\sigma^{\dagger}\sigma)^2 + \lambda_{\sigma h}H^{\dagger}H \sigma^{\dagger}\sigma\,,
    \\
    V_{{6\mathrm{D}}}(H,\sigma) &= \frac{\delta_0}{\Lambda^2}(H^{\dagger}H)^3 + \frac{\delta_2}{\Lambda^2}(H^{\dagger}H)^2\sigma^{\dagger}\sigma + \frac{\delta_4}{\Lambda^2}H^{\dagger}H(\sigma^{\dagger}\sigma)^2
    +
    \frac{\delta_6}{\Lambda^2}(\sigma^{\dagger}\sigma)^3\,,
    \\
    V_{\mathrm{soft}} (\sigma) &= \frac{1}{2}\mu_b^2 \left(\sigma^2 + \sigma^{\ast 2} \right)\,.
    \label{eq:V-components}
    \end{aligned}
\end{equation}
In the real field basis, $H$ and $\sigma$ are found as follows
\begin{equation}
\begin{aligned}
H = \dfrac{1}{\sqrt{2}} 
\begin{pmatrix}
\omega_1 + i \omega_2  \\
\phi_h + h + i \eta
\end{pmatrix}\,,	
\qquad
\sigma = \dfrac{1}{\sqrt{2}} \left( \phi_\sigma + h^\prime + i J \right)\,,	
\end{aligned}
\end{equation}	
in terms of the classical-field configurations $\phi_h$ and $\phi_\sigma$, the corresponding radial fluctuations $h$ and $h^\prime$, and Goldstone bosons $\omega_{1,2}$, $\eta$ effectively ``eaten up'' by $W^\pm$ and $Z$ bosons, and Majoron $J$ staying physical in the spectrum. The latter acquires its pseudo-Goldstone mass via a soft U(1) $\to \mathbb{Z}_2$ breaking mass term $V_{\mathrm{soft}} (\sigma)$ and may play a role of Dark Matter. Further details of the mass spectrum, parameter space and the finite-temperature effective potential in this model can be found in Ref.~\cite{Addazi:2023ftv}.

As was mentioned above, the strong first-order EW phase transitions originate due to a sizeable trilinear coupling of the SM-like Higgs boson $h_1$ determining a potential barrier between the false and true Higgs vacua. At one-loop level, it receives contributions from the top quark $t$, heavy neutrinos $N$ and the second CP-even Higgs boson $h_2$ loops, i.e.
\begin{equation}
    \lambda_{h_1 h_1 h_1} = \lambda_{h_1 h_1 h_1}^{(0)} + \frac{1}{16 \pi^2} \left(\lambda_{h_1 h_1 h_1}^t + \lambda_{h_1 h_1 h_1}^N + \lambda_{h_1 h_1 h_1}^{h_2}\right)
    \label{eq:tri-Higgs}
\end{equation}
with explicit expression found in Ref.~\cite{Addazi:2023ftv}. 
\begin{figure}[htb!]
	\centering
	\includegraphics[width=0.48\textwidth]{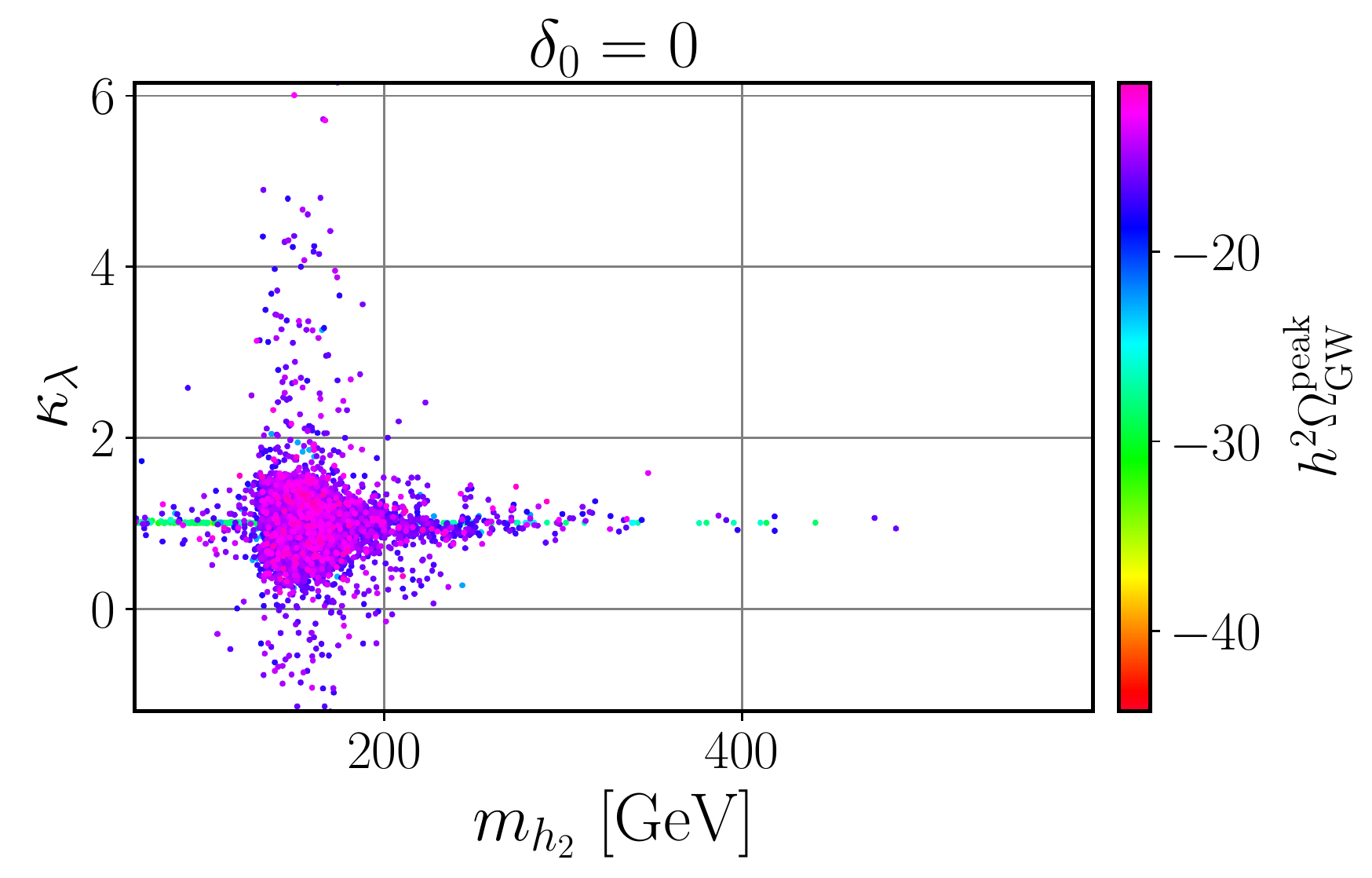}
	\includegraphics[width=0.48\textwidth]{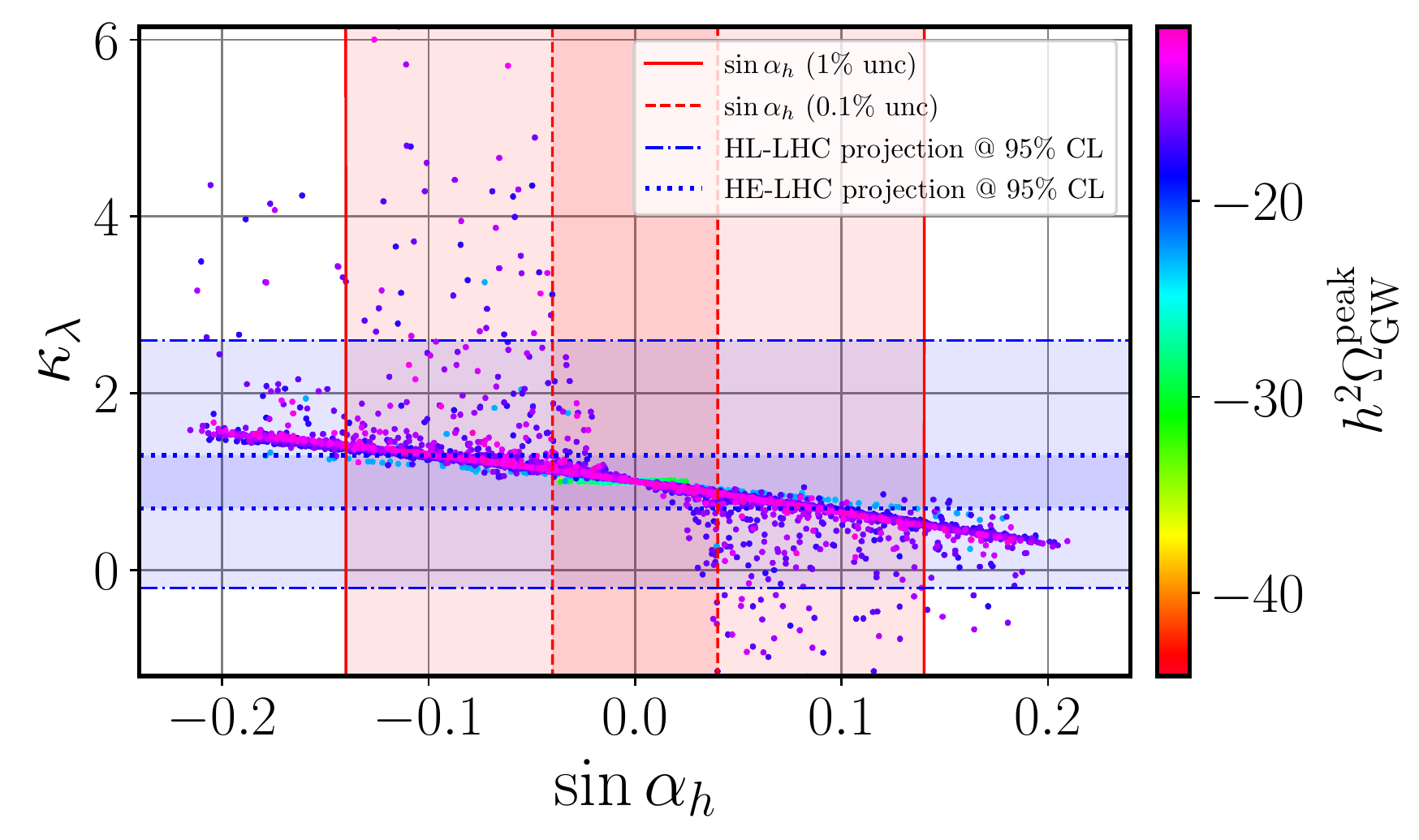}
 \caption{\footnotesize Left panel: correlation between the trilinear Higgs coupling modifier $\kappa_\lambda$ and the mass of the second CP-even Higgs boson mass $h_2$. Right panel: correlation between $\kappa_\lambda$ and the scalar CP-even mixing $\sin \alpha_h$. Only points featuring a strongly first-order phase transition are shown. In both panels, the peak-amplitude of the resulting primordial gravitational wave spectrum is presented in the color scale, and $\delta_0 = 0$ has been imposed in a restrictive parameter scan. From Ref.~\cite{Addazi:2023ftv}. }
	\label{fig:kappa-Omega}
\end{figure}

In a general parameter scan of Ref.~\cite{Addazi:2023ftv}, one extracts the Lagrangian parameters $\lambda_{\sigma h}$, $\lambda_\sigma$, $\lambda_h$, $\mu_b$ and $\delta_6$ in terms of five physical observables serving as input parameters -- two CP-even scalar masses, $m_{h_1}\simeq 125$ GeV and $m_{h_2}=[60\dots 1000]$ GeV, the scalar mixing angle $\alpha_h$ (bounded by $|\sin \alpha_h|<0.24$), the mass of the pseudo-Goldstone Majoron state $m_J$, and the Higgs branching ratio into Majorons $\mathrm{Br}\left( h_1 \to J J \right)$. Additional parameters varied in the scan are the EFT cut-off scale $\Lambda=[10\dots 1000]$ TeV, the Majoron VEV $v_\sigma=[100\dots 1000]$ GeV, light neutrino mass scale $m_{\nu}<0.1$ eV, as well as the magnitudes of the remaining dimension-6 operators $\delta_{0,2,4}$ chosen to be within wide ranges consistent with perturbativity of the corresponding quartic couplings in the EFT potential (\ref{eq:V-components}). In the original work of Ref.~\cite{Addazi:2023ftv}, the parameter ranges were chosen to generically comply with the LHC constraints on $\kappa_\lambda \lesssim 6.5$ \cite{CMS:2022dwd} (see Table~\ref{tab:sensitivity_kappa_lambda_hh} for the existing CMS and ATLAS bounds in various channels) and on the branching ratio of invisible Higgs decays ${\rm BR}(h_1 \to JJ) < 0.18$ from Ref.~\cite{CMS:2022qva}. The current ATLAS bound on the latter observable is somewhat tighter, ${\rm BR}(h_1 \to JJ) < 0.107$ \cite{ATLAS:2023tkt}, which may not impact the conclusions of Ref.~\cite{Addazi:2023ftv} in a significant manner. Besides, the case of a light Majoron has been considered, with $m_J < 100$ keV, that might, under certain conditions, play a role of Dark Matter candidate.

The main results of the restricted parameter scan are highlighted in Fig.~\ref{fig:kappa-Omega} where the correlations of the the trilinear Higgs coupling modifier $\kappa_\lambda$ with the mass of the second CP-even Higgs boson mass $h_2$ (left) and with the scalar CP-even mixing parameter $\sin \alpha_h$ (right) are shown for parameter space points that provide a strongly first-order phase transition~\cite{Addazi:2023ftv}. In this restrictive scan, the dimension-six Higgs operator has been omitted $\delta_0 = 0$ to showcase the importance of Higgs-singlet portal interactions encoded in $\delta_2$ and $\delta_4$. For such points, the strength of the phase transition takes maximal values $\alpha \gtrsim 0.1$ while the inverse duration is minimised $10\lesssim \beta/H \lesssim 100$. The color scale represents the maximal value of the induced cosmic gravitational wave spectrum $h^2 \Omega_\mathrm{GW}^\mathrm{peak}$. Here, the magenta points, with the ballpark localised at $100 \lesssim m_{h_2}/{\rm GeV} \lesssim 250$ and $0\lesssim \kappa_\lambda \lesssim 2$, represent gravitational wave signals strong enough to be potentially detectable at the future LISA \cite{LISA:2017pwj}, BBO \cite{Crowder:2005nr,Corbin:2005ny} and DECIGO \cite{Seto:2001qf,Isoyama:2018rjb} facilities.

\subsubsection{Conclusions}

The same operators that enhance the strength of the phase transition also contribute to the Higgs-pair, triple-Higgs and the associated (with an additional scalar) Higgs production channels. These observables of the Higgs sector provide a direct access to the Higgs vacuum structure and, thus, to dynamics of the phase transitions. The future measurements of these channels at the LHC, its forthcoming upgrades and future colliders can be cross-correlated with possible observations of primordial gravitational waves enabling to efficiently constrain New Physics at a TeV scale \cite{Basler:2016obg,Huang:2016cjm,Basler:2017uxn,Hashino:2018wee,Biekotter:2022kgf}. Moreover, correlations with other measurements such as the searches for $A\to ZH$ (with $H\to t\bar t$) in the case of 2HDMs \cite{Biekotter:2023eil} or for deviations from the SM prediction in the $W$ boson mass in the case of scalar-triplet extensions \cite{Addazi:2022fbj} provide other possible probes for parameter space domains in multi-scalar models that feature strong first-order EW phase transitions.

Fig.~\ref{fig:kappa-Omega} (right) provides a good example illustration of the interplay between collider and gravitational wave measurements. In particular, the power of such an interplay can be understood considering the measurements of the CP-even scalar mixing angle $\alpha_h$ (red vertical lines) \cite{Papaefstathiou:2022oyi} and the Higgs trilinear coupling $\kappa_\lambda$ (blue horizontal lines) \cite{Goncalves:2018qas} at the forthcoming LHC upgrades and future colliders. Hence, already in the example of Majoron EFT discussed above, one notices that correlations between collider and gravitational-wave observables may pose significant constraints on New Physics models that feature portal-type couplings between the Higgs boson and BSM scalars.

\clearpage
\section{{\bfseries Simulation and parton level Monte Carlo/ simplified models} \label{sec:simp}}
{\sl A. Papaefstathiou, T. Robens, R. Zhang}

{
In section \ref{sec:qcd}, we discussed the current state of the art and open points for the simulation of triple Higgs final states following SM predictions, also adressing available Monte Carlo tools for both signal and background. Here, in turn, we describe the available tools for double resonance enhanced triple Higgs production in new physics scenarios. We briefly elaborate on of the currently available tools that focusses on a simple production and decay chain. In addition, we point to possible pitfalls that can arize from an oversimplified approach in scenarios where additional features of a particular UV realization are important.

}

\subsection{Introduction}
We here focus on resonance-enhanced triple scalar production, i.e. scenarios that allow for onshell production of the form

\begin{\eqn}\label{eq:procdom}
p\,p\,\rightarrow\,h_3\,\rightarrow\,h_1\,h_2\,\rightarrow\,h_1\,h_1\,h_1,
\end{\eqn}
where we assumed mass a hierarchy $M_3\,\geq\,M_2\,\geq\,M_1$. Basically all new physics scenarios that contain at least 4 scalar fields in the gauge eigenbasis (we here count complex degrees of freedom separately) can provide such scenarios. The simplest realization is however the extension of the scalar potential by a complex or, equivalently, two real singlets, see e.g. \cite{Barger:2008jx} for early work. 

In case the above decay chain is dominant, the following parameters are minimally needed to describe the above process:
\begin{\eqn*}
M_1,\,M_2,\,M_3;\,\Gamma_1,\,\Gamma_2,\,\Gamma_3;\,g_{h_3\,t\,\bar{t}},\,\lam_{123},\,\lam_{112},
\end{\eqn*}
where $\lam_{ijk}$ denote triple scalar couplings. If decays of the triple scalar states are specified, in principle also coupling modifications of the $h_{125}$ to the respective final states are still allowed within a range of $\kappa\,\geq\,0.96$, see e.g.  \cite{Robens:2023pax}, and should be included in the parameter list. Note that for a specific UV complete realization the widths are not free parameters.
\subsection{Current model version}

The phenomenological Lagrangian described by the \texttt{MadGraph5\_aMC@NLO} model, found at~\cite{gitlab}, contains scalar interactions of the form:

\begin{equation}\label{eq:scalarcouplings}
V\,\supset\,\sum_{i,j,k}\,\lambda_{ijk}\,h_i h_j h_k + \sum_{i,j,k,l}\,\lambda_{ijkl}\,h_i h_j h_k h_l \;,
\end{equation}
where $i,j,k,l=1,2,3$ correspond to the three physical scalar particles present in the model, $h_1$, $h_2$, $h_3$. 

The couplings of the scalars to the rest of the SM particles are SM-like and are each re-scaled by a single parameter $\kappa_i$: 

\begin{equation}\label{eq:rescaledcouplings}
g_{h_i XX} = g_{hXX}^{\mathrm{SM}} \kappa_i\;,
\end{equation}
where $i=1,2,3$. 

If we identify $h_1$ with the SM-like Higgs boson, then $m_1 \simeq 125$~GeV and $\kappa_1 \lesssim 1$, and $\kappa_{2,3} \simeq 0$, following experimental constraints steming from the SM-like Higgs boson signal strength measurements. 

This model has been implemented into {\texttt{MadGraph5\_aMC@NLO}} with the input parameters defined by Eqs.~(\ref{eq:scalarcouplings}) and ~(\ref{eq:rescaledcouplings}), where the correspondence to the input parameters of the  \texttt{MadGraph5\_aMC@NLO} is given in Table~\ref{table:params}.

\begin{table}[ht!]
    \centering
    \begin{tabular}{c|c}
        Parameter & \texttt{MadGraph5\_aMC@NLO} \\ \hline 
        $\lambda _{ijk}$ & \texttt{Kijk} \\
        $\lambda _{ijkl}$ & \texttt{Kijkl} \\
        $\kappa_i$ & \texttt{ki}\\\hline
        Particle & \\ \hline
        $h_1$ & \texttt{h} \\
        $h_2$ & \texttt{eta0} \\
        $h_3$ & \texttt{iota0} \\\hline
    \end{tabular}
    \caption{The correspondence between the parameters defined by  Eqs.~(\ref{eq:scalarcouplings}) and ~(\ref{eq:rescaledcouplings}) and those in the \texttt{MadGraph5\_aMC@NLO} model.}
    \label{table:params}
\end{table}

In addition, the masses of the new particles can be entered as free parameters. Note that this requires to use autowidth for consistency reasons, otherwise the rates given by the Monte Carlo run can be non physical\footnote{While using an arbitrary width might be corrected by rescaling in case the process of Eq. (\ref{eq:procdom}) is dominant, it immediately starts to fail once more physical descriptions of the production process are used. See discussion below for details.}.

As a caution, we want to emphasize that a randomly set of parameters does not guarantee that a realistic UV-complete model exists that can be mapped on the corresponding values. The above model also does not allow for non-uniform rescaling of couplings to SM particles, as e.g. present in new physics scenarios with additional doublets. In the current implementation, the important sum rule \cite{Gunion:1990kf} should always be obeyed:

\begin{\eqn*}
\sum_i\,\kappa^2_i\,=\,1.
\end{\eqn*}

\subsection{Enhancement of number of free parameters/ more physical description}

In general, if a process in the form of Eq. (\ref{eq:procdom}) is targeted, it is clear that various channels can contribute to the final state $h_i\,h_i\,h_i$. In particular, all possible intermediate indices can appear
\begin{\eqn}\label{eq:other}
p\,p\,\rightarrow\,h_i\,\rightarrow\,h_j\,h_1\,\rightarrow\,h_1\,h_1\,h_1,
\end{\eqn}
where $\left\{i,j,k\right\}\,\in\,\left\{1,2,3\right\}$. In such a scenario, in principle all $\kappa_i$ and $\lam_{ij1}/\,\lam_{j11}$ are needed as input parameters. While the above model allows for an inclusion of all these choosing arbitrary values, the total number of free parameters that needs to be taken into account is increased by 12, including the additional couplings $g_{h_{1/2}t\bar{t}}$.

\subsection{Possible pitfalls}
Clearly, possible pitfalls can occur when the process specified in eqn. (\ref{eq:procdom}) is no longer dominant. In this case, obviously a process can still be generated that corresponds to the one above, and used for cut optimization and, if applicable, training of neural networks or similar tools, but the correct mapping to a UV complete theory might then require recasting methods. One way to identify such scenarios is e.g. the case where 
\begin{\eqn*}
p\,p\,\rightarrow\,h_1\,\rightarrow\,h_1\,h_1\,\rightarrow\,h_1\,h_1\,h_1
\end{\eqn*}
with intermediate off-shell particles is large.

We have considered such a scenario in the TRSM, with the following input parameters \cite{trzhang}:
\begin{eqnarray}\label{eq:bphhhmax}
M_2\,=\,550\,\GeV&M_3\,=\,700\,\GeV& \nonumber\\
\theta_{hs}\,=\,-0.002826,&\theta_{hx}\,=\,0.04424,&\theta_{sx}\,=\,0.8908, \nonumber\\
v_s\,=\,739.2\,\GeV,&v_x\,=\,152.3\,\GeV.&
\end{eqnarray}

For this parameter point, the cross section for the total process without intermediate state specification is given by $\sigma_\text{tot}\,=\,0.06031 (4)\,\fb$; specifying the intermediate state as given above leads to 
$\sigma_{h_3}\,=\,0.02151 (2)\,\fb$ 
at 13 \TeV; in this scenario, although onshell production is possible, the double resonance enhanced process only contributes to about $40\,\%$ of the total cross section.

\begin{center}
\begin{figure}
\begin{center}
\includegraphics[width=0.45\textwidth]{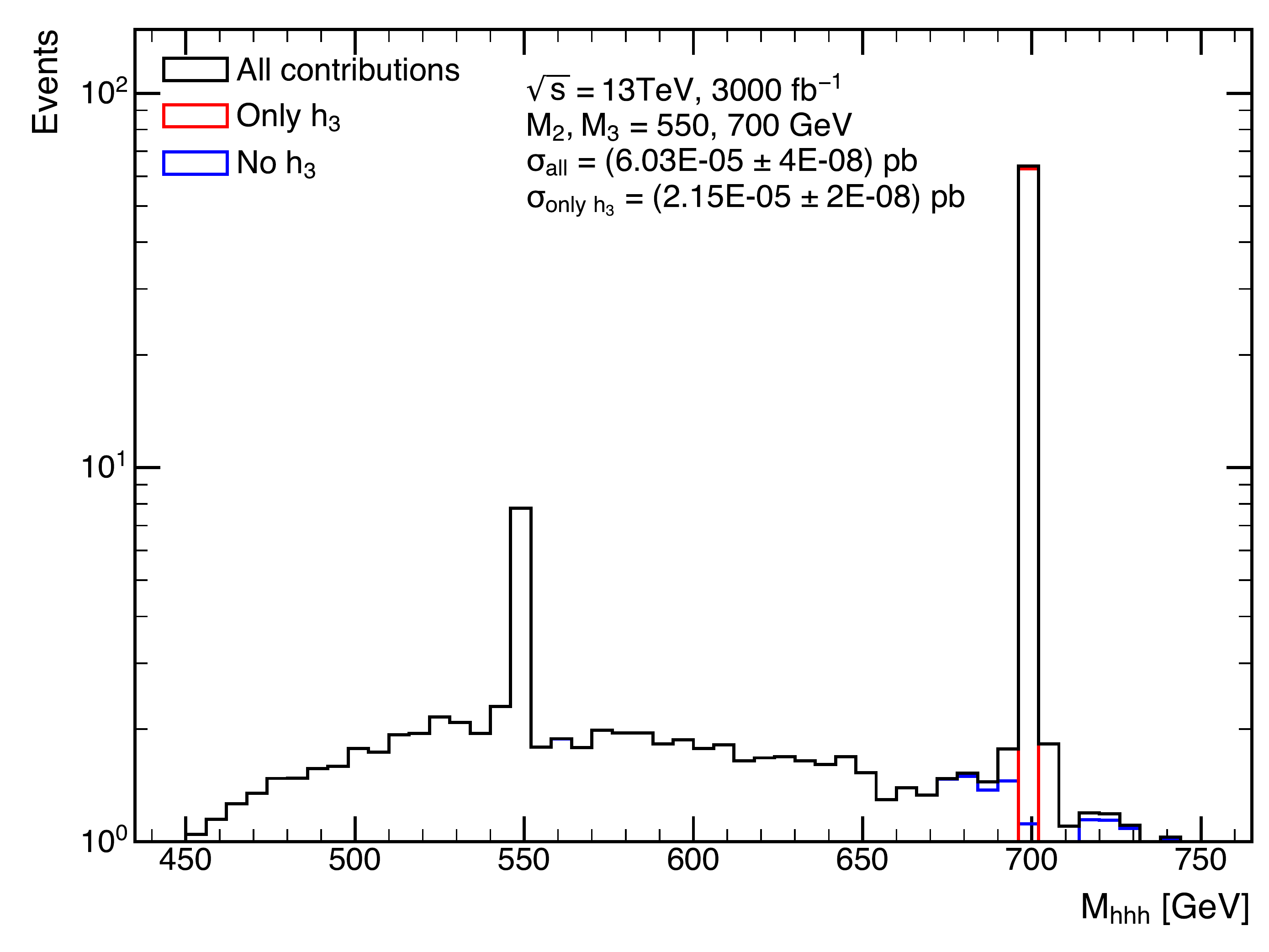}
\includegraphics[width=0.45\textwidth]{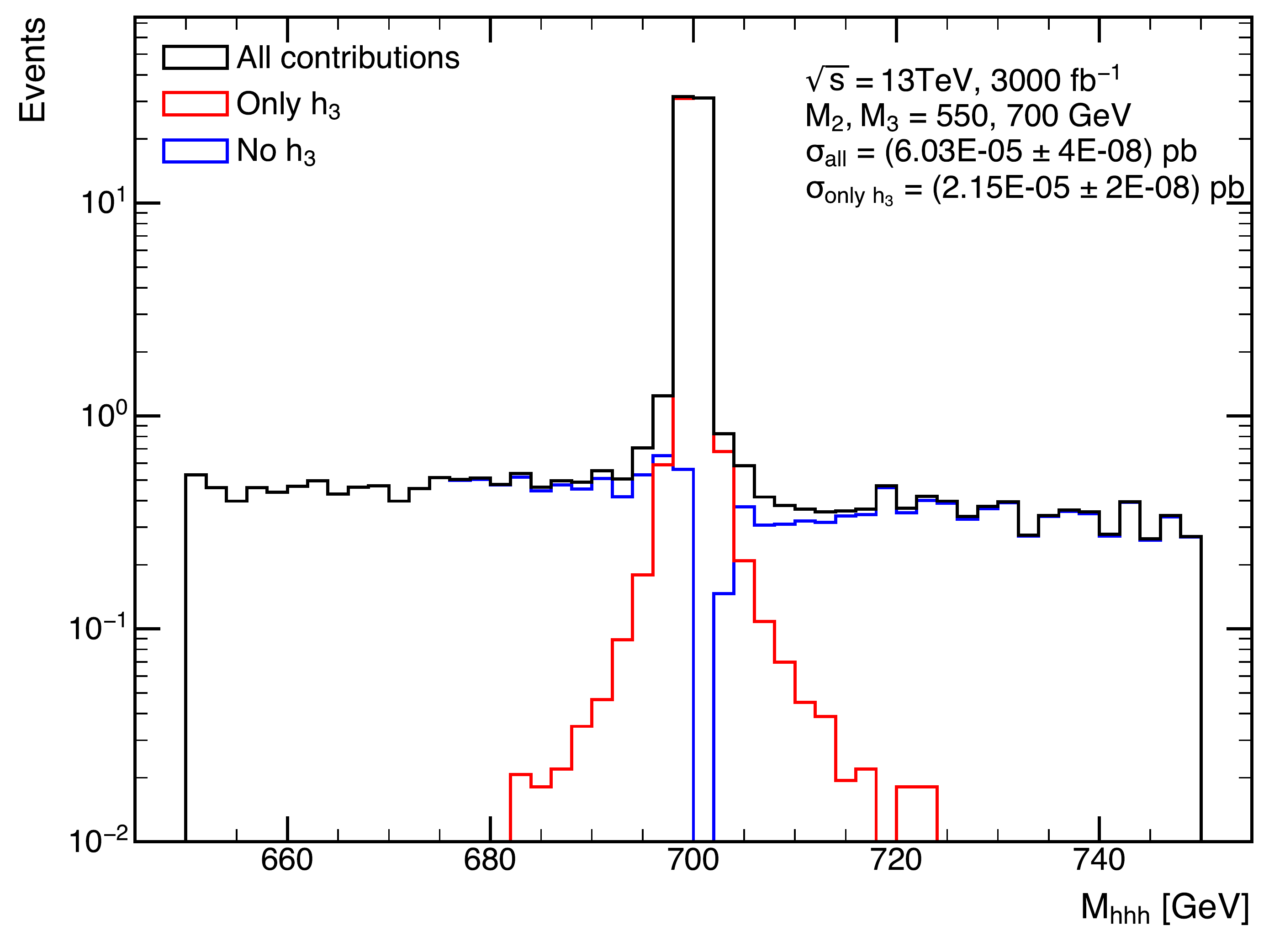}
\caption{\label{fig:mhhh} Comparison of $M_{hhh}$ invariant mass distributions for the point in eqn.(\ref{eq:bphhhmax}), for full process {\sl (black)}
, process excluding contributions from (\ref{eq:procdom}) {\sl(blue)}, 
and contributions via the dominant process (\ref{eq:procdom}) only {\sl(orange)}, for the whole mass region {\sl (left)} as well as zoomed in into the relevant mass region {\sl (right)}. Events are shown for 13 \TeV and $3000\,\fb^{-1}$. Bin size is 6 \GeV and 2 \GeV, respectively.}
\end{center}
\end{figure}
\end{center}

\begin{center}
\begin{figure}
\begin{center}
\includegraphics[width=0.45\textwidth]{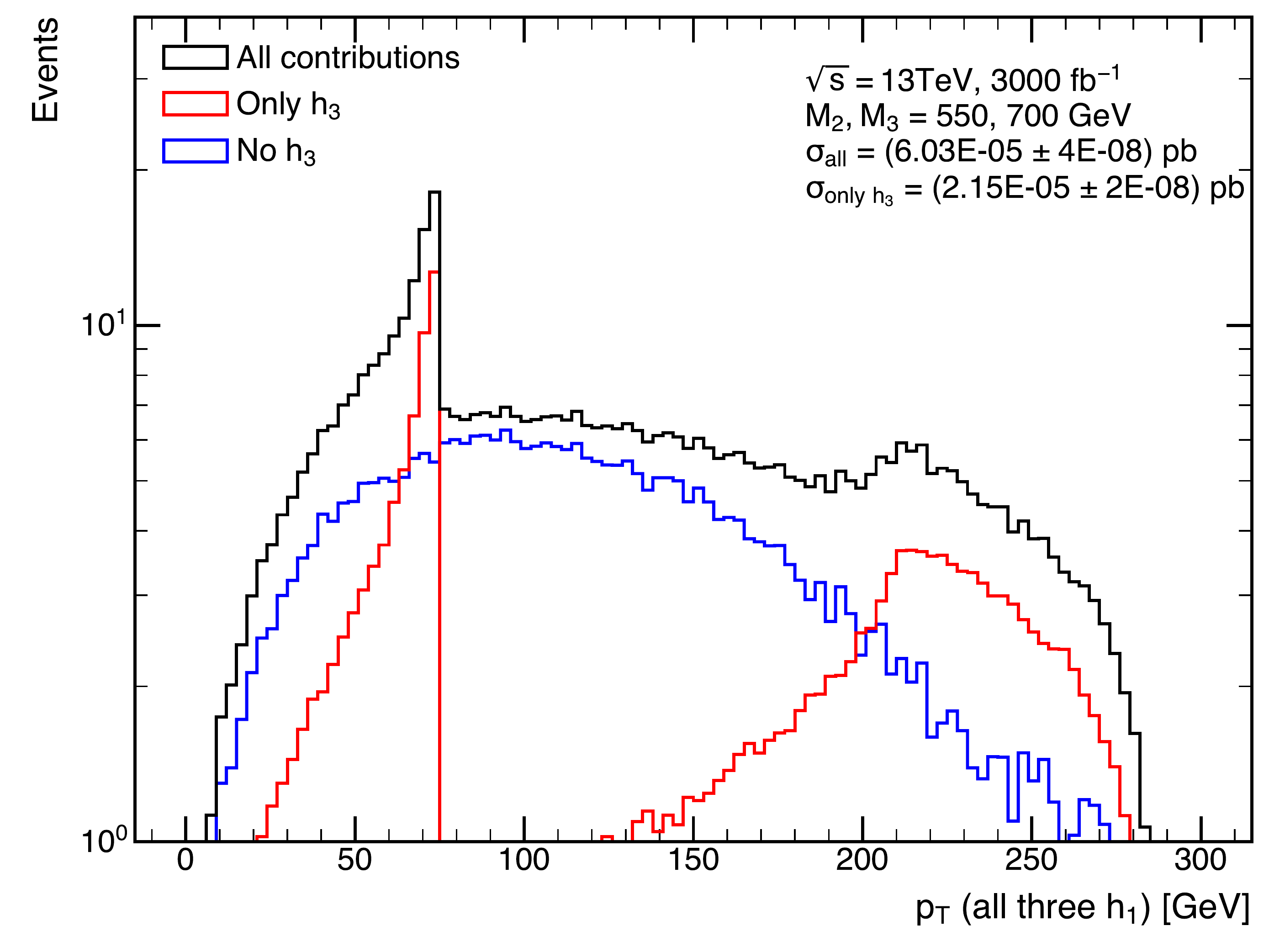}
\caption{\label{fig:pth} Comparison of $p_\perp$ distributions for all scalars for the point in eqn.(\ref{eq:bphhhmin}), for full process {\sl (black)}
, process excluding contributions from (\ref{eq:procdom}) {\sl(blue)}, 
and contributions via the dominant process (\ref{eq:procdom}) only {\sl(orange)}. Events are shown for 13 \TeV and $3000\,\fb^{-1}$.
Bin size is 3 \GeV.
}
\end{center}
\end{figure}
\end{center}

We display the respective distributions in the triple scalar invariant mass as well as the $p_\perp$ of all scalars in figures \ref{fig:mhhh} and \ref{fig:pth}, respectively.

For the mass distribution, we see that in the region around the peak the full process and the target process render the same distributions. However, off the mass peak the second obviously neglects additional contributions. Similarly, the $p_\perp$ distributions also differ between the two descriptions.

\begin{center}
\begin{figure}
\begin{center}
\includegraphics[width=0.45\textwidth]{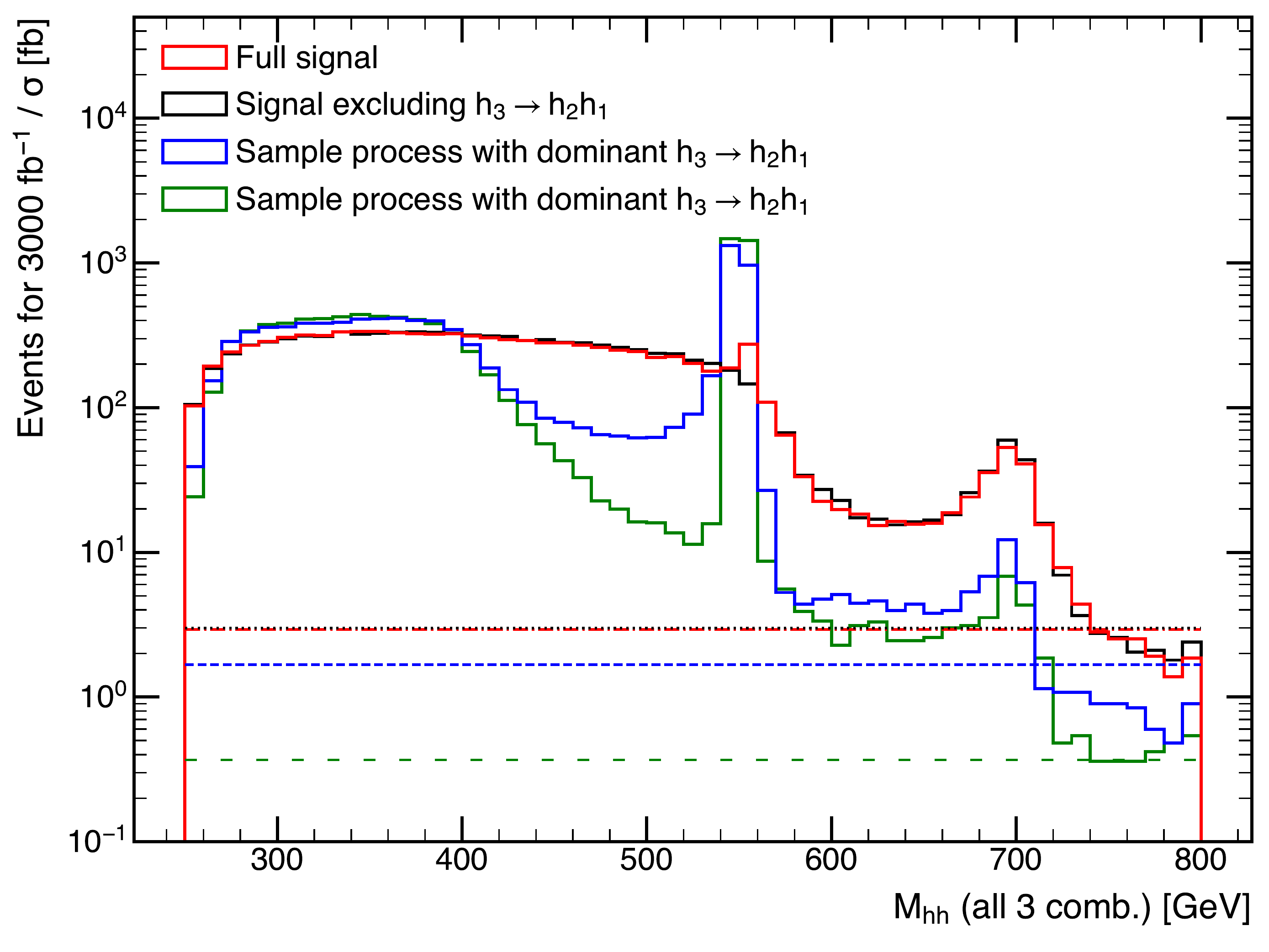}
\caption{\label{fig:mhh} Comparison of $M_{hh}$ invariant mass distributions for the point in eqn.(\ref{eq:bphhhmin}), for full process {\sl (red)}, process excluding contributions from (\ref{eq:procdom}) {\sl(black)}, and two other parameter point where $\gtrsim\,90\,\%$ stem from (\ref{eq:procdom}) {\sl (blue, green)}, normalized to the respective cross sections.
The lines signify 1 event at $\mathcal{L}_\text{int}\,=\,3000\,\fb^{-1}$ for $13\,\TeV$.
Bin size is 10 \GeV.
}
\end{center}
\end{figure}
\end{center}

Another interesting parameter point is a scenario where the target process only contributes about $6\,\%$ of the total cross section, and dominant contributions instead stem from
\begin{\eqn*}
p\,p\,\rightarrow\,h_3\,\rightarrow\,h_3\,h_1\,\rightarrow\,h_1\,h_1\,h_1
\end{\eqn*}
where again the intermediate particles can be offshell. An example of such a scenario is the point specified by
\begin{eqnarray}\label{eq:bphhhmin}
M_2\,=\,550\,\GeV&M_3\,=\,700\,\GeV,&\nonumber\\
\theta_{hs}\,=\,0.06232,&\theta_{hx}\,=\,0.2773,&\theta_{sx}\,=\,0.1150,\nonumber\\
v_s\,=\,269.2\,\GeV,&v_x\,=\,173.1\,\GeV.&
\end{eqnarray}
For this scenario, the $M_{hhh}$ distribution will feature a peak at $\sim\,700\,\GeV$, as around $90\%$ of the process is mediated via $h_3$ production. However, the $hh$ invariant mass distribution will look different. As an example, in figure \ref{fig:mhh} we show the corresponding contributions for the above point with and without the original target process given by eqn.~ (\ref{eq:procdom}), and in addition compare it to two sample processes where $h_3$ clearly dominates.

It is obvious that the above points are just examples, and it could very well be that in regions of parameter space with large rates the decay chain given by eqn.(\ref{eq:procdom}) is always dominant. However, depending on the models parameter space many other processes of the form given by eqn. (\ref{eq:other}) can appear that lead to different distributions in the parameter space of a UV complete model.

\clearpage
\section{{\bfseries Summary and open questions}}\label{sec:summ}
{\sl  J. Konigsberg, G. Landsberg, T. Robens
}

After the discovery of a particle with properties consistent with those of the Higgs boson of the electroweak Standard Model, all parameters of the scalar sector are now known in principle. However, to determine whether the scalar sector that is realized in Nature indeed corresponds to the one of the SM, both the triple ($\kappa_3$) and quartic ($\kappa_4$) scalar couplings of this Higgs boson need to be determined with sufficient precision. While $\kappa_3$ can be determined in the observation of processes leading to di-Higgs final states (see e.g. \cite{DiMicco:2019ngk} for a concise overview), $\kappa_4$ is only accessible at leading order in processes with triple Higgs final states.

Di-Higgs production is nearly exclusively sensitive to $\kappa_3$, while triple Higgs production is sensitive to both $\kappa_3$ and $\kappa_4$. A full determination of the Higgs potential is only possible through the combined measurement of these couplings. 
The cross section for triple Higgs production at hadron colliders, within the Standard Model, is known up to NNLO in QCD and ranges from $\mathcal{O}\lb 10^{-1}\fb\rb$ at 14 \TeV to $\mathcal{O}\lb \fb \rb$ at 100 \TeV colliders \cite{deFlorian:2019app}. These cross sections predict very small signal yields even at the highest energy colliders, and are several hundred times smaller than those for di-Higgs production. Therefore, a first attempt at discovering triple Higgs processes could be made by considering new physics scenarios with significantly enhanced production rates with respect to SM predictions. Even though these processes could include final states with different kinematic distributions from SM ones, they provide a very good testing ground for various experimental analysis strategies. But importantly, the possibility of discovering new physics in itself merits deliberate searches for such scenarios.

In this workshop, we addressed various such new physics models that extend the SM scalar sector and that can potentially result in observable signal yields at hadron colliders. Importantly, we also discussed in depth the experimental challenges, and the corresponding tools and techniques needed, in searching for these processes.
Such new physics models include additional scalars produced in conjunction with Higgs bosons, resulting in multi-Higgs processes with additional resonances that can be searched for experimentally due to their striking, and diverse, topologies. An example of such models in which, in addition to the discovered Higgs boson, two new real singlet scalars are introduced is the TRSM model, for which different regions of mass phase-space (for the three scalars involved) can yield signals large enough for potential discovery at the HL-LHC. In general, the extended scalar sector is broadly interesting and, for example, has potential connections to dark matter, hierarchies in the quark mass sector, or even electroweak baryogenesis. The message is that such multi-scalar searches should be pursued, both in model-dependent and model-independent fashions.

In terms of how to purse triple Higgs searches, it is illustrative to see where we are with the di-Higgs pursuits. The current expected sensitivity to di-Higgs production, by CMS and ATLAS, using each about 140~fb$^{-1}$ of data from Run~2, is in the range of 2.5-2.9 times the predicated rate of the SM. The main final state channels contributing to these searches are: $b\bar b b\bar b$, $b\bar b\tau^+\tau^-$, and $b\bar b\gamma\gamma$. With doubling of the Run 2 data in Run 3 and a possible combination of the results from the ATLAS and CMS experiments, it is possible that a 3-$\sigma$ evidence for di-Higgs production can be achieved at the LHC in the next couple of years. At the HL-LHC, with the techniques developed now and in future, and with the expected 3,000~fb$^{-1}$ of data, observation of di-Higgs production is achievable.

 There were extensive discussion of these techniques at the workshop, that are also applicable to triple Higgs searches. Because the Higgs decay to b-quark pairs is the largest ($\sim 58\%)$, focusing on final states that contain at least four b-quarks, from the decay of two of the three Higgs bosons, is the most promising path to larger signal yields and background reduction. A price in the efficiency
 in the detection of b-jets is nonetheless paid in order to reduce fakes while enhancing the signal. To alleviate that, the decay of the third boson, in addition to considering the $b\bar b$ mode, can also include non-b jets, such as hadronically decaying tau leptons, jets from the $W^+W^-$, or from $gg$ decay modes. 
 The main analysis components of these searches are: the triggers used to capture as much signal as possible, the b-tagging algorithms, the topological reconstruction of the event, the modeling of the multi-jet QCD background, and the algorithms for signal extraction. All of these elements have benefited significantly from the fast pace of machine learning developments that have now become integral to advanced analyses.

Implementing b-tagging and full event reconstruction algorithms in the triggers is paramount to maintaining trigger rates reasonably low, while recording as much signal as possible. Access to low level detector information at the earliest levels of the trigger paths is, and will continue to be, critical so that ML techniques can be implemented in fast hardware such as field-programmable gate array (FPGA) and GPU based engines. Flavor-tagging algorithms have evolved tremendously from the early days in which silicon tracking detectors were first used in hadron colliders. The continuous evolution of silicon tracker technology, combined with ML techniques, nowadays results in algorithms that can identify b-jets with an efficiency of $\sim 80\%$ and a fake rate at the $10^{-2}$ level. Similar ML algorithms have also been extremely useful in charm-tagging and in the identification of tau lepton hadronic decays, all relevant in triple Higgs searches.

The complexity of these multi-object final states, with potentially multiple resonances, lends itself further to a broad usage of ML techniques. Event-wide ML algorithms have proven to be extremely effective in discerning signal from background at the event selection stage and, very importantly, in modeling QCD background using data-driven techniques, as this is an important component of the systematic uncertainty in the searches. At the end stage of the analyses, dedicated ML algorithms focus on signal extraction in specific signal regions.
In di-Higgs and in triple Higgs searches, be them from SM processes or not, there are regions of phase space in which, due to the large boost of the particles involved, the final state jets are close together and merge. Experiments use ML algorithms to discern whether jet event activity in large fiducial cones is from decays of boosted objects such as bosons or top quarks. ML algorithms have also been developed to identify whether the merged jets originate from heavy flavor. In any given multi-jet process events can be categorized as fully resolved --with all final state jets far away enough from each others, as semi-merged  --with some of the jets closely merged, and as fully merged in which every jet is merged with another. Techniques to optimize the categorization and consequent utilization of the appropriate analysis techniques for each case are important, with the ultimate sensitivity being the deciding figure of merit. A complication in signal event reconstruction stems from additional jets from pileup interactions, and from initial and final state radiation, and these need to be included in the event reconstruction ML algorithms.

Another important area addressed in the workshop was the intricacy of theoretical challenges and recent advancements in reaching higher orders of precision in QCD calculations of hadronic processes. These are important in establishing predictions of signal cross sections, including accurate kinematical differential distributions.
Related to this, the Monte Carlo generators modeling of multi-jet processes has become more and more critical as ML algorithms need to be trained using very large simulated samples of background multi-jet events that are modeled as accurately as possible. 

In conclusion, the HHH workshop held in Dubrovnik last Summer brought together theorists and experimentalists to a dedicated forum in which the discussions proceeded with intense focus and great camaraderie. As reflected in this white paper, the projection of everyone's knowledge and experience onto the triple Higgs landscape has helped establish an initial stage of common understandings and a road map on how to pursue the detection of HHH final states at hadron colliders. It is clear at the moment that this is no easy feat, but history is such that over a decade ago just the observation of $H\to b\bar b$ decay was considered very far from achievable. That was established. Nowadays evidence for di-Higgs production may already be within reach in the next couple of years. In the near future the HL-LHC will deliver enormous amounts of data to experiments over a long period of time. We should remain confident that continuous, and significant, improvements on every aspect of the ingredients that  make up these quests will be made. Beyond that, there is the promise of a future 100~TeV collider, where the SM HHH production cross section increases by a factor of 60 or so, and rough estimates, at this stage, predict almost reaching evidence for this process with 30~ab$^{-1}$ of data.  History shows this is bound to get better and better. It is an exciting time to push on this frontier.

\clearpage
\section*{{Acknowledgements}}
AP acknowledges support by the National Science Foundation under Grant No.\ PHY 2210161. TR is supported by the Croatian Science Foundation under grant IP-2022-10-2520. BF has received partial support from the French \emph{Agence Nationale de la Recherche} (ANR) through Grant ANR-21-CE31-0013 (DMwithLLPatLHC). The organizers also want to thank the IUC staff for creating a fantastic workshop atmosphere. B.X. Liu is supported in part by NSERC (Canada), and Sun Yat-sen University Shenzhen Campus under project 74140-12240013. The work of P. Ferreira and R. Santos~is supported in part
by the Portuguese Funda\c{c}\~{a}o para a Ci\^{e}ncia e Tecnologia under Contracts UIDB/00618/2020, UIDP/00618/2020, CERN/FIS-PAR/0025/2021, CERN/FIS-PAR/0021/2021 and CERN/FIS-PAR/0037/2021. MM acknowledges support by the BMBF project 05H21VKCCA. TR is supported by the Croatian Science Foundation (HRZZ) under project IP-2022-10-2520. The organizers also acknowledge partial financial support for some participants from the Inter University Center Dubrovnik. The work of G. Landsberg and M. Stamenkovic is partially supported by the US Department of Energy Grant no. DE-SC001001. 
\addcontentsline{toc}{section}{References}
\bibliography{all}


\end{document}